\documentclass[aps,prc,superscriptaddress,showpacs,floatfix,twocolumn]{revtex4}

\usepackage{graphicx}	
\usepackage{xspace} 
\usepackage{amsmath}

\newcommand{\sigmappe}{\sigma_{pp}^e\xspace}
\newcommand{\sigmappin}{\sigma_{pp}^{\rm in}\xspace}
\newcommand{\pte}{p_T^e\xspace}
\newcommand{\pt}{p_T\xspace}
\newcommand{\mt}{m_T\xspace}

\newcommand{\snn}{\sqrt{s_{\rm NN}}\xspace}
\newcommand{\npart}{N_{\rm part}\xspace}
\newcommand{\ncol}{N_{\rm coll}\xspace}

\newcommand{\gevc}{GeV/$c$\xspace}

\newcommand{\naae}{N_{\rm AuAu}^e\xspace}

\newcommand{\nppe}{N_{pp}^e\xspace}
\newcommand{\raa}{R_{\rm AuAu}\xspace}

\newcommand{\taa}{T_{\rm AuAu}\xspace}
\newcommand{\rg}{R_\gamma\xspace}
\newcommand{\rcn}{R_{\rm CN}\xspace}
\newcommand{\rnp}{R_{\rm NP}\xspace}
\newcommand{\rkp}{R_{\rm KP}\xspace}

\newcommand{\aee}{(N_{e^+} + N_{e^-})/2\xspace}
\newcommand{\anee}{(N_{e^{+}} + N_{e^{-}})/2\xspace}
\newcommand{\axa}{A+A\xspace}

\newcommand{\auau}{Au+Au\xspace}

\newcommand{\pp}{$p$+$p$\xspace}

\begin{document}


\title{Heavy Quark Production in $p+p$ and Energy Loss and Flow of Heavy Quarks \\
in Au+Au Collisions at $\snn = 200$ GeV}

\newcommand{\abilene}{Abilene Christian University, Abilene, Texas 79699, USA}
\newcommand{\banaras}{Department of Physics, Banaras Hindu University, Varanasi 221005, India}
\newcommand{\bnlcoll}{Collider-Accelerator Department, Brookhaven National Laboratory, Upton, New York 11973-5000, USA}
\newcommand{\bnlphys}{Brookhaven National Laboratory, Upton, New York 11973-5000, USA}
\newcommand{\caucr}{University of California - Riverside, Riverside, California 92521, USA}
\newcommand{\charlesczech}{Charles University, Ovocn\'{y} trh 5, Praha 1, 116 36, Prague, Czech Republic}
\newcommand{\ciae}{China Institute of Atomic Energy (CIAE), Beijing, People's Republic of China}
\newcommand{\cns}{Center for Nuclear Study, Graduate School of Science, University of Tokyo, 7-3-1 Hongo, Bunkyo, Tokyo 113-0033, Japan}
\newcommand{\colorado}{University of Colorado, Boulder, Colorado 80309, USA}
\newcommand{\columbia}{Columbia University, New York, New York 10027 and Nevis Laboratories, Irvington, New York 10533, USA}
\newcommand{\czechtech}{Czech Technical University, Zikova 4, 166 36 Prague 6, Czech Republic}
\newcommand{\dapnia}{Dapnia, CEA Saclay, F-91191, Gif-sur-Yvette, France}
\newcommand{\debrecen}{Debrecen University, H-4010 Debrecen, Egyetem t{\'e}r 1, Hungary}
\newcommand{\elte}{ELTE, E{\"o}tv{\"o}s Lor{\'a}nd University, H - 1117 Budapest, P{\'a}zm{\'a}ny P. s. 1/A, Hungary}
\newcommand{\fit}{Florida Institute of Technology, Melbourne, Florida 32901, USA}
\newcommand{\fsu}{Florida State University, Tallahassee, Florida 32306, USA}
\newcommand{\gsu}{Georgia State University, Atlanta, Georgia 30303, USA}
\newcommand{\hiroshima}{Hiroshima University, Kagamiyama, Higashi-Hiroshima 739-8526, Japan}
\newcommand{\ihepprot}{IHEP Protvino, State Research Center of Russian Federation, Institute for High Energy Physics, Protvino, 142281, Russia}
\newcommand{\illuiuc}{University of Illinois at Urbana-Champaign, Urbana, Illinois 61801, USA}
\newcommand{\instpasczech}{Institute of Physics, Academy of Sciences of the Czech Republic, Na Slovance 2, 182 21 Prague 8, Czech Republic}
\newcommand{\isu}{Iowa State University, Ames, Iowa 50011, USA}
\newcommand{\jinrdubna}{Joint Institute for Nuclear Research, 141980 Dubna, Moscow Region, Russia}
\newcommand{\kaeri}{KAERI, Cyclotron Application Laboratory, Seoul, Korea}
\newcommand{\kek}{KEK, High Energy Accelerator Research Organization, Tsukuba, Ibaraki 305-0801, Japan}
\newcommand{\kfki}{KFKI Research Institute for Particle and Nuclear Physics of the Hungarian Academy of Sciences (MTA KFKI RMKI), H-1525 Budapest 114, POBox 49, Budapest, Hungary}
\newcommand{\korea}{Korea University, Seoul, 136-701, Korea}
\newcommand{\kurchatov}{Russian Research Center ``Kurchatov Institute", Moscow, Russia}
\newcommand{\kyoto}{Kyoto University, Kyoto 606-8502, Japan}
\newcommand{\labllr}{Laboratoire Leprince-Ringuet, Ecole Polytechnique, CNRS-IN2P3, Route de Saclay, F-91128, Palaiseau, France}
\newcommand{\lawllnl}{Lawrence Livermore National Laboratory, Livermore, California 94550, USA}
\newcommand{\losalamos}{Los Alamos National Laboratory, Los Alamos, New Mexico 87545, USA}
\newcommand{\lpc}{LPC, Universit{\'e} Blaise Pascal, CNRS-IN2P3, Clermont-Fd, 63177 Aubiere Cedex, France}
\newcommand{\lund}{Department of Physics, Lund University, Box 118, SE-221 00 Lund, Sweden}
\newcommand{\muenster}{Institut f\"ur Kernphysik, University of Muenster, D-48149 Muenster, Germany}
\newcommand{\myongji}{Myongji University, Yongin, Kyonggido 449-728, Korea}
\newcommand{\nagasaki}{Nagasaki Institute of Applied Science, Nagasaki-shi, Nagasaki 851-0193, Japan}
\newcommand{\newmex}{University of New Mexico, Albuquerque, New Mexico 87131, USA }
\newcommand{\nmsu}{New Mexico State University, Las Cruces, New Mexico 88003, USA}
\newcommand{\ornl}{Oak Ridge National Laboratory, Oak Ridge, Tennessee 37831, USA}
\newcommand{\orsay}{IPN-Orsay, Universite Paris Sud, CNRS-IN2P3, BP1, F-91406, Orsay, France}
\newcommand{\peking}{Peking University, Beijing, People's Republic of China}
\newcommand{\pnpi}{PNPI, Petersburg Nuclear Physics Institute, Gatchina, Leningrad region, 188300, Russia}
\newcommand{\riken}{RIKEN Nishina Center for Accelerator-Based Science, Wako, Saitama 351-0198, JAPAN}
\newcommand{\rikjrbrc}{RIKEN BNL Research Center, Brookhaven National Laboratory, Upton, New York 11973-5000, USA}
\newcommand{\rikkyo}{Physics Department, Rikkyo University, 3-34-1 Nishi-Ikebukuro, Toshima, Tokyo 171-8501, Japan}
\newcommand{\saispbstu}{Saint Petersburg State Polytechnic University, St. Petersburg, Russia}
\newcommand{\saopaulo}{Universidade de S{\~a}o Paulo, Instituto de F\'{\i}sica, Caixa Postal 66318, S{\~a}o Paulo CEP05315-970, Brazil}
\newcommand{\seoulnat}{System Electronics Laboratory, Seoul National University, Seoul, Korea}
\newcommand{\stonybrkc}{Chemistry Department, Stony Brook University, Stony Brook, SUNY, New York 11794-3400, USA}
\newcommand{\stonycrkp}{Department of Physics and Astronomy, Stony Brook University, SUNY, Stony Brook, New York 11794, USA}
\newcommand{\subatech}{SUBATECH (Ecole des Mines de Nantes, CNRS-IN2P3, Universit{\'e} de Nantes) BP 20722 - 44307, Nantes, France}
\newcommand{\tenn}{University of Tennessee, Knoxville, Tennessee 37996, USA}
\newcommand{\titech}{Department of Physics, Tokyo Institute of Technology, Oh-okayama, Meguro, Tokyo 152-8551, Japan}
\newcommand{\tsukuba}{Institute of Physics, University of Tsukuba, Tsukuba, Ibaraki 305, Japan}
\newcommand{\vandy}{Vanderbilt University, Nashville, Tennessee 37235, USA}
\newcommand{\waseda}{Waseda University, Advanced Research Institute for Science and Engineering, 17 Kikui-cho, Shinjuku-ku, Tokyo 162-0044, Japan}
\newcommand{\weizmann}{Weizmann Institute, Rehovot 76100, Israel}
\newcommand{\yonsei}{Yonsei University, IPAP, Seoul 120-749, Korea}
\affiliation{\abilene}
\affiliation{\banaras}
\affiliation{\bnlcoll}
\affiliation{\bnlphys}
\affiliation{\caucr}
\affiliation{\charlesczech}
\affiliation{\ciae}
\affiliation{\cns}
\affiliation{\colorado}
\affiliation{\columbia}
\affiliation{\czechtech}
\affiliation{\dapnia}
\affiliation{\debrecen}
\affiliation{\elte}
\affiliation{\fit}
\affiliation{\fsu}
\affiliation{\gsu}
\affiliation{\hiroshima}
\affiliation{\ihepprot}
\affiliation{\illuiuc}
\affiliation{\instpasczech}
\affiliation{\isu}
\affiliation{\jinrdubna}
\affiliation{\kaeri}
\affiliation{\kek}
\affiliation{\kfki}
\affiliation{\korea}
\affiliation{\kurchatov}
\affiliation{\kyoto}
\affiliation{\labllr}
\affiliation{\lawllnl}
\affiliation{\losalamos}
\affiliation{\lpc}
\affiliation{\lund}
\affiliation{\muenster}
\affiliation{\myongji}
\affiliation{\nagasaki}
\affiliation{\newmex}
\affiliation{\nmsu}
\affiliation{\ornl}
\affiliation{\orsay}
\affiliation{\peking}
\affiliation{\pnpi}
\affiliation{\riken}
\affiliation{\rikjrbrc}
\affiliation{\rikkyo}
\affiliation{\saispbstu}
\affiliation{\saopaulo}
\affiliation{\seoulnat}
\affiliation{\stonybrkc}
\affiliation{\stonycrkp}
\affiliation{\subatech}
\affiliation{\tenn}
\affiliation{\titech}
\affiliation{\tsukuba}
\affiliation{\vandy}
\affiliation{\waseda}
\affiliation{\weizmann}
\affiliation{\yonsei}
\author{A.~Adare} \affiliation{\colorado}
\author{S.~Afanasiev} \affiliation{\jinrdubna}
\author{C.~Aidala} \affiliation{\columbia}
\author{N.N.~Ajitanand} \affiliation{\stonybrkc}
\author{Y.~Akiba} \affiliation{\riken} \affiliation{\rikjrbrc}
\author{H.~Al-Bataineh} \affiliation{\nmsu}
\author{J.~Alexander} \affiliation{\stonybrkc}
\author{A.~Al-Jamel} \affiliation{\nmsu}
\author{K.~Aoki} \affiliation{\kyoto} \affiliation{\riken}
\author{L.~Aphecetche} \affiliation{\subatech}
% ...
\collaboration{PHENIX Collaboration} \noaffiliation

\date{\today}

\begin{abstract}

 
Transverse momentum ($\pte$) spectra of electrons from semileptonic weak 
decays of heavy flavor mesons in the range of $0.3 < \pte < 9.0$ GeV/$c$ 
have been measured at midrapidity ($|\eta|<0.35$) by the PHENIX 
experiment at the Relativistic Heavy Ion Collider in \pp~and 
\auau~collisions at $\snn = 200$ GeV.  In addition, the azimuthal 
anisotropy parameter $v_2$ has been measured for $0.3 < \pte < 5.0$ 
GeV/$c$ in \auau~collisions.  The nuclear modification factor $R_{\rm 
AA}$ with respect to $p+p$ collisions indicates substantial energy loss 
of heavy quarks in the produced medium.  Comparisons of $R_{\rm AA}$ and 
$v_2$ are made to various model calculations.

\end{abstract}

	
\maketitle


 
\section{INTRODUCTION}\label{sec:introduction}

Numerous experimental results from the Relativistic Heavy Ion Collider 
(RHIC) at Brookhaven National Laboratory (BNL) have firmly established 
that the matter created in central \auau~collisions at $\snn=200$ GeV 
cannot be explained by the common expectation of a weakly interacting 
gas of quarks and gluons at very high densities 
\cite{Adcox:2004mh,wp_brahms,wp_phobos,wp_star}.  Strong suppression 
observed in measurements of the nuclear modification factor $\raa(\pt)$ 
for light-flavor hadrons shows that high-$\pt$ scattered partons suffer 
a significant energy-loss in the medium \cite{ppg003,ppg014,sup_star}.  
The large magnitude of the differential elliptic flow parameter 
$v_2(\pt)$ and its $\pt$ and mass dependencies measured in a limited 
phase space approximately agree with the theoretical predictions based 
on an ideal hydrodynamic fluid \cite{hydro,parton_v2,ppg022,star_v2_1}.  
In addition, the observation of universal scaling of $v_2$ for various 
hadrons suggests that the flow pattern is established at the partonic 
level before hadronization \cite{Adare:2007a}.  These experiments 
indicate that the medium is strongly interacting and exhibits 
hydrodynamic behavior.

However, it has been pointed out that qualitative evidence for a 
near-perfect hydrodynamic fluid is not sufficient in terms of 
thermodynamic and transport concepts \cite{Arnold:2005a,Molnar:2003a}.  
The issue is how perfect is the near-perfect fluidity observed at RHIC.  
The validation of the perfect fluidity requires the dimensionless ratio 
of shear viscosity $\eta$ to entropy density $s$ to be small.

Heavy quarks (charm and bottom) are important probes of the dense matter 
formed at RHIC.  Because of their large masses, their dominant 
production mechanism is restricted to parton-parton collisions in the 
initial stage of the reaction.  They can interact with the medium 
differently than light quarks and gluons due to their heavy mass.  It 
was predicted that the energy loss of heavy quarks would be smaller than 
that of light quarks and gluons due to suppression of small angle gluon 
radiation, called the ``dead cone effect" \cite{dk,armesto}.

Recently, heavy quark measurements along with models for heavy quark 
interaction have opened new possibilities to investigate other 
interaction mechanisms such as collisional energy-loss and in-medium 
fragmentation.  In some models, the heavy quark diffusion coefficient 
$D^{\rm HQ}$ controls the extent to which the initial power-law $p_T$ 
spectrum approaches the thermal spectrum and the extent to which the 
heavy quark will follow the underlying flow of the medium.  
Simultaneous measurement of heavy-quark $\raa(\pte)$ and $v_2(\pte)$ can 
provide an estimate of $D^{\rm HQ}$.

Heavy quark measurements in \pp~collisions serve as a testing ground of 
QCD.  Because of their large mass, it is expected that next-to-leading 
order perturbative QCD (NLO pQCD) can describe the production cross 
section of charm and bottom at high energy, particularly at high $p_T$.  
At the Tevatron with $\sqrt{s} = 1.9$ TeV, bottom production is well 
described by NLO pQCD~\cite{Cacciari:2004ur}.  Charm production cross 
sections at high $\pt$ are found to be higher than the theory by 
$\approx$ 50\%, but are compatible within the theoretical 
uncertainties~\cite{Acosta:2003ax}.


There are several ways to measure heavy quark production.  The most 
direct method is to reconstruct $D$ or $B$ mesons from their decay 
products, such as $D \rightarrow K \pi$ or $B \rightarrow J/\psi K$.  
PHENIX is currently not capable of measuring the decay vertex of the 
heavy meson, which makes this method very difficult.  The method which 
is employed by this analysis is to measure single leptons from heavy 
flavor decay.  Both the charm and bottom quarks have relatively large 
branching ratios ($\sim 10\%$) to single electrons or single muons.  
Production yields and momentum distributions of the parent charm or 
bottom hadrons can be inferred from the invariant spectrum of single 
leptons from the decay.  This method has an advantage in heavy-ion 
collisions at RHIC where the first method suffers from a very large 
combinatorial background due to the high multiplicity of the event.

Single electrons in hadronic collisions were first observed in the early 
1970's in $\sqrt{s}=52.7$ GeV \pp~collisions at the CERN 
ISR~\cite{ISR01} before the discovery of charm.  Subsequently, several 
experiments reported single electron production at the ISR.  In 
$p\bar{p}$ collisions at $\sqrt{s}$=630 GeV, UA1 \cite{UA1_muon} 
measured bottom production via single muons and UA2 \cite{UA2_electron} 
reported the charm cross section from single electron measurements.  At 
the Tevatron collider, the CDF \cite{CDF_lepton} and D0 \cite{D0_lepton} 
experiments measured bottom production via single muons or single 
electrons.

At RHIC, PHENIX first measured charm production in Au+Au collisions at 
$\sqrt{s_{\rm NN}} = 130$ GeV via measurement of single electrons 
\cite{Adcox:2002cg}.  Subsequently, PHENIX reported results of single 
electron measurements in \pp \cite{Adler:2005fy} and \auau~collisions 
\cite{Adler:2004ta,ppg040,ppg056} at midrapidity from the 2002 data set.  
In Au+Au collisions, the total yield of heavy-flavor decay electrons was 
found to scale with the number of binary nucleon-nucleon collisions 
\cite{Adler:2004ta} as expected for a point-like process.  As already 
mentioned, the energy loss of heavy quarks was expected to be reduced 
due to the dead-cone effect.  Consequently, it was expected that high 
$p_T$ suppression and $v_2$ of heavy-flavor electrons would be much 
weaker than those of light mesons.  In contrast, a strong suppression of 
heavy-flavor electrons for $2<p_T<5$ GeV/$c$ \cite{ppg056} and a 
non-zero electron $v_2$ for $p_T < 2$ GeV/$c$ \cite{ppg040} were 
discovered.  The STAR experiment has also measured the yield of 
electrons from heavy flavor decays at RHIC \cite{star_e}.

This article presents measurements of single electrons $\aee$ from 
semileptonic decays of heavy quarks (charm and bottom) at midrapidity 
($|\eta|<0.35$) in $p+p$ and Au+Au collisions at $\sqrt{s_{\rm NN}}$=200 
GeV.  The data were collected in 2004 (Au+Au), and 2005 ($p+p$).  We 
extend the previous PHENIX analyses of electron measurements in \pp~and 
\auau~collisions at $\snn = 200$ GeV to a broader $\pte$ range: 
$0.3<\pte<9.0$ GeV/$c$ and with a much higher precision.  Part of the 
results have been published in \cite{ppg065,ppg066}.


This article is organized as follows:~Section~\ref{sec:detector} 
presents an overview of the PHENIX detector system related to the 
analysis.  Section~\ref{sec:analysis} presents the details of the data 
analysis.  Section~\ref{sec:results} shows the $\pte$ distribution of 
the invariant spectrum in the 2005 \pp~and invariant yields and elliptic 
flow in 2004 \auau~collisions.  The results are compared with 
theoretical predictions and discussed in Sec.~\ref{sec:discussion}.  
Finally, Sec.~\ref{sec:summary} gives a summary and conclusions of the 
analysis.

 
\section{PHENIX DETECTOR}\label{sec:detector}

\begin{figure}[htbp]
  \includegraphics[width=1.0\linewidth]{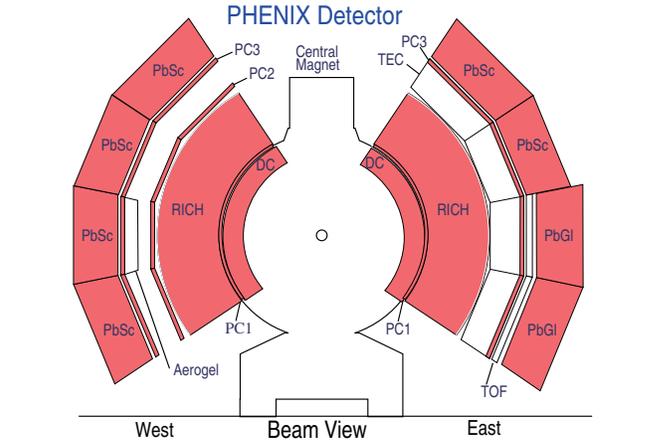}
\caption{(Color online) Beam view (at $z=0$) of the PHENIX central arm
detector in 2004 (\auau) and 2005 (\pp).  The detectors used in the
present analysis are the drift chamber (DC) and the multiwire
proportional pad chambers (PC1, PC2, PC3) for charged particle tracking,
the ring-imaging \v{C}erenkov detector (RICH) for electron
identification, and the electromagnetic calorimeters, which are Lead
Glass (PbGl) and Lead Scintillator (PbSc), for energy measurement.
 \label{fig:Phenix_2004}}
\end{figure}

A detailed description of the complete PHENIX detector system can be 
found elsewhere 
\cite{Adcox:2003zm,Aronson:2003a,Adcox:2003a,Aizawa:2003a,Aphecetche:2003a,Allen:2003a}.  
Here we describe the parts of the detector system that are used in this 
analysis, namely, two global detectors and two central arm 
spectrometers.  The global detectors are the beam-beam counter (BBC) and 
the zero-degree calorimeter (ZDC).  Each of the central arms covers a 
pseudorapidity of $|\eta| < 0.35$ and an azimuthal angle of $\pi/2$.  
They contain the drift chamber (DC) and the multiwire proportional pad 
chamber (PC) for charged particle tracking, the ring-imaging 
\v{C}erenkov detector (RICH) for electron identification, and the 
electromagnetic calorimeter (EMCal) for energy measurement.  
Figure~\ref{fig:Phenix_2004} shows the beam view of the PHENIX 
detector.

\subsection{Global detectors}

The BBC and the ZDC measure the collision time and the collision vertex 
position $z_{{\rm vtx}}$ along the beam.  In \auau~collisions, they are 
used to determine the centrality of the collision \cite{Allen:2003a}.  
They also provide the first level, minimum bias (MB) trigger information 
for beam-beam collisions.

Two sets of BBCs are placed at $\pm 1.44$ m from the nominal interaction 
point along the beam line (one on the north side and one on the south).  
Each BBC consists of 64 \v{C}erenkov counter modules, arranged radially 
around the beam pipe.  Each module is made of a 3 cm long quartz 
radiator and a photomultiplier tube (PMT).  The PMTs can operate in a 
high magnetic field ($\sim 0.3$ T) that lies mostly parallel to the beam 
axis.  The BBC is used to estimate the number of charged particles in 
the pseudorapidity region $3.0<|\eta|<3.9$.  The vertex position 
resolution of the BBC is $\sim 0.6$ cm for \auau~collisions and a few cm 
in \pp collisions.

The ZDC 
measures the total energy carried by the forward neutrons produced along 
the beam direction.  These neutrons are produced either by Coulomb 
dissociation of the beam particles or by evaporation from beam 
spectators.  The ZDCs are placed at $\pm 18$ m from the interaction 
point along the beam line.  The angular acceptance of each ZDC is 
$|\theta|<2$ mrad ($|\eta|>6$).  Each ZDC consists of three modules of 
two-interaction-length deep tungsten-quartz \v{C}erenkov sampling 
calorimeters.  The energy resolution of the ZDC is $\delta E/E \sim 
218/\sqrt{E~{\rm (GeV)}}~\%$ \cite{Adler:2001a}.

\subsection{Central magnet}

The transverse momentum of each charged particle is determined by its 
bending curvature in the magnetic field provided by the PHENIX central 
magnet (CM) system \cite{Aronson:2003a}.  The CM is energized by two 
pairs of concentric coils and provides an axial magnetic field parallel 
to the beam pipe.  The coils can be run with the fields for the two coil 
sets adding (the ``$++$'' and ``$--$'' configurations) or canceling 
(``$+-$'' configuration).  During the Au+Au measurement in 2004 and the 
$p+p$ measurement in 2005, the CM was operated in the ``$++$'' and 
``$--$'' configurations.  In the ``$++$'' and ``$--$'' configurations, 
the field component parallel to the beam axis has an approximately 
Gaussian dependence on the radial distance from the beam axis, dropping 
from $0.9$ T at the center to $0.096$ T ($0.048$ T)  at the inner 
(outer) radius of DCs.  The total field integral is $\int B\cdot 
dl=1.15$ T$\cdot$m.

\subsection{Tracking detectors}

The DC and PC~\cite{Adcox:2003a} in the central arms measure charged
particle trajectories in the azimuthal direction to determine the
transverse momentum ($\pt$) of each particle.  By combining the polar
angle information from the innermost PC (PC1) with the $\pt$, the total
momentum ($p$) is determined.

The DC is positioned between 2.02 and 2.46 m in radial distance from the
$z$ axis for both the West and East arms.  Each DC occupies a
pseudorapidity of $|\eta|<0.35$ and $\pi/2$ in azimuth.  Each DC volume
consists of 20 sectors, each of which covers $4.5$ degrees in azimuth.
Each sector has six types of wire modules stacked radially, named X1,
U1, V1, X2, U2 and V2.  Each module is further divided into 4 drift
cells in the $\phi$ direction.  A plane of sense wires is at the center
of a drift cell, with 2 to 2.5 cm drift space.  The X wires run parallel
to the beam axis and measure the particle trajectory in the $r$-$\phi$
plane.  The U and V wires have a stereo angle of about 6.0 degrees
relative to the X wires in order to measure the $z$-coordinate of the
track.  The single X wire resolution is $\sim 150~\mu$m.  The intrinsic
tracking efficiency of the X modules is greater than $99~\%$.  Helium
bags were placed between the beam pipe and the DCs to reduce the photon
conversions and multiple scattering.

The PC determine the spacepoints along the straight line particle
trajectories outside the magnetic field.  They are multiwire
proportional chambers that form three separate layers of the central
tracking system.  The first PC layer (PC1) is located between the DC and
RICH at $2.47$-$2.52$ m in radial distance from the interaction point,
while the third layer (PC3) is located in front of EMCal occupying
$4.91$-$4.98$ m from the interaction point.  The second layer (PC2) is
placed behind RICH occupying $4.15$-$4.21$m in radial distance in the
West arm only.  Position information from the PC1 and the DC, along with
the vertex position measured by the BBC, are used in the global track
reconstruction to determine the polar angle of each charged track.

\subsection{Ring-imaging \v{C}erenkov detector (RICH)}

\begin{figure}[htbp]
  \includegraphics[width=1.0\linewidth]{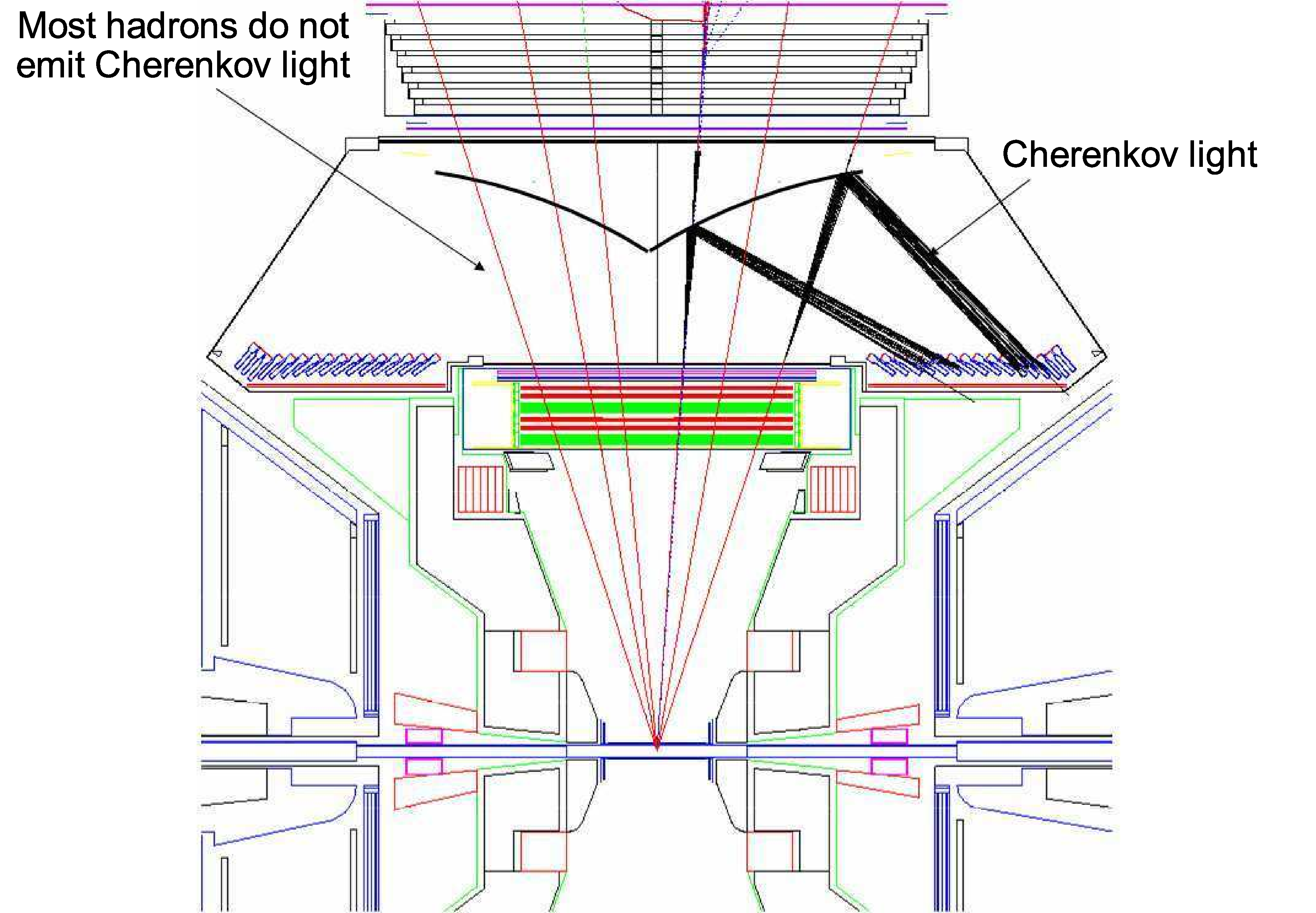}
  \caption{(Color online) Top view (at $y=0$) of the 
ring-imaging \v{C}erenkov detector (RICH) in the PHENIX East arm.  
  	 \label{fig:RICH_cherenkov}}
\end{figure}

The RICH is a threshold-type gas \v{C}erenkov counter and the primary 
detector used to identify electrons in PHENIX \cite{Aizawa:2003a,RICH}.  
It is located in the radial region of $2.5$-$4.1$m just outside PC1.  
The RICH in each central arm covers a pseudorapidity range of 
$|\eta|<0.35$ and $\pi/2$ in azimuthal angle.  Each volume contains 
mirror panels (0.53~\% of a radiation length thick), forming two 
intersecting spherical surfaces, with a total reflecting area of 20 
m$^2$.  The spherical mirrors focus \v{C}erenkov light onto two arrays 
of $80(\phi) \times 16(z) = 1280$ PMTs, each located on either side of 
the RICH entrance window as shown in Fig.~\ref{fig:RICH_cherenkov}.  
Each PMT has a magnetic shield that allows it to operate in a magnetic 
field up to 100 Gauss.  CO$_2$ gas at atmospheric pressure 
($n=1.000410$) was used as the \v{C}erenkov radiator.  The RICH has a 
\v{C}erenkov threshold of $\gamma=35$, which corresponds to $p>20$ 
MeV/$c$ for an electron and $p>4.9$ GeV/$c$ for a pion.

The average number of hit PMTs per electron track is about 5, and the 
average number of photo-electrons detected is about 10.  The $e/\pi$ 
separation is $\sim 10^4$ in $p+p$.

\subsection{Electromagnetic calorimeter (EMCal)}

The EMCal \cite{Aphecetche:2003a} consists of two arms, each covering a 
pseudorapidity region of $|\eta|<0.35$ and $\pi/2$ in azimuthal angle.  
Each of the two arms consists of four rectangular sectors.  Two adjacent 
sectors of the East arm are based on lead-glass (PbGl) calorimetry, 
whereas the remaining sectors are based on lead-scintillator (PbSc) 
calorimetry.  The radial distance from the $z$ axis is $5.10$ m for PbSc 
and $5.50$ m for PbGl.

The PbSc is a Shashlik-type sampling calorimeter made of alternating 
tiles of lead and scintillator.  It consists of $10.5 \times 10.5 \times 
37$ $\textrm{cm}^3$ ($18.2~X_0$) rectangular modules, each of which is
constructed from alternating layers of $1.5$ mm thick lead and
$4$ mm polystyrene-based scintillator layers.  The light from the 
scintillator is collected by optical fibers that run longitudinally 
through the module volume and are brought to four PMTs in the back 
end.  This structure divides each module into four equal square cross section
towers, from which the light is collected separately by the fibers.  A 
PbSc sector is subdivided into 18 ($3~(\phi)\times 6~(z)$) super-modules 
made of $12 \times 12$ towers.  Thus a PbSc sector consists of $36 (\phi) 
\times 72 (z) = 2592$ towers.  The nominal energy resolution is $\delta 
E/E \sim 4.5~\% \oplus 8.3/\sqrt{E~{\rm (GeV)}}~\%$.

The PbGl consists of $4\times4\times40$ $\textrm{cm}^3$ rectangular modules,
each assembled from 4 lead-glass crystals.  Light emitted by particles
depositing energy in the crystals is collected by a PMT at the back of each
module.  Twenty four of them ($4~(\phi) \times 6~(z)$) are combined 
together to form a PbGl-module.  Six PbGl-modules ($3~(\phi) \times 
2~(z)$) comprise a PbGl-super-module and 32 of them ($4~(\phi) \times 
8~(z)$) constitute a sector.  This means a PbGl sector consists of $48 
(\phi) \times 96 (z) = 4608 $ PbGl counters.  The PbGl has a nominal 
energy resolution of $\delta E/E \sim 4.3~\% \oplus 
7.7/\sqrt{E~{\rm (GeV)}}~\%$.

The EMCal served as the electron trigger device in the 2005  
\pp~data taking.  The trigger (PH) is fired whenever there is any hit which
produces an energy of more than $1.4$ GeV in an EMCal trigger tile unit.  The 
trigger tile consists of $4 \times 4$ non-overlapping towers.  All 
trigger bits output from the tiles are summed in each super module 
online.  Since there are 172 EMCal super-modules (108 PbSc + 64 PbGl), 
the PH trigger has 172 output channels.  $95~\%$ of them were live in the 
2005 \pp~data.  All the output bit information is summed and used 
for the event-trigger decision.  The trigger efficiency is evaluated in
section \ref{sub:trig}.
 
 
\section{ANALYSIS}\label{sec:analysis}
%
 
 
\subsection{Data set and event selection (\auau)}\label{sub:auau_data}

The data for \auau~collisions at $\snn = 200$ GeV were taken during 
the 2004 run of RHIC.  The minimum bias (MB) trigger was defined as:

\begin{eqnarray}
  {\rm MB} \equiv ({\rm BBC} \ge 2) \&\&
(|z|<38~\textrm{cm}).
\end{eqnarray}

Here, ${\rm BBC} \ge 2$ means that at least two hits are required in 
both the North and South BBCs.  The offline trigger also requires at 
least one hit in one of the ZDCs.  The vertex position along the beam 
line is determined online from the timing difference between the two 
BBCs.  The MB trigger efficiency for inelastic \auau~collisions is 
evaluated as $92.2^{+2.5}_{-3.0}$\%.

In the offline analysis, a tighter vertex cut ($|{\bf bbcz}|<20$~cm) is 
required to eliminate conversion electrons from forward materials.  
Here {\bf bbcz} is the vertex position determined by BBC in the offline 
analysis.  Any abnormal data due to unusual beam or detector conditions 
are removed.  Table~\ref{tab:01-04-01} summarizes the number of MB 
events used for the analysis of electron spectra.  For the $v_2$ 
analysis, only those events with a good reaction plane calibration were 
used.  The number of MB events used for the $v_2$ analysis was $7.1 
\times 10^8$ events.

The MB events are divided into five centrality classes of 0--10\%, 
10--20\%, 20--40\%, 40--60\%, and 60--92\%.  See section~\ref{sub:cent} 
for the centrality determination.

There are two groups of runs: normal runs and converter runs.  In the 
converter runs, an additional photon converter was inserted around the 
beam pipe.  See section~\ref{sub:converter} for the converter runs and 
their use for the analysis.

\begin{table}[tbh]
  \caption{Number of events for each centrality class for
runs with and without additional converter material
installed.\label{tab:01-04-01}}
  \begin{ruledtabular} \begin{tabular}{ccc}
  Centrality class & Normal runs  & Converter runs  \\
 \hline
	0--92\%  		& $7.48\times 10^8$ & $5.79\times 10^7$ \\
  	0--10\%		& $8.06\times 10^7$ & $6.31\times 10^6$ \\
	10--20\%		& $8.07\times 10^7$ & $6.29\times 10^6$ \\
	20--40\% & $1.61\times 10^8$ & $1.25\times 10^7$ \\  
  	40--60\%	 & $1.61\times 10^8$ & $1.24\times 10^7$ \\
  	60--92\%	 & $2.65\times 10^8$ & $2.04\times 10^7$ \\
  \end{tabular} \end{ruledtabular}
\end{table}
 
 
\subsection{Data set and event selection $(p+p)$}\label{sub:auau_pp}

The data for $p+p$ collisions at $\sqrt{s}=200$ GeV were collected 
during the polarized $p+p$ collisions in 2005.  The proton beams had 
approximately 50\% longitudinal polarization with alternating spin 
orientations in successive bunches.  The polarization of the protons has
negligible effect on the cross section.

Two data sets are used for the electron analysis: (1) the minimum bias 
(MB) data set and (2) a ``photon'' (PH) trigger data set.  The MB trigger 
for $p+p$ required at least one hit in both the North and South BBC 
detectors in coincidence with the beam bunch crossing and the event 
vertex position within $|{\bf bbcz}|<$ 30 cm from the nominal 
collision point along the beam axis.  The PH trigger required a minimum 
energy deposit of 1.4 GeV in an overlapping tile of $4 \times 4$ EMCal 
towers in coincidence with the MB trigger.  The PH trigger had nearly 
100\% trigger efficiency for electrons with $p_T$ above 2 GeV/$c$ in the 
active trigger tiles.

The MB trigger cross section is $\sigma_{\rm BBC} = 23.0 \pm 2.2$ mb.  
Since only $\simeq$ 50\% of inelastic $p+p$ collisions satisfy the MB 
trigger condition, only a fraction of the inclusive electron production 
events were triggered by the PH trigger.  This fraction is assumed to be 
momentum and process independent, and is determined to be $\epsilon_{\rm 
bias} = 0.79 \pm 0.02$ \cite{Adler:2005ph} from the yield ratio of high 
$p_T$ $\pi^0$'s with and without the BBC trigger.  The value of 
$\epsilon_{\rm bias}$ is slightly higher than that of 2002 
(0.75)~\cite{Adler:2005fy}, due to a lower BBC threshold.

After the selection of good runs and the vertex cut, an 
integrated luminosity ($\mathcal{L}$) of 45 nb$^{-1}$ in the MB data set 
and 1.57 pb$^{-1}$ in the PH data set are used for the analysis.  During 
a part of the 2005 $p+p$ run, the same photon converter that was used in the 
2004 Au+Au run was inserted around the beam pipe.  The integrated 
luminosity in the converter run period is approximately 7\% of the 
total luminosity.

 
\subsection{Centrality (Au+Au)}\label{sub:cent}

\begin{figure}[htbp]
  \includegraphics[width=1.0\linewidth]{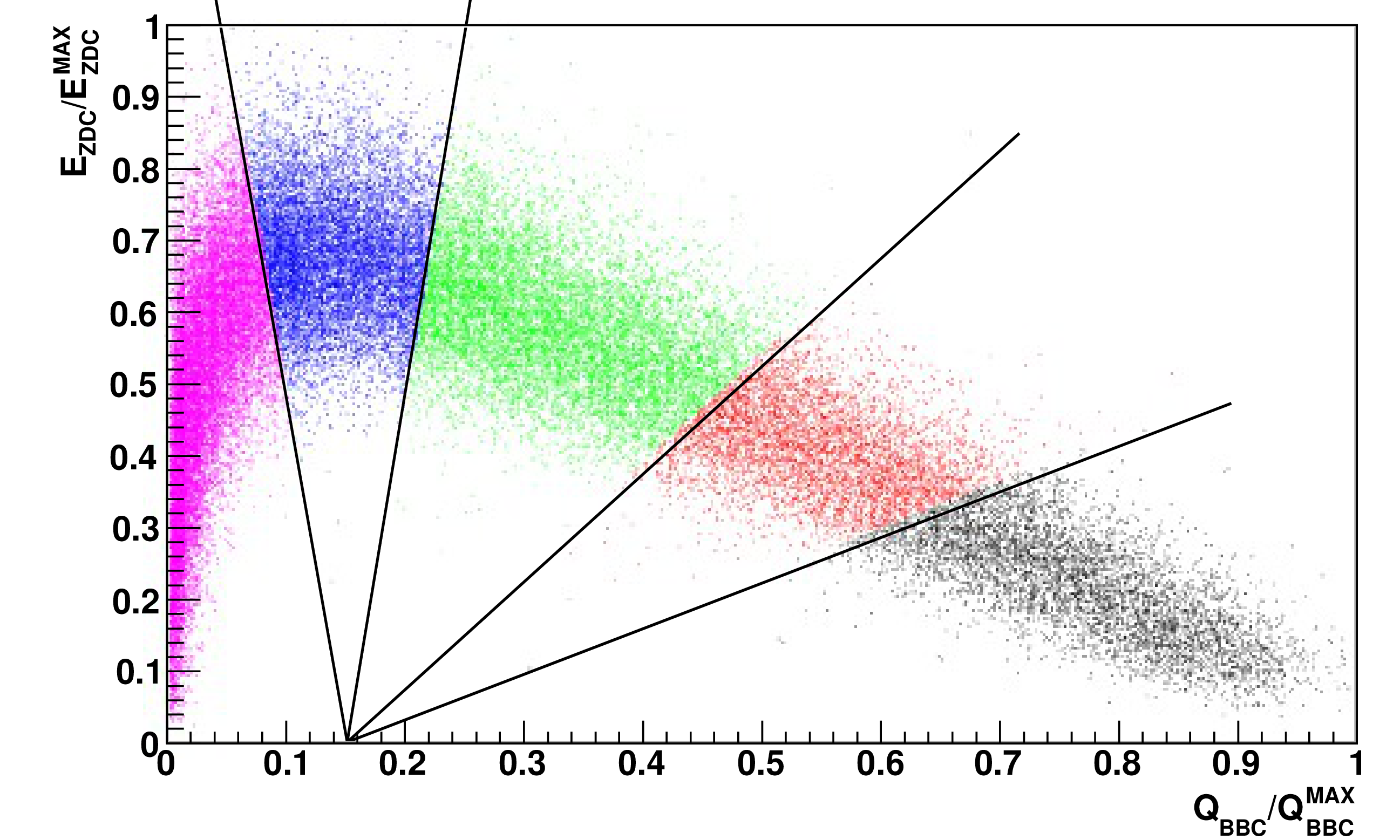}
  \caption{(Color online) The correlation between fractional BBC charge 
Q/Q$_{\rm max}$ and fractional of ZDC energy E/E$_{\rm max}$.  The sections
from right to left correspond to centrality classes 0--10~\%, 
10--20~\%, 20--40~\%, 40--60~\%, and 60--92~\%.
\label{fig:bbc_zdc}}
\end{figure}

The centrality of each \auau~collision is determined 
by the measurements of BBC charge and ZDC energy.  
Figure~\ref{fig:bbc_zdc} shows the correlation between BBC total charge 
and ZDC total energy for real data, each of which is normalized by the maximum
value.  The amount of BBC charge is proportional to the particle
multiplicity and is correlated to the overlapping area of two colliding nuclei.  
In contrast, the total ZDC energy is smaller in peripheral collisions than it
is in central collisions.  In the most peripheral collisions, a large fraction
of the nuclei become spectators.  The number of free neutrons reaching the ZDC
becomes small, due to the formation of deuterons or heavier charged nuclei from
the spectator neutrons and protons, which are swept by the magnets.  The
centrality determination from the BBC and ZDC energy is schematically shown in
Fig.~\ref{fig:bbc_zdc}.  Each centrality class contains approximately the same
number of events .  The most peripheral class is 
normalized to be 92\%.

The relation between centrality, the number of binary collisions 
($\ncol$), and the number of participants ($\npart$) is obtained from a 
Glauber Monte Carlo simulation~\cite{Glauber}.  The systematic 
uncertainties in $\ncol$ and $\npart$ are calculated from the 
uncertainty in the Glauber parameters, the centrality determination, and 
BBC and ZDC responses.  Table \ref{tab:ncoll_npart} summarizes the average 
$\ncol$, $\npart$, the associated nuclear overlap function ($\taa$) and 
their systematic errors for each centrality class.


\begin{table}[hbtp]
  \caption{$\npart$, $\ncol$ and $\taa$ \cite{Glauber} by Glauber calculation
for \auau~collisions at $\snn$ = 200
	 GeV.~\label{tab:ncoll_npart}}
\begin{ruledtabular} \begin{tabular}{cccc}
 Centrality class & $\langle \npart \rangle$ (syst) & $\langle \ncol \rangle$
 (syst) & $\langle \taa \rangle (\textrm{\rm MB}^{-1})$ (syst) \\
  \hline
 ~0--92\% & 109.1 (4.1)  & 257.8 (25.4) & 6.14  (0.45)  \\  
 ~0--10\% & 325.2 (3.3)  & 955.4 (93.6) & 22.75 (1.56)  \\
 10--20\% & 234.6 (4.7)  & 602.6 (59.3) & 14.35 (1.00)  \\
 20--40\% & 140.4 (4.9)  & 296.8 (31.1) & 7.07  (0.58)  \\
 40--60\% & 59.95 (3.6)  & 90.70 (11.8) & 2.15  (0.26)  \\
 60--92\% & 14.50 (2.5)  & 14.50 (4.00) & 0.35  (0.10)  \\
\end{tabular} \end{ruledtabular} 
\end{table}

 
\subsection{Track Reconstruction} \label{sub:track} 

\begin{figure}[htbp]
  \includegraphics[width=1.0\linewidth]{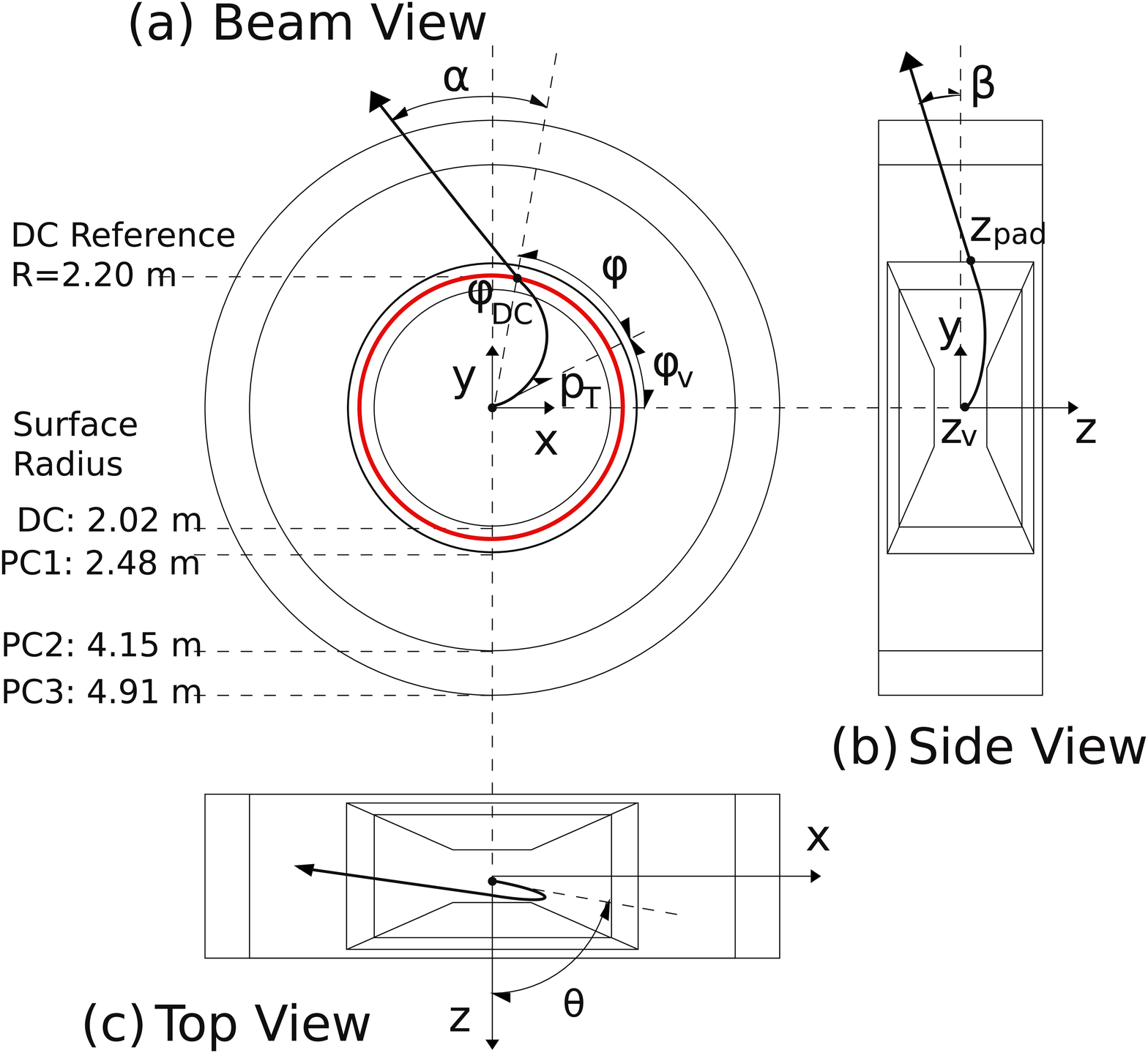}
\caption{(Color online) A charged particle trajectory and the kinematic
parameters 
are shown in (a) beam view, (b) side view, and (c) top view of PHENIX 
central region.\label{fig:parameters}}
 \end{figure}

Charged particles are reconstructed by the DC and PCs.  
Figure~\ref{fig:parameters} shows a trajectory up to PC3 and associated
kinematic parameters.  Assuming that the track originates on the beam axis, the
intersections of the trajectory with each detector plane are uniquely determined
by four initial kinematic parameters: $z_{{\rm vtx}}$, $\theta$,
$\phi_{\rm v}$, and the transverse momentum $\pt$.  Here, $z_{{\rm vtx}}$
is the interaction vertex along the $z$-axis, $\theta$ is the angle between the
initial direction of the particle track and $z$-axis, and $\phi_{\rm v}$ is
the initial azimuthal angle of the trajectory.  These initial parameters are
reconstructed frommeasured variables:~$\alpha$, $\beta$, $z_{\rm pad}$, and
$\phi_{\rm DC}$.  $\alpha$ is the angle between the projection of the
trajectory in the $x-y$ plane and the radial direction, at the intersection
point of the trajectory with the circle of DC reference radius 
($R_{\rm DC}=2.2~m$).  $z_{\rm pad}$ is the intersection point of 
the trajectory with PC1 surface radius ($R_{\rm PC1}=2.48~m$).  
$\beta$ is obtained by considering the plane which includes the $z$-axis and 
$z_{\rm pad}$.  It is defined as the angle between the projection of 
the trajectory to the plane and the line connecting the vertex to the position
of the pad chamber hit.  $\phi_{\rm DC}$ is the azimuthal angle of
intersection point of the trajectory with the circle of radius with
$R_{\rm DC}$.

The tracking process starts by collecting hits in DC X1 and X2 wires.  The 
hits are projected to the $x-y$ plane at $z=0$.  Then, a Hough transform is 
applied to find the sets of hit points based on the assumption of the 
track having a straight line trajectory inside the DC 
volume~\cite{Ben90,Ohlsson92}.  The technique is performed using all 
possible X1 and X2 hit combinations taking $\alpha$ and 
$\phi_{\rm DC}$ as the parameters in Hough space.  After this 
process, the direction of the found track line is specified by $\alpha$ 
and $\phi_{\rm DC}$.  The hits in the U and V wires and PC1 hits are 
associated with the track in order to obtain $z_{\rm pad}$ and 
$\beta$.  After the pattern recognition and track reconstruction, the initial
momentum vector of the track at the collision vertex is calculated.  A
look-up-table technique is used for fast
processing~\cite{Myrhiem79,Chikanian96}.  After the initial kinematic parameters
are obtained, each reconstructed track is associated with hit information from
the outer detectors (PC2, PC3, RICH, and EMCal).

 
\subsection{Electron Identification}\label{sub:eid}

Electron candidates are selected from reconstructed charged tracks based 
on information from the RICH and EMCal.  For each of the reconstructed 
tracks, the following variables are calculated and used for electron 
identification (eID);

\begin{itemize}
\item {\bf quality} --
Tracks are required to have hits in both the X1 and X2 sections of the DC and
be uniquely associated with hits in the U or V stereo wires.  At least one
matching PC1 hit is also required.

\item {\bf emcsdphi\_e} --
Displacement in $\phi$ of the electron hit position of the associated 
EMCal cluster from the projected position of the track in units of 
standard deviations.  For example, ${\bf emcsdphi\_e} < 2$ means that 
the position of the associated EMCal cluster in $\phi$ is within 2 
$\sigma$ of the projected track position.  This variable was calibrated
specifically for electrons.

\item {\bf emcsdz\_e} --
Same as {\bf emcsdphi\_e}, but for the $z$ coordinate.

\item {\bf prob} --
The probability that the associated EMCal cluster is an electromagnetic 
shower.  This variable is calculated from the $\chi^2$ value between the 
actual tower energy distribution of the cluster and the expected 
distribution for an electromagnetic shower.  For example, a cut, 
{\bf prob} $>0.01$, has 99\% efficiency for a photon or electron 
shower, while it rejects a large fraction of hadrons.

\item {\bf n0} --
Number of hit RICH PMTs in an annulus region with inner radius of 3.4 cm 
and outer radius of 8.4 cm around the the track projection on the RICH.  
The expected radius of a \v{C}erenkov ring emitted by an electron is 5.9 cm.

\item {\bf n1} --
The number of hit RICH PMTs within 11 cm around the projection point of
the track.

\item {\bf chi2/npe0} --
A $\chi^2$-like shape variable of the RICH ring associated with the track.

\item {\bf disp} --
A variable representing the displacement of the RICH ring center
from the projected track position.  Units are cm.

\item {\bf dep} --
A variable of energy momentum matching.  This variable is calculated as 
{\bf dep} $= (E/p - 1)/\sigma_{E/p}$, where $E$ is the energy 
measured by EMCal; $p$ is the momentum of the track; and $\sigma_{E/p}$ 
is the standard deviation of Gaussian-like $E/p$ distribution.  $E/p$ is less
than 1 for hadrons since hadrons do not deposit their full energy in the
calorimeter.  $\sigma_{E/p}$ depends on the momentum of the electron.
\end{itemize}

\begin{table}
\caption{Electron ID cuts used in Au+Au analysis}
\begin{ruledtabular} \begin{tabular}{c}
eID cuts for Au+Au\\
\hline
$\sqrt{\mbox{\textsf{emcsdphi\_e}}^2 + \mbox{\textsf{emcsdz\_e}}^2} <$ 2.0\\
\textsf{n0} $\geq$ 2\\
\textsf{n1} $\geq$ 3 if $p_T <$ 5.0 GeV/$c$\\
\textsf{n1} $\geq$ 5 if $p_T >$ 5.0 GeV/$c$\\
\textsf{disp} $<$ 5.0\\
\textsf{chi2/npe0} $<$ 10.0\\
\textsf{prob} $>$ 0.01 if $p_T <$ 5.0 GeV/$c$\\
\textsf{prob} $>$ 0.2 if $p_T >$ 5.0 GeV/$c$\\
\textsf{dep} $>$ -2.0\\
\end{tabular} \end{ruledtabular}
\label{tab:RGeid}
\end{table}

\begin{table}
\caption{Electron ID cuts used in $p+p$ analysis}
 \begin{ruledtabular}  \begin{tabular}{c}
eID cuts for $p+p$\\
\hline
$|\textsf{emcsdphi\_e}|<4$\\
$|\textsf{emcsdz\_e}|<4$\\
\textsf{n0} $\geq$ 2\\
\textsf{n1} $\geq$ 2\\
\textsf{prob} $>$ 0.01\\
$0.50 < E/p < 1.3 (0.2<p<0.3$ GeV/$c$)\\
$0.55 < E/p < 1.3 (0.3<p<0.4$ GeV/$c$)\\
$0.60 < E/p < 1.3 (0.4<p<0.6$ GeV/$c$)\\
$0.65 < E/p < 1.3 (0.6<p<0.8$ GeV/$c$)\\
$0.70 < E/p < 1.3 (0.8<p<5.0$ GeV/$c$)\\
\hline
\end{tabular} \end{ruledtabular}
\label{tab:RGeid_pp}
\end{table}

Tables \ref{tab:RGeid} and \ref{tab:RGeid_pp} summarize the eID cuts used for 
the Au+Au and $p+p$ analyses, respectively.  For 
tracks with $p_T$ below 4.8 GeV/$c$, the RICH is fired only by 
electrons.  Thus the RICH is the primary means of electron detection in 
this $p_T$ region.  The $E/p$ cuts remove remaining 
background.  The main cause of electron mis-identification below 5 
GeV/$c$ is that a charged hadron track is accidentally associated with 
hits in the RICH detector.  Since the particle multiplicity in $p+p$ is 
very small, typically few tracks in each of the central 
arm spectrometers, the probability of such an accidental overlap is much 
smaller.  This permits us to use looser electron cuts in the $p+p$ analysis.  
Note that although hadrons could fire the RICH above 4.8 GeV/$c$, the number of
photoelectrons is small for hadrons below 5 GeV/$c$, and such hadrons are easily
distinguishable from electrons.  Thus we tighten our cuts above 5 GeV/$c$
without worrying about extra hadron contamination.

In $p+p$ collisions the hadron contamination after the eID cuts is very small.
The contamination level is estimated by reversing the {\bf prob} cuts to enhance
the hadron background.  The estimated hadron contamination is 3\% at
$p_T=0.3$ GeV/$c$ and less than 1\% for $0.8 < p_T < 5$ GeV/$c$
with eID efficiency of approximately 90\%.  In Au+Au, there still remains hadron 
background due to accidental overlap between a track and RICH hits.  
This background is estimated and subtracted as described below.

\begin{figure}
  \includegraphics[width=0.8\linewidth]{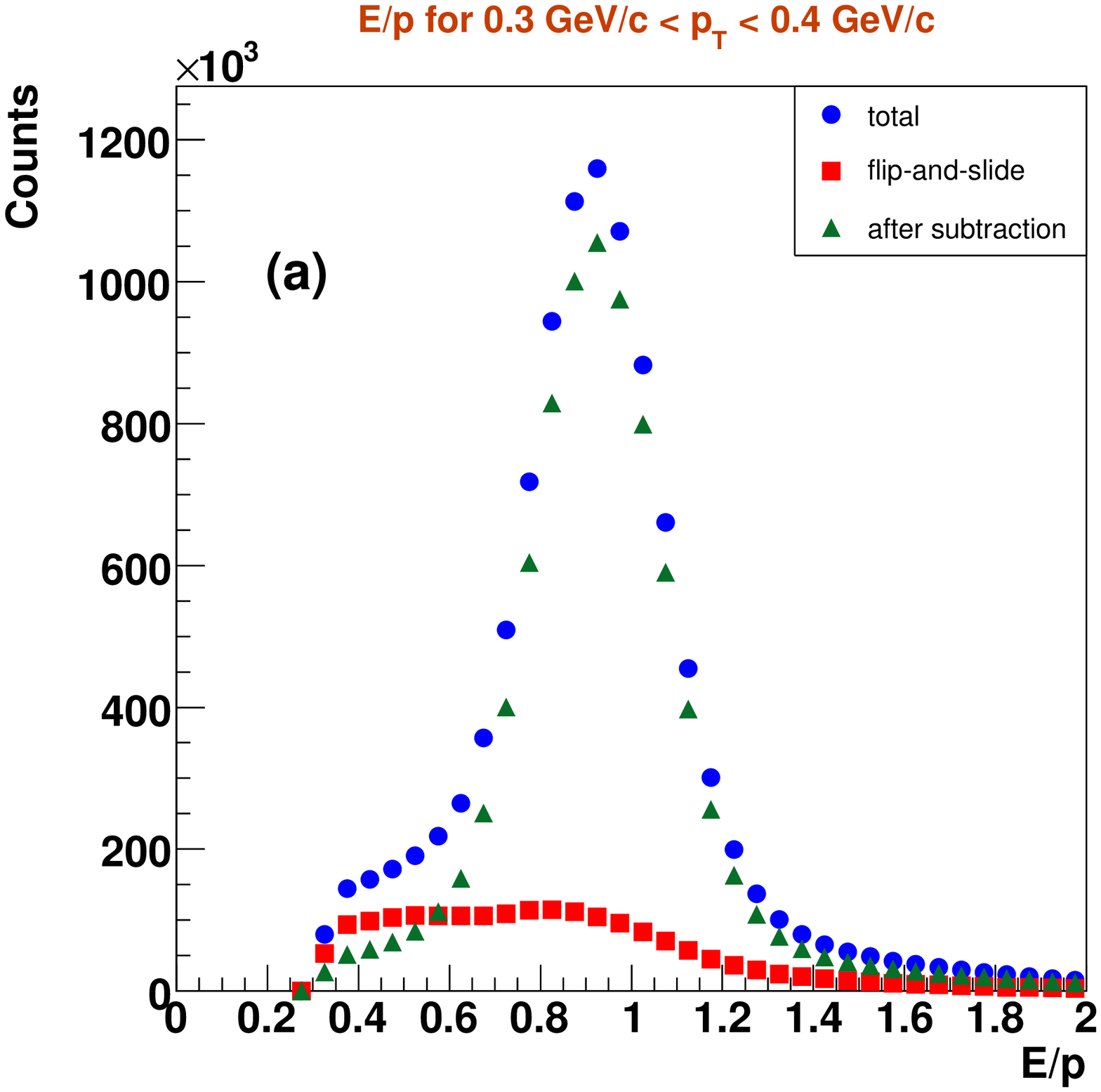}
  \includegraphics[width=0.8\linewidth]{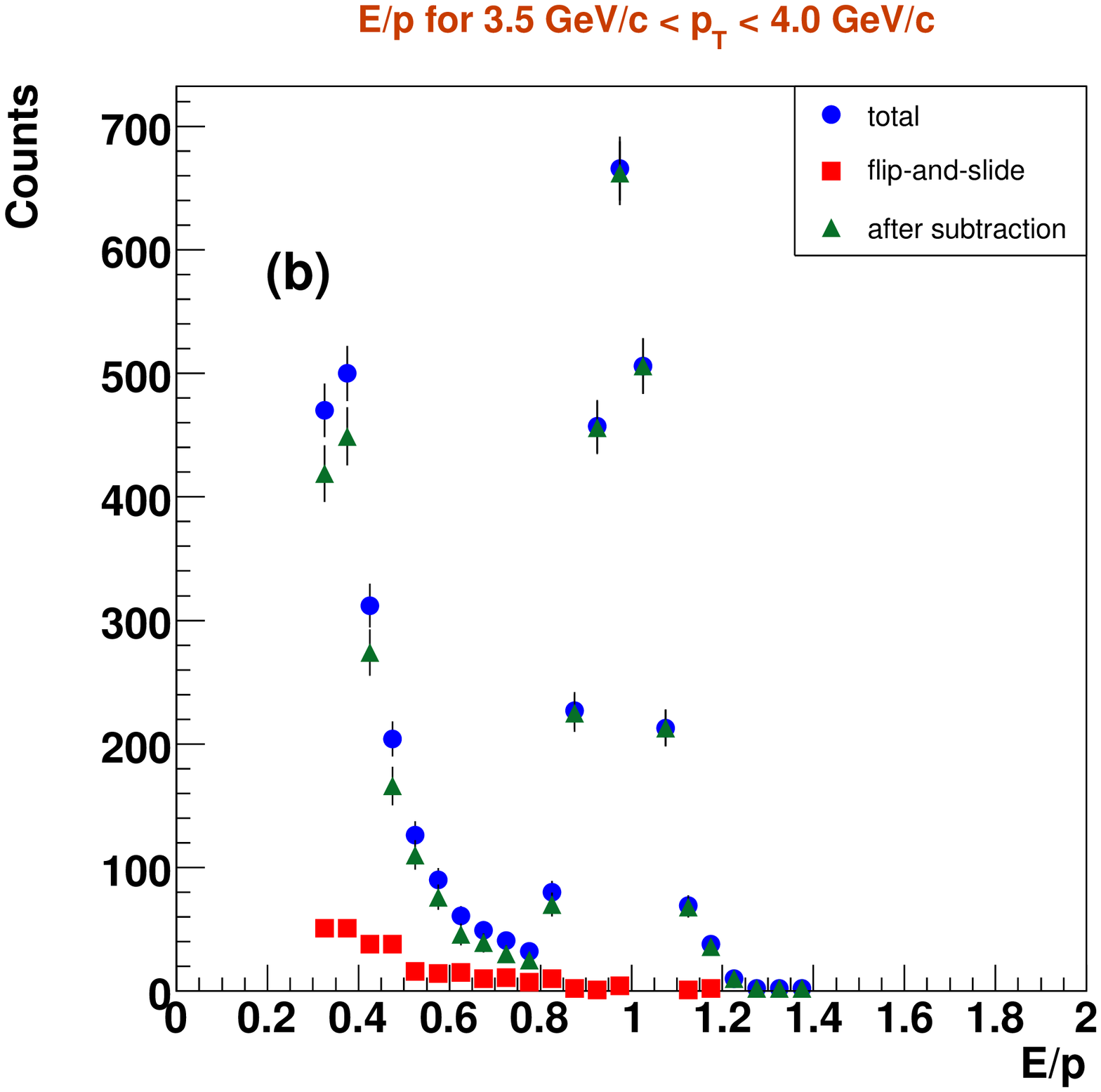}
  \includegraphics[width=1.0\linewidth]{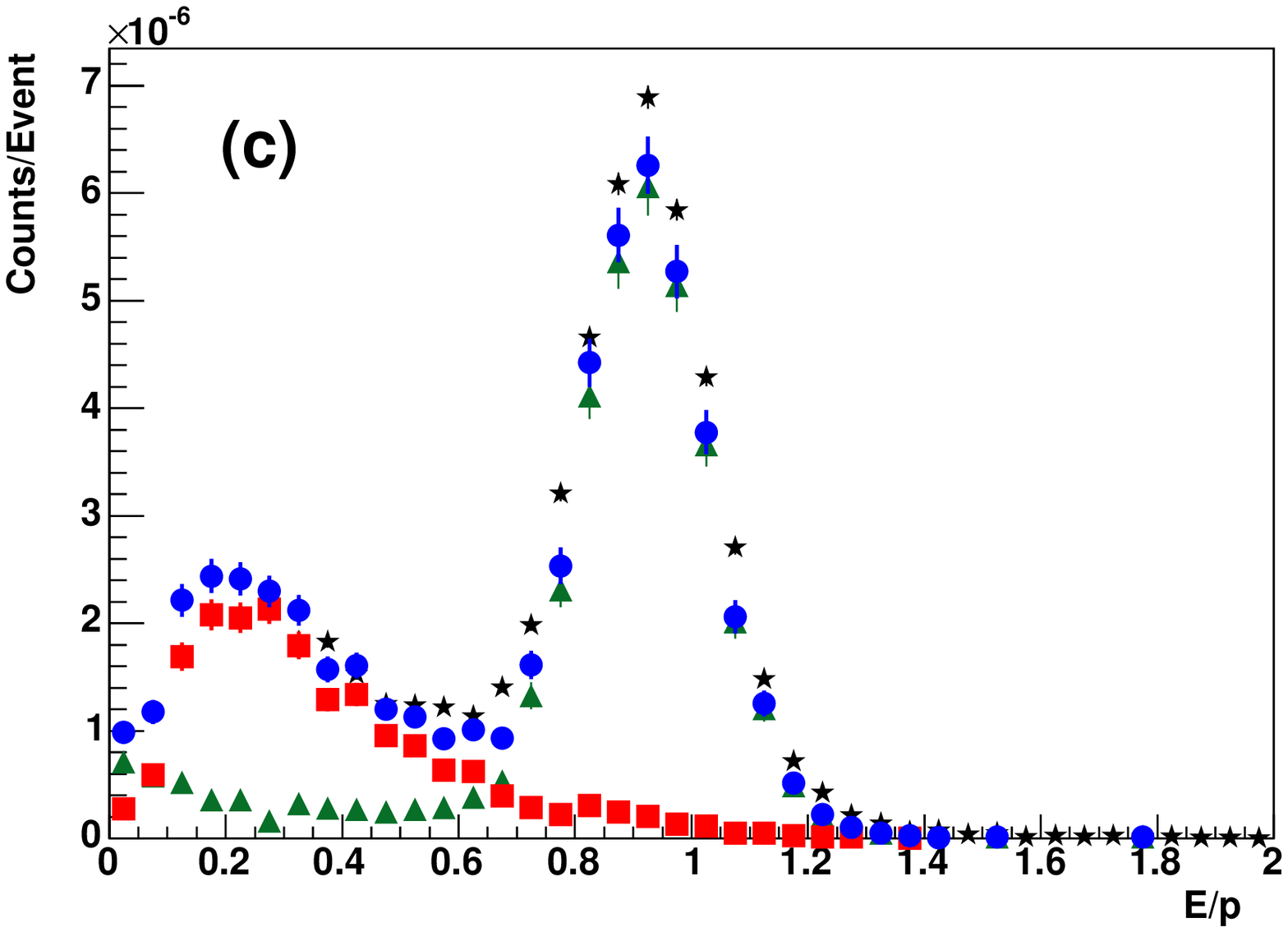}
\caption{(Color online) E/p distributions for various $\pt$ ranges.
(a) and (b) are for Au+Au collisions.  (c) is for $p+p$ collisions,
where the black stars are the data points and the solid circles show the
GEANT~\protect\cite{GEANT01:W5013} simulation.  The triangles and
squares show the contributions from simulation of $\pi^0$ and $K_{e3}$,
respectively.  See text for more details.
\label{fig:EoP1}}
\end{figure}

Figure~\ref{fig:EoP1} (a) and (b) show $E/p$ distributions for two 
typical $p_T$ ranges in Au+Au collisions.  All cuts in 
table~\ref{tab:RGeid} are applied except for the {\bf dep} cut since it 
is just a $p_T$-dependent cut in $E/p$.  The square points are the 
estimated distributions of the remaining hadron background which are 
randomly associated with a ring in the RICH.  This estimation is 
performed by swapping the north and south sides of the RICH in the same 
event in software, and reconstructing the track matching to RICH.  That 
is, DC tracks from the south are matched with RICH hits in the north, 
and vice-versa.  Since the north and the south sides of the RICH are 
identical and there was only $\sim$ 1 percent of dead channels in the 
RICH PMTs, this method gives a proper statistical estimate of the random 
hadron background in the electron sample.

After subtraction of the random hadron associations (shown as squares in 
Fig.~\ref{fig:EoP1} (a) and (b)), an additional low $E/p$ tail remains 
in the distribution at high $p_T$, as shown in Fig.  \ref{fig:EoP1} (b).  
This tail is due to electrons from kaon decay ($K_{e3}$) and photon 
conversions that occur far from the collision vertex.  These background 
electrons are reconstructed with a momentum higher than the actual 
momentum of the electrons, and as such have a low $E/p$.

A full GEANT \cite{GEANT01:W5013} simulation of the PHENIX detector was 
performed to determine the background from $K_{e3}$ decays and from 
photon conversions.  Figure~\ref{fig:EoP1}(c) shows the $E/p$ 
distribution in $p+p$ for $0.8 < p_T < 1.0$ GeV/$c$ compared with the 
GEANT simulation.  The black stars are the data points and the circles 
show the simulation.  The triangles and squares show the contributions 
from simulation of $\pi^0$ and $K_{e3}$, respectively.  The simulated 
$\pi^0$ and $K_{e3}$ decays went through the same offline analysis chain 
as in the real data, and identical electron identification cuts were 
applied.  In the $p+p$ data, the contamination of hadron 
mis-identification is negligible.  In the $\pi^0$ simulation, electrons 
are mainly produced either by the Dalitz decay or by photon conversion 
in the beam pipe.  These conversion electrons produce the peak around 
$E/p = 1$.  Note that in Fig.~\ref{fig:EoP1}(c) the data is above the 
simulation since the data also contain nonphotonic electrons.  A small 
number of conversions in the helium bag contribute to the tail in the 
low $E/p$ region.  The main cause of the low $E/p$ tail is $K_{e3}$ 
decay, shown as squares.  Since the simulation reproduces the low $E/p$ 
tail of the data very well, we conclude that the $K_{e3}$ background 
under the Gaussian peak, which will remain after the {\bf dep} or $E/p$ 
cuts, can be determined from the simulation.

There is no adjusted normalization in this simulation/data comparison.  
In the simulation, we assume $d\sigma/dy(pp\rightarrow \pi^0)$ = 46.0 
mb, $d\sigma/dy(pp\rightarrow K^L)$ = $d\sigma/dy(pp\rightarrow K^\pm)$ 
= 4.0 mb at midrapidity.  The momentum distribution of $\pi^0$ and kaons 
are based on PHENIX measurements in $p+p$.

For $p_T$ above 4.8 GeV/$c$, charged pions begin to radiate 
\v{C}herenkov light in the RICH.  Thus tighter electron selection cuts 
are applied to reject pion background.  In both the Au+Au and $p+p$ 
analyses, tighter cuts of {\bf n1} $\geq$ 5 and {\bf prob} $>$ 0.2 are 
added.  In the $p+p$ analysis, the $E/p$ cut is tightened to 
$0.8<E/p<1.3$.  With these cuts, the electron measurement is extended to 
9 GeV/$c$ in $p_T$.  The tighter cuts are applied to tracks above 5 
GeV/$c$, as the \v{C}herenkov radiation is very weak between 4.8 and 5 
Gev/$c$, and the data were binned in multiples of 0.5 GeV/$c$.

\begin{figure*}[th]
  \includegraphics[width=0.48\linewidth]{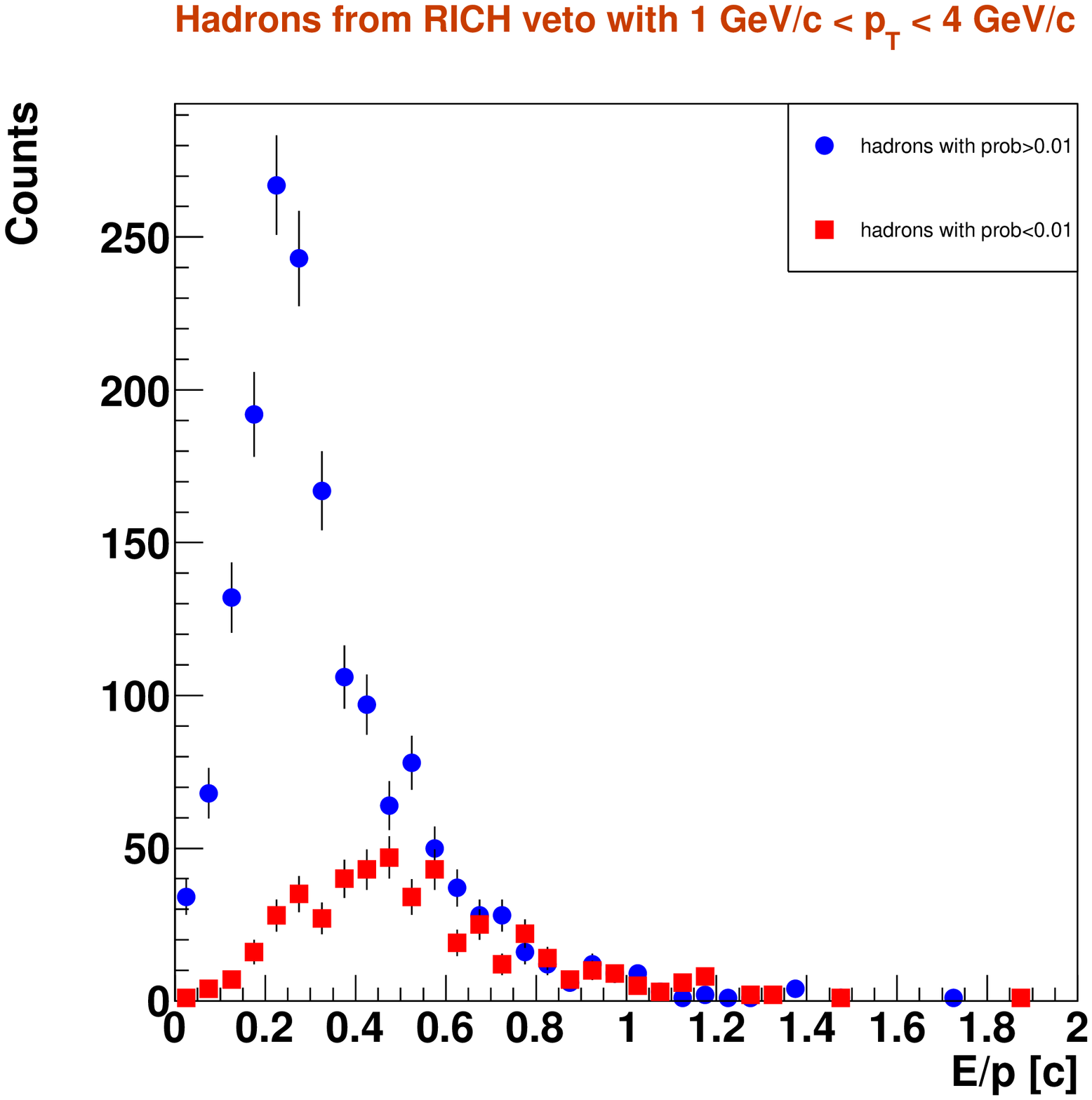}
  \includegraphics[width=0.48\linewidth]{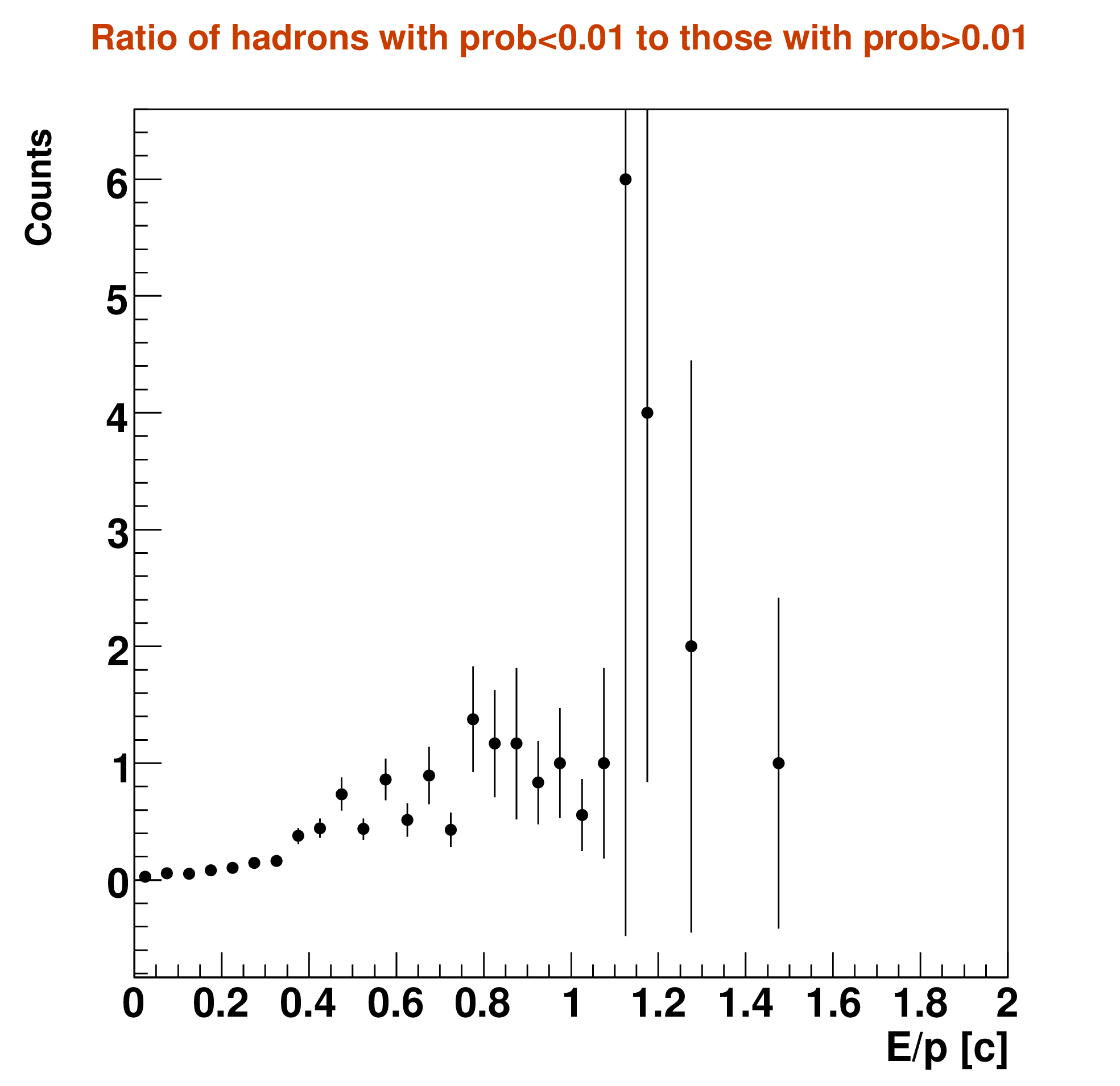}
  \includegraphics[width=0.48\linewidth]{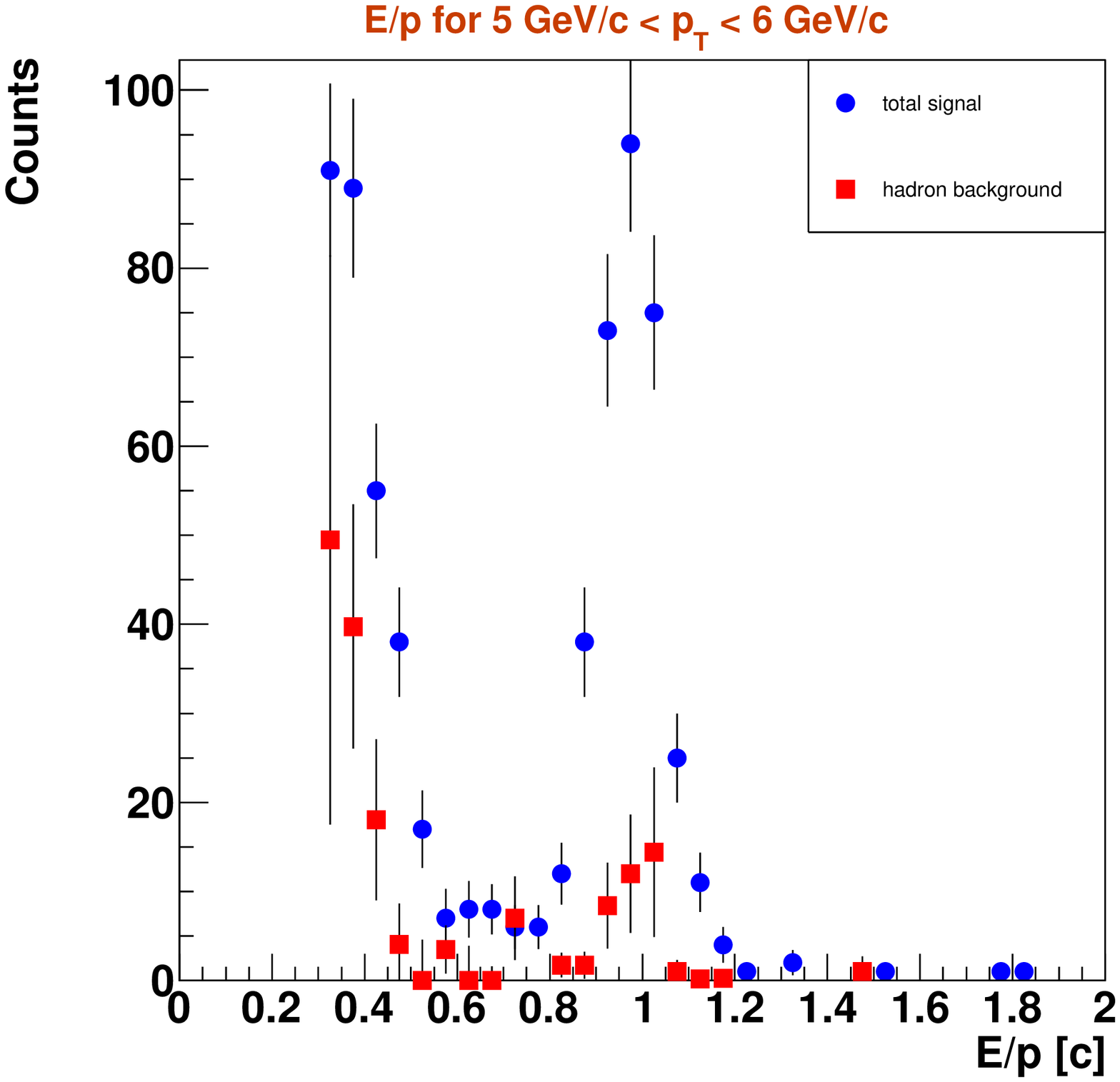}
  \includegraphics[width=0.48\linewidth]{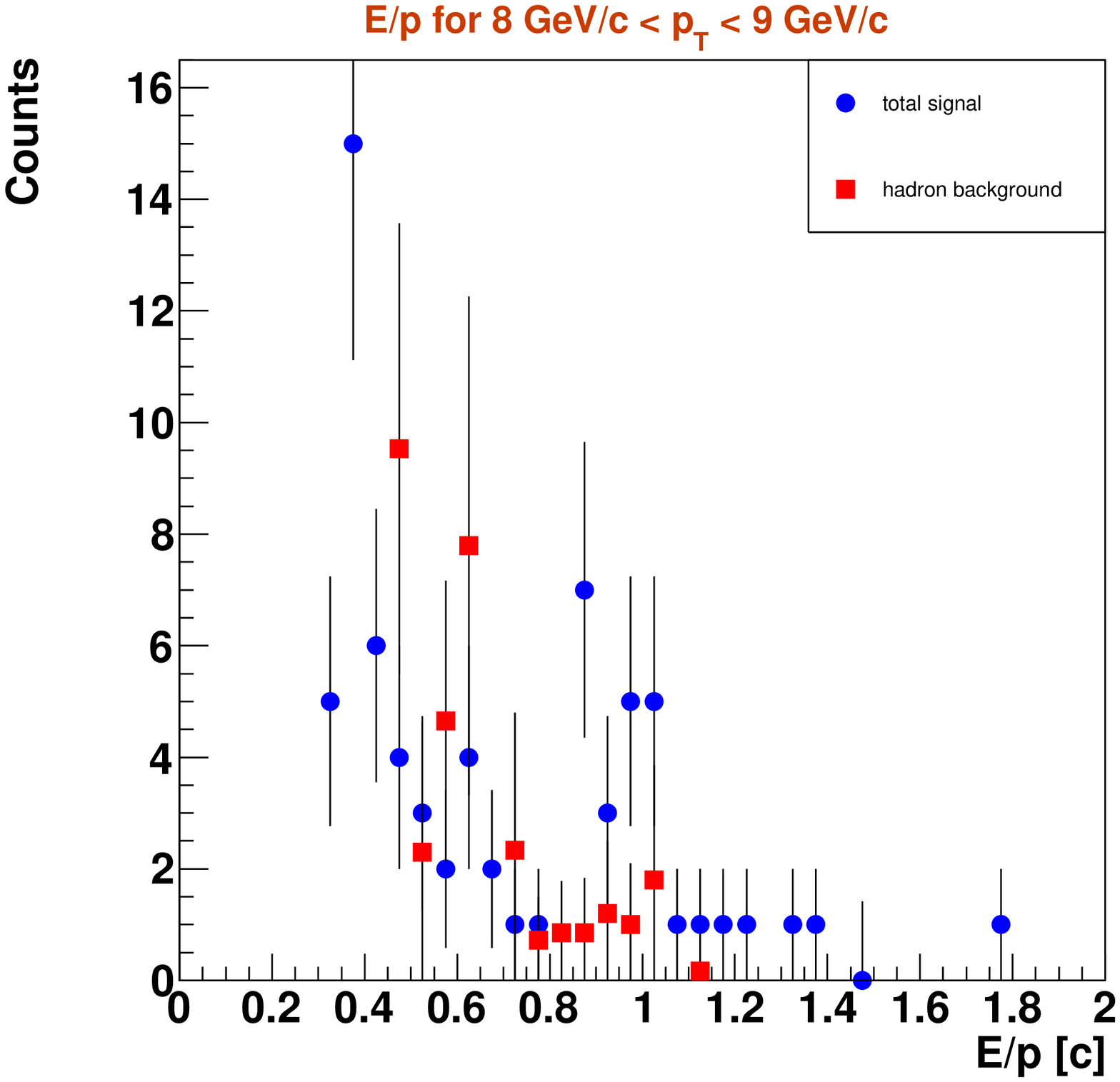}
\caption{(Color online) E/p distributions for (upper) hadron samples from a 
RICH veto cut which are used in the contamination study and (lower) the
indicated high-$\pt$ ranges, shown along with the estimation of the
hadronic contamination.
\label{fig:had1}}
\end{figure*}

The remaining hadron background with the tighter cuts is studied using 
the shape of the $E/p$ distribution.  Here we rely on the fact that the 
distribution of the {\bf prob} and $E/p$ variables are roughly 
independent of $p_T$ for hadrons at high $p_T$, and that a cut on {\bf 
prob}$<$0.01 eliminates the vast majority of electrons.  First, we 
obtain a sample of hadrons in the $p_T$ range of 1-4 GeV/c by imposing a 
veto on the RICH.  The hadron sample is then divided into two samples, 
one with {\bf prob}$>$0.01 and the other with {\bf prob}$<$0.01.  The 
ratio of these two hadron samples is taken.  The upper plots in 
Fig.~\ref{fig:had1} show the two hadron samples and their ratio.  The 
$E/p$ distributions for $p_T$ above 5 GeV/$c$ are then estimated from 
the data with cuts identical to those described above, except with the 
{\bf prob} cut reversed to {\bf prob}$<$0.01.  These $E/p$ distributions 
are then divided by the ratio shown in the upper right plot of 
Fig.~\ref{fig:had1} to obtain an estimate of the $E/p$ distributions of 
hadrons passing the {\bf prob}$>$0.01 cut.

The lower plots of Fig.~\ref{fig:had1} show the estimate for the 
hadronic background along with the total $E/p$ distribution.  This 
procedure actually gives an overestimation of the background, as some 
electrons do pass the {\bf prob}$<$0.01 cut.  This can be seen in the 
lower left plot as a peak in the hadronic sample.  From these plots, we 
determine that any hadronic background below 8 GeV/c is negligible.  
Between 8-9 GeV/c in $p_T$, we estimate a background of 20\%, with an 
uncertainty of 10\%. The eID efficiency of the tighter cuts is $p_T$ 
independent.  In $p+p$, it is determined to be 57\% of that for $p_T<5$ 
GeV/$c$ by applying the same tighter cuts for $p_T<5$ GeV/$c$.

\subsection{Acceptance correction}\label{sub:acc}

The acceptance correction in this analysis covers the following three
components:  the geometrical acceptance correction
($\epsilon^{\rm geo}$), the eID efficiency
($\epsilon^{\rm eID}$) and the reconstruction efficiency
($\epsilon^{\rm reco}$).  $\epsilon^{\rm geo}$ accounts for the
fraction of electrons that do not hit any detectors due to the finite
solid angle covered by the detectors.  $\epsilon^{\rm eID}$ is the
correction factor for signal loss by eID cuts.  $\epsilon^{\rm reco}$
takes into account that the measured electron spectrum in the detector
is different from the real spectrum due to detector responses and the
track reconstruction method.  All the values are computed at once by the
full detector simulation for single $e^{\pm}$ events.  Single $e^+$ and
$e^-$ events are generated uniformly in phase space ($0 < \pte < 15$
GeV/$c$, $|y| < 0.5$, and $0< \phi <2\pi$).  They are processed by the
full GEANT simulation program of the PHENIX
detector\cite{GEANT01:W5013}.  The output simulation data files are
processed by the event reconstruction chain of PHENIX.  The acceptance
correction factor (${\epsilon}^{\rm acc}_{\Delta y}$) is calculated as
follows:

\begin{eqnarray}
  {\epsilon}^{\rm acc}_{\Delta y}
  	\equiv {\epsilon}^{\rm geo} 
	\cdot  {\epsilon}^{\rm reco} 
	\cdot  {\epsilon}^{\rm eID} 
 	= \frac{dN_e^{\rm out}/d\pte}{dN_e^{\rm in}/d\pte},
\end{eqnarray}

The $\pte$ distributions of input and output electron yields are 
$dN_e^{\rm in}/d\pte$ and $dN_e^{\rm out}/d\pte$, respectively.  The 
same eID cuts and fiducial cuts used in real data processing are applied 
to the output.  Since the input electrons are generated flat in $\pte$, 
we must take into account that the smearing of the spectrum due to 
imperfect momentum reconstruction affects a steeply falling spectra 
differently than our simulated spectra.  Each track reconstructed in the 
simulation is weighted by the $\pte$ of the corresponding input 
electron.

\begin{figure}[hbtp]
  \includegraphics[width=1.0\linewidth]{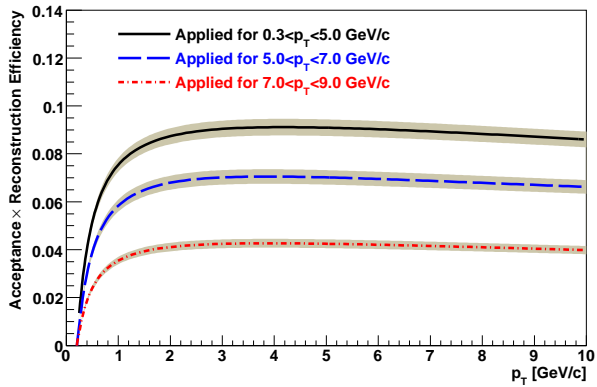}
  \caption{(Color online) $\epsilon^{\rm acc}_{\Delta y}$ for $\anee$ as a
function of $p_T$.  The top, 
  	 middle and bottom curves are calculated with eID cuts applied
for 
	 $0.3<p_T<5.0$, $5.0<p_T<7.0$, and $7.0<p_T<9.0$ GeV/$c$ respectively.
\label{fig:acc}}
\end{figure}

Figure~\ref{fig:acc} shows three curves of $\epsilon^{\rm acc}_{\Delta 
y}$ as a function of $p_T$ for $\anee$.  The top, middle and bottom 
curves are calculated with each set of eID cuts applied for 
$0.3<p_T<5.0$, $5.0<p_T<7.0$, and $7.0<p_T<9.0$ GeV/$c$ respectively.

\subsection{Trigger efficiency ($p+p$)}\label{sub:trig}

The trigger efficiency of the PH trigger in \pp~ collisions is 
determined from the MB data set as the fraction of electrons that 
satisfies the PH trigger.  Since the trigger condition determined by the 
online level-1 trigger processor was recorded, we can apply the same PH 
trigger to the events recorded in the MB trigger in the offline 
analysis.  We require that the PH trigger bits of the event are set, and 
that the trigger tile that fires the PH trigger is hit by the electron 
candidate track selected by the offline analysis.

\begin{figure}
\includegraphics[width=1.0\linewidth]{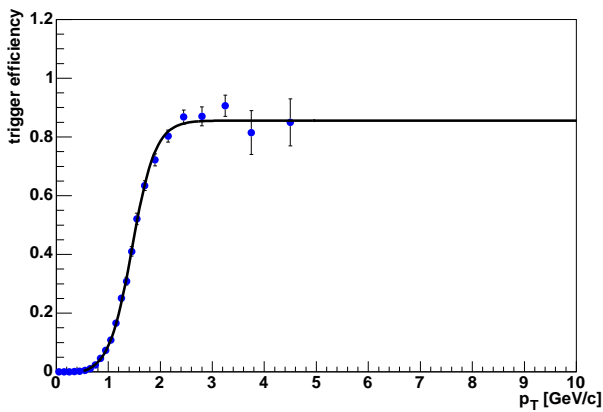}
\caption{(Color online) \label{fig:trig_eff}
Efficiency of the PH trigger in $p+p$ determined from the minimum-bias
data set.  
}
\end{figure}

Figure~\ref{fig:trig_eff} shows the trigger efficiency of the PH trigger 
thus determined.  The effective trigger threshold is about 1.4 GeV, and 
the efficiency saturates above 2 GeV.  The trigger efficiency at the 
plateau is $\approx$ 86\%, consistent with the fraction of active 
trigger tiles.  The curve is a Fermi-like function that is fitted to the 
data.  The fitted function is used as the trigger efficiency in the 
analysis.

\subsection{Occupancy correction (Au+Au)}\label{sub:embed}

In addition to the efficiency for a single electron passing through the 
detector to pass the fiducial and eID cuts due to detector dead area and 
the efficiencies described above, there is also a finite efficiency loss 
for particle detection due to the presence of other particles nearby.  
To get a quantitative understanding of the multiplicity-dependent 
efficiency loss, we embed simulated single electrons and positrons into 
data files containing detector hits from real collisions.  The same 
simulated particles are used in this method as those used to estimate 
the single particle efficiency loss.  The simulated particles are 
embedded into events such that the $z$ vertex of the simulated particle 
and the event are similar.  The simulated $e^\pm$ are run through the 
GEANT simulator of PHENIX, and the hits are added to the data files 
containing hits from a real Au+Au event.  Next, these new files 
containing the embedded $e^\pm$ are run through the entire 
reconstruction software to produce track candidates containing the 
variables upon which we make identification cuts.  We then define the 
embedding efficiency as
\begin{equation}
\epsilon_{\textrm{embed}} = \frac{\# \ \textrm{reconstructed} \ e^\pm \
\textrm{from embedded data}}{\# \ \textrm{reconstructed} \ e^\pm \ \textrm{from
single track data}}
\end{equation}
where a reconstructed particle from embedded data has most of its 
detector hits associated with hits from the simulated particle.

As a systematic cross-check on this method for determining the 
multiplicity-dependent efficiency loss, a data-driven method was 
employed.  The general strategy of this data-driven method is to select 
a very pure sample of electron conversion pairs from the data with an 
invariant mass cut.  We apply tight eID cuts to one of the tracks in a 
given pair to increase the chance that the pair really is an electron 
pair.  Then, we measure the efficiency loss as a function of collision 
centrality of tighter cuts (cuts actually used in the analysis) relative 
to that of the loose cuts.  We then still need simulation to determine 
the multiplicity-dependent loss due to the loose eID cuts, but 
simulation is more reliable when we use loose cuts.  In order to get a 
pure sample of electrons, we isolate $e^\pm$ pairs from 
photon-conversions in the beampipe and from $\pi^0$ Dalitz decays.  We 
assume, reasonably, that the multiplicity-dependent efficiency loss for 
these electrons is the same as for all electrons.

\begin{figure}[htb]
  \includegraphics[width=1.0\linewidth]{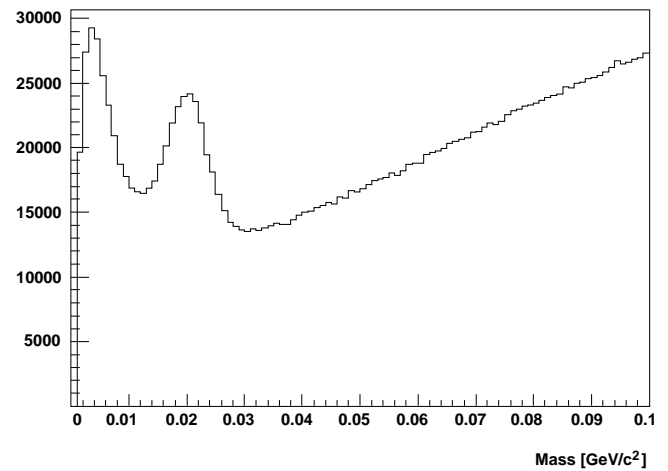}
  \caption{Invariant mass distribution of $e^\pm$ pairs from min-bias Au+Au data
 used for the multiplicity-dependent efficiency loss estimate.
\label{fig:e_pm_mass}}
\end{figure}

Figure~\ref{fig:e_pm_mass} shows the invariant mass distribution of 
$e^\pm$ pairs from the data.  The peak at $\sim$ 20 MeV/$c^2$ 
corresponds to the apparent mass of electrons from photon conversions in 
the beampipe.  The peak at $\sim$ 5 MeV/$c^2$ corresponds to $e^\pm$ 
from $\pi^0$ Dalitz decays.  As mentioned above, we would like to have 
electrons identified by using both loose cuts and tight cuts.  However, 
if loose cuts are applied to both the electron and the positron in the 
pair, then the sample will contain contamination from hadrons.  In order 
to make sure that the pair is really an $e^\pm$ pair, we apply tight eID 
cuts to one of the particles in the pair.  We restrict ourselves to 
pairs with one of the particles in the acceptance of the Time of Flight 
detector \cite{Aizawa:2003a} so that we can use the TOF for electron 
identification.

The cuts used for the analysis for $p_T <$ 5 GeV/$c$ are then applied 
to the other particle in the pair, and the relative efficiency loss is 
calculated.  As described in the previous section, for $p_T >$ 5 
GeV/$c$ we obtain the efficiency by the relative electron yield 
between the samples with the cuts used for the two $p_T$ ranges.  
This relative yield includes multiplicity-dependent effects, so we 
only explicitly calculate the multiplicity-dependent efficiency loss 
for the cuts used for $p_T <$ 5 GeV/$c$.  Some of the relative 
efficiency loss using the method just described is due to the single 
particle efficiency.  To obtain an estimate of the 
multiplicity-dependent component of this relative loss, we divide the 
relative loss by the loss for peripheral collisions (60\%-93\% 
centrality), since the multiplicity-dependent loss in peripheral 
collisions is very low.  So we have defined the multiplicity-dependent 
efficiency loss from the data-driven method for centrality $C$ as
\begin{equation}
\epsilon_{\textrm{mult}}^{\textrm{data}, C} =
\frac{\epsilon_{\textrm{eID}}^{\textrm{data},
C}}{\epsilon_{\textrm{eID}}^{\textrm{data}, 60--93\%}}
\end{equation}
with
\begin{equation}
\epsilon_{\textrm{eID}}^{\textrm{data}, C} = \frac{\epsilon_{\textrm{tight
cuts}}^{C}}{\epsilon_{\textrm{loose cuts}}^{C}}.
\end{equation}
Ideally we would measure the above efficiency with no bias caused by making
the initial loose cuts.  We correct the multiplicity-dependent efficiency
loss by using the efficiency from the embedding simulation for the loose cuts,
$\epsilon_{\textrm{loose}}^{\textrm{embed}, C}$ to obtain the more accurate
estimate of the multiplicity-dependent loss from the otherwise data-driven
method.

\begin{table}[htbp]
\caption{Multiplicity-dependent efficiency loss calculated by two
methods.  $\epsilon^{\textrm{embed},C}$ is the efficiency
loss calculated from the embedding
simulation.$\epsilon_{\textrm{loose}}^{\textrm{embed}, C}$ is the same as above
but with  the loose cuts.  ~\label{tab:embedding}}
\begin{ruledtabular} \begin{tabular}{cccc}
 Centrality  &  $\epsilon^{\textrm{embed},C}$ & 
$\epsilon_{\textrm{mult}}^{\textrm{data},C}
\epsilon_{\textrm{loose}}^{\textrm{embed}, C}$ \\ 
  \hline
 00--93\% 	&  0.852 & 0.882\\
 0--10\%  	&  0.771 & 0.769\\
 10--20\% 	&  0.835 & 0.856\\
 20--40\% 	&  0.900 & 0.924\\
 40--60\% 	&  0.952 & 0.977\\
 60--93\% 	&  0.982 & 0.997\\
\end{tabular} \end{ruledtabular}
\end{table}

Table~\ref{tab:embedding} displays the various embedding efficiencies 
described above.  The difference between the left and right columns 
gives an idea of the systematic error involved in this estimation.  
The use of the sample of conversion and Dalitz $e^\pm$ pairs will be 
discussed further in the section pertaining to systematic error 
analysis.
 
 
\subsection{Electron background cocktail}\label{sub:cocktail}

The inclusive electron spectra consists primarily of three components:
(1) ``nonphotonic" electrons from heavy-flavor decays, 
(2) ``photonic" background from Dalitz decays of light neutral mesons
  and photon conversions (mainly in the beam pipe), and 
(3) ``nonphotonic" background from $K \rightarrow e\pi\nu$ ($K_{e3}$), 
  and dielectron decays of vector mesons.

At high electron $p_T$, Drell-Yan processes also account for a small 
but non-negligible contribution to the electron spectrum.  Also, heavy 
quarkonia decays contribute at high electron $p_T$, which, although 
arguably can be included in electrons from ``heavy-flavor" decays, 
should be distinguished from open heavy flavor decays for the purpose 
of interpretating the measurement.

The photonic background is much larger than the nonphotonic 
background.  The signal of electrons from heavy-flavor decays is small 
compared to the photonic background at low $\pt$ ($S/B < 0.2$ for $p_T < 
0.5$~\gevc) but rises with increasing $\pt$ ($S/B > 1$ for $p_T > 
2$~\gevc).  In order to extract the heavy-flavor signal, the background 
has to be subtracted from the inclusive electron spectra.  One 
technique to accomplish this task is the so-called ``cocktail 
subtraction" method described in detail here.  

A cocktail of electron spectra from background sources is calculated using
a Monte Carlo event generator of hadron decays and then subtracted from 
the inclusive electron spectra.  This technique requires that the phase space
distributions of the relevant background sources be well known.  The PHENIX
measurements of the relevant electron sources are precise enough to constrain
the background within a systematic uncertainty better than 15~\% for all $\pt$.
This uncertainty is of the same order as the signal to background ratio at 
the lowest $\pt$ and, therefore, it is not sufficiently small to 
extract the heavy-flavor signal via the cocktail subtraction over the 
full $\pt$ range.  Hence, at low $\pt$, a complementary technique to 
subtract the background, the so-called ``converter subtraction" method, 
is used to extend the heavy-flavor measurement to the lowest 
$\pt$ with good precision.  Consequently, the converter subtraction is 
the key to extracting the total heavy-flavor yield or cross section since 
most of the electrons from heavy-flavor decays have low $\pt$.  
However, towards high $\pt$, {\it e.g.} for $\pt > 2$~\gevc where the 
converter subtraction starts to suffer from a lack of statistical 
precision, it is beneficial to apply the cocktail subtraction since 
the signal to background ratio is large, statistics is irrelevant due 
to the Monte Carlo nature of the cocktail subtraction, and the 
cocktail input is known with small systematic uncertainties as 
discussed in the following.

\begin{figure}[tbh]
\includegraphics[width=1.0\linewidth]{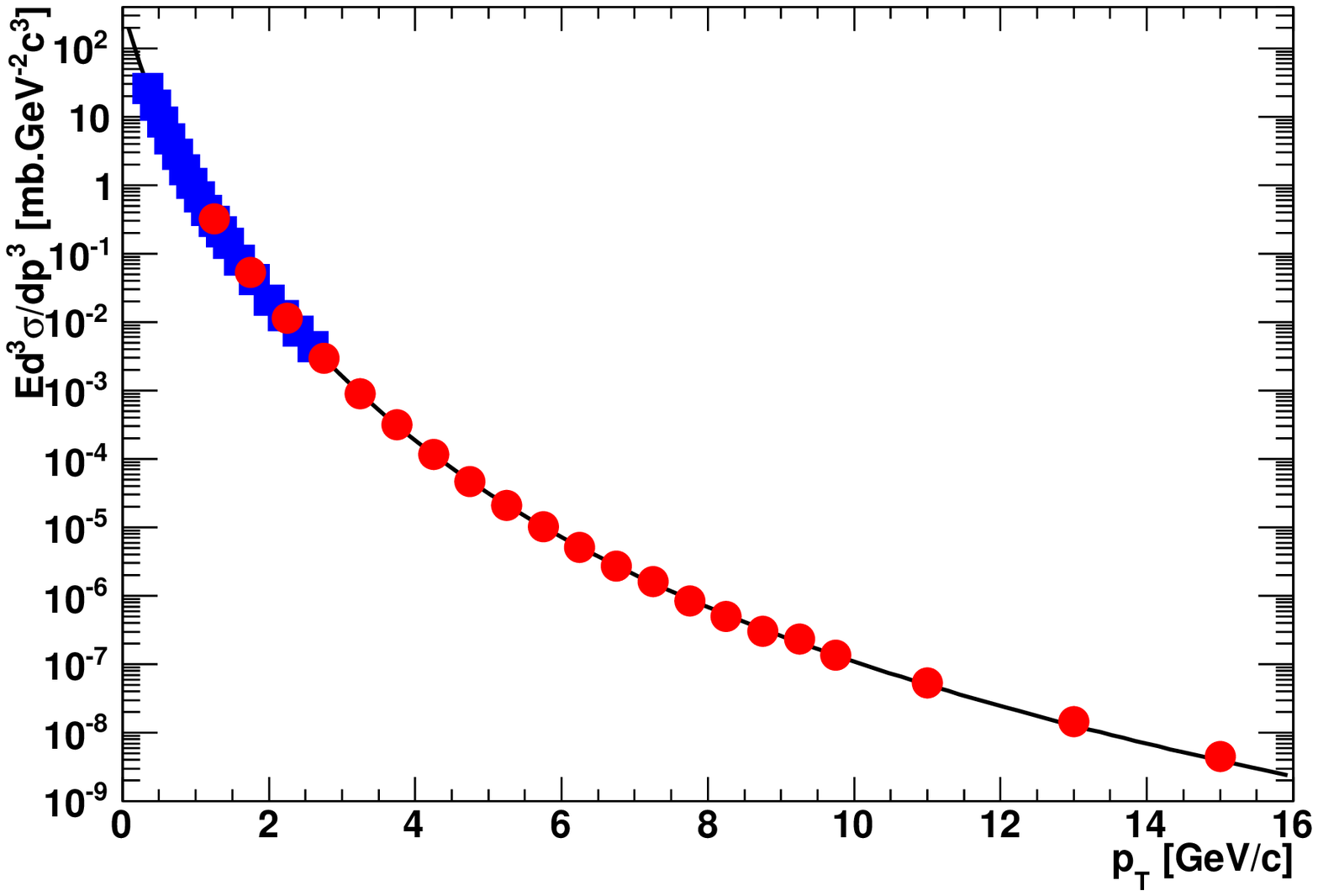}
\includegraphics[width=1.0\linewidth]{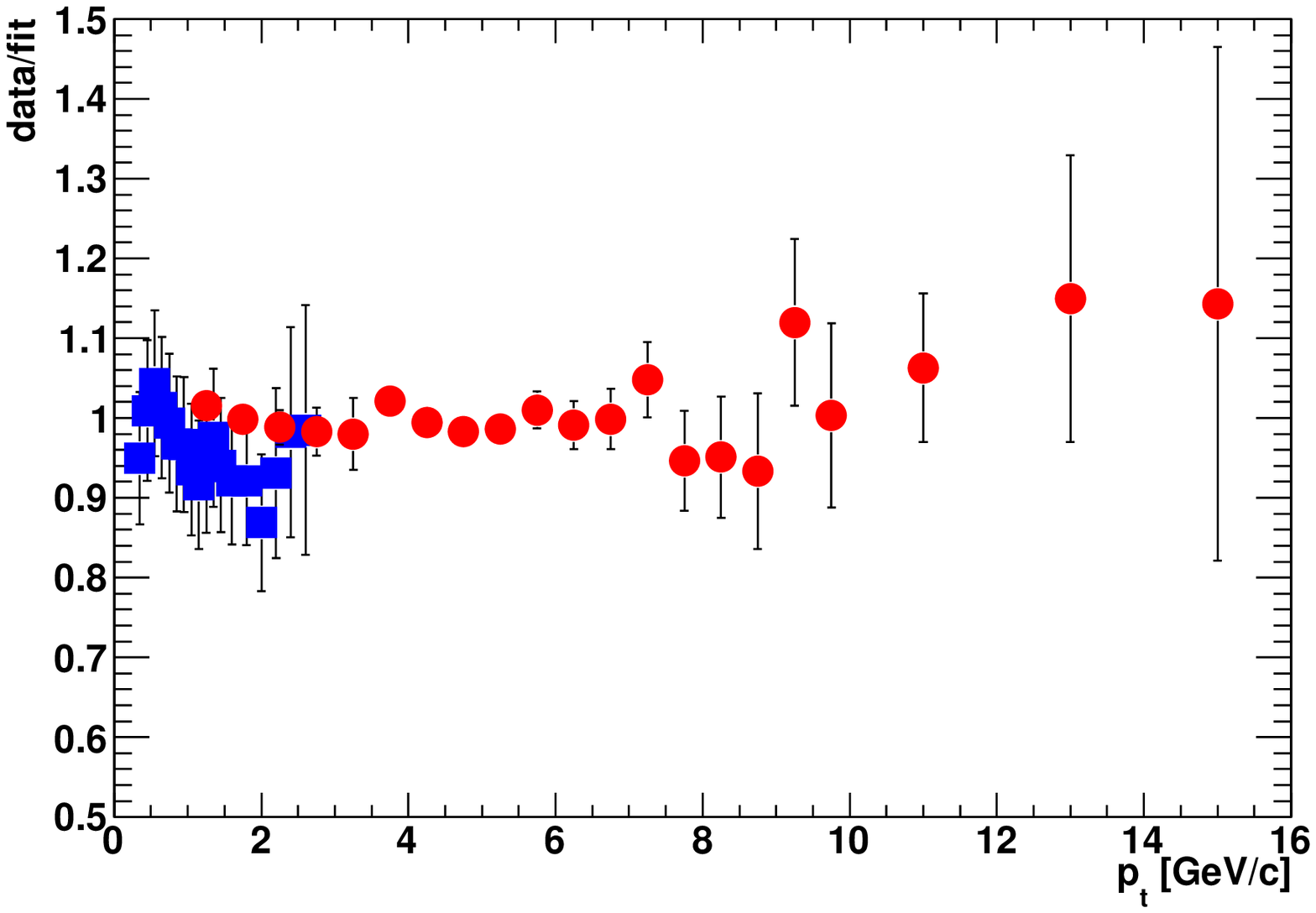}
\caption{(Color online) (upper) Invariant differential cross section of charged
pions (squares at low $p_T$) and neutral pions (circles) for $p+p$ collisions, 
together with a fit according to Eq.~\ref{fitfunc}.  (lower) Ratio of the data
to the fit.
\label{fig:pion_input_pp}}
\end{figure}

\begin{figure}[tbh]
\includegraphics[width=1.0\linewidth]{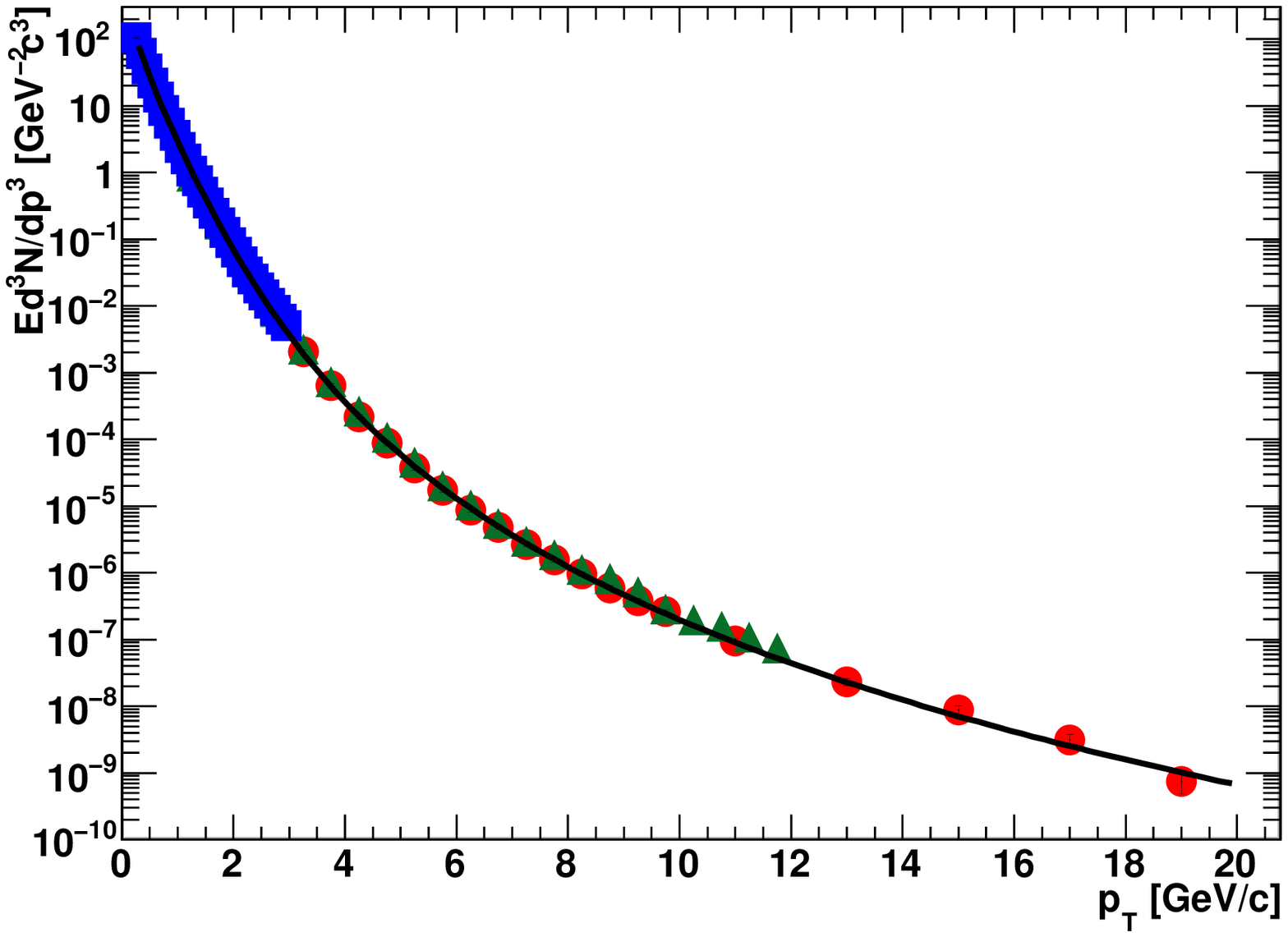}
\includegraphics[width=1.0\linewidth]{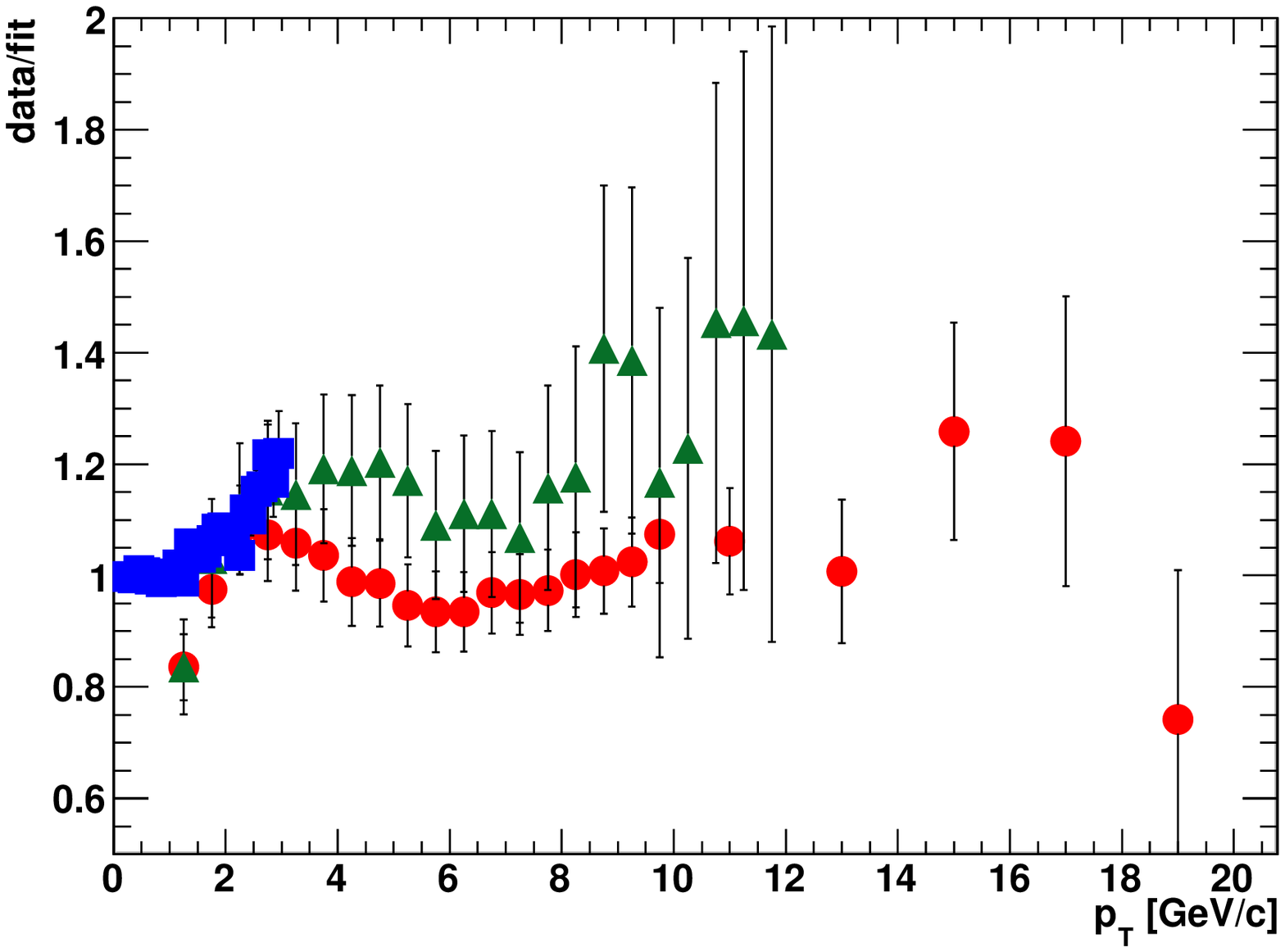}
\caption{(Color online) (upper) Invariant multiplicity of charged pions (squares
at low $p_T$) and neutral pions for Au+Au collisions from 2002 (triangles) and
2004 (circles) together with a fit according to Eq.~\ref{fitfunc}.  (lower) Ratio
of the data to the fit to the 2004 data.}
\label{fig:pion_input_auau}
\end{figure}

The most important background source, except for the highest electron 
$\pt$ where contributions from direct radiation become dominant (see 
below), comes from the neutral pion.  The contribution of $\pi^0$ decays to
the photonic background is twofold.  First, the Dalitz decay of neutral 
pions ($\pi^0 \rightarrow e^+e^-\gamma$) is a primary source of 
electrons from the collision vertex.  Second, the conversion of 
photons from the decay $\pi^0 \rightarrow \gamma\gamma$ in material in 
the PHENIX central arm aperture (mainly in the beam pipe) gives rise to a 
secondary source of electrons originating away from the original 
collision vertex.  It is crucial to note that the contribution from 
photon conversions is smaller than the contribution from Dalitz 
decays.  This is due to the carefully minimized material budget in the 
PHENIX central arms.  Apart from the beam pipe, which is made out of 
Beryllium and contributes less than 0.3~\% of a radiation length to 
the material budget, helium bags constitute the only material between 
the beam pipe and the tracking and electron identification detectors 
in PHENIX.  As was verified in a full GEANT simulation of $\pi^0$ 
decays, the ratio of electrons from the conversion of photons from 
$\pi^0 \rightarrow \gamma\gamma$ decays to electrons from $\pi^0$ 
Dalitz decays is 0.403 with a systematic uncertainty of about 10~\%.  This ratio
is independent of $\pt$ in the range relevant here, {\it i.e.} for $\pt > 
1$~\gevc.  For heavier mesons this ratio is rescaled in the cocktail to 
properly account for the fact that the branching ratio for the Dalitz 
decay relative to the $\gamma\gamma$ decay grows slightly with 
increasing parent meson mass.  Consequently, once the phase space 
distributions of $\pi^0$ and heavier mesons are available it is 
straight forward to determine both the Dalitz decay contribution to 
the background electron spectrum as well as the corresponding 
contribution from the conversion of photons from the same parent 
mesons.

The phase space distributions of $\pi^0$ are obtained via simultaneous 
fits to the charged and neutral pion spectra.  This approach is 
only valid under the assumption that the invariant $\pi^0$ spectra and 
the averaged charged pion spectra $(\pi^+ + \pi^-)/2$ are the same.  
While this assumption in general is well justified, at low $\pt$, {\it 
i.e.} for $p_T < 1$~\gevc, the decay of $\eta$ mesons into three 
$\pi^0$ creates a tiny charge asymmetry.  However, according to a 
PYTHIA \cite{Sjostrand:2001} calculation and consistent with data, this
asymmetry is only about 2\% and can be safely ignored in this context.

Figure~\ref{fig:pion_input_pp} shows the comparison of the neutral and
charge-averaged invariant differential cross sections of pions in
\pp~collisions at $\sqrt{s} = 200$~GeV in comparison with a simultaneous fit to
the data with a modified Hagedorn parametrization:
\begin{equation}
E \frac{d^3\sigma}{d^3p} = 
\frac{c}{(exp(-a p_T - b p_T^2) + p_T / p_0)^n}
\label{fitfunc}
\end{equation}
where $a$, $b$, $c$, $p_0$, and $n$ are fit parameters.  Both an absolute
comparison as well as the ratio of the data to the fit are shown to demonstrate
the excellent quality of the parametrization.

For \auau~collisions the $\pi^0$ invariant differential
multiplicity distributions are obtained by equivalent fits to the measured
$\pi^0$ \cite{ppg014,ppg080} and $\pi^\pm$ \cite{ppg026} spectra independently
for various centrality selections.  Figure~\ref{fig:pion_input_auau} shows the
comparison of data and parametrization on absolute and relative scales for
minimum bias \auau~collisions as an example.

It is obvious that the neutral pion spectrum from 2004 is systematically
different from the 2002 result.  The difference between the parametrization and
both sets of neutral pion spectra gives rise to an additional systematic
uncertainty in the resulting electron cocktail.  The difference between the
parametrizations is assumed to be one standard deviation of error due to this
systematic difference.  This systematic uncertainty reaches its maximum value
for electrons in the $\pt$ range between 3 and 5~\gevc and is significantly less
for lower as well as higher $\pt$.

Given that pion decays are the most important cocktail ingredient at 
low and intermediate $\pt$ it is obvious that the cocktail systematic 
uncertainty is largely dominated by the uncertainty in the pion spectra as
well.  To evaluate this uncertainty the full cocktail calculation is
repeated with the pion cross section moved up or down by one standard
deviation in the systematic uncertainty, propagating the uncertainty in the pion
spectra to the electron cocktail.  With a systematic uncertainty of ~10~\% almost
independent of $\pt$, some of which originates from the difference between the
Run 2 and Run 4 $\pi^0$ measurements, the pion input represents the largest
contributor to the electron cocktail uncertainty, except for at the highest
$\pt$ where direct radiation becomes important.

Other light mesons contributing to the electron cocktail are the 
$\eta$, $\rho$, $\omega$, $\eta'$, and $\phi$ mesons via their Dalitz 
decays and/or the conversion of photons from their decays.  However, 
only the $\eta$ meson is of any practical importance here.

For the cocktail calculation, the shape of the invariant $\pt$ distributions 
and the relative normalizations to the $\pi^0$ are required as input parameters.
The $\pt$ spectra are derived from the pion spectrum by $m_T$ scaling, 
{\it i.e.} the same modified Hagedorn parametrizations are used 
(Eq.~\ref{fitfunc}), but with $\pt$ replaced by $\sqrt{\pt^2 + m_{meson}^2 
- m_{\pi^0}^2}$.  
The resulting $\eta / \pi^0$ ratios agree as a function of $\pt$ well within 
experimental uncertainties for $\pt > 2$~\gevc with corresponding PHENIX data 
for \pp~and \auau~collisions.

Since the chosen approach of $m_T$ scaling ensures that at high $\pt$ the 
spectral shapes of all meson distributions are the same, the normalization of 
the meson spectra relative to the pion spectrum can be given by the ratios of 
mesons to pions at high $\pt$ (5 \gevc is used here).  
The values used for \pp~collisions are shown in Table~\ref{tab:m_to_p0_pp}.

\begin{table}
\caption{Ratios of mesons to neutral pions in $p+p$ collisions}
 \begin{ruledtabular} \begin{tabular}{c}
meson to pion ratios\\
\hline
$\eta / \pi^0 = 0.48 \pm 0.03$  \cite{ppg051}\\
$\rho / \pi^0 = 1.00 \pm 0.30$  \cite{an089}\\
$\omega / \pi^0 = 0.90 \pm 0.06$  \cite{Ryabov:1998rt}\\
$\eta'  / \pi^0 = 0.25 \pm 0.075$ \cite{an089}\\
$\phi / \pi^0 = 0.40 \pm 0.12$  \cite{an089}\\
\end{tabular}  \end{ruledtabular}
\label{tab:m_to_p0_pp}
\end{table}

For \auau~collisions the same central values are used, but the 
uncertainties of the ratio $\eta / \pi^0$ and $\omega / \pi^0$ are, 
conservatively, increased to 0.10 and 0.27, respectively, since 
precision measurements are available for the production of these 
mesons in \pp~collisions but not in \auau~collisions at all 
centralities.

The contribution from the $K_{e3}$ decay and the semi-leptonic decay of $K^0_S$
can only be determined via a full GEANT simulation, taking into account the 
exact electron identification cuts.  The electron cocktail includes
parametrizations based on such GEANT simulations using the charged kaon spectra
measured in \pp~and \auau~ collisions as input \cite{ppg026}.  Systematic
uncertainties are fully propagated from the kaon spectra to the spectra of
reconstructed electrons from $K_{e3}$ decays.  In any case, the contribution
from kaon decays is only relevant ({\it i.e.} larger than 10~\%) for electrons
with $\pt < 1$~\gevc.  

Contributions to the electron cocktail from direct radiation are 
twofold.  First, real photons produced in initial hard scattering 
processes, {\it i.e.} so-called direct photons, convert in material in 
the PHENIX aperture exactly as photons from light neutral meson 
decays.  Second, every source of real photons also presents a source of 
virtual photons.  
Consequently, direct real photon production is accompanied by direct 
virtual photon production, {\it i.e.} the emission of $e^+e^-$ pairs.  In the
case of the neutral pion these two sources are the the $\gamma\gamma$ decay of
the $\pi^0$ and the corresponding Dalitz decays, which are also called internal
conversions.  The measured real direct photon spectra are parametrized and the 
conversion electron spectra of these are added to the electron cocktail.  

\begin{figure}
\includegraphics[width=1.0\linewidth]{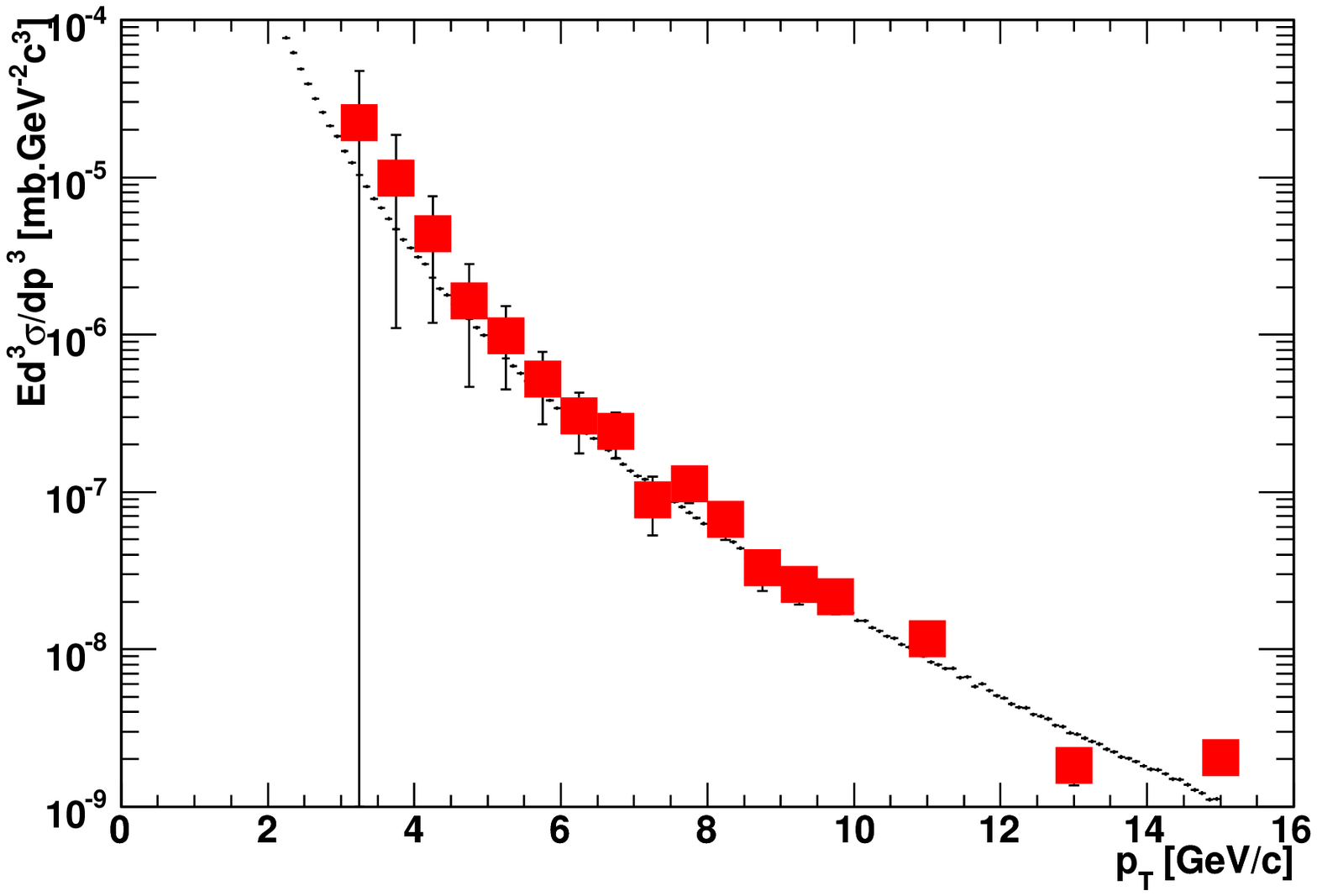}
\includegraphics[width=1.0\linewidth]{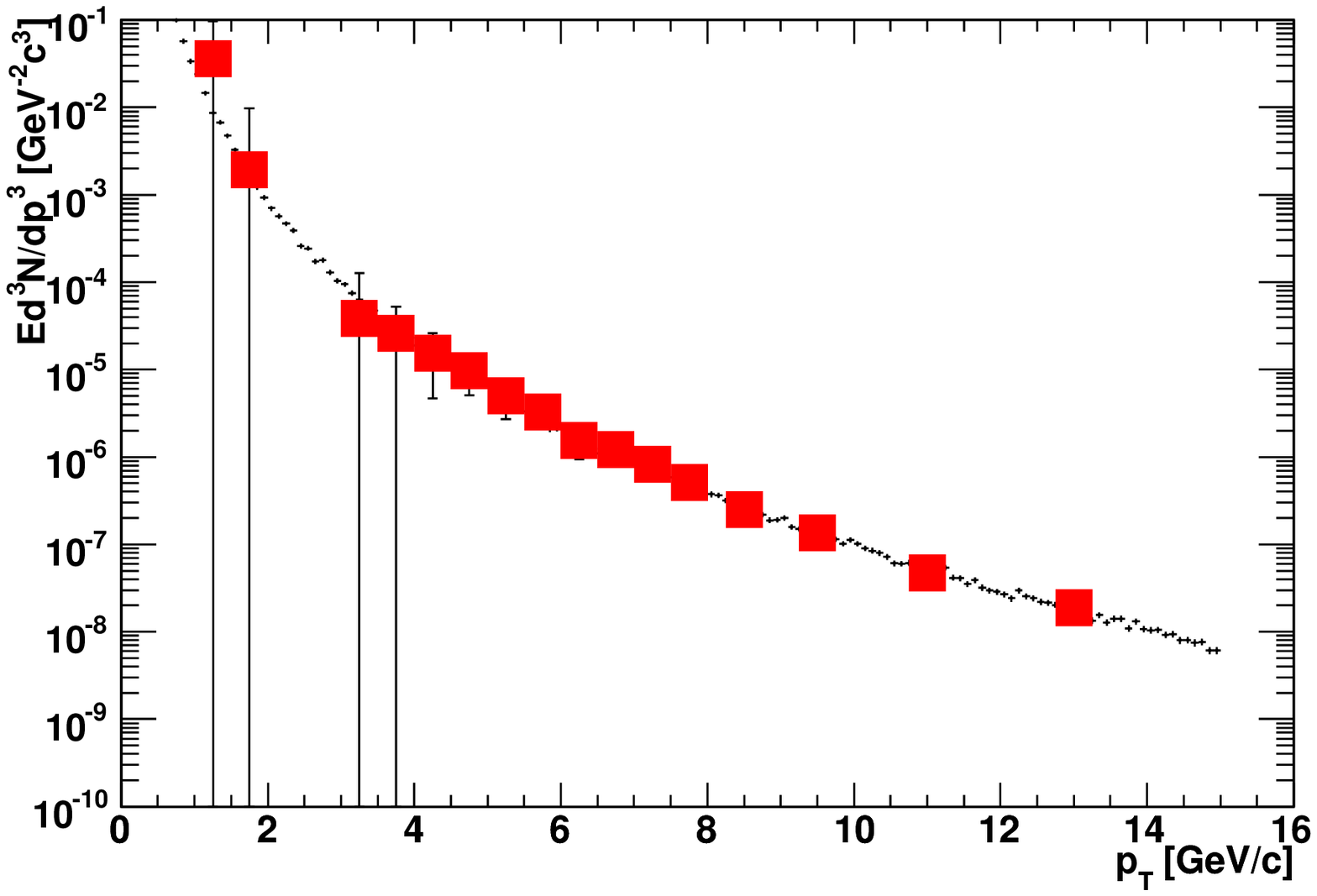}
\caption{(Color online) Measured direct photon spectrum (red squares)
compared with the cocktail parametrization 
(histogram) for (upper) \pp and (lower) Au+Au collisions.
\label{fig:direct_photon}
}
\end{figure}

Figure~\ref{fig:direct_photon} shows comparisons of the measured 
direct photon spectrum with the cocktail parametrization
for (upper)~\pp \cite{Adler:2005qk} and (lower) minimum bias \auau~collisions
\cite{ppg042}.  In accordance with direct photon measurements in
\auau~collisions, the direct photon yield is assumed to scale with the number
of binary collisions as a function of the centrality for $p_T$ below 12 GeV/$c$
in the \auau~electron cocktail.  Note that although the parametrization used
underestimates the central value of the direct photon measurement below 5
GeV/$c$, the contribution from direct photons in this $p_T$ range is small
enough that effects of this slight underestimation can be neglected.

The ratio of virtual direct photons to real direct photons depends on 
$\pt$ because the phase space for dielectron emission increases with 
increasing $\pt$ \cite{ppg088}.  The very same effect is seen in the Dalitz
decays of light neutral mesons.  For instance, the Dalitz decay branching ratio
relative to the two photon decay branching ratio is larger for the $\eta$ meson
than for the $\pi^0$.  Consequently, the ratio of virtual and real direct photon
emission increases with $\pt$.

\begin{figure*}
\includegraphics[width=0.48\linewidth]{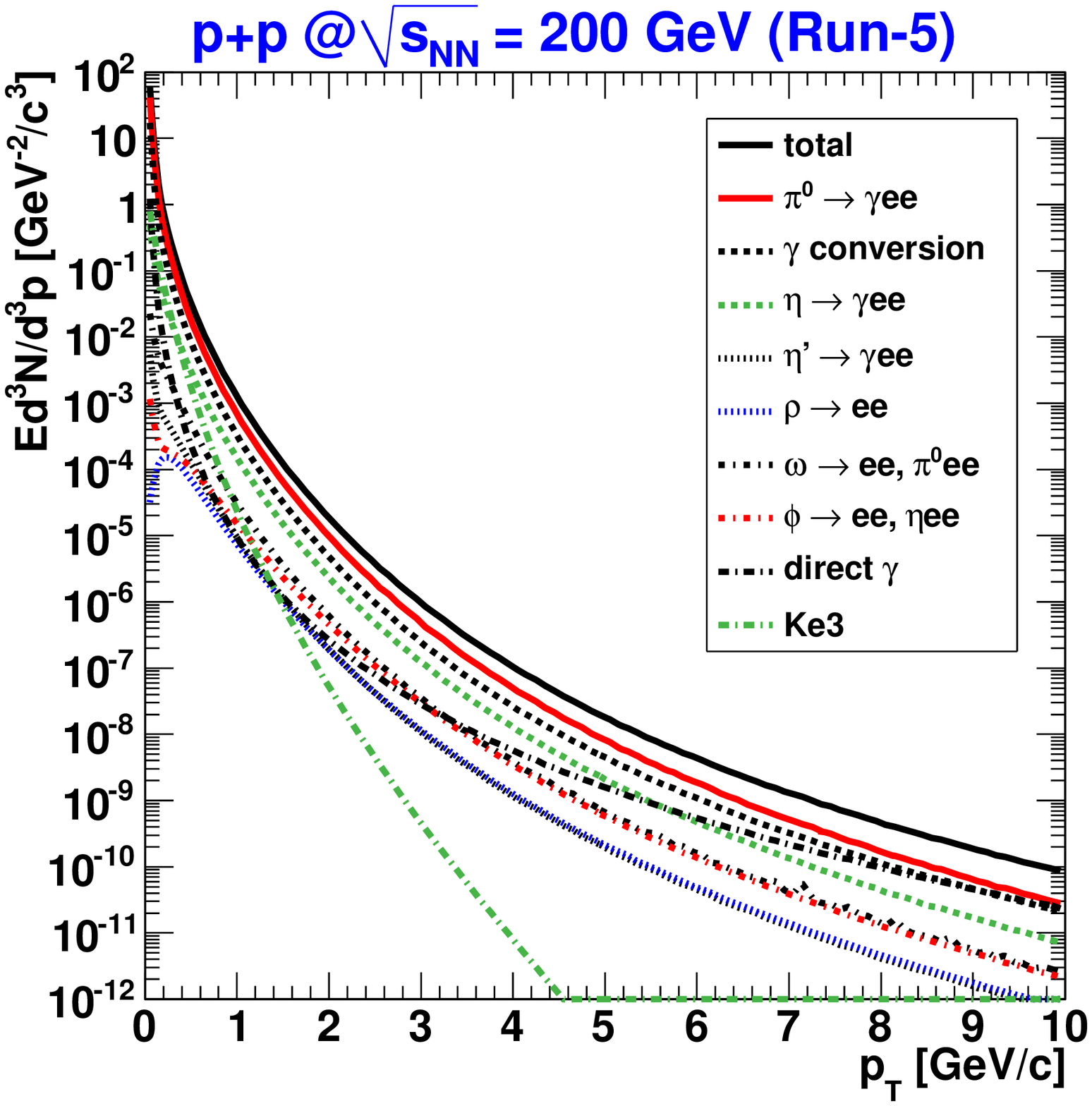}
\includegraphics[width=0.48\linewidth]{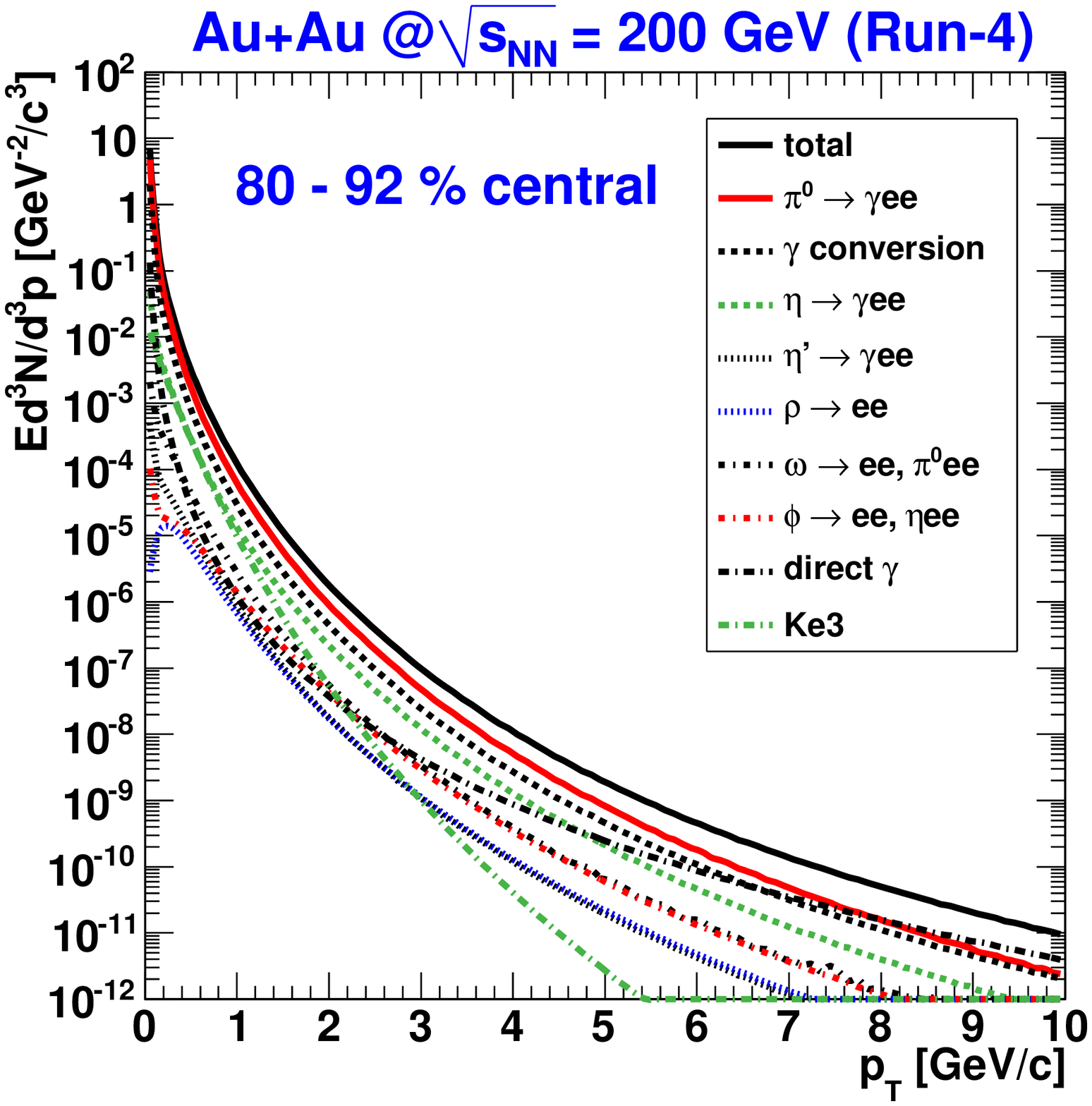}
\includegraphics[width=0.48\linewidth]{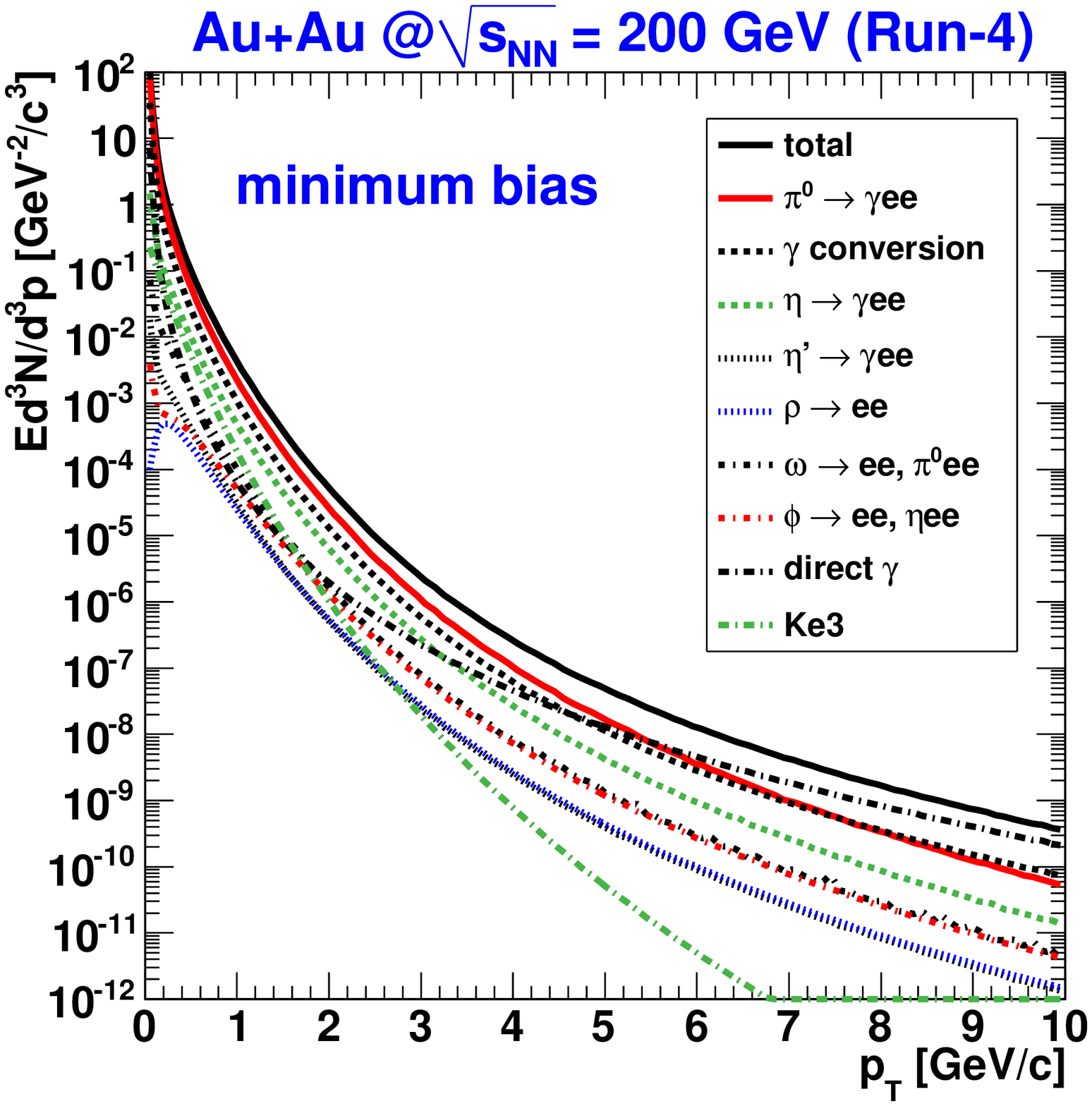}
\includegraphics[width=0.48\linewidth]{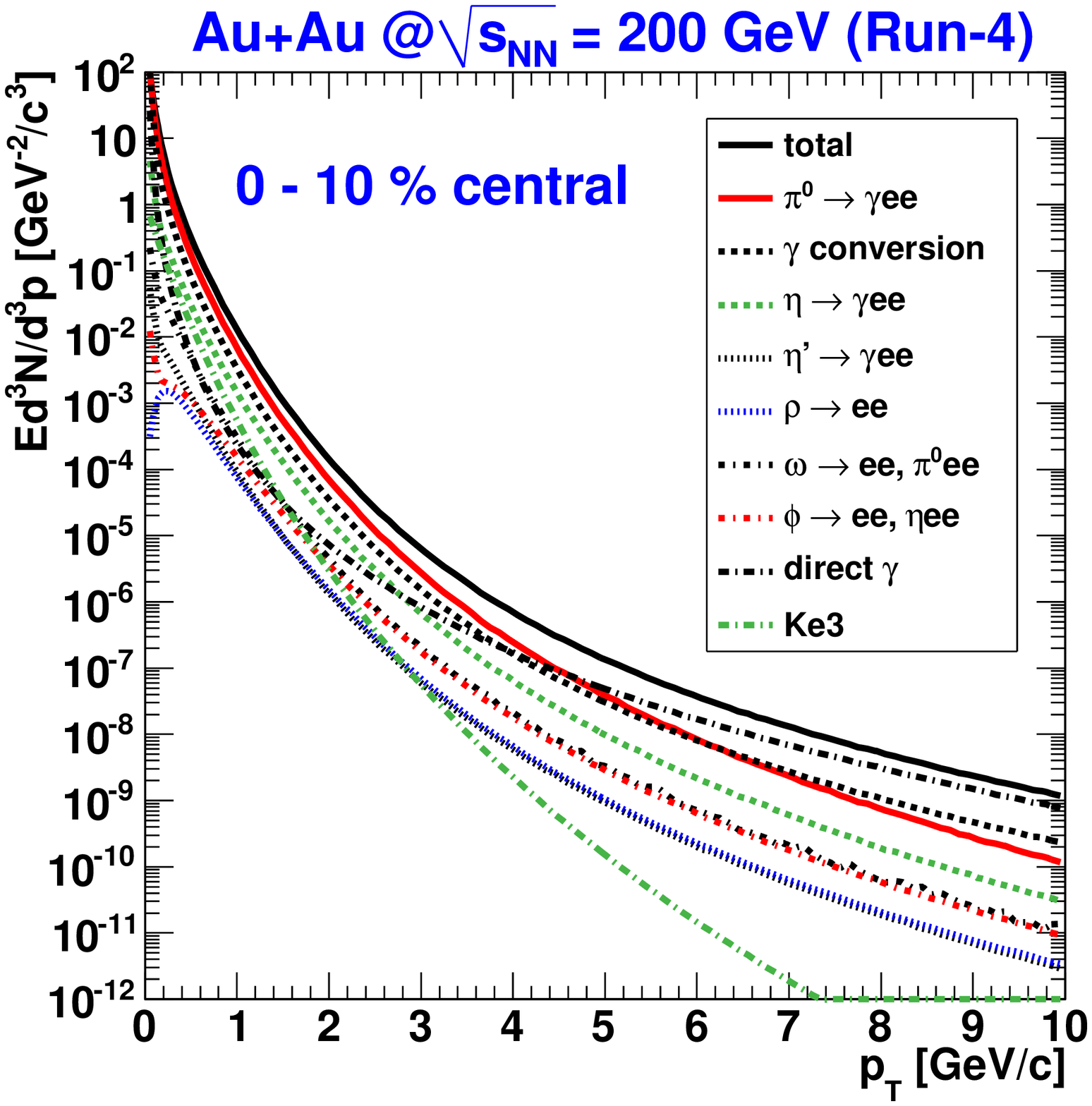} 
\caption{(Color online) Invariant differential cross sections and 
multiplicity of electrons for electron cocktails in \pp and 
\auau~collisions at 200~GeV for the indicated centrality ranges.  
\label{fig:cocktails}
}
\end{figure*}

Figure~\ref{fig:cocktails} shows the 
resulting electron cocktails for \pp and \auau~collisions at 200 GeV.
Systematic errors are estimated for all cocktail ingredients, 
propagated to the corresponding electron spectra, and then added in 
quadrature to determine the total cocktail systematic error.
The following systematic errors are assigned to the various inputs:

\begin{itemize}
\item pion spectra: obtained via full cocktail calculations using pion 
  spectra moved up (down) by the systematic uncertainty of the pion 
  spectra as input (almost no $\pt$ dependence).  This is the dominant
  systematic uncertainty, except for the case of high $\pt$ electrons
  (above 5~\gevc) in central \auau~collisions, where direct radiation
  dominates the electron background and, at the same time, dominates
  the systematic uncertainty in the electron cocktail.
\item meson to pion ratios: the systematic uncertainties are listed in Table
\ref{tab:m_to_p0_pp}.
  Since the contributions from all other mesons are much smaller than the 
  contribution from $\eta$ decay only the $\eta$ is of any practical
  relevance.  This contribution is small compared to the uncertainty in
  the pion spectra and it depends only slightly on $\pt$.
\item conversion material in the aperture: the contribution from photon
  conversions obviously depends on the material present in the aperture.
  A careful analysis of fully reconstructed dielectrons from photon
  conversions suggests that this uncertainty is not larger than 10~\%.
  It is crucial to note that the contribution from photon conversion
  to the background electron spectra is less than half of the contribution
  from direct Dalitz decays as a consequence of the careful minimization
  of the material budget in the PHENIX central arm aperture.  For a 
  reliable measurement of nonphotonic electrons this is an essential
  prerequisite.  
\item $K_{e3}$ decay: this contribution is estimated via a full GEANT 
  simulation.  Given the limited statistics of this calculation a 50~\% 
  systematic error is assigned.
\item direct radiation: this contribution is directly propagated from the 
  systematic error quoted for the direct photon measurement.  It is
  relevant only at high $\pt$ in central \auau~collisions.
\end{itemize}

In previous analyses, the single electron background cocktail has not included 
contributions originating from quarkonia decay ($J/\psi$ and 
$\Upsilon$) or Drell-Yan processes.  Each of these processes has a 
small total cross section relative to the electron cocktail 
background.  However, with increasing $p_T$ these processes begin to 
contribute meaningfully and are included in the cocktail.  

PHENIX has measured the $J/\psi$ $p_T$ spectrum from 0 to 9 GeV/$c$ in $p+p$
collisions \cite{ppg097}.  To determine the appropriate input to the Monte
Carlo event generator used to calculate the components of the electron 
cocktail, two functional fits to the $J/\psi$ data are performed.  The 
first uses the so-called Kaplan function $p_0 \left( 1 + \left( 
p_T/p_1 \right)^2 \right)^{-n}$, and the second assumes the $m_T$ 
scaling function used in \cite{ppg085}.  Since both functions may 
provide valid representations of the true $J/\psi$ spectral shape, the average
of these two functional fits to the data determines the central values used in
the cocktail.  The upper (lower ) systematic uncertainties of the spectral shape
are determined by normalizing each of the functional forms up (down) by 10\% and
using the largest (smallest) result for the systematic error.  In
practice, the lower bound of the spectral shape is determined by the smaller
Kaplan function normalized down a further 10\%, while the upper bound is set by
the larger $m_T$ scaling function normalized up 10\%.

While measured $J/\psi$ $p_T$ spectra exist at RHIC energies, the 
$\Upsilon$ $p_T$ spectra have not been measured.  Since the overall 
production cross section for $\Upsilon$ is estimated to be roughly 1\% of that
of the $J/\psi$, contributions from $\Upsilon$ decay contribute much less to 
the single electron cocktail.  A NLO calculation for $\Upsilon$ 
production in the color evaporation model provides the input to the 
Monte Carlo calculation of the central value 
estimates~\cite{RHICII_heavy}.  Without a measured $\Upsilon$ $p_T$ 
spectrum to characterize the systematic uncertainty to the Monte Carlo 
input, the same relative systematic uncertainties derived from the 
$J/\psi$ data are applied to the $\Upsilon$ cocktail estimate.  Compared to the
quoted uncertainties in the NLO calculation, the use of data-driven systematic
uncertainties for the $\Upsilon$ provides a more conservative estimate.

A leading-order (LO) Drell-Yan calculation \cite{Vogelsang_private} of 
the single-inclusive lepton $p_T$ spectrum, \pp~$\rightarrow (e^{+} 
+e^{-})/2 + X$, is also included in the updated electron cocktail.  The 
calculation is for $|y|<0.5$ and uses the CTEQ6M parton distributions
\cite{cteq6m}.  The scale used is $p_T$, no cut is placed on the lepton pair
mass, and a K-factor of 1.5 is applied.

\begin{figure}
\includegraphics[width=1.0\linewidth]{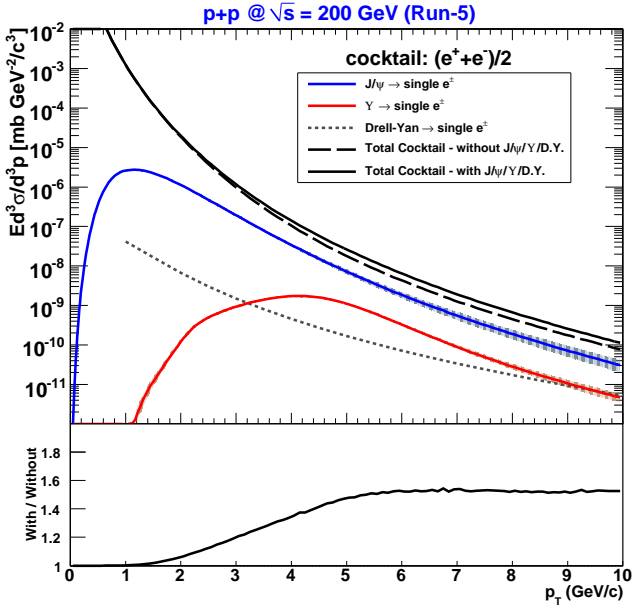}
\caption{(Color online) Electron background cocktail with quarkonium and
Drell-Yan contributions for $p+p$ collisions.\label{fig:cocktail_additions}}
\end{figure}

Figure~\ref{fig:cocktail_additions} shows the cocktail for \pp collisions with
the quarkonium and Drell-Yan contributions.  The bottom section of the plot
shows the ratio of the cocktail with the quarkonium and Drell-Yan to that
without.  

\begin{figure}
\includegraphics[width=1.0\linewidth]{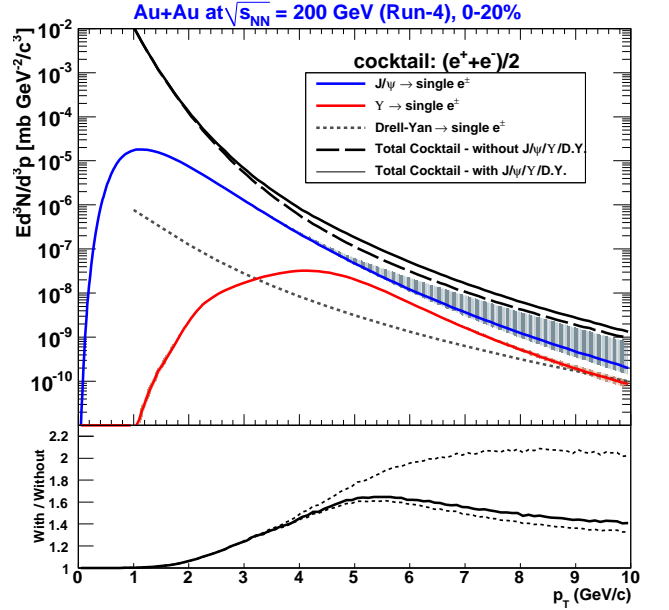}
\caption{(Color online) Electron background cocktail with quarkonium and
Drell-Yan contributions for 0-20\% centrality Au+Au collisions.
\label{fig:cocktail_additions_auau}}
\end{figure}

PHENIX has measured the $J/\Psi$ $p_T$ spectrum out to 5 GeV/$c$ in Au+Au
collisions at 200 GeV \cite{ppg068}.  In order to extrapolate out to higher
$p_T$, the measured $J/\Psi$ spectrum from $p+p$ collisions is scaled by
$R_{\rm AA}$ and $N_{{\rm coll}}$.  Above 5 GeV/$c$, two extreme scenarios are
considered, and the difference in the two resulting estimates for the
extrapolation is assumed as a systematic error.  In the first scenario, the
$R_{\rm AA}$ is kept constant from its value at 5 GeV/$c$.  In the second scenario,
the $R_{\rm AA}$ is assumed to increase linearly from its value at 5 GeV/$c$ up to
a value of 1 at 10 GeV/$c$, above which it is assumed to be constant.  Figure
\ref{fig:cocktail_additions_auau} shows the cocktail with quarkonium and
Drell-Yan contributions for 0-20\% centrality Au+Au collisions.

The $J/\Psi$ spectrum in Au+Au collisions has not been measured separately for
0--10\% and 10--20\% centralities.  In order to estimate the electron background
in these centralities separately, we assume that the $R_{\rm AA}$ of $J/\Psi$ is
the same in 0--10\% and 10--20\% centrality collisions.

\subsection{Converter subtraction method}\label{sub:converter}  

\begin{figure}[htbp]
  \includegraphics[width=1.0\linewidth]{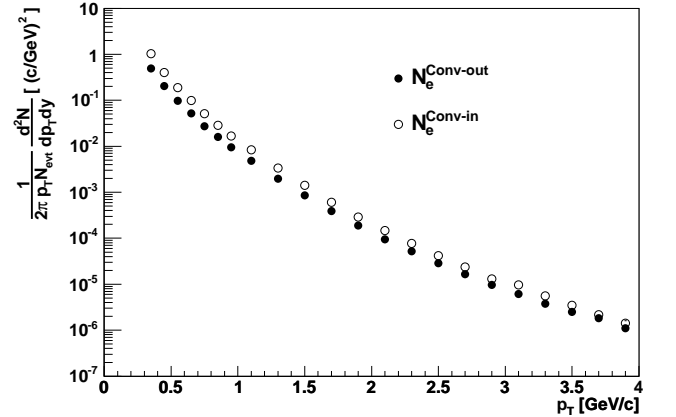}
\caption{Invariant yields of inclusive electrons with (open circles) 
and without (filled circles) the photon converter for MB events.
\label{fig:inc_wwo_conv_mb}}
\end{figure}

The yields of photonic and nonphotonic electrons are obtained by 
measuring the difference between inclusive electron yields with and without 
a photon converter of precisely known thickness: a brass sheet of 
$1.680~\%$ radiation length ($X_0$).  Figure~\ref{fig:inc_wwo_conv_mb} 
shows the corresponding $\pte$ spectra for MB events.

These yields can be expressed as the following relations:
\begin{eqnarray}
  N^{\rm Conv\--out}_e &=& N^{\gamma}_e + N^{\rm Non\--\gamma}_e, \\
  N^{\rm Conv\--in}_e &=& R_{\gamma} N^{\gamma}_e + (1-\epsilon)
N^{\rm Non\--\gamma}_e.  
\end{eqnarray}

Here, $N^{\rm Conv\--in}_e$ ($N^{\rm Conv\--out}_e$) is the 
measured electron yield with (without) the converter.  $N^{\gamma}_e$ 
($N^{\rm Non\--\gamma}_e$) is the photonic (nonphotonic) electron 
yield.  $\epsilon$ represents a small loss of 
$N^{\rm Non\--\gamma}_e$ due to the converter.  This blocking factor 
has been already evaluated in a previous measurement ($\epsilon \sim 
2.1\%$) \cite{Adler:2004ta}.  The main issue in this calculation is to 
determine $\rg$, i.e, to understand how much the photonic electron 
yield is increased by the converter.  The main source of photonic 
electrons is a mixture of mesons 
($\pi^0,~\eta,~\eta^{\prime},~\omega$, and $\phi$) decaying into real 
or virtual photons with their different $\pt$ slopes.

\begin{table}[htbp]
  \caption{Radiation length ($L$) of material near the interaction point.
The conversion probability ($P^{\rm Conv}$) is calculated for the case of
electrons emitted at $\pte\ =1.0$ GeV/$c$.~\label{tab:conv_prob01}}
\begin{ruledtabular} \begin{tabular}{ccc}
   Material & $L$ ($X_0$) & $P^{\rm Conv}$ \\ 
 \hline
  Beam pipe ($Be$) & 0.288\%  	  & 0.201\% \\ 
  Air ($r<30~cm$)  & 0.099\%  & 0.069\% \\ 
  Total	 & 0.387\%  & 0.270\% \\ 
  Converter (brass) & 1.680\%  	  & 1.226\% \\ 
\end{tabular} \end{ruledtabular}
\end{table}

To calculate $\rg$, it is necessary to know exactly the material 
amounts near the interaction point.  Table~\ref{tab:conv_prob01} shows 
a list of each material thickness in units of radiation length.  To 
reduce the air contribution, a helium bag was installed in the space 
surrounded by the DC ($0.3<r<1.0$ m).  The converter sheet was rolled 
just around the beam pipe in converter runs.  The conversion 
probability ($P^{\rm Conv}$) in Tab.~\ref{tab:conv_prob01} is 
calculated for the case of electrons emitted at $\pte =1.0$ GeV/$c$.  
The equivalent conversion probability of a virtual photon in $\pi^0$ 
Dalitz decay ($P^{\rm Dalitz}$) is 0.598\%~\cite{Tsai}.  $\rg$ can 
be estimated with these values at $\pte = 1.0$ GeV/$c$.

\begin{eqnarray}
  \rg = \frac{P^{\rm Conv} + P^{\rm Dalitz}~({\rm with~converter})}
 {P^{\rm Conv} + P^{\rm Dalitz}~({\rm without~converter})} 
	 \sim 2.41.
\end{eqnarray}

To obtain a more realistic value of $\rg$ considering geometrical 
effects, GEANT-based Monte-Carlo simulations \cite{GEANT01:W5013} for 
photon conversions were performed with and without the converter.  
$\rg$ is determined for $\pi^0$ and $\eta$ separately.  We use the 
$\pi^0$ spectrum measured by PHENIX as the input for the $\pi^0$ 
simulation and assume $\mt$ scaling ($\pt \rightarrow \sqrt{\pt^2 + 
m_{\eta}^2 - m_{\pi^0}^2}$, normalized at high $\pt$ to $\eta/\pi^0 = 
0.48 \pm 0.1$) to obtain the input for the $\eta$ simulation.  Since the 
$\eta$ mass is larger than the $\pi^{0}$ mass, the phase space of $\eta$ 
Dalitz decay is slightly larger than that of $\pi^{0}$ Dalitz decay.  
The relative branching ratio (Dalitz decay)/(two $\gamma$ decay)  is 
1.2\% for $\pi^{0}$ and 1.5\% for $\eta$ \cite{PDG}.  This difference 
makes $R^{\eta}_{\gamma}$ smaller than $R^{\pi^{0}}_{\gamma}$.  
Contributions from other mesons which undergo Dalitz decay 
($\eta^{\prime}, \rho,~\omega,$~and~$\phi$) are small (6\% at $\pt=3$ 
GeV/$c$, and smaller at lower $\pt$).  The particle ratios at high $\pt$ 
($\eta^{\prime}/\pi^0=0.25 \pm 0.13,~\rho/\pi^0=\omega/\pi^0 = 1.0 \pm 
0.5,~\phi/\pi^0=0.4 \pm 0.2$) are used to determine $\rg$.  The 
$\phi/\pi^0$ ratio used here is consistent with our $\pi^0$ and $\phi$ 
measurement.  The uncertainties in the particle ratios are included in 
the systematic uncertainties of $\rg$.  For this method, it is essential 
that the amount of material is accurately modeled in the simulation.  
We compared the yield of identified photon conversion pairs in the data 
and in the simulation, and concluded that the simulation reproduces 
$\rg$ within $\pm 2.7\%$.  This uncertainty is included in the overall 
systematic uncertainty.

\begin{figure}[htbp]
  \includegraphics[width=1.0\linewidth]{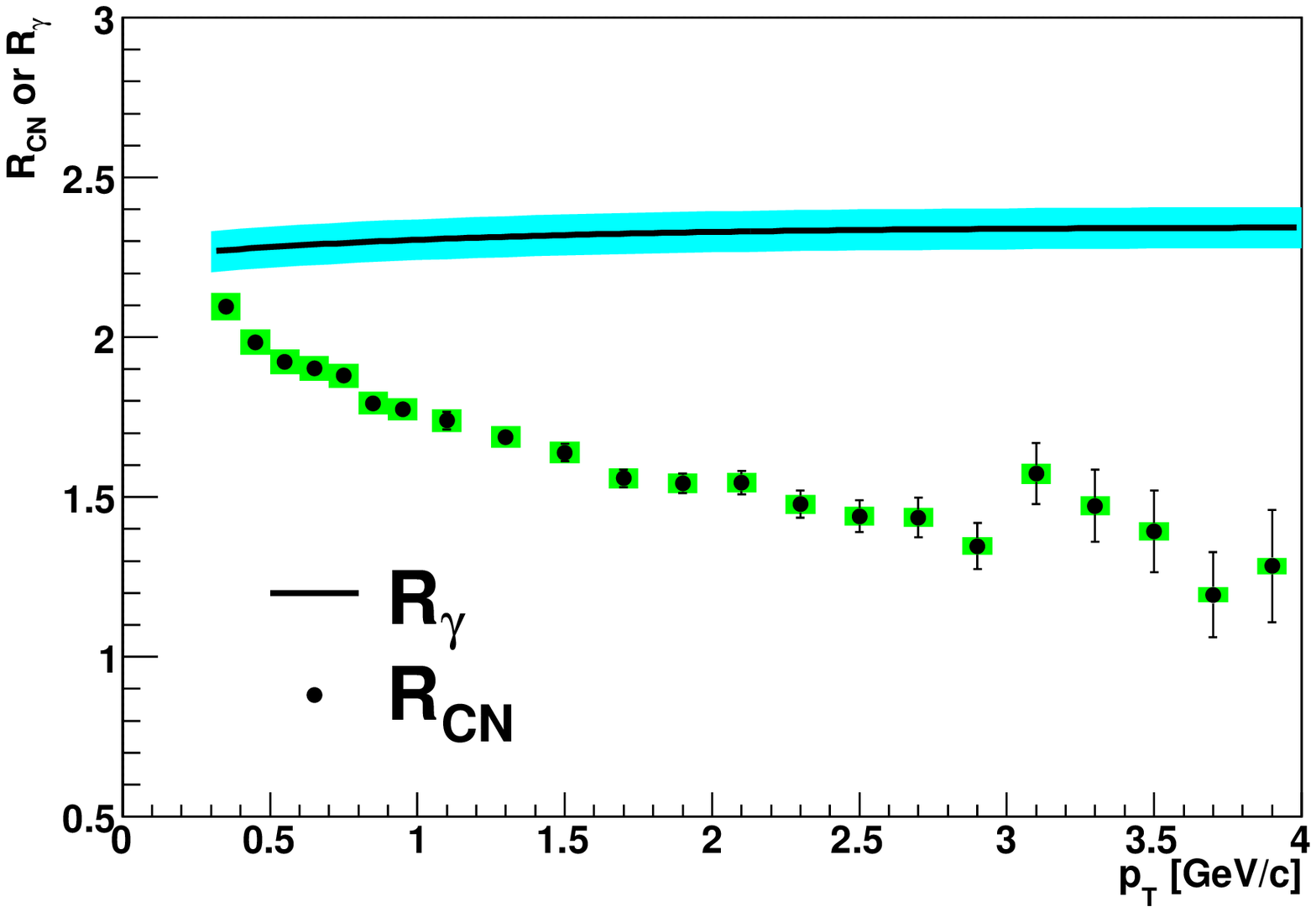}
  \includegraphics[width=1.0\linewidth]{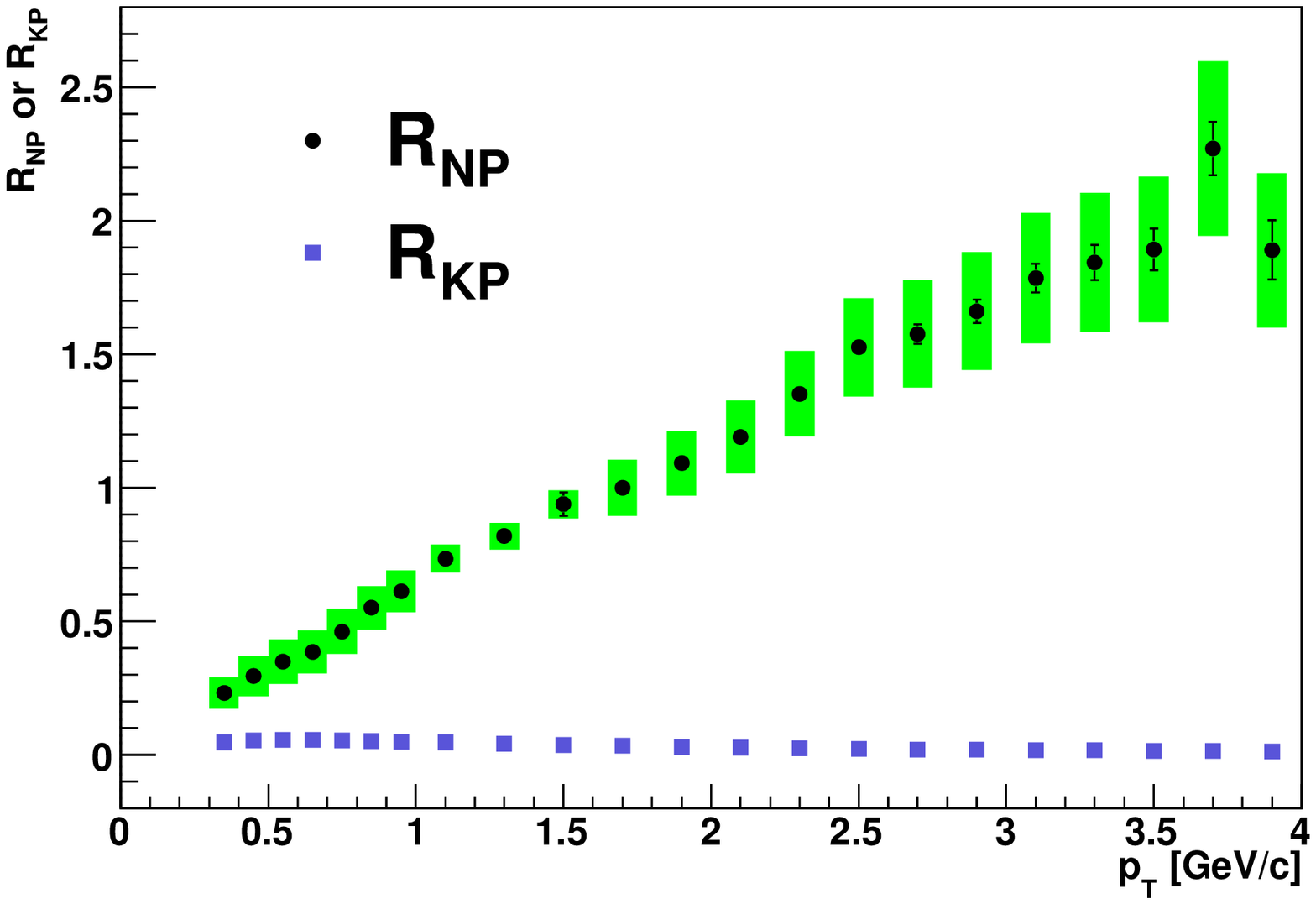}
\caption{(Color online) (upper) Ratio of inclusive electrons 
with/without the converter ($\rcn$, points with systematic error boxes) 
and ratio of photonic electrons with/without the converter ($\rg$, 
solid line with a systematic error band).  
(lower) The ratio of nonphotonic yield to photonic electron yield 
($\rnp$, filled circles) and the ratio of electron yield from kaon 
decays to photonic electron yield ($\rkp$, filled squares) as a 
function of $\pte$ for MB.  Filled circles with statistical error bars 
indicate $\rnp$ produced by both the converter and cocktail methods.
\label{fig:rcn_rg_rnp_mb}
}
\end{figure}

In the upper plot in Fig.~\ref{fig:rcn_rg_rnp_mb}, $\rg$ is indicated 
as a solid curve, which is compared with the ratio of the inclusive 
electron yield with/without the photon converter ($\rcn$):
\begin{eqnarray}
  \rcn \equiv \frac{N^{\rm Conv\--in}_e}{ N^{\rm Conv\--out}_e} 
 =  \frac{\rg + (1-\epsilon)\rnp}{1+\rnp}.  
\end{eqnarray}
Here, $\rnp$ is the ratio of nonphotonic/photonic electron yields 
($N^{\rm Non\--\gamma}_e/N^{\gamma}_e$).  If there were no 
nonphotonic contributions ($\rnp=0$), then we would have $\rcn=\rg$.  
The lower plot in Fig.~\ref{fig:rcn_rg_rnp_mb} shows that $\rcn$ 
gradually decreases with increasing $\pte$, while $\rg$ slightly 
increases with $\pte$.  The difference between $\rcn$ and $\rg$ proves 
the existence of nonphotonic electrons.


The converter method is applied for $0.3\le \pte < 1.6$ GeV/$c$, and the 
cocktail method is applied for $\pte \ge 1.6$ GeV/$c$.  Boxes in the 
lower plot in Fig.~\ref{fig:rcn_rg_rnp_mb} are systematic errors of 
$\rnp$ from each method.  $\rnp$ increases with $\pte$ and is more than 
$1.0$ in $\pte \gtrsim 1.6$ GeV/$c$.  This plot gives an important 
demonstration that the amount of conversion material is relatively small 
in PHENIX.

\begin{figure*}[htbp]
  \includegraphics[width=0.48\linewidth]{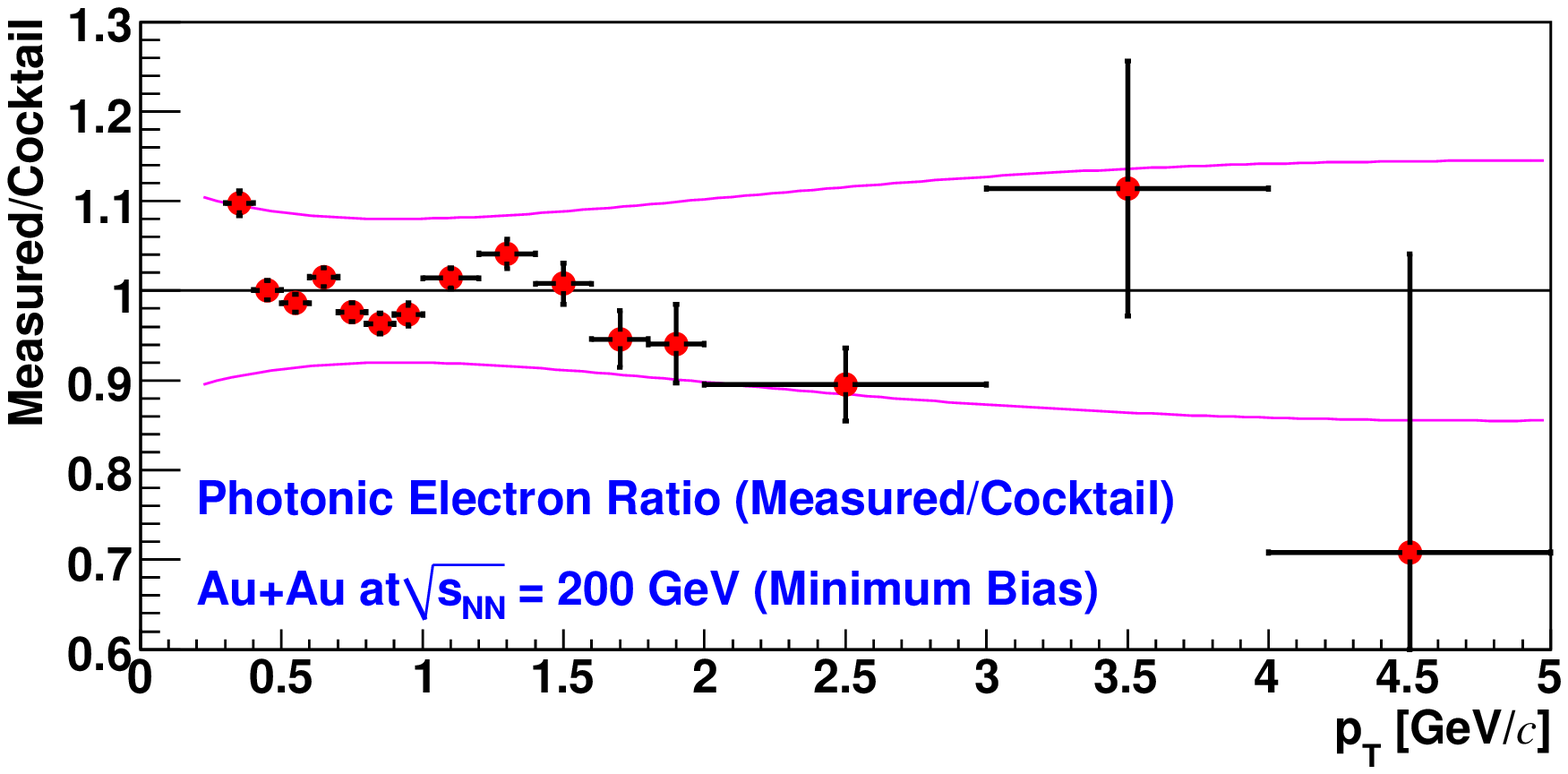}
  \includegraphics[width=0.48\linewidth]{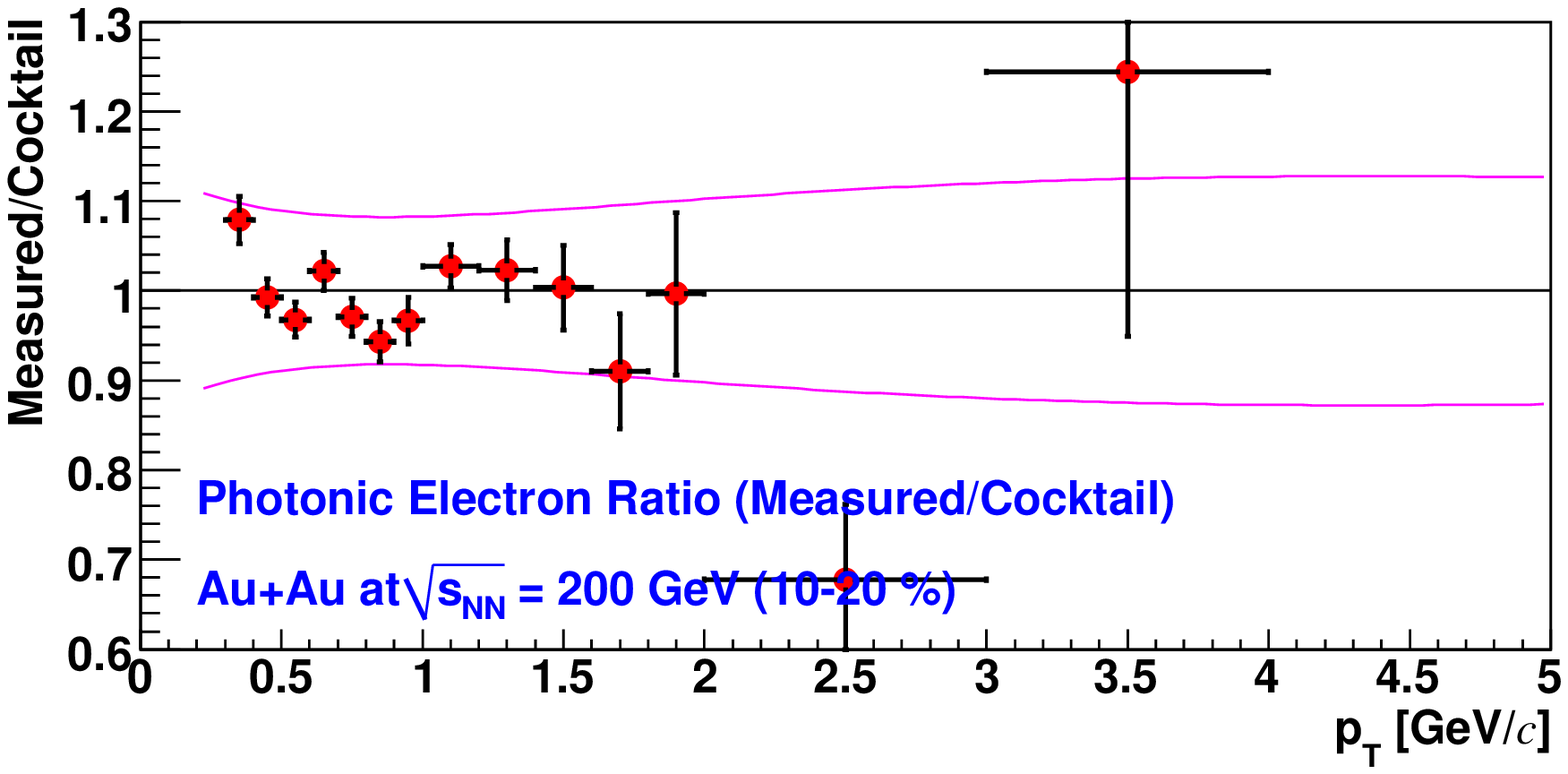}
  \includegraphics[width=0.48\linewidth]{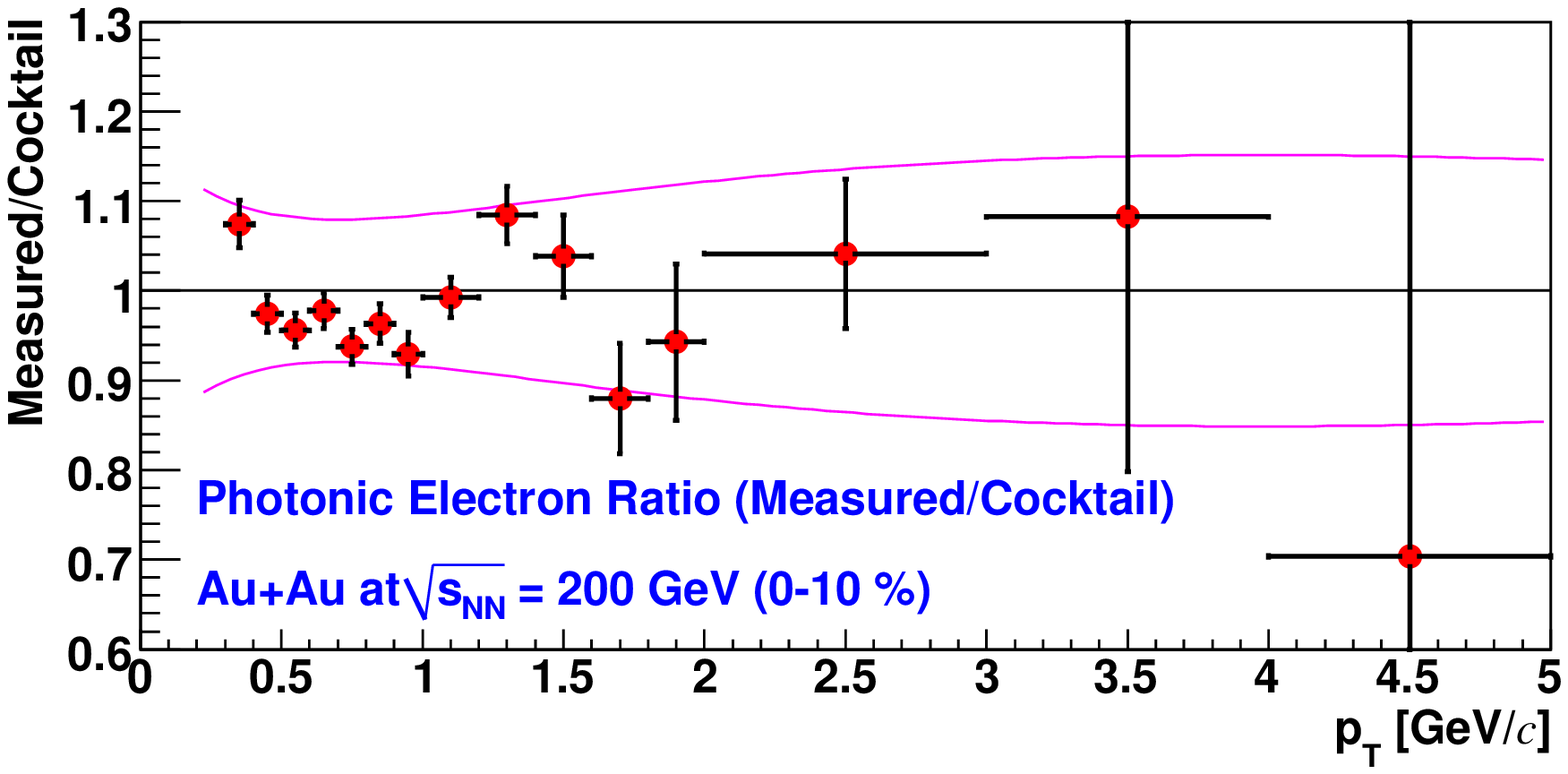}
  \includegraphics[width=0.48\linewidth]{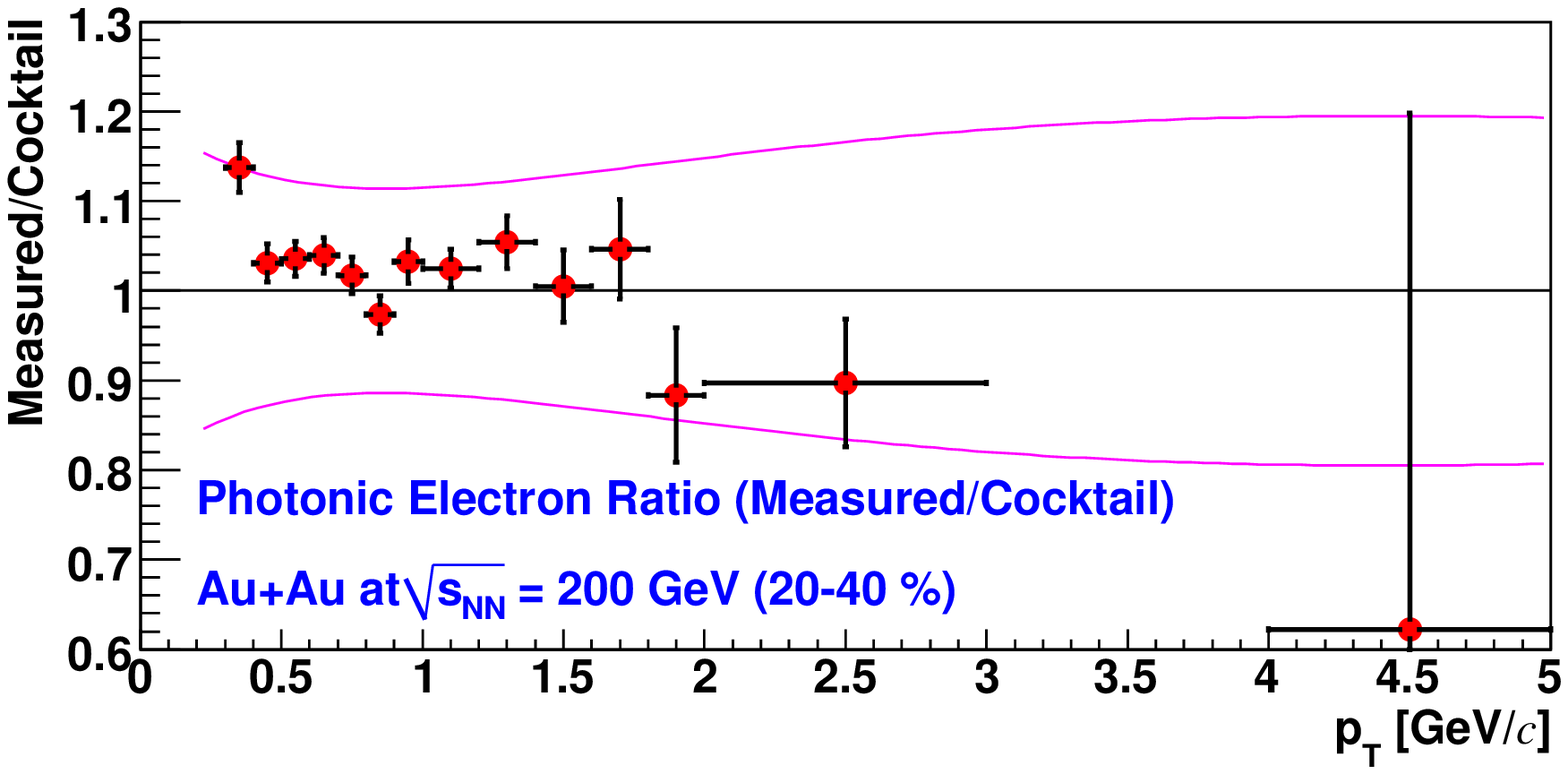}
  \includegraphics[width=0.48\linewidth]{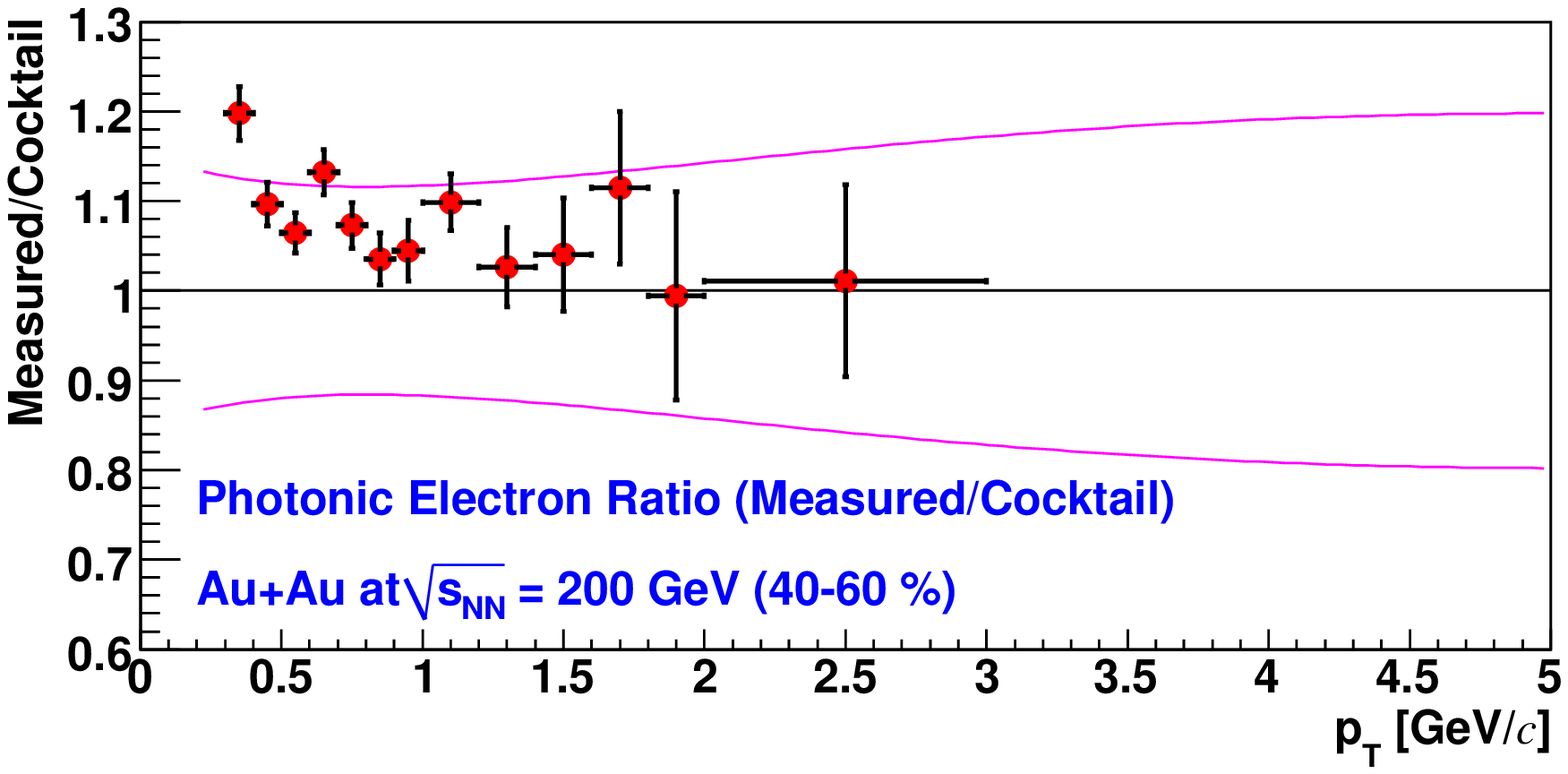}
  \includegraphics[width=0.48\linewidth]{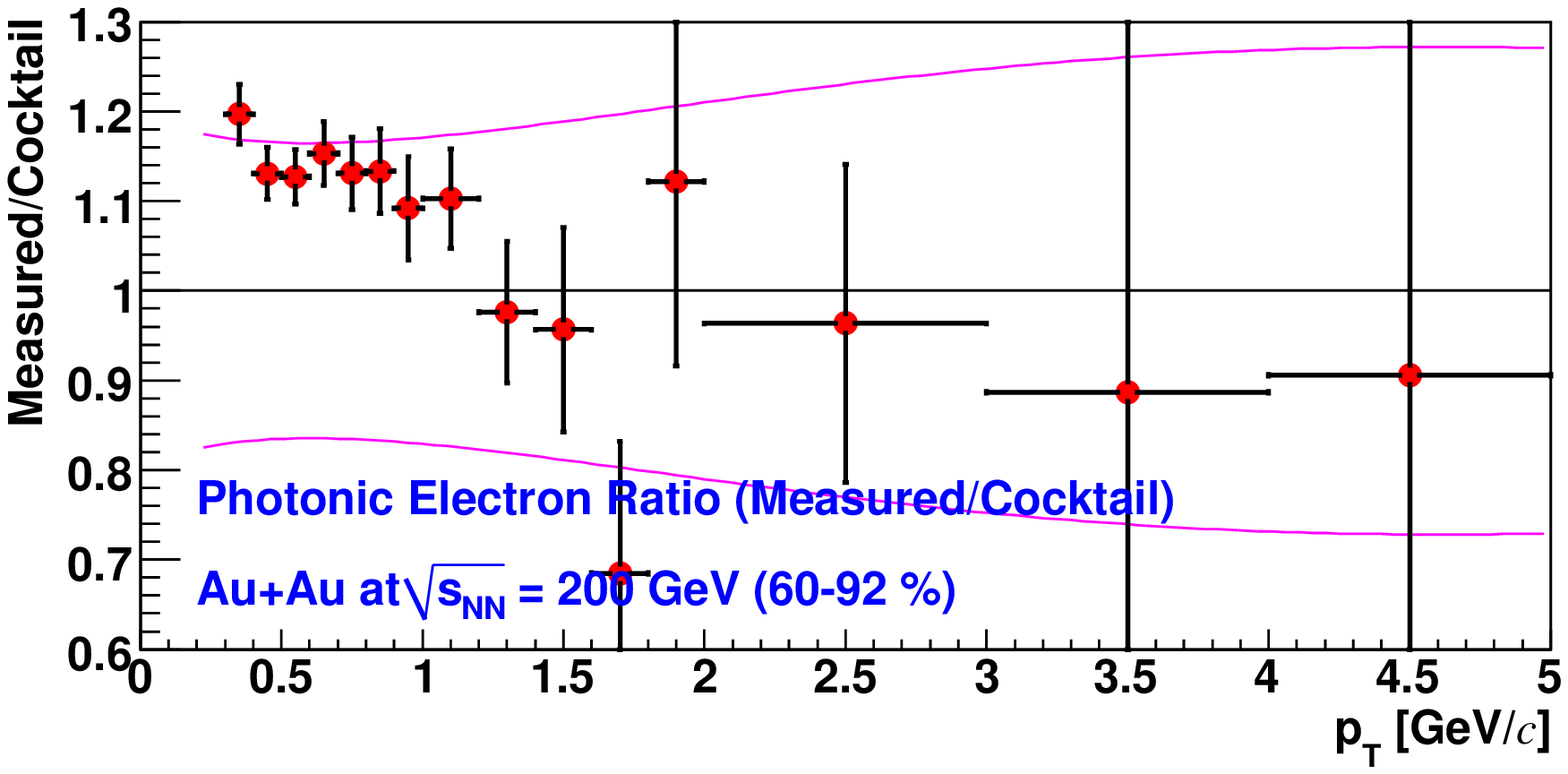}
  \caption{(Color online) The ratio of photonic electrons measured by the
converter 
method to the cocktail calculation in \auau~collisions for the indicated
centralities.
The upper and lower curves show the systematic error of the cocktail.  
Error bars are statistical only.  
\label{fig:conv_cock_ratio_auau}}
\end{figure*}

\begin{figure}[htbp]
  \includegraphics[width=1.0\linewidth]{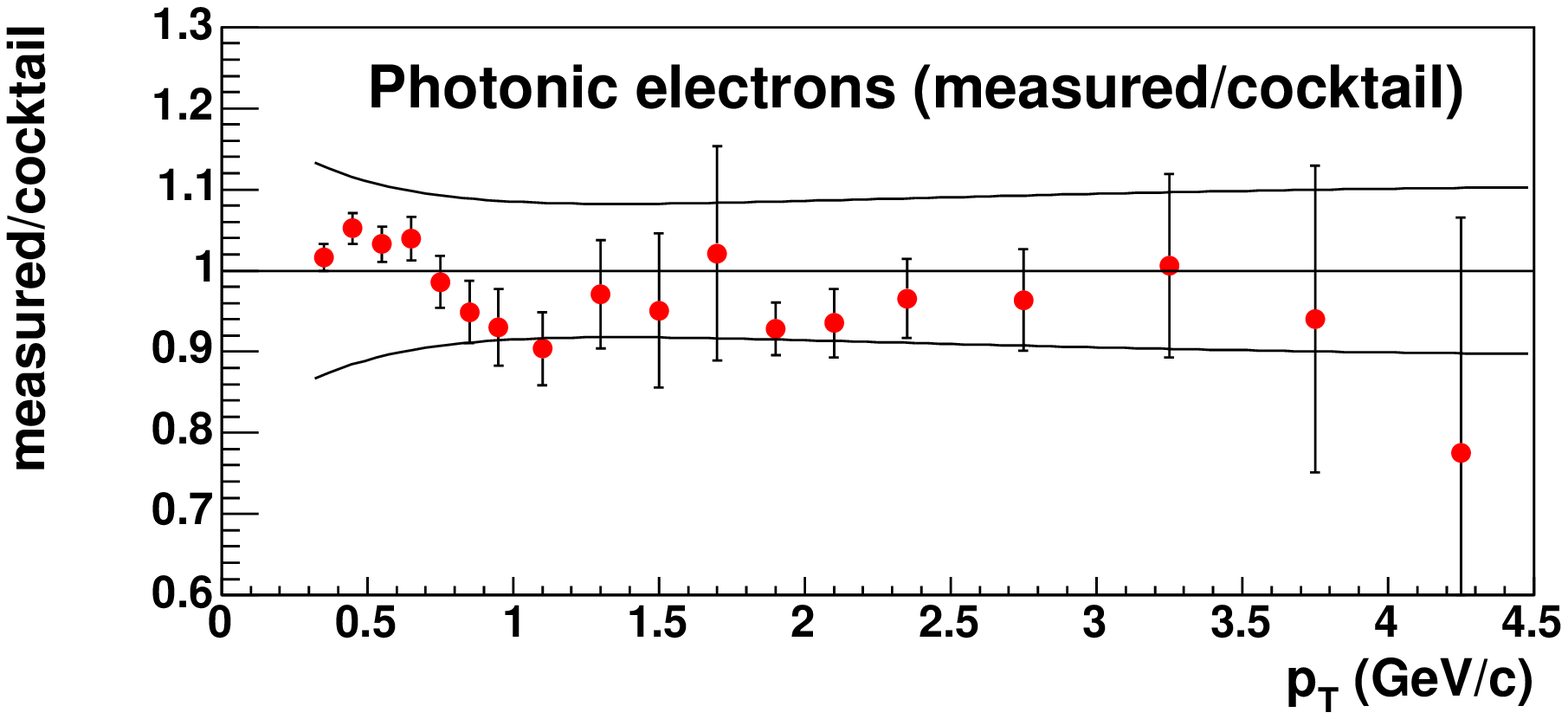}
\caption{(Color online) The ratio of photonic electrons measured by the
converter 
method to the cocktail calculation in 2005 \pp~collisions.  The data 
from the MB (PH) data set are shown below (above) 1.8 GeV/$c$.  The 
upper and lower curves show the systematic error of the cocktail.  
Error bars are statistical only.  
\label{fig:conv_cock_ratio_pp}}
\end{figure}

To tune the normalization of the photonic electron spectra in the 
cocktail, the converter/cocktail ratio of photonic electrons is 
calculated for all centrality classes in \auau~collisions.  The ratio in 
MB is fitted by a constant (converter/cocktail = 1.182).  The constant 
is applied to correct the cocktail spectra of photonic electrons for all 
centrality classes in \auau~collisions.  The ratios of photonic electron 
yield from the converter method to that from the corrected cocktail 
method for all centrality classes in 2004 \auau~collisions are shown in 
Fig.~\ref{fig:conv_cock_ratio_auau}.  The curves in each plot indicate 
the systematic errors of the cocktail method.  
Figure~\ref{fig:conv_cock_ratio_pp} shows the converter/cocktail ratio 
(after scaling by 1.182) of photonic electrons in 2005 \pp~collisions.

\subsection{Kaon decay background}\label{sub:Ke3}

After the subtraction of photonic electrons, almost all of the remaining 
background is from kaon decays.  Kaons contribute to the nonphotonic 
electron spectra via $K_{\rm e3}$ decay and are only substantial at low 
$\pte$ ($<1$ GeV/$c$).  Since kaons have a long lifetime, the decay 
electrons do not originate from the primary collision vertex.  
Therefore, their contribution can only be determined reliably in a full 
detector simulation.  This simulation was performed similarly to the 
$\pi^0$ simulation.  It is done for charged kaons and $K^0_L$ to 
estimate the final electron rate.  The input spectra have a 
parametrization obtained by the PHENIX kaon measurements~\cite{ppg026}.

To compare kaon contributions with photonic and nonphotonic 
electrons, the ratio of the electron yield from kaon decays over the 
photonic electron yield ($\rkp$) is shown as filled squares in the 
lower plot of Fig.~\ref{fig:rcn_rg_rnp_mb}.  The background from kaons 
becomes negligible above a $\pte$ of 1 GeV/$c$.
 
 
\subsection{Invariant spectra}\label{sub:invariant}

The invariant cross section for single electrons from heavy-flavor decay in
$p+p$ collisions is calculated using the following formula,
\begin{equation}
E\frac{d^3\sigma^{\rm HF}}{dp^3} = \frac{1}{\mathcal{L}}\frac{1}{2\pi p_{\rm
T}}\frac{N_e^{\rm HF}}{\Delta p_T \Delta y}\frac{1}{A\epsilon_{\rm
rec}}\frac{1}{\epsilon_{\rm bias}},
\end{equation}
where ${\mathcal{L}}$ is the integrated luminosity; $N_e^{\rm HF}$ is the 
electron yield from heavy-flavor decay after subtraction of photonic and 
nonphotonic background contributions; $A\epsilon_{\rm rec}$ is the 
product of the geometrical acceptance and reconstruction efficiency 
described in \ref{sub:acc}.  For the PH data set, $A\epsilon_{\rm 
rec}$ also includes the PH trigger efficiency described in \ref{sub:trig}.
The cross sections from the MB and the PH data sets are consistent with each
other in overlapping $p_T$ regions.

 
\subsection{Systematic uncertainties in invariant spectra}\label{sub:systematic}
%
%

In the $p+p$ analysis, systematic uncertainties are categorized into (a) 
inclusive electron spectra, (b) cocktail subtraction, and (c) converter 
subtraction.  Category (a) is common to both the converter and cocktail 
methods, and includes the uncertainties in luminosity (9.6\%), 
geometrical acceptance (4\%), eID efficiency (3\%) (from comparing to 
conversion electrons and simulation), and the PH trigger efficiency (3\% 
at the plateau).  Uncertainties in cocktail subtraction (category (b)) 
include the normalization (8\%) and $p_T$ dependent shape uncertainty 
(2\% at $p_T \simeq$ 2 GeV/$c$, increasing to 6\% at 9 GeV/$c$) which 
originates from increased measurement errors on the background sources.  
In the converter analysis (category(c)) the dominant uncertainties are 
in $R_\gamma$ (2.7\%) and in the relative acceptance in the converter 
and the normal runs (1.0\%).

In the Au+Au analysis there is no trigger efficiency correction, but 
there is additional uncertainty from the correction due to 
multiplicity-dependent efficiency loss.  A conservative systematic 
uncertainty of 4\% was added, ascertained from the differences of the 
two columns in Table \ref{tab:embedding}.

These uncertainties are propagated into the uncertainties in the 
heavy-flavor electron yields and added in quadrature.  At low (high) 
$p_T$ the uncertainties are amplified (reduced) to large (small) 
uncertainties in the heavy-flavor electron signal by approximately a 
factor of $1/(S/B)$, where $S/B$ is the ratio of the nonphotonic 
electron yield to the photonic electron yield.
 
 
\subsection{Inclusive electron $v_2$}\label{sub:inv_v2}

\begin{figure}[htbp]
\includegraphics[width=\linewidth]{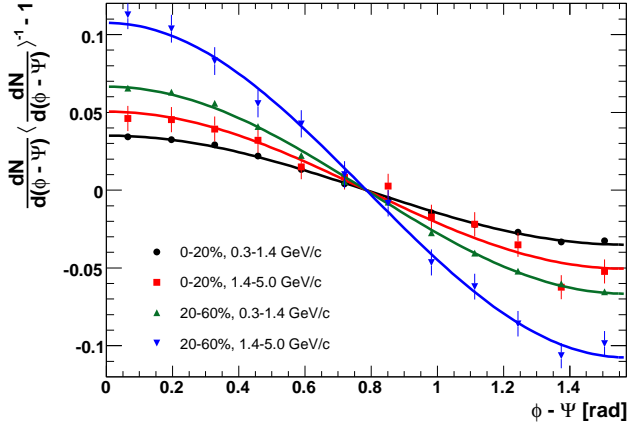}
\caption{(Color online) Azimuthal distributions relative to the reaction plane
for 
various centrality and transverse momenta.  The distributions have been 
normalized to their average value and centered around 0.
\label{fig:dNdphi_norm}}
\end{figure}

The value of the inclusive electron $v_{2}$ was measured with the reaction plane
method which can be written as
\begin{equation}
\frac{dN}{d\phi}=N_{0}\left\{1+2v_{2}\cos(2(\phi-\Psi))\right\},
\label{eq:Eqrpmethod}
\end{equation}
where $N_{0}$ is a normalization constant, $\phi$ is the azimuthal angle 
of electrons, and $\Psi$ is the reaction plane angle.  The reaction 
plane was determined from the multiplicity in each segment of the 
beam-beam counters.  As the measurement of the reaction plane is 
sensitive to non-flow effects such as jets, resonance decays and HBT 
correlations, the reaction plane was measured in two well-separated 
rapidity intervals.  Since each BBC is roughly three units of 
pseudorapidity away from the central arms, non-flow effects in the 
reaction plane measurement are expected to be small.  
Figure~\ref{fig:dNdphi_norm} shows the azimuthal distribution of 
inclusive electrons for various centrality and $p_T$ ranges, fit to the 
form of Eqn.~\ref{eq:Eqrpmethod}.  One can see from the agreement of the 
curves and data points that higher order Fourier coefficients do not 
contribute much to the measured azimuthal electron distribution.  Using 
the reaction plane measured with the BBCs, the inclusive electron 
$v_{2}$ was calculated as $v_{2}=\langle\cos(2(\phi-\Psi_{\textrm{\rm 
meas}}))\rangle.  $ The true $v_2$ with respect to the true reaction 
plane can be expressed in terms of the observed $v_2$ with respect to 
the measured reaction plane as \cite{ollitrault, rpm}

\begin{equation}
v_2 = \frac{v_2^{\textrm meas}}{\langle \textrm{cos}(2(\Psi_{\textrm{\rm meas}}
- \Psi_{\textrm{\rm true}})) \rangle}.
\end{equation}
Two methods were used to extract $\langle 
\textrm{cos}(2(\Psi_{\textrm{\rm meas}} - \Psi_{\textrm{\rm true}})) \rangle$.  
The first is an analytical calculation \cite{ollitrault,rpm}
\begin{align}
&\langle \textrm{cos}(2(\Psi_{\textrm{\rm meas}} - \Psi_{\textrm{\rm true}})) \rangle =
\\ \nonumber &\frac{\sqrt{\pi}}{2\sqrt{2}} \chi \textrm{exp}(\frac{-\chi^2}{4})
[I_0(\frac{\chi^2}{4}) + I_1(\frac{\chi^2}{4})]
\end{align}
where $\chi$ 
is equal to $v_2^{\textrm{\rm meas}} \sqrt{2N}$, and $N$ is the BBC 
multiplicity.  This method was verified to be well approximated by 
\cite{rpm}
\begin{equation}
\langle \textrm{cos}(2(\Psi_{\textrm{\rm meas}} - \Psi_{\textrm{\rm true}})) \rangle
\sim \sqrt{2 \langle \textrm{cos}(2(\Psi_{\textrm{\rm meas}}^{N} -
\Psi_{\textrm{\rm meas}}^{S}))\rangle} \label{eq:RPres}
\end{equation}
with $\Psi_{\textrm{\rm meas}}^{N(S)}$ is the measured reaction plane using 
only the north (south) BBC.

\begin{figure}
 \includegraphics[width=1.0\linewidth]{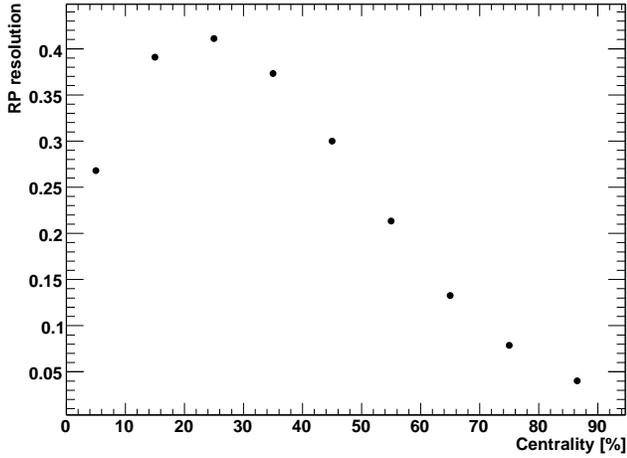}
  \caption{Reaction plane resolution as a function of centrality.}
\label{fig:RPres}
\end{figure}

\begin{figure*}[tb]
\includegraphics[width=0.48\linewidth]{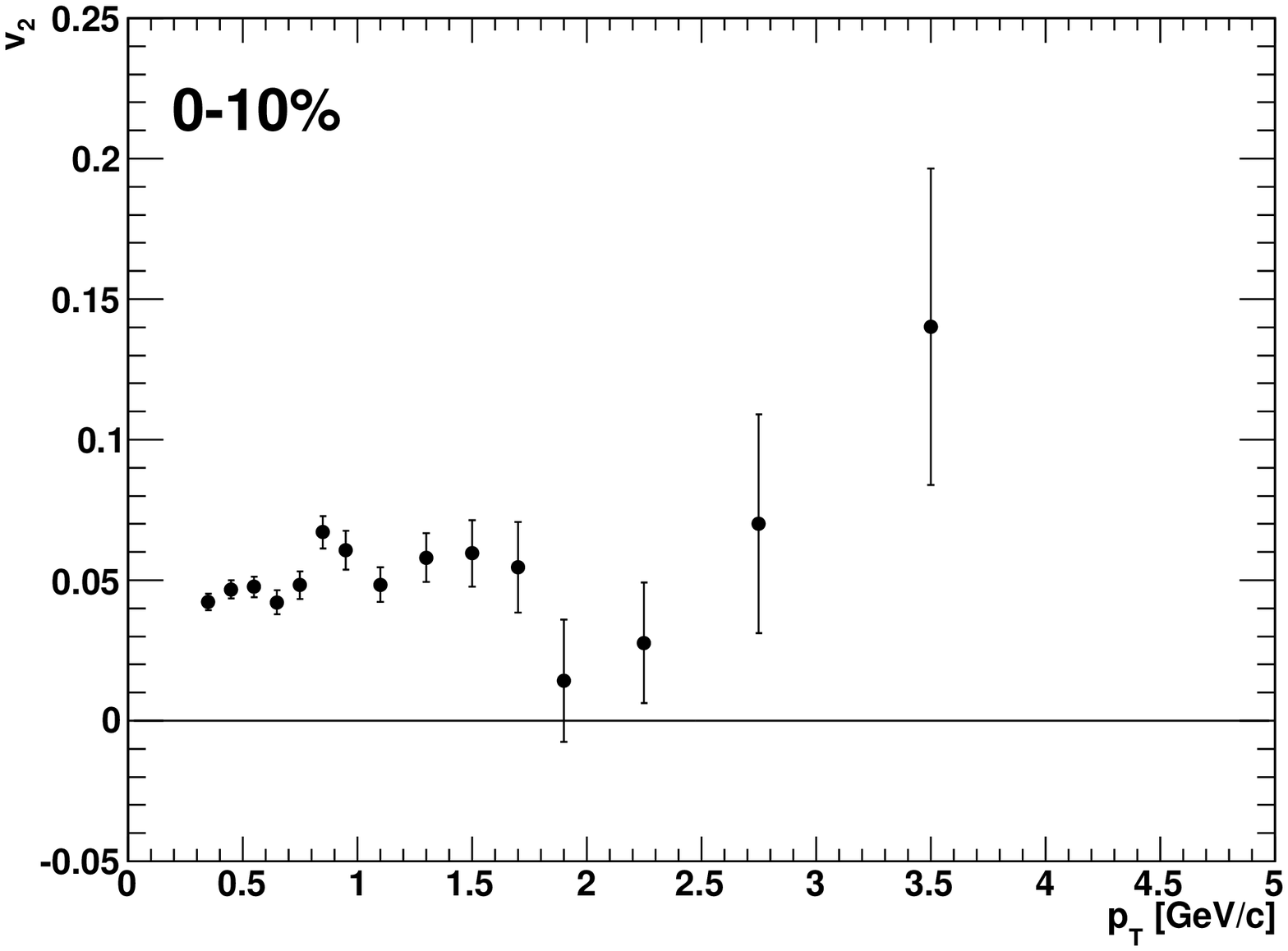}
\includegraphics[width=0.48\linewidth]{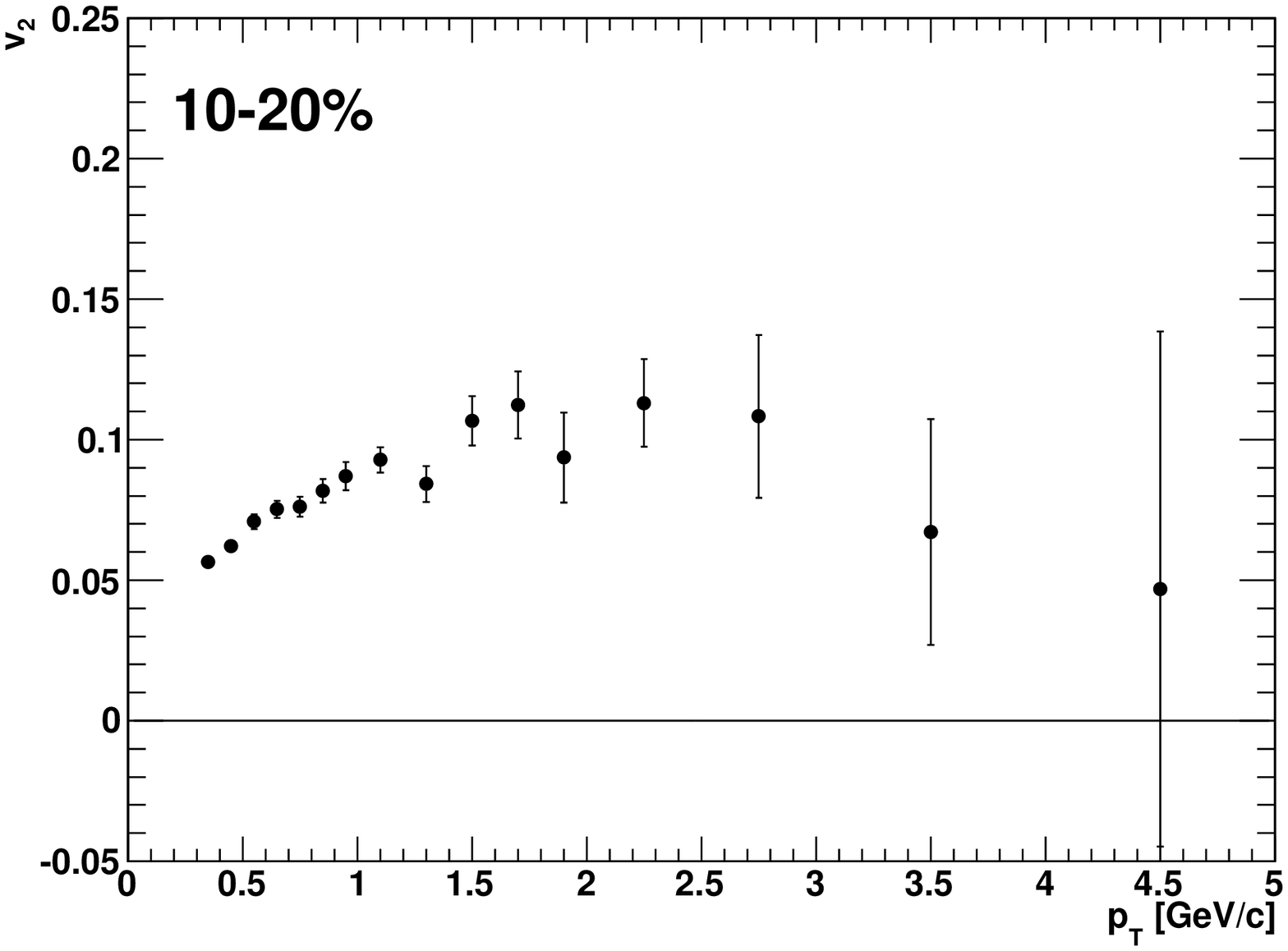}
\includegraphics[width=0.48\linewidth]{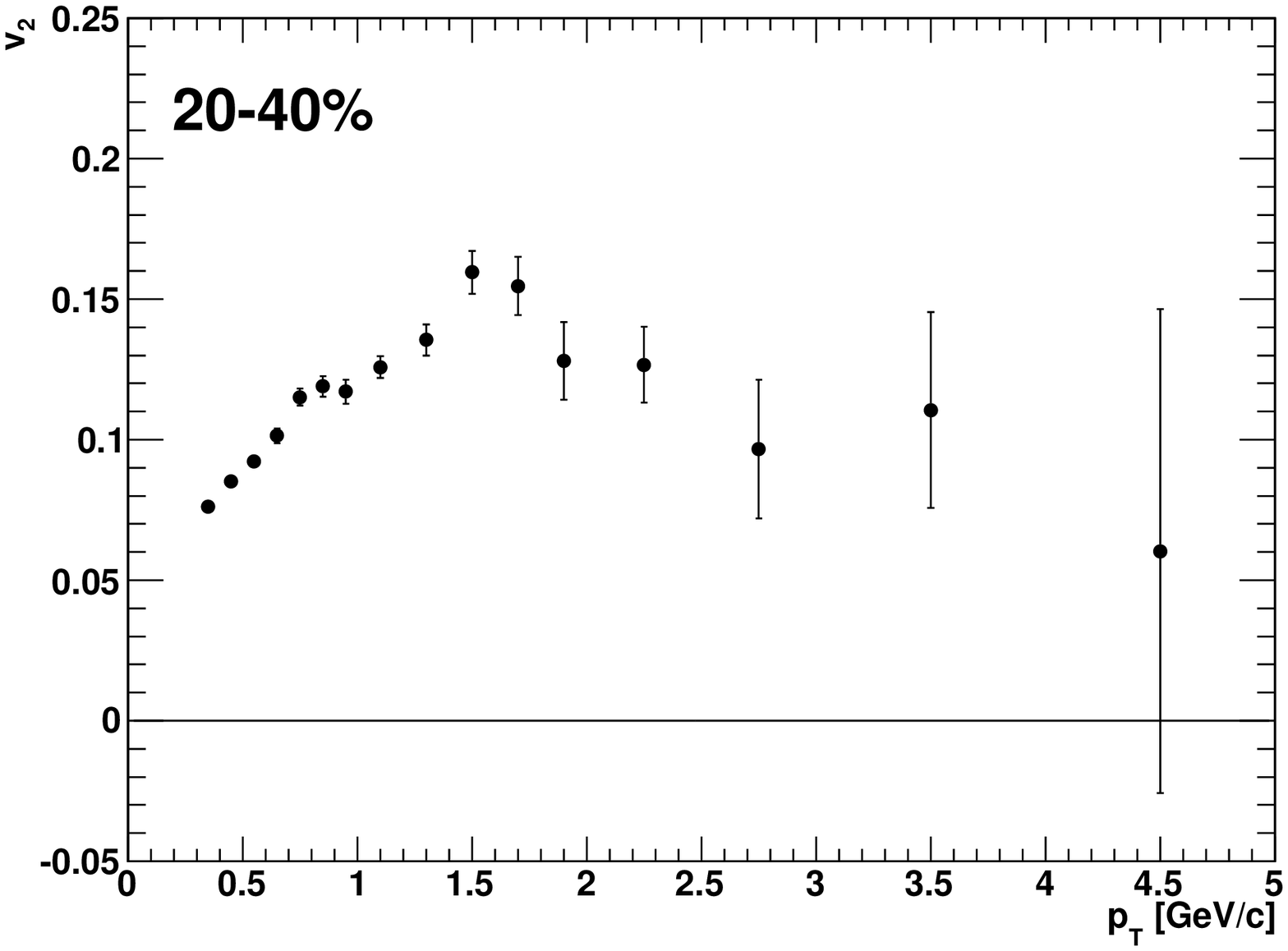}
\includegraphics[width=0.48\linewidth]{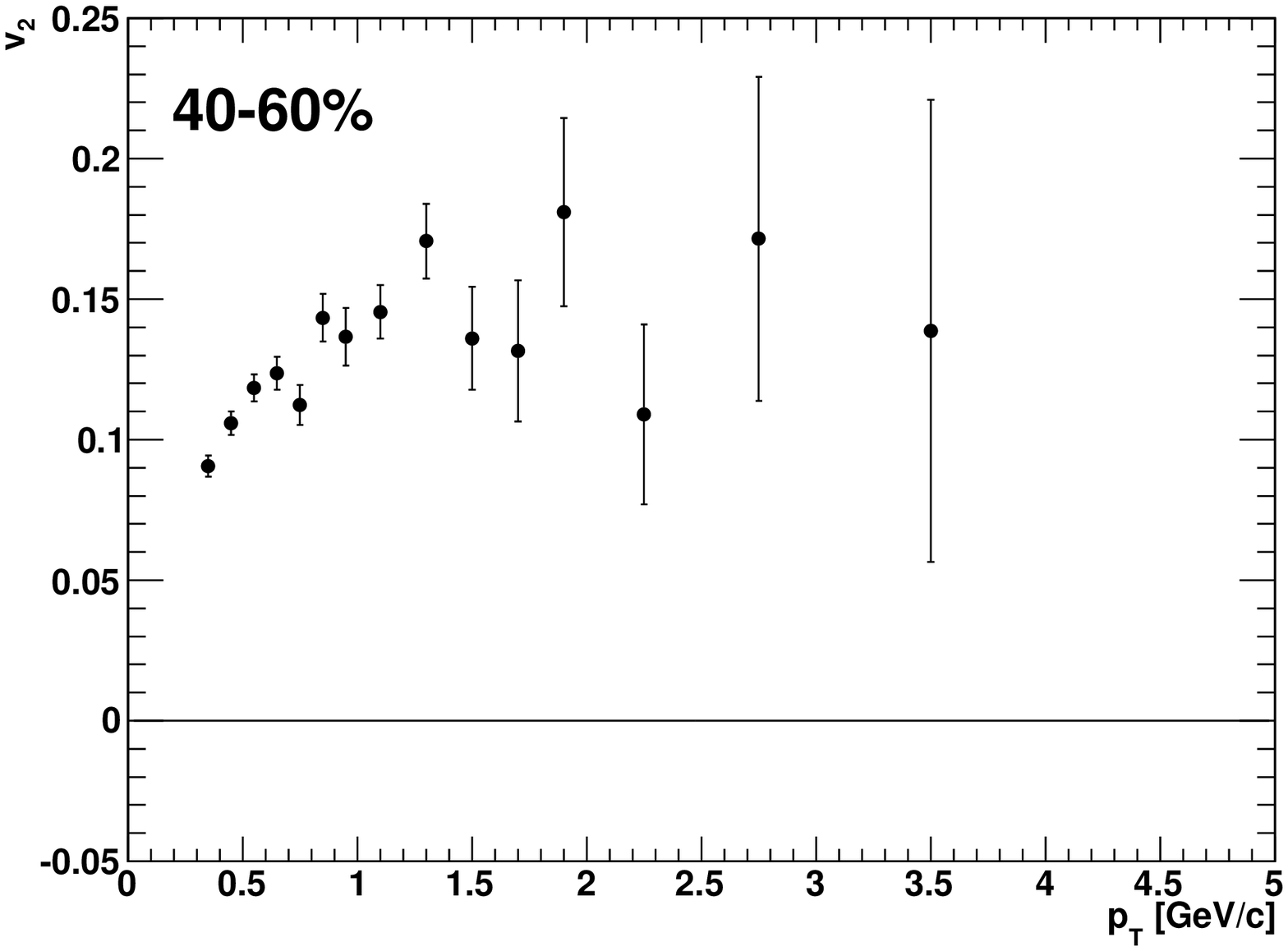}
\includegraphics[width=0.48\linewidth]{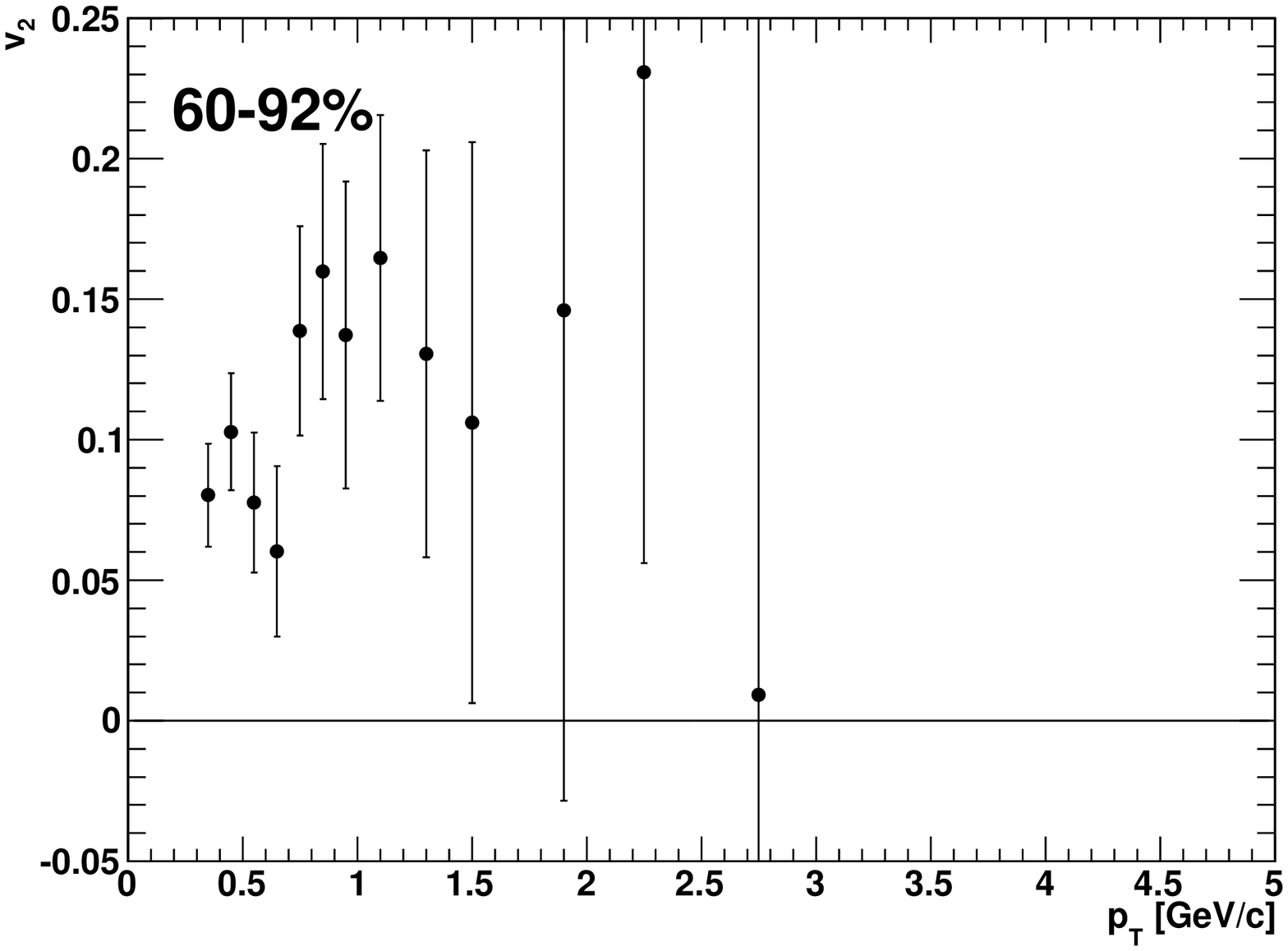}
\caption{$p_T$ dependence of inclusive electron $v_2$, for different centrality
bins.  A 5\% systematic scale uncertainty is not shown.}
\label{fig:incv2_cent}
\end{figure*}

\begin{figure}[tb]
\includegraphics[width=1.0\linewidth]{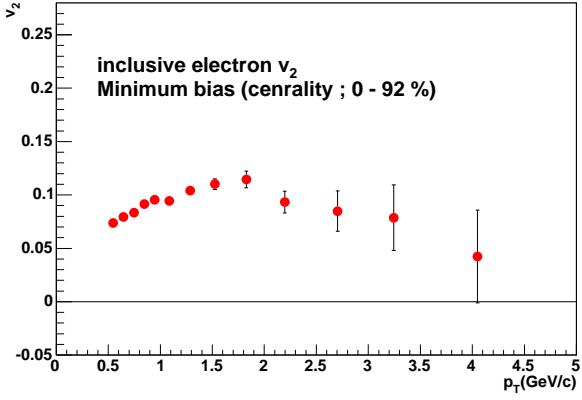}

\caption{(Color online) Inclusive electron $v_{2}$ for minimum bias events as a 
function $p_{T}$ measured in Au+Au collisions at $\sqrt{s_{\rm NN}}$ 
= 200 GeV.
\label{fig:incv2_MB}
}
\end{figure}

Figure~\ref{fig:RPres} shows the reaction plane resolution determined 
from Equation \ref{eq:RPres}.  Less than 10\% of background remains due 
to accidental RICH associations of hadrons.  Contributions from such 
background sourced was subtracted as:

\begin{equation}
\frac{dN}{d(\phi-\Psi_{R.P.})} = \frac{dN^{e}_{cand}}{d(\phi-\Psi_{R.P.})} -
\frac{dN^{e}_{back}}{d(\phi-\Psi_{R.P.})}, 
\end{equation}

\noindent where $N^{e}_{cand}$ is the number of electrons identified by 
the RICH and $N^{e}_{back}$ is the number of the background particles.  
The number of random associations are obtained by replacing the north 
and the south sides of the RICH in the software, as described 
previously.  The $v_2$ for a given centrality bin $[a,b]$ can be 
expressed as

\begin{equation}
v_2^{\textrm{bin}} = \dfrac{\displaystyle \int_a^b \frac{v_2^{\textrm{\rm meas}}(C)
\dfrac{dN_{e^+,e^-}}{dC}}{\textrm{res}(C)} dC}{\displaystyle \int_a^b
\dfrac{dN_{e^+,e^-}}{dC} dC}.  \label{eq:v2int}
\end{equation}

\noindent where $res(C)$ is the reaction plane resolution for a given 
centrality $C$.  Due to lack of statistics, Equation \ref{eq:v2int} was 
approximated by a Riemann sum over 10\% centrality bins.  It was 
verified that $v_2$ is not sensitive to changing the bins for the 
Riemann sum.  This is because the integrand in the numerator of Equation 
\ref{eq:v2int} is quite flat with centrality, as the resolution grows 
with both the measured $v_2$ and the particle multiplicity.  
Figure~\ref{fig:incv2_cent} shows the transverse momentum and centrality 
dependence of the inclusive electron $v_{2}$.  The minimum-bias $v_{2}$ 
is shown in Fig.~\ref{fig:incv2_MB}.

\subsection{Background $v_2$ cocktail}\label{sub:v2_cocktail}

The azimuthal distribution of electrons ($dN^{e}/d\phi$) is given as the 
sum of the azimuthal distributions of photonic electrons 
($dN^{\gamma}/d\phi$) and nonphotonic electrons ($dN^{{\rm non}-\gamma}/d\phi$):

\begin{eqnarray}
\frac{dN_{e}}{d\phi}&=&\frac{dN_{e}^{\gamma}}{d\phi}+\frac{dN_{e}^{{\rm non}-\gamma}}{
d\phi}.
\label{eq:EQ_v2_bg_0}
\end{eqnarray}

Expanding Eq.  \ref{eq:EQ_v2_bg_0} into the Fourier expansion, 
the nonphotonic electron $v_{2}$ can be expressed as
\begin{equation}
v_{2_{e}}^{{\rm non}-\gamma} = \frac{(1+R_{\rm NP})v_{2_{e}}-v_{2_{e}}^{\gamma}}{R_{\rm NP}}.
\end{equation}
where $v_{2_{e}}$ is the $v_{2}$ of inclusive electrons, 
$v_{2_{e}}^{\gamma}$ is the $v_2$ of the photonic electrons, and 
$R_{\rm NP}$ is the ratio of nonphotonic to photonic electron yields 
which was defined in section \ref{sub:converter}.  The photonic electron $v_{2}$
can be estimated from the $v_{2}$ of electrons from various photonic sources as
\begin{equation}
v_{2}^{\gamma} = \sum R_{i}v_{2_{e}}^{i}
\label{eq:EQ_v2_bg_1}
\end{equation}
where $R_{i}$ is the relative contribution of electron source $i$ to 
the background and $v_{2}^{i}$ is electron $v_{2}$ from the electron 
source.

\begin{figure}[tb]
\includegraphics[width=1.0\linewidth]{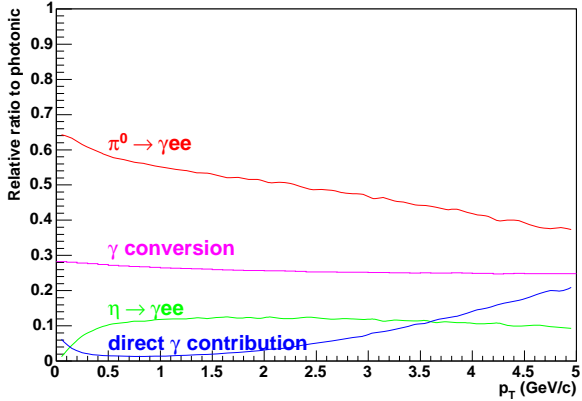}
\caption{(Color online) Relative contributions of electron from $\pi^0$, $\eta$
and 
direct $\gamma$ to the background.}
\label{fig:contribution}
\end{figure}

\begin{figure}[tb]
\includegraphics[width=1.0\linewidth]{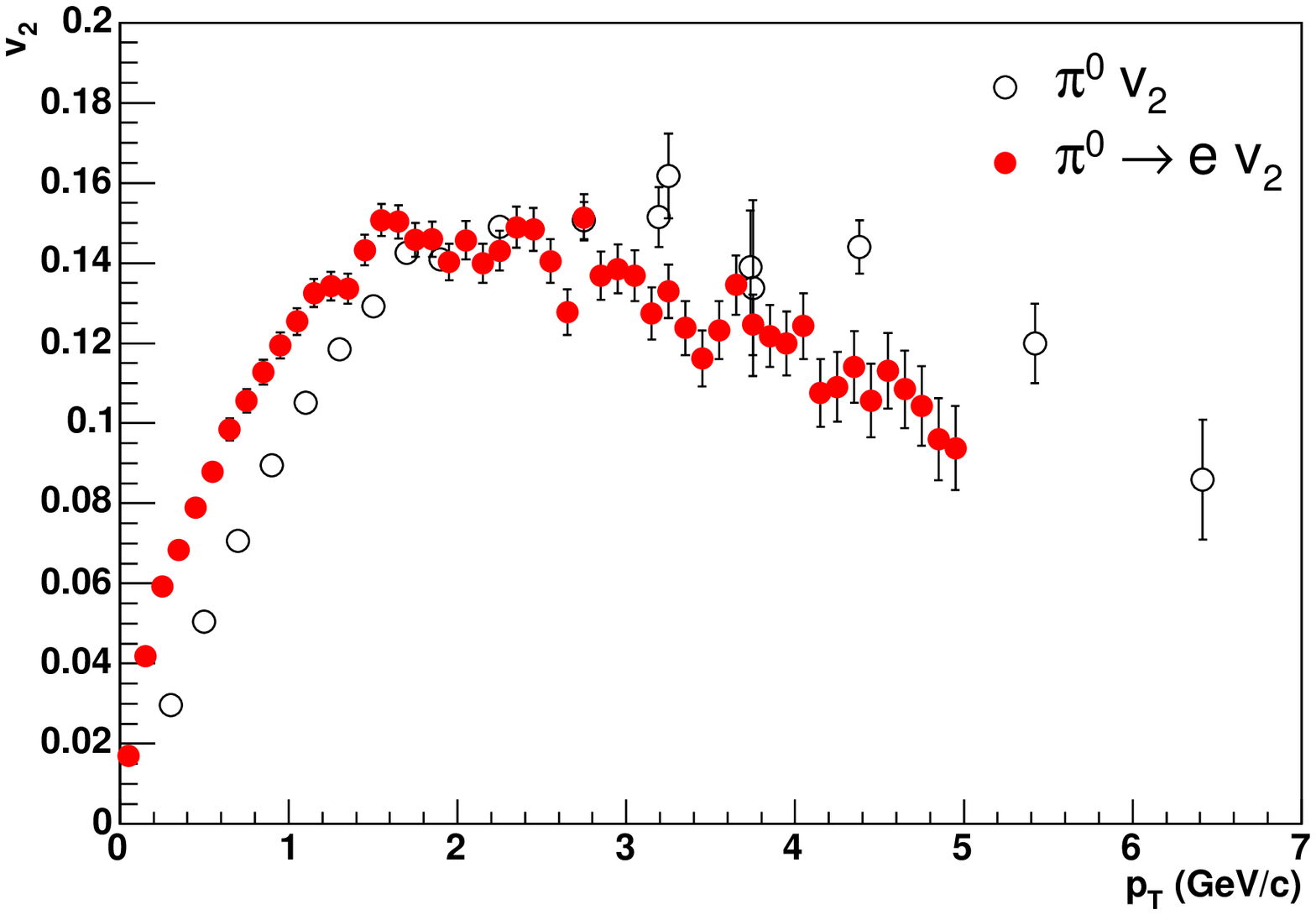}
\caption{(Color online) Electron $v_2$ from $\pi_0$ decay as a function of
$p_T$, from a simulation.}
\label{fig:pi0decayv2}
\end{figure}

\begin{figure}[tb]
\includegraphics[width=1.0\linewidth]{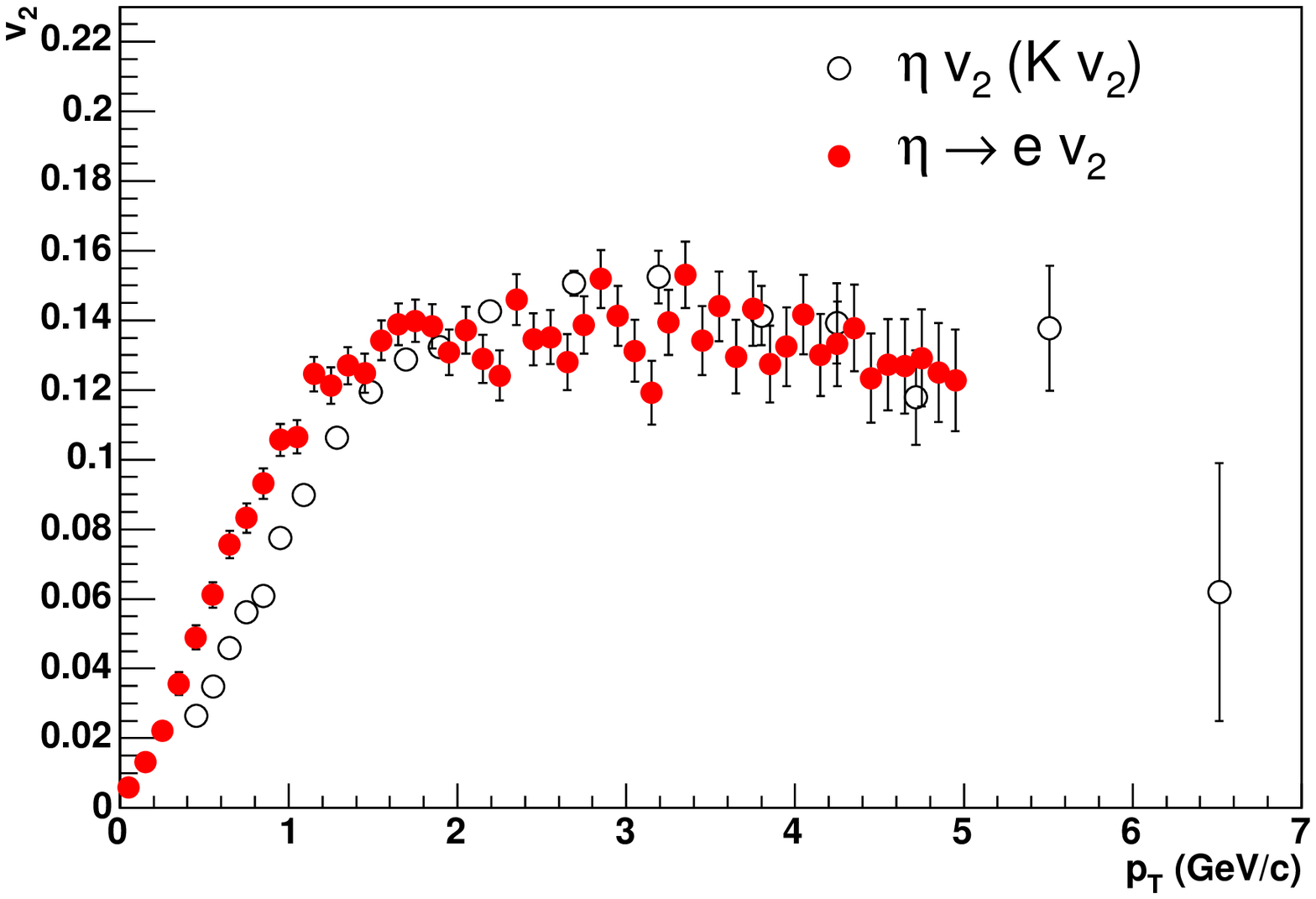}
\caption{(Color online) Electron $v_2$ from $\eta$ decay as a function of
$p_T$, from a simulation.
}
\label{fig:etadecayv2}
\end{figure}

\begin{figure}[tb]
\includegraphics[width=1.0\linewidth]{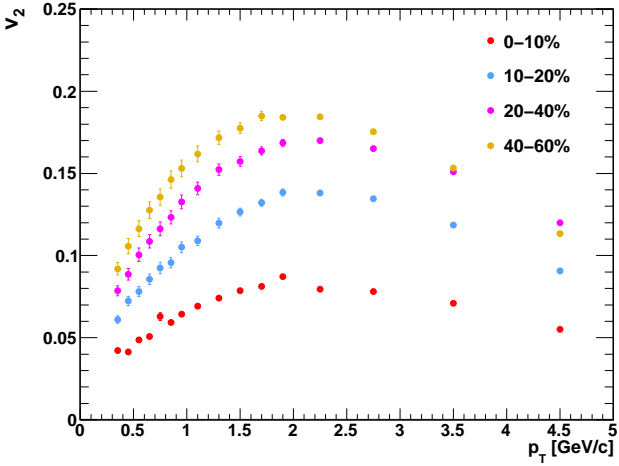}
\caption{(Color online) Electron $v_2$ from photonic sources as a function of 
$p_T$ and centrality.}
\label{fig:phov2}
\end{figure}

The dominant sources of photonic electrons are photon conversions and 
Dalitz decays from $\pi^{0}$.  In addition, we also took into account 
electrons from $\eta$ and direct photon decays when calculating the 
photonic electron $v_{2}$.  The relative contributions of electrons 
from those sources are shown in Fig.  \ref{fig:contribution}.  The other 
sources were ignored when calculating the photonic electron $v_{2}$ 
due to their small contribution.  The measured $v_{2}$ for
$\pi^{\pm}$ and $\pi^0$ was used as input for a simulation of $\pi^0
\rightarrow e^{\pm} v_2$,  and transverse kinetic energy scaling
\cite{Adare:2007a} was assumed for the $\eta$, $\rho$, and $\omega$.  The
direct photon $v_2$ was assumed to be zero.  The decay electron $v_{2}$ from
$\pi^{0}$ and $\eta$ from Monte Carlo simulations are shown in
Fig.~\ref{fig:pi0decayv2} and Fig.~\ref{fig:etadecayv2}.  The photonic electron
$v_{2}$ that was calculated from the results is shown in Fig.~\ref{fig:phov2}.
The middle solid line is the mean value of the photonic electron $v_{2}$ and the
upper and lower dashed lines show one standard deviation of the systematic
uncertainty.  The uncertainty mainly comes from the uncertainty of the measured
input $v_{2}$.  The $v_2$ from photonic sources for various centralities is shown
in Fig.~\ref{fig:phov2}.

\subsection{Photonic $v_2$ converter method}\label{sub:v2_converter}

\begin{figure}[tb]
\includegraphics[width=1.0\linewidth]{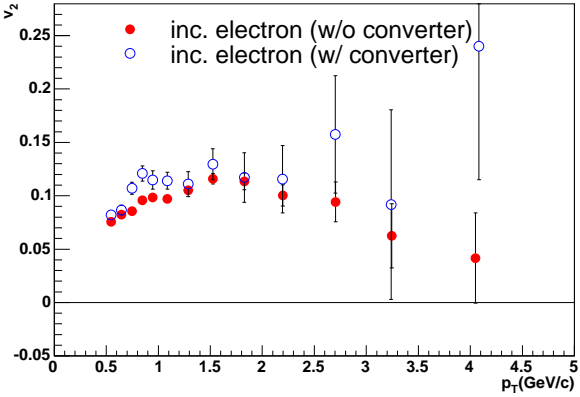}
\caption{(Color online) Inclusive electron $v_{2}$ with and without the
converter 
installed.
\label{fig:incv2_w_o_conv}
}
\end{figure}

\begin{figure}[tb]
\includegraphics[width=1.0\linewidth]{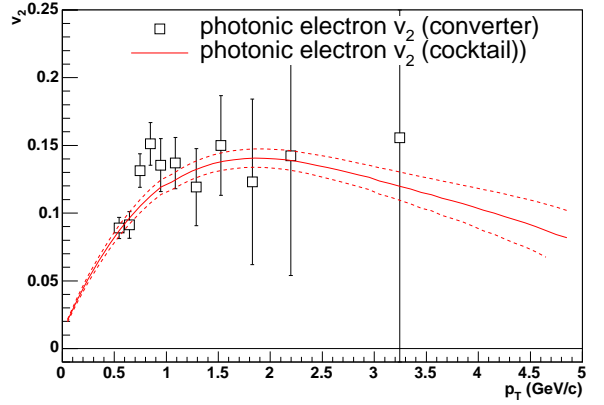}
\caption{(Color online) Photonic electron $v_{2}$ determined by two independent
methods, the cocktail and the converter method.
The lines are determined by the cocktail method and the squared points are
obtained by the converter method.
\label{fig:phov2_phov2}
}
\end{figure}

The photonic electron $v_{2}$ can be also determined by the converter 
method.  Nonphotonic and photonic electron $v_{2}$ can be separated by 
using the inclusive electron $v_{2}$ measured with 
($v_{2_{e}}^{conv-in}$) and without ($v_{2_{e}}^{conv-out}$) the 
converter as
\begin{align}
\nonumber &v_{2_{e}}^{{\rm non}-\gamma} = \\ &\frac{
R_{\gamma}(1+R_{\rm NP})v_{2_{e}}^{conv-out} -
(R_{\gamma}+R_{\rm NP})v_{2_{e}}^{conv-in}}{R_{\rm NP}(R_{\gamma}-1)} \\
\nonumber &v_{2_{e}}^{\gamma} = \\ &\frac{ (1+R_{\rm NP})v_{2_{e}}^{conv-out} -
(R_{\gamma}+R_{\rm NP})v_{2_{e}}^{conv-in}}{(1-R_{\gamma})}.
\end{align}
Figure~\ref{fig:incv2_w_o_conv} shows the inclusive electron $v_{2}$ 
with and without the converter.  If the photonic electrons and 
nonphotonic electrons have the same $v_{2}$, the $v_{2}$ measured 
with and without the converter would be the same.  Due to the small 
statistics of the converter runs, $v_{2}$ measured with the converter 
has a large statistical uncertainty.  The photonic electron $v_{2}$ 
obtained by the converter method is shown as open squares in 
Fig.~\ref{fig:phov2_phov2}.  The result is consistent with the photonic 
electron $v_{2}$ determined by the cocktail method within statistical 
uncertainty.

\subsection{Heavy flavor $v_2$ and systematic uncertainties}\label{sub:v2_sys}

The remaining background after subtracting photonic background is 
kaon decay as described previously.  The electron $v_{2}$ from kaon 
decays was also calculated by a Monte Carlo simulation assuming transverse
kinetic energy scaling and it was removed from the nonphotonic electron sample.
The electron $v_{2}$ from kaon decays was subtracted from the nonphotonic
electron $v_{2}$ as \begin{equation} v_{2_{e}}^{heavy} = 
\frac{v_{2_{e}}^{{\rm non}-\gamma}-R_{KNP}v_{2_{e}}^{K}}{1-R_{KNP}}.  
\end{equation} Here $R_{KNP}$ is the ratio of the yield of 
electrons from kaon decays and from all other nonphotonic sources 
($R_{KNP} = N_{e}^{K}/N_{e}^{{\rm non}-\gamma}$), and $v_{2_{e}}^{heavy}$ is the
$v_2$ of electrons from heavy flavor decays.  After subtracting kaon 
decays, the main source of nonphotonic electrons remaining is heavy 
flavor decays.

Systematic uncertainties of the heavy flavor electron $v_{2}$ are 
summarized below.

\begin{itemize}

\item Reaction plane determination: The systematic uncertainty of the reaction
plane determination was estimated by measuring the inclusive electron $v_{2}$
separately with the north BBC, the south BBC, and both the north and south 
BBCs combined.  The maximum difference is about 5 $\%$ and we apply it 
as the uncertainty due to the reaction plane determination.

\item Electron identification: The systematic uncertainty from electron
identification was estimated by measuring the inclusive electron $v_{2}$ while 
changing the electron identification cuts from the standard cuts 
which were described in previous section.  In this analysis we 
changed the cut parameters of $E/p$, $n0$ and $\chi^{2}/npe0$.  The relative
change in $v_2$ from varying the cuts on $E/p$ is about $2\%$, while for $n0$ it
is about 2$\%$ and for $\chi^{2}/npe0$ it is 1$\%$.  The total systematic 
uncertainty assigned from electron identification is 3 $\%$.

\item Background $v_{2}$: As described in the previous section, the
uncertainties from the photonic electron $v_{2}$ come from the $\pi^{0}$
$v_{2}$ and kaon $v_{2}$, which have about 5 $\%$ uncertainties.  We applied
these values to the uncertainty of the photonic electron $v_{2}$.  We also 
applied 5 $\%$ for the $v_{2}$ from three-body kaon decays.

\item $R_{\rm NP}$: The systematic uncertainty of $R_{\rm NP}$ comes from the
uncertainties of inclusive electron spectra and the subtracted background 
spectra.  The systematic uncertainty of the inclusive electron 
spectra includes the uncertainties in the geometrical acceptance, 
the reconstruction efficiency and the occupancy correction, as 
described previously.

\end{itemize}

The total systematic uncertainty was obtained by a quadratic sum of 
the above uncertainties.

 
\section{RESULTS}\label{sec:results}
%
%

 
\subsection{Heavy-flavor electron cross sections $(p+p)$}

\begin{figure}[htbp]
  \includegraphics[width=1.0\linewidth]{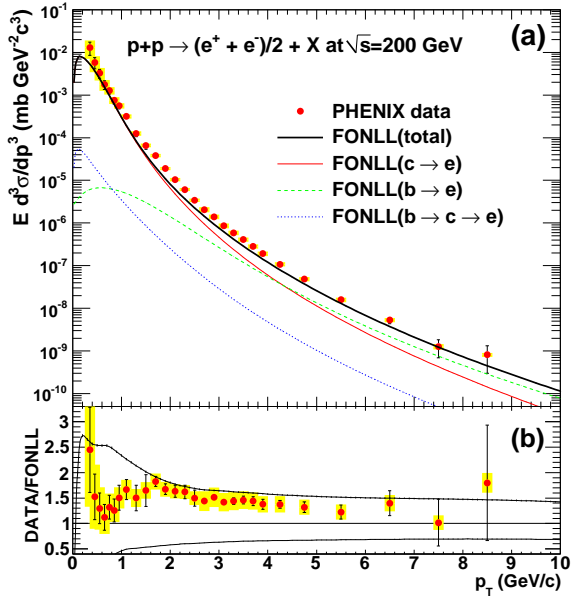}
  \caption{(Color online) (a)~Invariant differential cross sections of single 
electrons as a function of $\pt$ in $p+p$ collisions at $\sqrt{s} = 
200$ GeV.  The error bars (bands) represent the statistical (systematic) errors.
The curves are the FONLL calculations \cite{fonll}.  (b)~The ratio of FONLL/Data
as a function of $\pt$.  The upper (lower) curve shows the theoretical upper 
(lower) limit of the FONLL calculation.  In both panels, a 10\% 
normalization uncertainty is not shown.  \label{fig:PP_Fig3}}
\end{figure}

Fig.~\ref{fig:PP_Fig3}~(a) shows the invariant differential cross 
section of electrons from heavy-flavor decays.  Fig.~\ref{fig:PP_Fig3}~(b) 
shows the yield divided by a FONLL calculation \cite{fonll}.  All background has
been subtracted.  The data from the two analysis methods (converter and
cocktail) are combined: at low $p_T$ ($p_T$ $< 1.6$ GeV/$c$) 
the converter subtraction method is applied to the MB data set; at intermediate
$p_T$ ($1.6< p_T < 2.6$~GeV/$c$) the converter method is applied to
the PH data set; and at high $p_T$ ($p_T>2.6$~GeV/$c$) the cocktail
method is applied to the PH data set.

\begin{figure}[htbp]
  \includegraphics[width=1.0\linewidth]{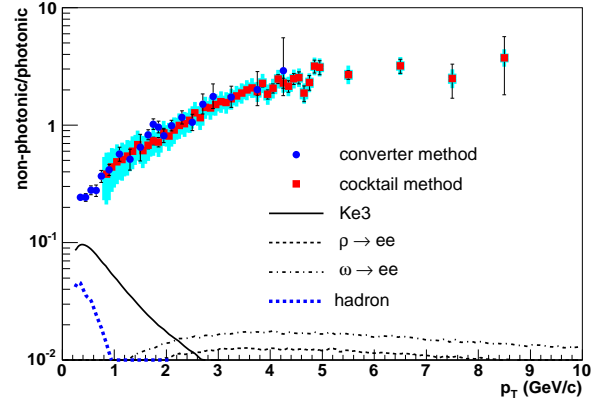}
  \caption{(Color online) Ratio of nonphotonic electrons to photonic back- 
ground.  Error bars are statistical errors and the error bands show 
the cocktail systematic errors.  The solid, dashed, dot- dashed, and 
dotted curves are the remaining nonphotonic back- ground from Ke3 , 
$\rho \rightarrow e^{+}e^{-}$, $\omega \rightarrow e^{+}e^{-}$ , 
and hadron contamination, respectively.
 \label{fig:PP_Fig2}}
\end{figure}

Figure~\ref{fig:PP_Fig2} shows the ratio of nonphotonic electrons 
(including charmonium, bottomonium, and Drell-Yan) to the photonic background.

\subsection{Heavy flavor electron invariant yield (\auau)}
The differential invariant yield spectra as a function of $\pte$ 
for the measured signal electrons from heavy flavor decays are 
calculated as in Eq.~\ref{eq:inv_eq_auau}.  $N^{\rm evt}_i$ is 
the number of events in a centrality class $i$ ($i=$~0--10~\%, 
10--20~\%, 20--40~\%, 40--60~\%, and 60--92~\%).  
$\epsilon^{\rm Hadron}(\pte)$ is the hadron contamination factor 
that is mentioned in \ref{sub:eid}.  $\epsilon^{\rm acc}_{\Delta 
y}(\pte)$ is the acceptance correction in \ref{sub:acc}.  
$\epsilon^{\rm embed}_i$ is the embedding efficiency in 
\ref{sub:embed}.  $\Delta \pte$ is the $\pte$ bin width.  $\Delta y$ is the
rapidity range ($|y|<0.5$) where the input $e^{\pm}$ are distributed at the
first stage of single particle simulation for the acceptance calculation (see 
\ref{sub:acc}).  $N^{\rm HF}_i(\pte , e^-)$ and 
$N^{\rm HF}_i(\pte , e^+)$ are the resulting counts of signal 
electrons and positrons from heavy flavor decays by the converter 
method or cocktail method.

\begin{align}
\label{eq:inv_eq_auau} &\frac{1}{2\pi\pte}\frac{d^2 N^{\rm HF}_i}{d\pte dy} 
  \equiv \frac{1}{2\pi\pte N^{\rm evt}_i} 
  \times \frac{1-\epsilon^{\rm Hadron}(\pte)}
  {\epsilon^{\rm acc}_{\Delta
y}(\pte)\cdot\epsilon^{\rm embed}_i} 
  \\ \nonumber &\times \frac{1}{\Delta \pte \Delta y}
  \times \frac{N^{\rm HF}_i(\pte , e^-) + N^{\rm HF}_i(\pte , e^+)}{2}.
\end{align}

\begin{figure*}[htbp]
  \includegraphics[width=0.48\linewidth]{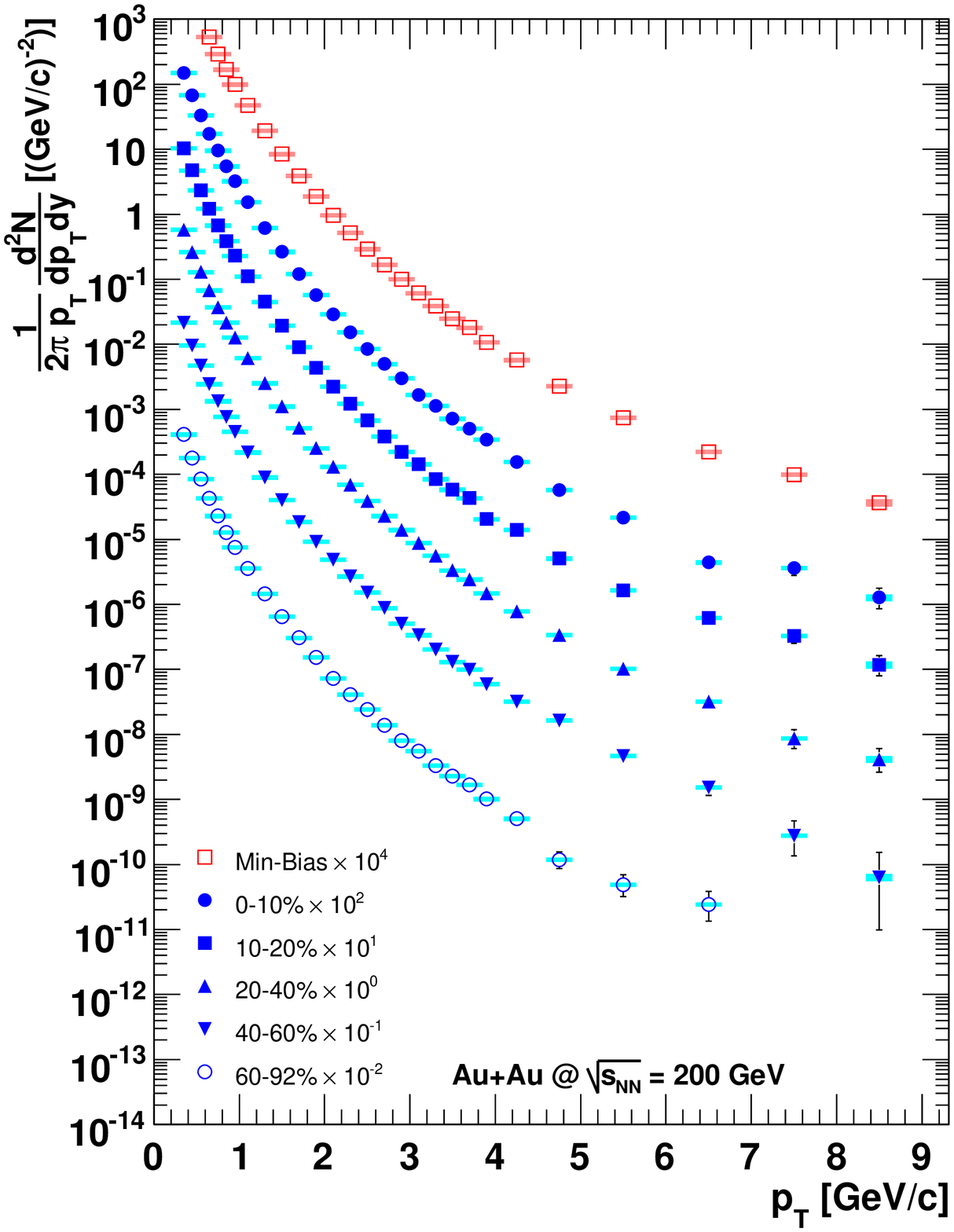}
  \includegraphics[width=0.48\linewidth]{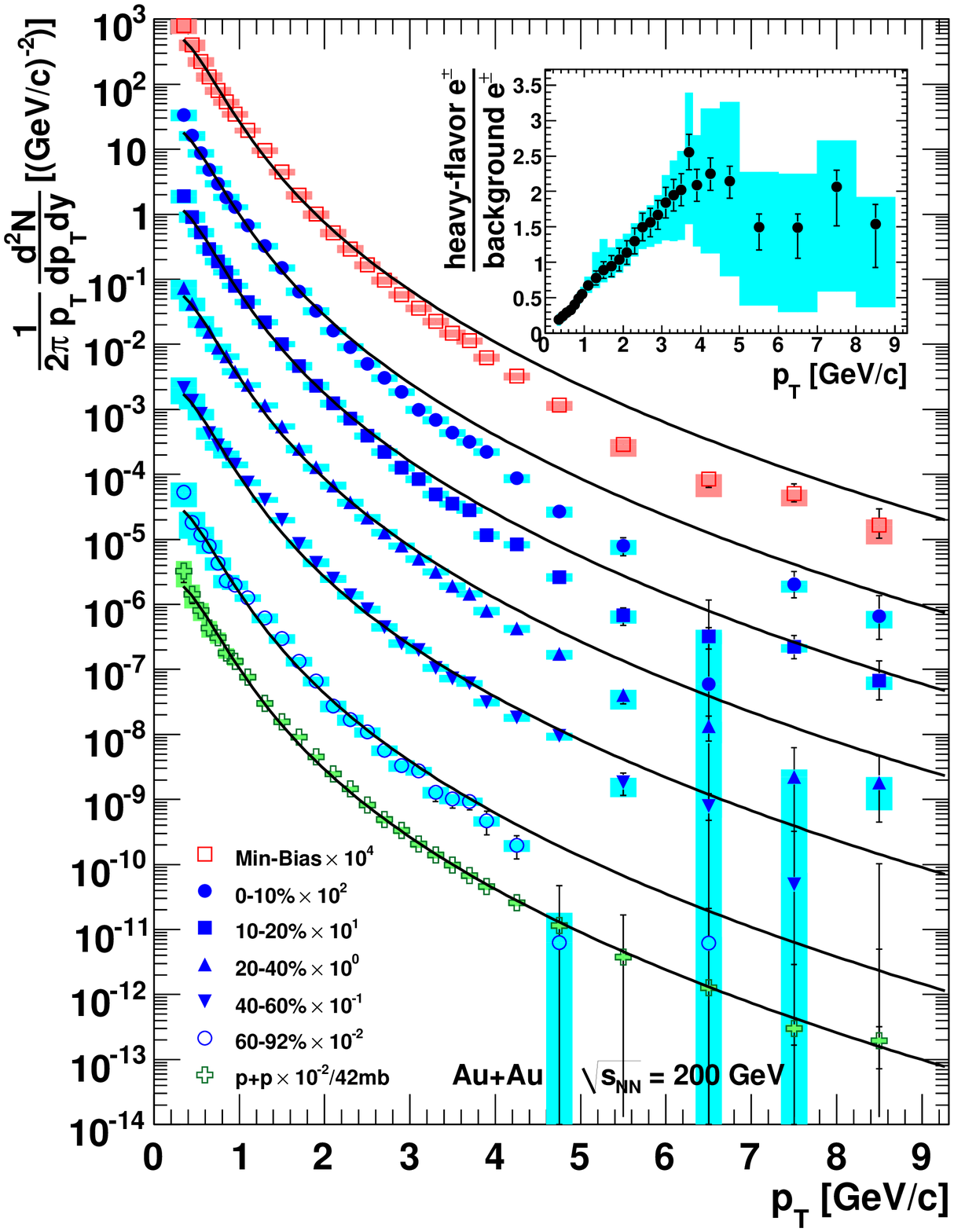}
  \caption {(Color online)
Invariant yields of (left) inclusive electrons and (right)
open heavy flavor electrons for different Au+Au centrality classes, 
scaled by powers of ten for clarity.  Error bars (boxes) 
depict statistical (systematic) uncertainties.
The insert shows the ratio of these electrons to those 
from background sources in minimum bias events.  The curves are scaled fits to
the $p+p$ spectrum.
\label{fig:AuAuinc}}
\end{figure*}

Figure~\ref{fig:AuAuinc} shows the invariant yield of inclusive 
and nonphotonic electrons in Au+Au collisions for various ranges in centrality.
From top to bottom the spectra correspond to data in MB events, the five
centrality classes, and the 2005 \pp~ data.  The various curves are fits to the
\pp data scaled by powers of ten for clarity (each factor is shown in the legend
of Fig.~\ref{fig:AuAuinc}).  These spectra are produced by the converter method
for $0.3< \pt<1.6$ GeV/$c$ and by the cocktail method for $1.6<\pt<9.0$
GeV/$c$.  The boxes and bars are systematic and statistical errors for each data
point, respectively.  Most of the nonphotonic electron yield is from the decay
of open heavy flavor mesons.  The curves overlaid in the right panel of
Fig.~\ref{fig:AuAuinc} are the fit to the corresponding data from \pp~
collisions with the spectral shape taken from a FONLL calculation~\cite{fonll}
and scaled by the nuclear thickness function $\taa$ for each centrality class.  
They are also scaled by powers of ten for clarity.  Each scaled FONLL curve
almost agrees with the measured data in the low-$\pt$ region.  On the other hand,
we can see the disagreement in the high-$\pt$ region between the measured data
and the FONLL curve in the minimum bias, 0--10\%, 10--20\%, and 20--40\% centrality
classes.  The data points are lower than each FONLL curve at high-$\pt$.  To
quantify the suppression, the nuclear modification factor is calculated in the
next subsection.

The insert box in the right panel of Fig.~\ref{fig:AuAuinc} shows 
the ratio of signal (electrons from heavy-flavor decays) to 
background as a function of $\pt$ for minimum bias events.  
This signal/background ratio ($R_{\rm SB}$) is calculated as:
\begin{eqnarray}
  R^{\rm SB} = \frac{N^{\rm HF}_e}{N^{\rm \gamma}_e + N^{\rm KV}_e}
  = \frac{N^{\rm Non\--\gamma}_e - N^{\rm KV}_e}{N^{\rm \gamma}_e +
N^{\rm KV}_e}.
\end{eqnarray}
Here, $N^{\rm Non\--\gamma}_e$ is the measured nonphotonic 
electron yield, $N^{\rm \gamma}_e$ is the yield of photonic 
electron background, and $N^{\rm KV}_e$ is the yield of electron 
backgrounds from kaons, vector mesons, quarkonia, and Drell-Yan.  Filled circles
with statistical error bars show $R^{\rm SB}$ produced by both the converter 
and cocktail methods.  Boxes are systematic errors of $R_{\rm SB}$.  
The yields from the converter and cocktail methods are combined at $\pt 
= 1.6$ GeV/$c$.  The signal to background increases rapidly with $\pt$, 
reaching one for $\pt \approx 1.5$~GeV/$c$, reflecting the small 
amount of conversion material in the detector acceptance.
 
 
\subsection{$\raa(\pt)$}

\begin{figure*}[htb]
  \includegraphics[width=0.48\linewidth]{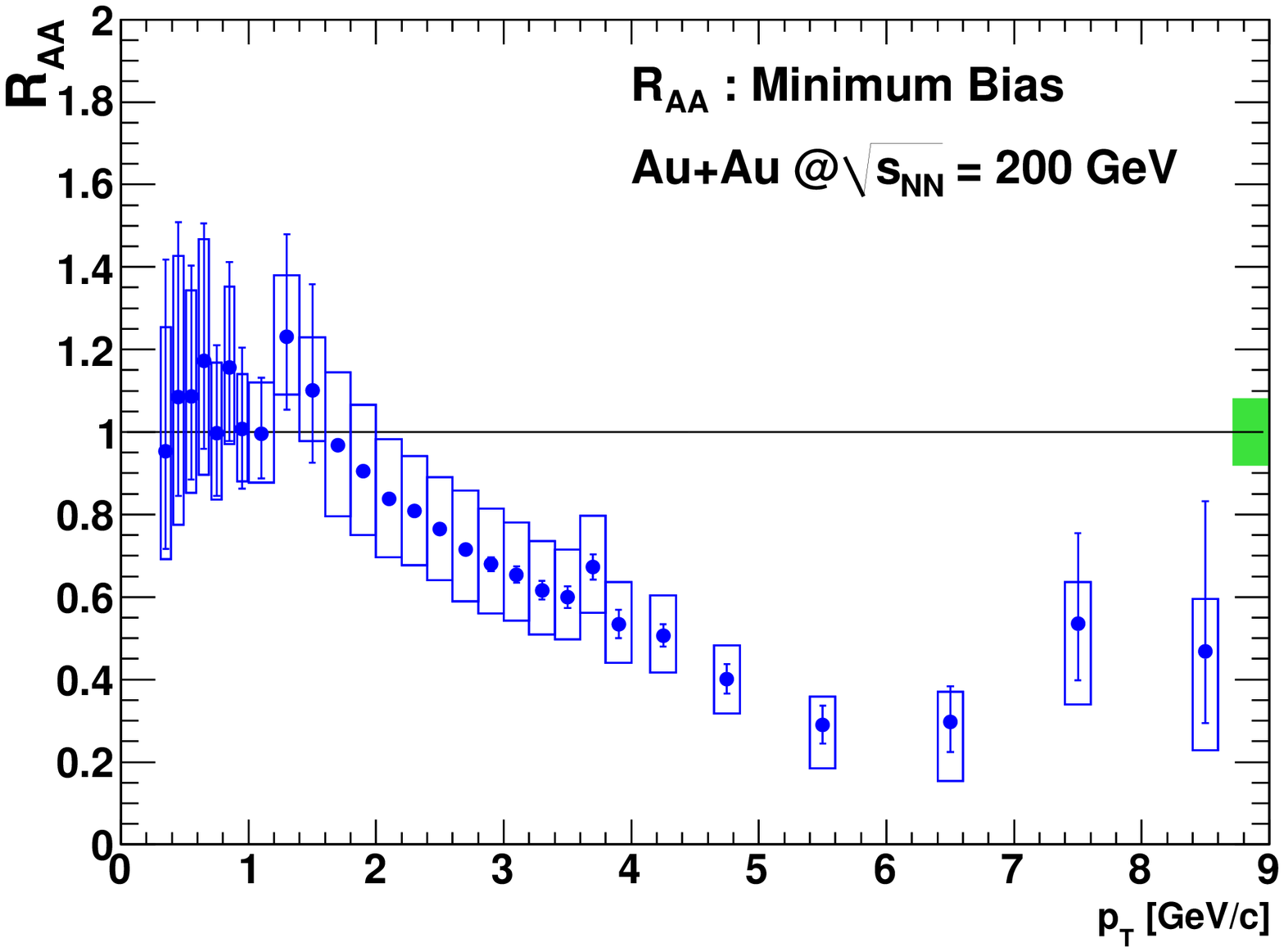}
  \includegraphics[width=0.48\linewidth]{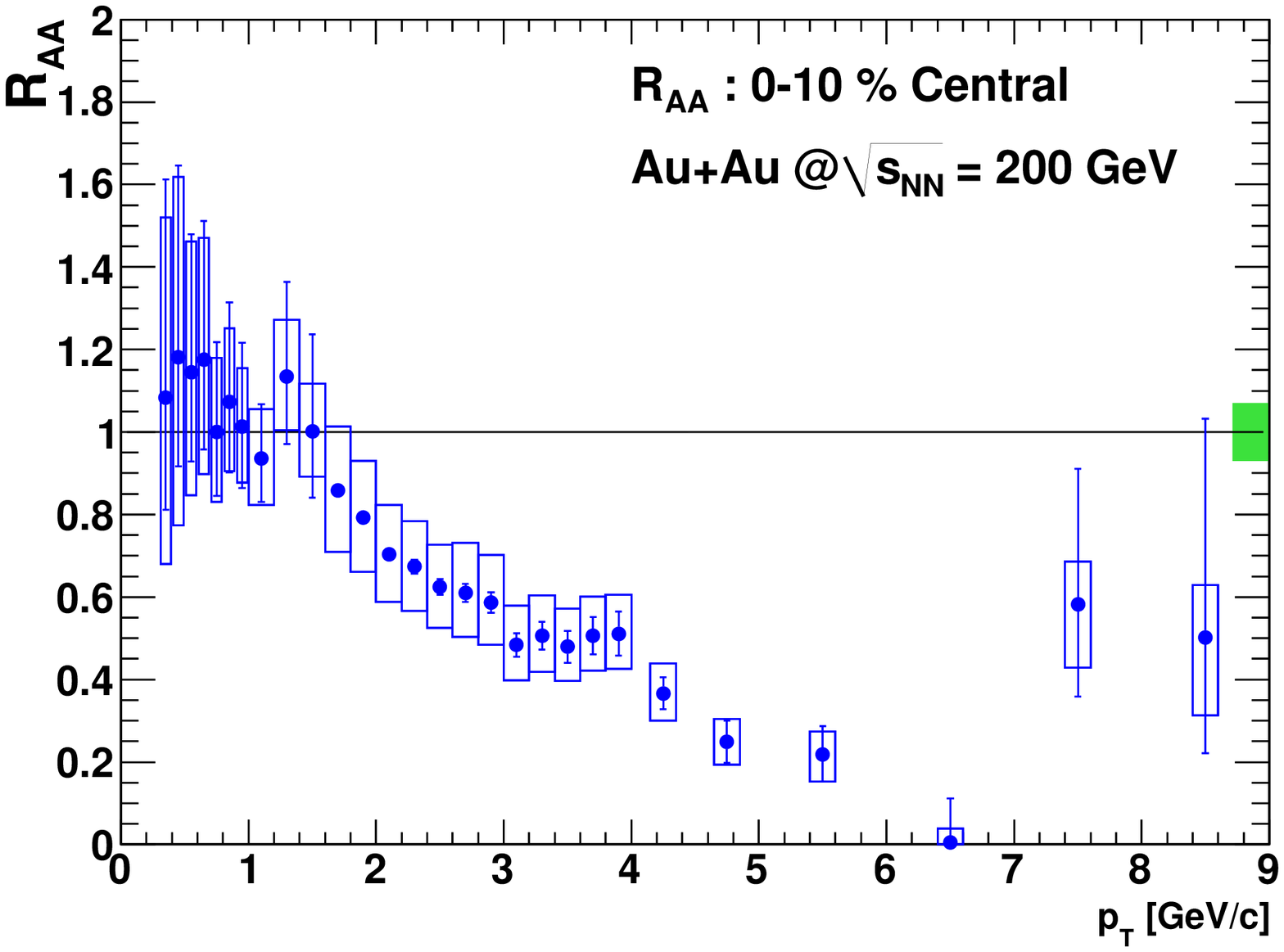}
  \includegraphics[width=0.48\linewidth]{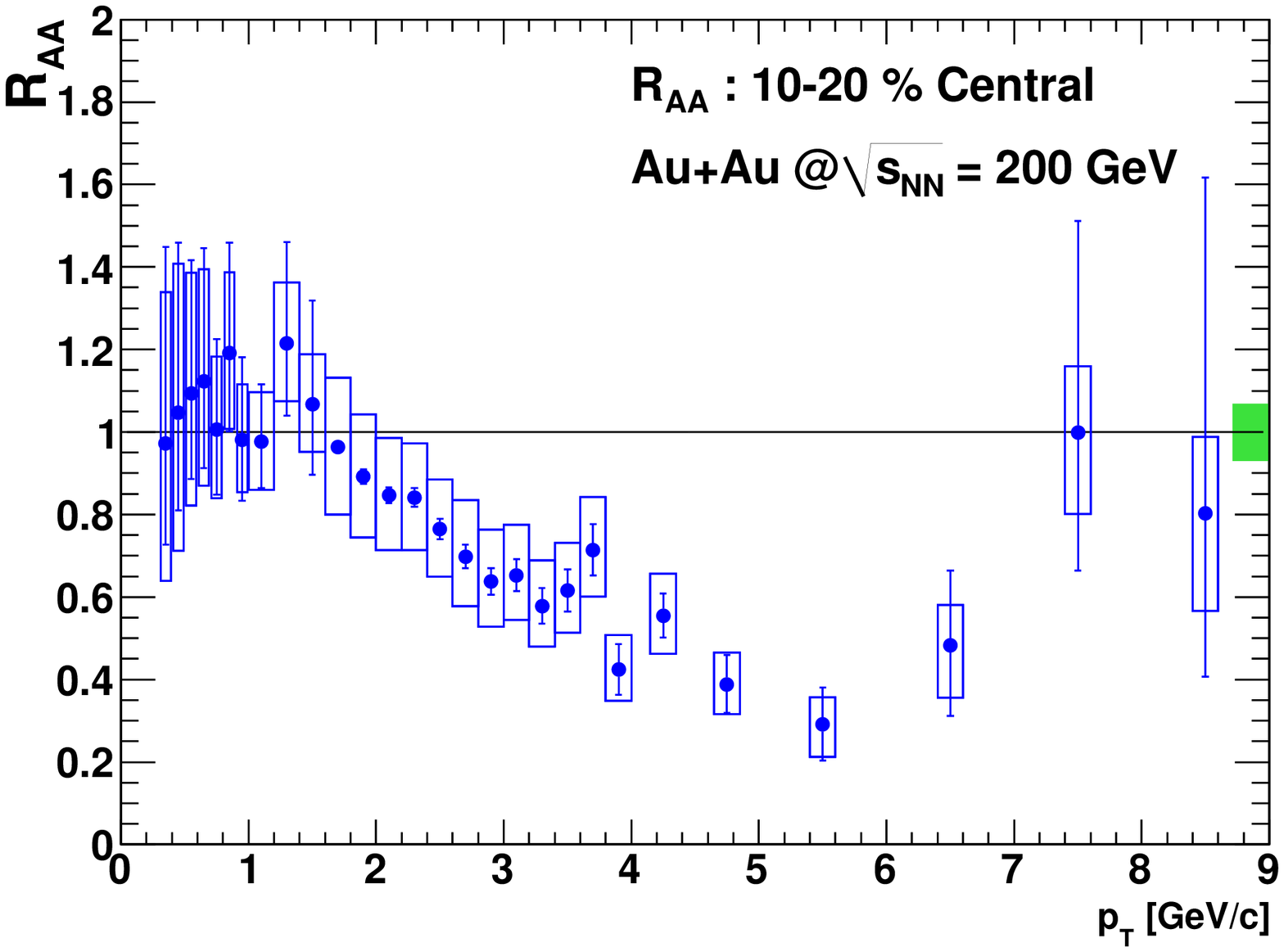}
  \includegraphics[width=0.48\linewidth]{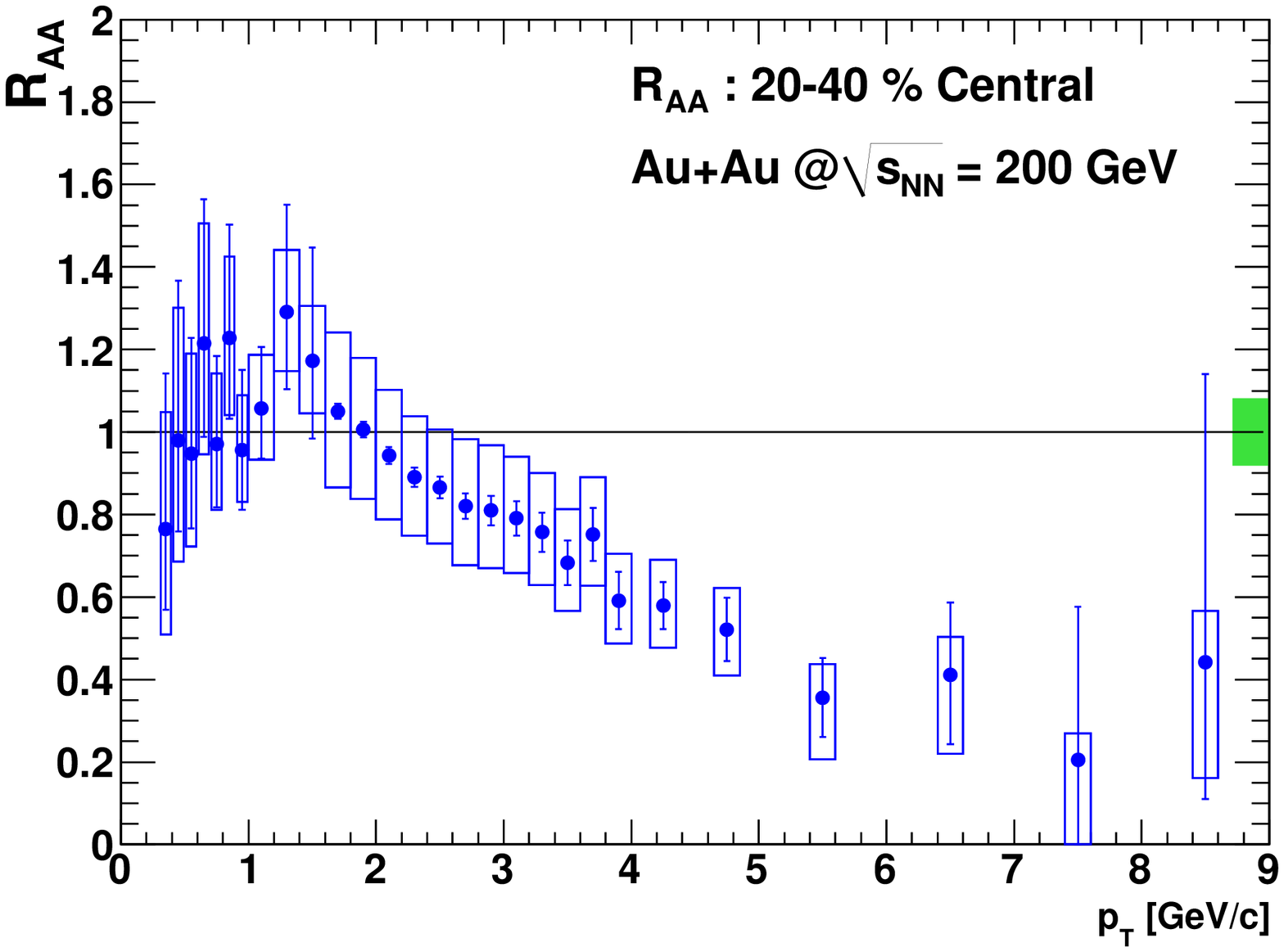}
  \includegraphics[width=0.48\linewidth]{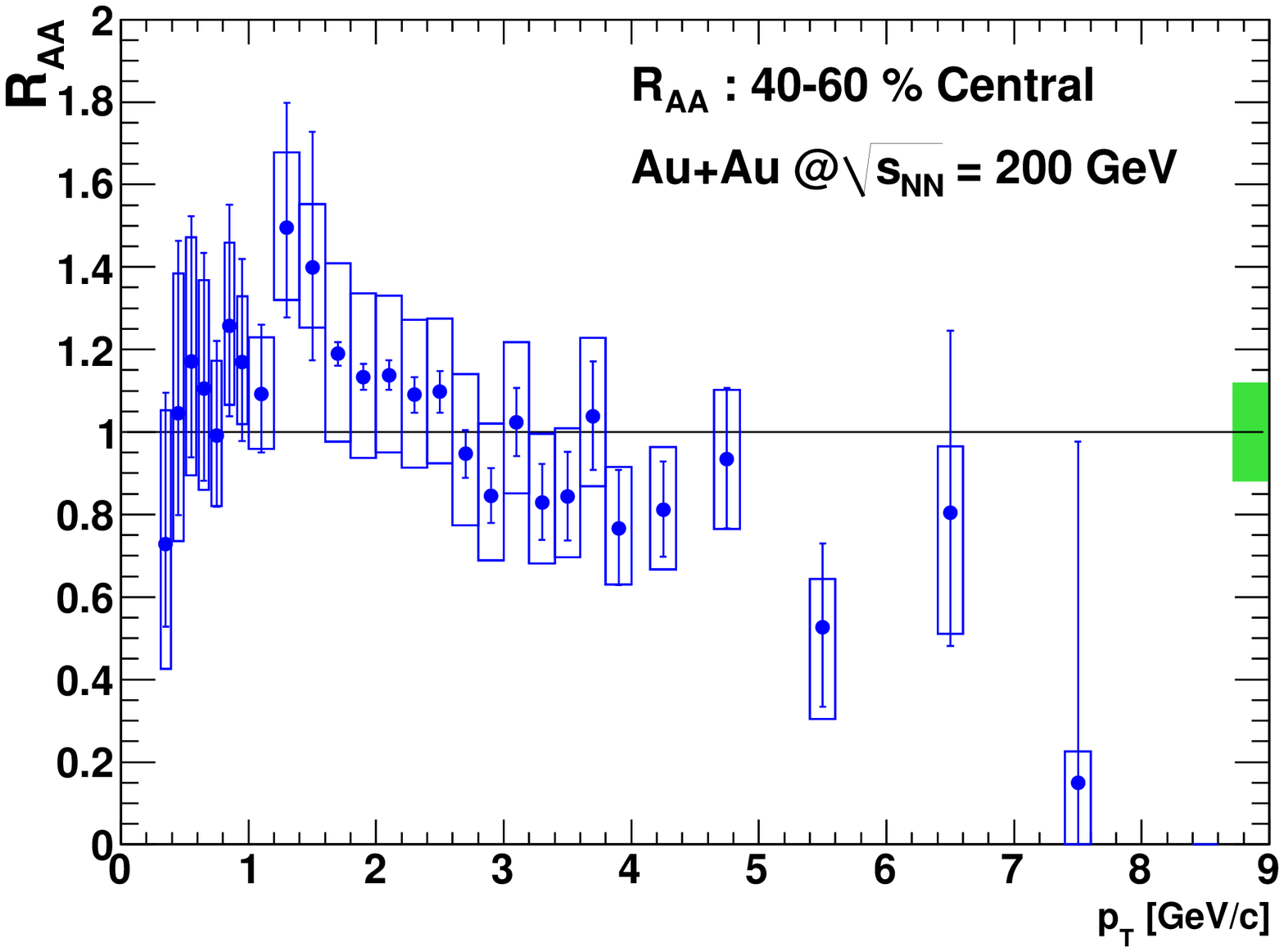}
  \includegraphics[width=0.48\linewidth]{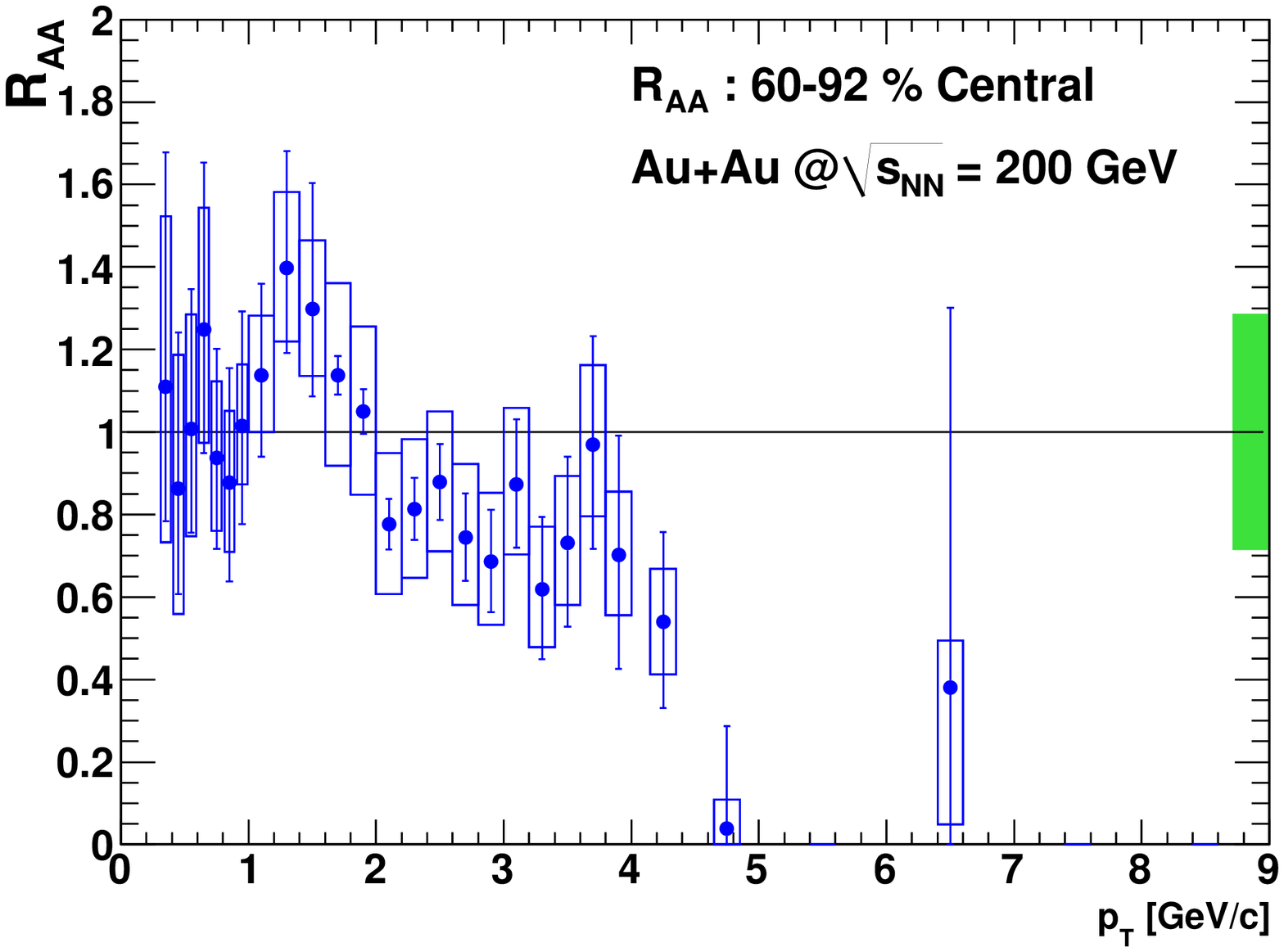}
\caption{(Color online) Open heavy flavor electron $\raa$ for the indicated
centralities.  
\label{fig:RAA_C}}
\end{figure*}

As seen in Fig.~\ref{fig:AuAuinc}, the \auau\ spectra agree well 
with the \pp\ reference at low $\pte$ for all centrality classes, but 
a large suppression with respect to \pp\ develops towards high $\pte$ 
in central collisions.  To quantify the suppression effect, the 
nuclear modification factors, $\raa(\pte)$ is calculated:
\begin{eqnarray}
  \raa(\pte) 
  &=& \frac{d\naae/d\pte}
		  {\langle \ncol \rangle \times d\nppe/d\pte} \\
	  &=& \frac{d\naae/d\pte}
		  {\langle \taa \rangle \times d\sigmappe/d\pte}
\end{eqnarray}
where $d\naae/d\pte$ is the differential invariant yield in \auau\ 
collisions and $d\nppe/d\pte$ ($d\sigmappe/d\pte$)  is the 
differential invariant yield (cross section)  in \pp\ collisions at a 
given $\pte$ bin.  For $\pte < 1.6$~GeV/$c$, $d\nppe/d\pt$ 
($d\sigmappe/d\pte$) is taken from the 2005 $p+p$ data.  At higher $\pt$, 
the  fits shown in Fig.~\ref{fig:AuAuinc} 
are used to remove statistical fluctuations, and the statistical error is
moved to a systematic error in the shape of $\raa$.  $\taa(b)$ is the nuclear
thickness function for \axa~ at an impact parameter ($b$) and 
$\ncol(b) =\taa(b) \times \sigmappin$.  Here, $\sigmappin$ is the inelastic
scattering cross 
section of \pp\ collisions (42 [mb]).  $\raa(\pte)$ for each 
centrality class is shown in Fig.~\ref{fig:RAA_C}.  Systematic errors 
consist of contributions from $\taa$ (a horizontal band around 
$\raa(\pte) = 1.0$), and from \pp\ and \auau\ data (vertical bands for each 
data point).  The statistical errors are shown as vertical bars for 
each data point.  Data points and errors for each centrality class are 
listed in the Appendix.

If no nuclear modification exists and binary scaling is correct, 
$\raa$ should be unity.  In all centrality classes, $\raa$ is consistent within
errors with unity for $\pt<2$ GeV/$c$.  However, we can 
see very clear suppression for 0--10\%, 10--20\%, 20--40\%, and minimum 
bias events in the high-$\pt$ region.  This observed strong suppression in 
the high-$\pt$ region indicates that the medium created by \auau\ 
collisions is so dense that not only light quarks, but also heavy 
quarks lose their energy, or merely that heavy quarks in \auau\ collisions 
are not produced as compared with scaled \pp\ collisions.  The total 
integrated yield in \auau\ collisions is compared with the 
binary-scaled yield in \pp\ collisions in the next subsection.
 
 
\subsection{Integrated $\raa (\npart)$}

\begin{figure*}[htbp]
  \includegraphics[width=0.48\linewidth]{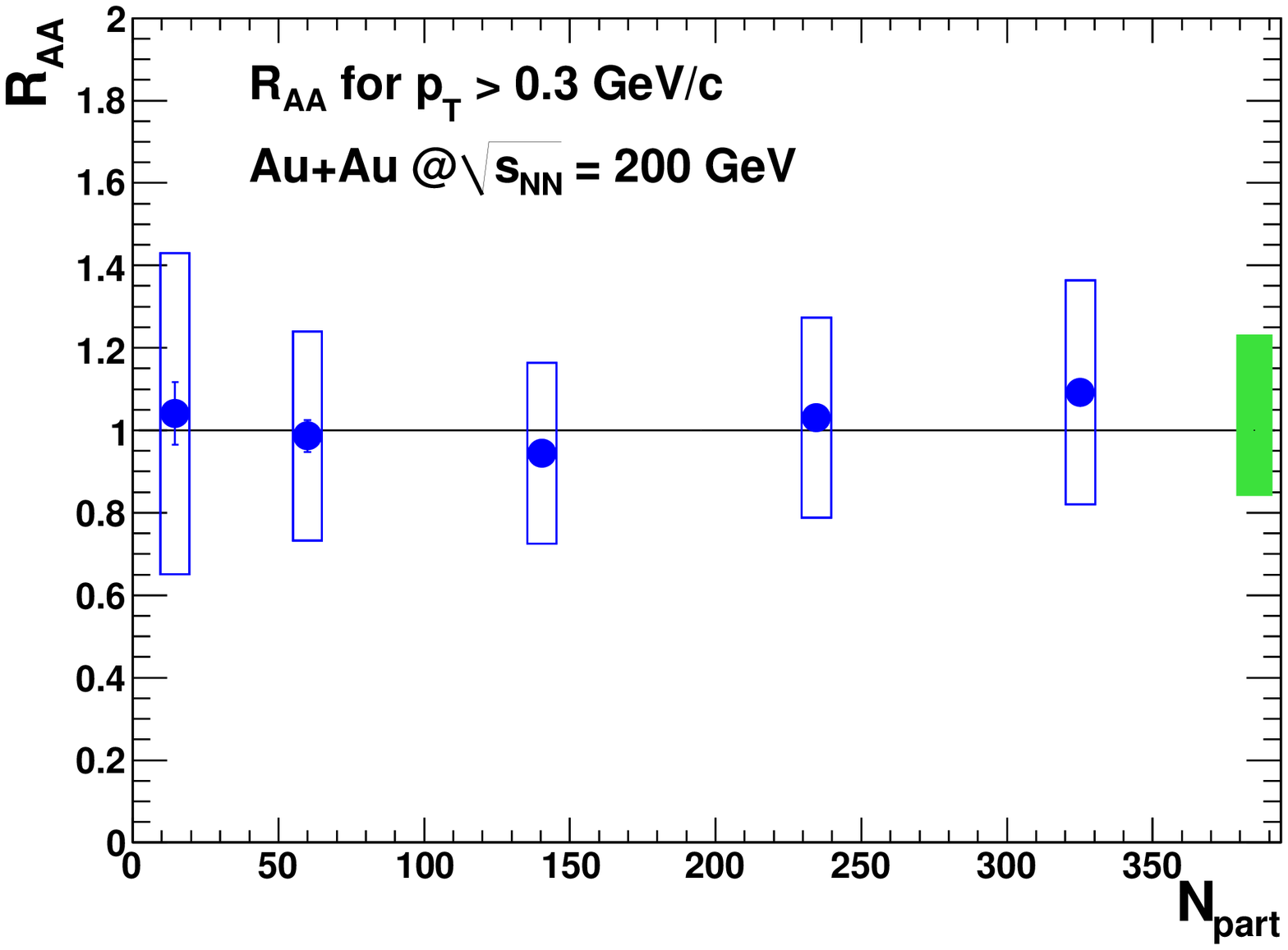}
  \includegraphics[width=0.48\linewidth]{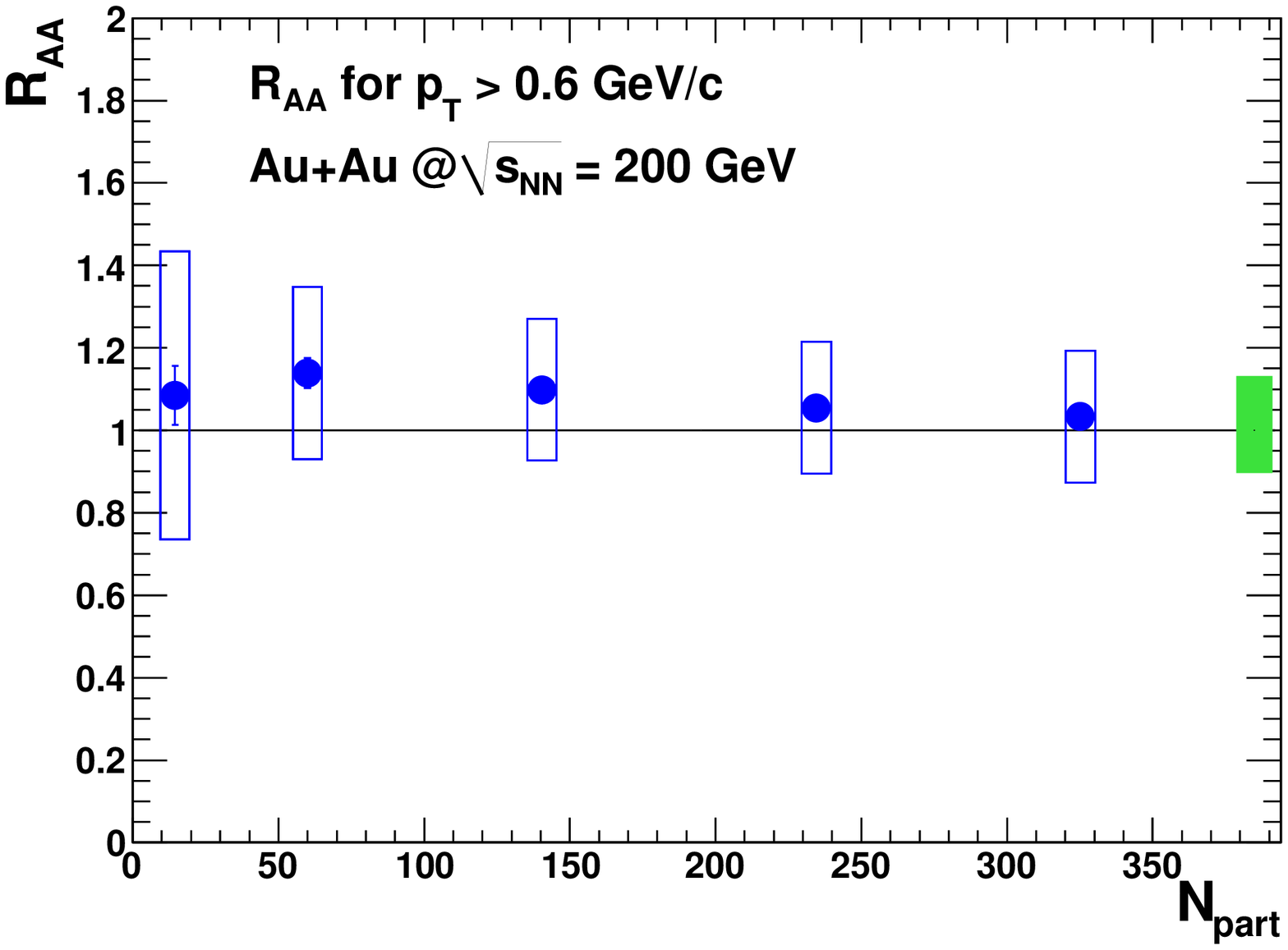}
  \includegraphics[width=0.48\linewidth]{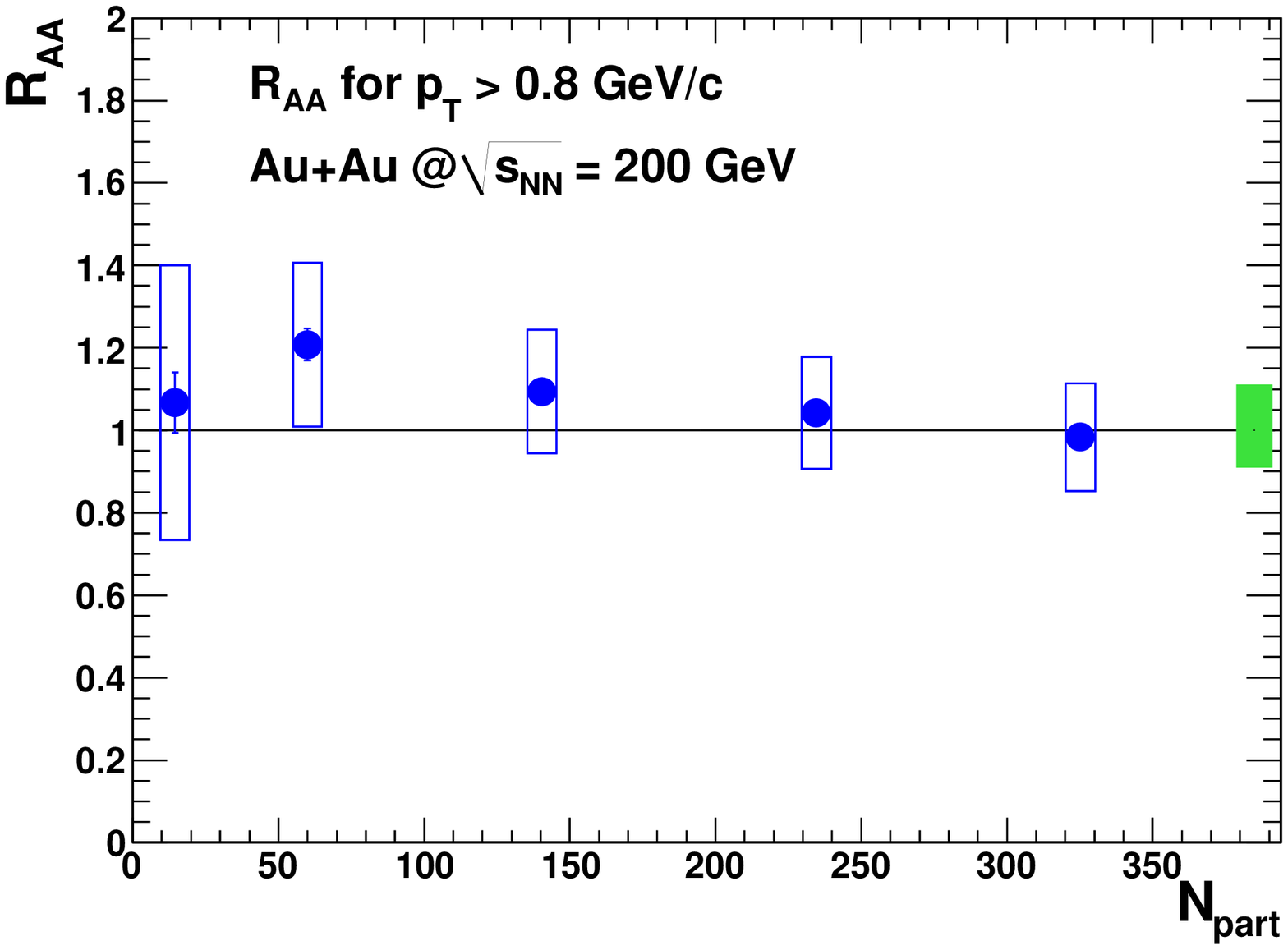}
  \includegraphics[width=0.48\linewidth]{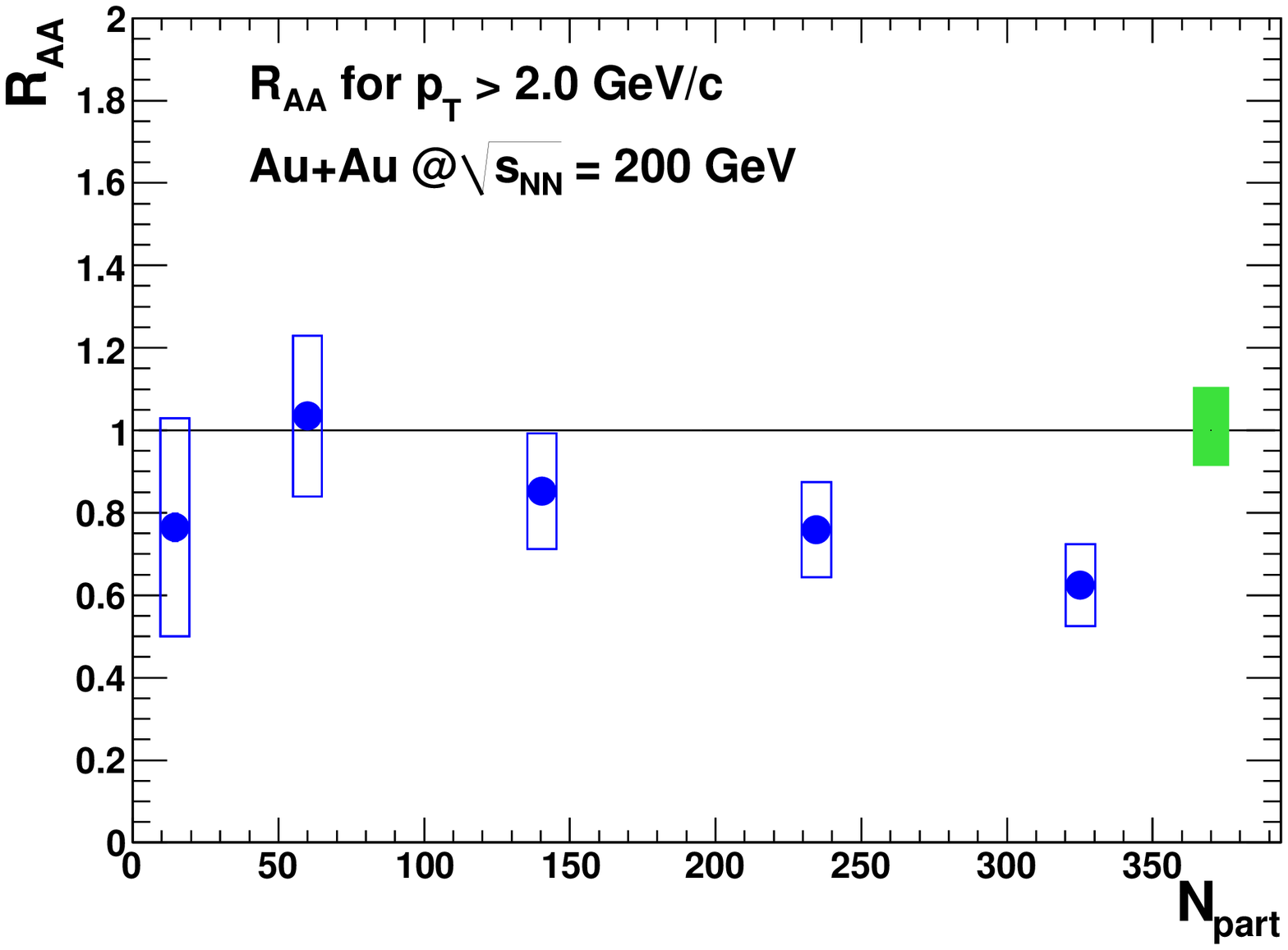}
  \includegraphics[width=0.48\linewidth]{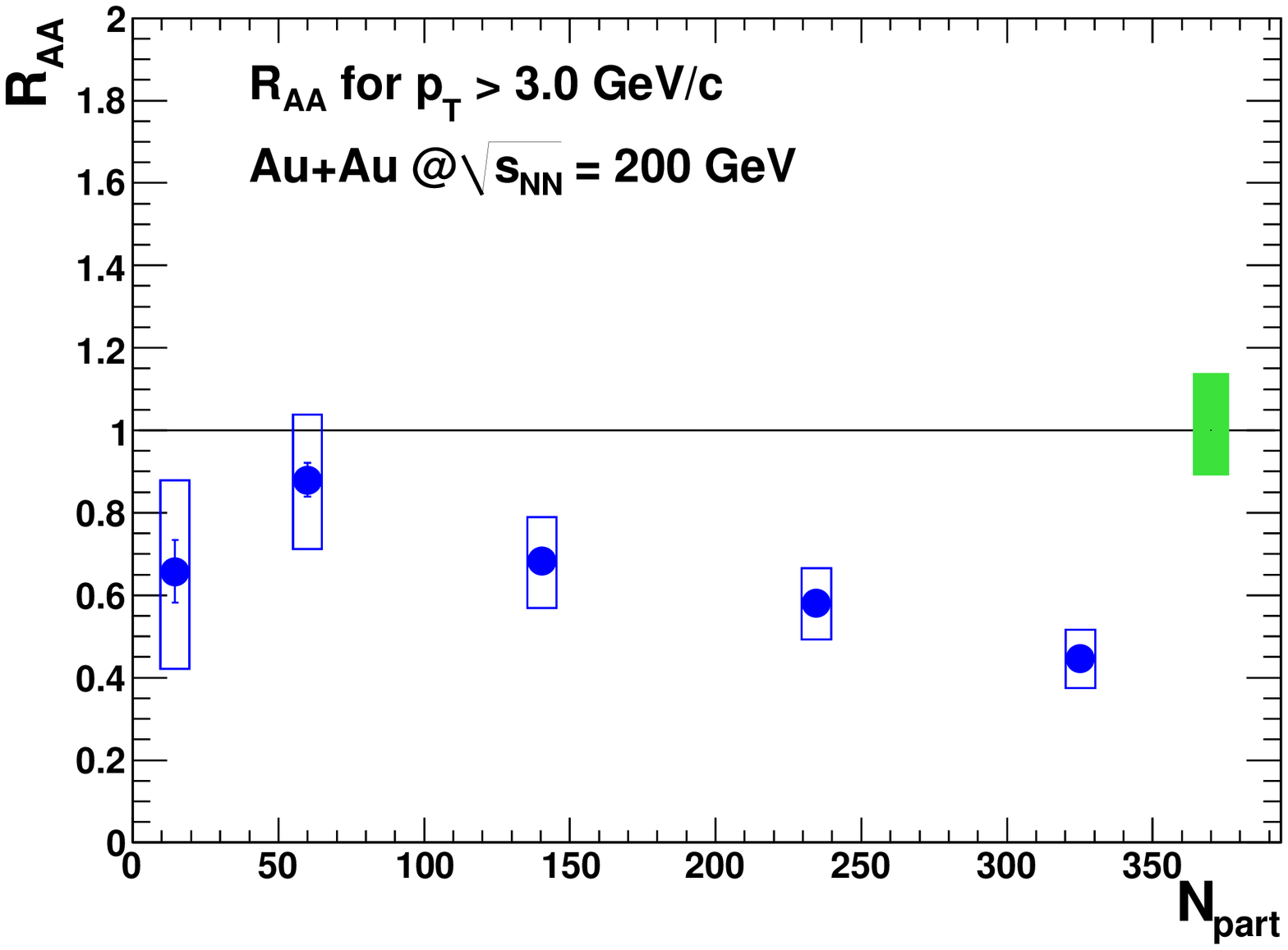}
  \includegraphics[width=0.48\linewidth]{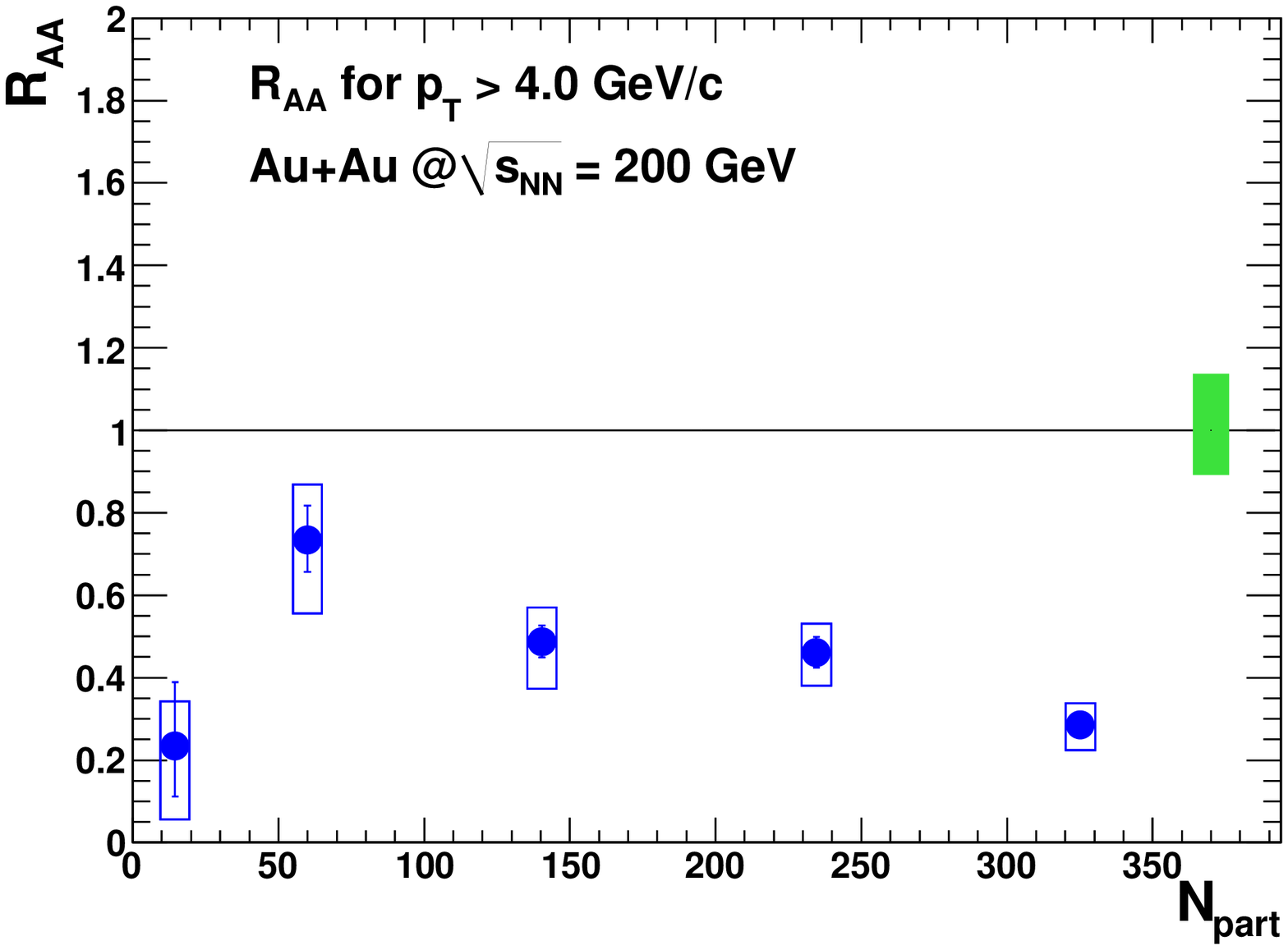}
\caption{(Color online) 
Nuclear modification factors $\raa$ for open heavy flavor electrons vs
centrality, integrated above the indicated $\pte$ ranges.  
\label{fig:Integrated_RAA}}
\end{figure*}


We calculate the $\pte$-integrated $\raa$ as a function of $\npart$ as:
\begin{eqnarray}
  \raa^{\pte}(\npart) = \frac{N^e_{\rm AuAu}(p_T)}{\langle \taa \rangle
\times
\sigma^e_{\rm pp}(p_T)}
  = \frac{N^e_{\rm AuAu}(p_T)}{\langle \ncol \rangle \times
N^e_{\rm pp}(p_T)},
\end{eqnarray}
where $N^e_{\rm AuAu}(p_T)$ is the total electron yield above a 
transverse momentum of $p_T$.  Figure~\ref{fig:Integrated_RAA} shows
$\raa^{\pte}(\npart)$ for electrons from heavy-flavor decays for six different 
integrated $\pte$ ranges as a function of the number of participant 
nucleons, $\npart$.  When the lower limit of integration is reduced to $\pte$ =
0.3 GeV/$c$, which includes more than half of the heavy-flavor decay 
electrons predicted by the FONLL calculation in \pp~collisions, $\raa$ 
is close to unity for all $\npart$.  This behavior suggests that the total yield
of electrons from heavy flavor decays in \auau~collisions is the same as the 
binary-scaled yield in \pp~collisions.  The observed strong 
suppression phenomenon can be interpreted as the result of softened 
heavy quarks due to energy loss in the created medium.  For $\pte 
> 0.6,~0.8,~2.0,~3.0$, and $4.0$~GeV/$c$, $\raa$ decreases 
systematically with centrality.  The behavior seems to be similar to 
the observed suppression for $\pi^0$ and $\eta$ mesons~\cite{ppg014,ppg051}.
However, quantitative comparison with the suppression of the light mesons
requires an understanding of the decay kinematics of open charm and bottom
mesons.

 
\subsection{Total charm cross section}

\begin{figure}[htbp]
\includegraphics[width=\linewidth]{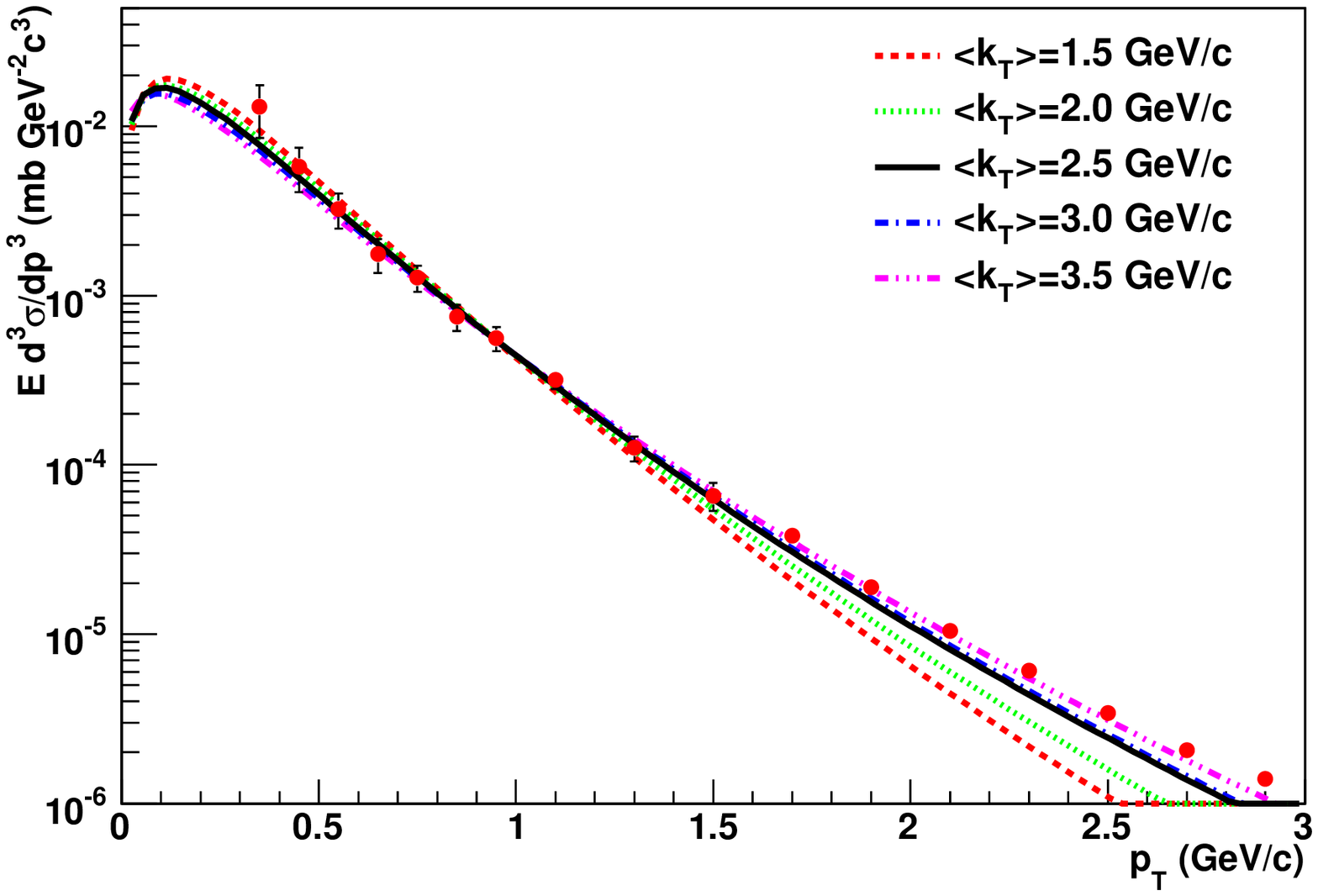}
\caption{(Color online)
Comparison of electron $p_T$ distribuion calculated by PYTHIA and the data
(filled circles).  The five curves are electron $p_T$ distribuions from charm
decay calculated using PYTHIA with different $<k_T>$ values shown in the legend.
All five curves are fitted to the data for $0.3 < p_T < 1.6$ GeV/$c$.
\label{fig:comp_pythia}}
\end{figure}

\begin{table*}[htbp]
\caption{Charm cross section per $N+N$ collision in centrality bins
(0--10\%,10--20\%,20--40\%,40--60\%, 60--92\% and Min Bias(MB)) in Au+Au and $p+p$.
The second column showS the nuclear overlap integral $T_{\rm AA}$ of the centrality.
\label{tab:charm_yield}}
\begin{ruledtabular}
\begin{tabular}{cccccc}
Centrality & $T_{\rm AA}$ (mb$^{-1}$) & $\frac{d\sigma^e}{dy}|_{p_T>0.4}$ ($\mu$b) &
$\frac{d\sigma^e_{c\bar{c}}}{dy}$ ($\mu$b) & $\frac{d\sigma_{c\bar{c}}}{dy}$
($\mu$b) & $\sigma_{c\bar{c}}$ ($\mu$b)\\
\hline
0--10\%      & $22.8\pm1.6$ & $6.50\pm0.12\pm0.92$ & $11.9\pm0.2\pm2.0$ & $134\pm
~2 \pm 27$ & $620\pm11\pm156$ \\
10--20\%     & $14.4\pm1.0$ & $6.30\pm0.13\pm0.96$ & $11.6\pm0.2\pm2.1$ & $130\pm
~3 \pm 27$ & $600\pm12\pm155$ \\
20--40\%     & $7.07\pm0.58$ & $6.15\pm0.13\pm1.07$ & $11.3\pm0.2\pm2.2$ &
$127\pm ~3 \pm 29$ & $586\pm12\pm159$ \\
40--60\%     & $2.16\pm0.26$ & $6.64\pm0.22\pm1.28$ & $12.2\pm0.4\pm2.6$ &
$137\pm ~5 \pm 33$ & $633\pm21\pm180$ \\
60--92\%     & $0.35\pm0.10$ & $5.98\pm0.42\pm1.68$ & $11.0\pm0.8\pm2.4$ &
$123\pm ~9 \pm 39$ & $570\pm40\pm201$ \\
0--92\% (MB) & $6.14\pm0.50$ & $5.96\pm0.08\pm0.96$ & $10.9\pm0.2\pm2.1$ &
$123\pm ~2 \pm 27$ & $568\pm~8\pm150$  \\
$p+p$       & NA   & $5.79\pm0.59\pm1.64$ & $10.7\pm1.1\pm3.2$    & $119 \pm 12 \pm 38 $ & $551\pm57\pm195$ \\
\end{tabular}
\end{ruledtabular}
\end{table*}

We determine the total charm cross section in $p+p$ from the
heavy-flavor electron invariant cross sections shown in Fig.~\ref{fig:PP_Fig3}.
The heavy-flavor electron cross section at low $p_T$ is domianted by charm, while
the bottom contribution is small, order of 1\%.

We follow the following steps to determine the total charm cross section.
\begin{enumerate}
\item\label{step:pt} The invariant cross section
is extraploated to $p_T$=0 to obtain the integrated
cross section $d\sigma^e/dy$ at y=0 over all $p_T$.
\item The cross section is convertered to a charm cross
section at $y=0$ $(d\sigma_{cc}/dy|_{y=0})$ using a
total electron branching ratio of charm (BR($c\rightarrow e$))
as
\begin{equation}
\frac{d\sigma_{cc}}{dy} = \frac{1}{BR(c\rightarrow e)} \frac{1}{C_{e/D}}\frac{d\sigma^e}{dy}\label{eq:dsdy}
\end{equation}
Here $BR(c\rightarrow e)$ is the total charm to electron branching ratio, and
$C_{e/D}=0.935$ is a kinematical correction factor to account for the difference
of rapidity distribution of electrons and $D$ mesons.
\item\label{step:y} Finally, the cross section is extrapolated to the entire
rapidity range to obtain the total charm production cross section
$\sigma_{c\bar{c}}$.
\end{enumerate}

We use two methods for the step~\ref{step:pt}.  In the first method, 
the total charm cross section in $p+p$ collisions is derived by 
numerically integrating the 
heavy-flavor electron cross section for $p_T>$ 0.4 GeV/$c$: 
$d\sigma_e(p_T>0.4)/dy = 5.79 \pm~0.59 \pm~1.64~\mu$b.  Here the 
systematic uncertainty is obtained by integrating the upper and the lower 
systematic uncertainty limits of the differential cross section, since the 
systematic uncertainties are essentially coherent.  
We don't include the lowest $p_T$ point, at $p_T=0.35$ GeV/$c$ since
its large individual uncertainty would make the uncertainty in the
numerical integral larger.  The cross section is then
extrapolated to $p_T$ = 0 using the spectrum shape predicted by 
FONLL: $d\sigma_e(p_T>0)/dy=10.6 \pm 1.1 \pm 3.2~\mu$b.  Here we have 
assigned 10\%
(a quarter of estimated cross section for $p_T<0.4$ GeV/$c$)
to the systematic uncertainty of the extrapolation, and
have subtracted contributions from $b \rightarrow e$ and
$b \rightarrow c \rightarrow e$ cascade decays ($0.1 \mu$b).

In the second  method for the step~\ref{step:pt}, 
we use PYTHIA to calculate the spectrum shape from charm decay electrons.
Here we use PYTHIA as an event generator for $D$-meson decays.
Since electrons are from $D$ and other charm
hadron decays, its momentum distribution in low $p_T$ part and the high $p_T$ part
are correlated.  If one can adjust the $p_T$ distributions of charm hadrons so that
the electron spectrum for just above the un-measured region, $ p_T >0.3$ GeV/$c$, is
reproduced, the $p_T$ distribution in the lower $p_T$ region should be well
described by the simulation.

Figure~\ref{fig:comp_pythia} shows comparisons between the PYTHIA 
calculations and the data.  The five curves in are from PYTHIA 
calculations, and the red points are from the data. The PYTHIA curves 
are fitted to the data for the fit range $0.3 < p_T < 1.6$ GeV/$c$, with 
their normalization to be the sole fitting parameter.  The shapes of 
PYTHIA curves are fixed. The PYTHIA curves underestimate the data for 
high $p_T$ ($p_T>1.6$ GeV/$c$).  However, in the $p_T$ ranges that is 
used for the fit, the PYTHIA curves can reasonably describe the data. 
The five curves corresponds to 5 different choices of $<k_T>$ ranging 
from 1.5 GeV/$c$ to 3.5 GeV/$c$ shown in the legend. The five curves are 
very similar in the low $p_T$ region below the data. This is because the 
electron spectra in the region is mainly determined by the decay 
kinematics.  Since the solid black curve ($<k_T>=2.5$ GeV/$c$) describes 
the data well, we use it as the central value.  We note that similar 
values of $<k_T>$ are found in our two hadron correlation analyses in 
$p+p$~\cite{ppg029}.

Using the second method, we obtained $d\sigma^{e}/dy(p_T>0) = 
10.9\pm0.6\pm2.8 \mu$b.  Here we have subtracted beauty contribution and 
added the systematic uncertainty of $\pm 1.4 \mu$b in quadrature as the 
systematic uncertainty of this extrapolation procedure. This systematic 
uncertainty is estimated as 50\% of the difference of the cross section 
with $<k_T>=3.5$ GeV/$c$ (minimum cross section) and with $<k_T>=1.5$ 
GeV/$c$ (maximum cross section). Since the second method gives a value 
of $d\sigma^e/dy$ very close to that of the first method, we use the 
value obtained from the first method in the charm cross section 
calculation.

We determine the charm production cross section, $d\sigma_{c\bar{c}}/dy$ 
= $119 \pm 12 \pm 38~\mu$b using Eq.~(\ref{eq:dsdy}). The branching 
ratio, $BR(c\rightarrow e) = 9.5 \pm 1.0$\%, is calculated using the 
following charmed hadron ratios: $D^+/D^0 = 0.45 \pm 0.10, D_s/D^0 = 
0.25 \pm 0.10$, and $\Lambda_c/D^0 = 0.1 \pm 0.05$.  In the step
~\ref{step:y}, we calculated the rapidity distribution of charm quark using 
HVQMNR~\cite{Mangano:1992kq} with CTEQ5M~\cite{Lai:1999wy} PDF.  The 
total charm cross section is determined to be $\sigma_{c\bar{c}} = 551 
\pm~57^{\rm stat} \pm~195^{\rm sys}~\mu$b.  Here we have assigned a systematic
uncertainty of 15\% to the extrapolation to the full rapidity.

The charm cross section obtained here is slightly smaller than that 
published in \cite{ppg065} ($567 \pm 57^{\rm stat} \pm 193^{\rm 
sys}~\mu$b) but agrees well within the systematic uncertainties.  The 
difference comes from the fact that we now subtract the contributions 
from $K^0_s$ and $J/\psi$ from the electron spectrum. The cross section 
is compatible with our previous measurement in the 2002 
run~\cite{Adler:2005fy} ($920 \pm 150^{\rm stat} \pm 540^{\rm 
sys}~\mu$b) and agrees well with the value derived from the yield of the 
dielectron continuum~\cite{ppg085} ($544 \pm 39^{\rm stat} \pm 142^{\rm 
sys} \pm 200^{\rm model}~\mu$b).

The FONLL cross section ($256^{+400}_{-146}~\mu$b) is compatible with 
the data within its uncertainty.  Although the data extends to high 
$p_T$ where the bottom contribution is expected to be dominant, the 
present analysis does not separate charm and bottom contributions.  The 
bottom cross section predicted by FONLL is $1.87^{+.99}_{-.67}~\mu$b, 
and the upper FONLL curve is consistent with the data.  The bottom cross 
section by FONLL is also consistent with the bottom cross section in 
$p+p$ we deduced from electron-hadron charge correlation~\cite{ppg094} 
and dielectorn continuum mass distribution~\cite{ppg085}.

%
%

Total charm yields in Au+Au in various centrality bins are evaluated in 
a similar way from the heavy-flavor electron spectra. For each 
centrality bin, the heavy-flavor electron spectrum is integrated for 
$p_T>$ 0.4 GeV/$c$ to obtain $dN^e/dy|_{p_T>0.4}$.  The electron cross 
section per $N+N$ collision, $d\sigma^e/dy|_{p_T>0.4}$ is then obtained as
\begin{equation}
\frac{d\sigma^e}{dy}|_{p_T>0.4} = \frac{1}{T_{\rm AA}}\frac{dN^e}{dy}|_{p_T>0.4}.
\end{equation}
The cross section is then extrapolated to $p_T=0$ and the contribution 
from $b$ decays and $b$ cascade decays are then subtracted to obtain the 
electron cross section from charm decay per $N+N$ collision 
$d\sigma^e_{cc}/dy$. Here we assume that $R_{\rm AA}=1.0$ for 
$p_T$-integrated $b$ production. The cross section $d\sigma_{cc}/dy$ is 
obtained using Eq.(\ref{eq:dsdy}). Finally, the cross section is 
extrapolated to the full rapidity range using the rapidtiy distribution 
of charm from HVQMNR to obtain total charm cross section 
$\sigma_{c\bar{c}}$ per $N+N$ collision.  We use the same correction 
factors $BR(c \rightarrow e)$, $C_{e/D}$, and rapidity distribution as 
in $p+p$ collisions.  Table~\ref{tab:charm_yield} summarizes the charm 
cross section per N+N collisions in each centrality bins of Au+Au 
collisions.  The charm cross sections per $N+N$ in Au+Au thus obtained 
agree well with the $p+p$ cross section for all centrality bins.  The 
data are consistent with the results from our previous measurement in 
Au+Au collisions in the 2002 run~\cite{Adler:2004ta} ($622 \pm 57^{\rm 
stat} \pm 160^{\rm sys}~\mu$b per $NN$).

\begin{figure}[htbp]
\includegraphics[width=1.0\linewidth]{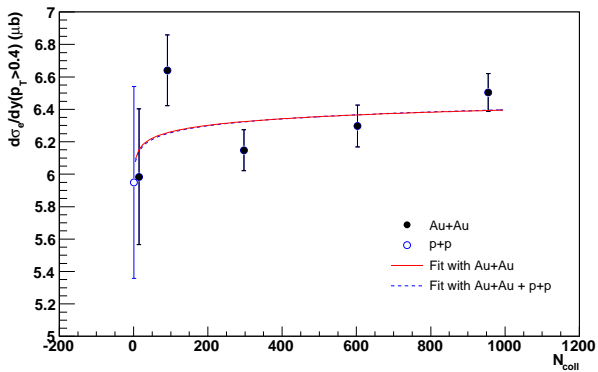}
\caption{(Color online) 
$N_{\rm coll}$ dependence of charm cross section per
$N+N$ collisions in Au+Au and $p+p$ collisions.
The error bar show the statistical errors only.  The curves in the figure
is the power-law fit $A\cdot N_{\rm coll}^\alpha$.
\label{fig:charm_vs_Ncoll}}
\end{figure}

Figure~\ref{fig:charm_vs_Ncoll} shows 
$d\sigma^e_{c\bar{c}}/dy|_{p_T>0.4}$ per $N+N$ collision in Au+Au and 
$p+p$ as a function of $N_{\rm coll}$. The cross section is more than 
half ($\simeq$ 54\%) of the total electron cross section from charm 
decay $d\sigma^e_{c\bar{c}}/dy$.  Fig.~\ref{fig:RAA_C} shows that 
$R_{\rm AA}$ is consistent with unity for $p_T<1.4$ GeV/$c$.  Thus the 
shape of the electron spectrum in Au+Au and $p+p$ is almost independent 
of $N_{\rm coll}$ in the low $p_T$ region. This indicates that the 
fraction of electron cross section below $p_T<0.4$ GeV/$c$ is almost 
independent of $N_{\rm coll}$.  In the following, we evaluate the 
$N_{\rm coll}$ dependence of the total charm cross section assuming that 
it is the same as the $N_{\rm coll}$ dependence of 
$d\sigma^e_{c\bar{c}}/dy|_{p_T>0.4}$.

To quantitatively evaluate the $N_{\rm coll}$ dependence of total charm 
cross section, we fit the data with a power-law function $A \cdot N_{\rm 
coll}^{\alpha}$.  We perform the fit with and without the $p+p$ data.  
The fitted curves are shown in Figure \ref{fig:charm_vs_Ncoll}, and the 
results of the fits are summarized in Table~\ref{tab:charm_alpha}.

Most of the sources of systematic uncertainties in the charm cross 
section are independent of $N_{\rm coll}$ and they do not affect the 
value of $\alpha$. For the systematic uncertainties in $\alpha$, we 
consider the following three remainig sources: (i) Multiplicity 
dependence of electron reconstruction efficiency (ii) systematic 
uncertainties in $R_\gamma$ (iii) systematic uncertainties in $T_{\rm 
AA}$. The systematic uncertainty from these sources are propagated to 
$\alpha$ and added in quadrature to obtain the total systematic 
uncertainty, which is shown as the second error of $\alpha$ in 
table~\ref{tab:charm_alpha}. The value of $\alpha$ for Au+Au and $p+p$ 
combined is $\alpha=1.0097\pm0.0094^{\rm stat}\pm0.0403^{\rm sys}$, 
consistent with unity, indicating that total charm cross section scales 
with $N_{\rm coll}$.

\begin{table}
\caption{ The results of power-law fits $A \cdot N_{\rm coll}^\alpha$
shown in Fig.~\ref{fig:charm_vs_Ncoll}.
In the fit, only statistical error is considered.
The uncertainty of $A$ and the first error of $\alpha$ are
the statistical erorrs obtained from the fit.  The second
error of $\alpha$ is systematical uncertainty.  
\label{tab:charm_alpha}}
\begin{ruledtabular} \begin{tabular}{cccc}
           & $A (\mu$b) & $\alpha$ & $\chi^2$/NDF \\
\hline
Au+Au      & $6.00\pm0.45$ & $1.0092\pm0.0120\pm 0.0506$ & 6.4/3\\
Au+Au + pp & $5.98\pm0.34$ & $1.0097\pm0.0094\pm 0.0403$ & 6.4/4\\
\end{tabular} \end{ruledtabular}
\end{table}

 
 
\subsection{Heavy flavor electron $v_2(p_T)$}

\begin{figure}[htpb]
\includegraphics[width=1.0\linewidth]{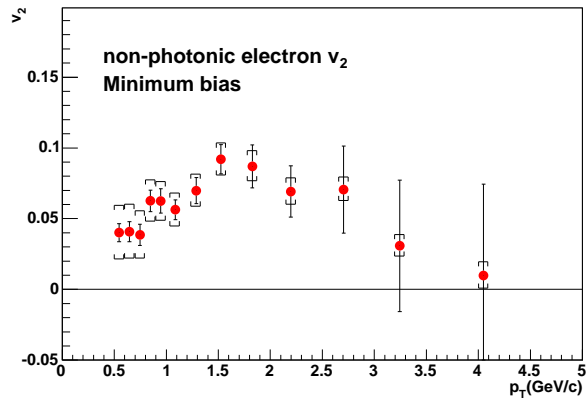}
\caption{(Color online) Nonphotonic electron $v_{2}$ as a function of $p_{T}$
in Au+Au collisions at $\sqrt{s_{\rm NN}}$ = 200 GeV.}
\label{fig:hqelectronv2}
\end{figure}

\begin{figure*}[htpb]
\includegraphics[width=0.48\linewidth]{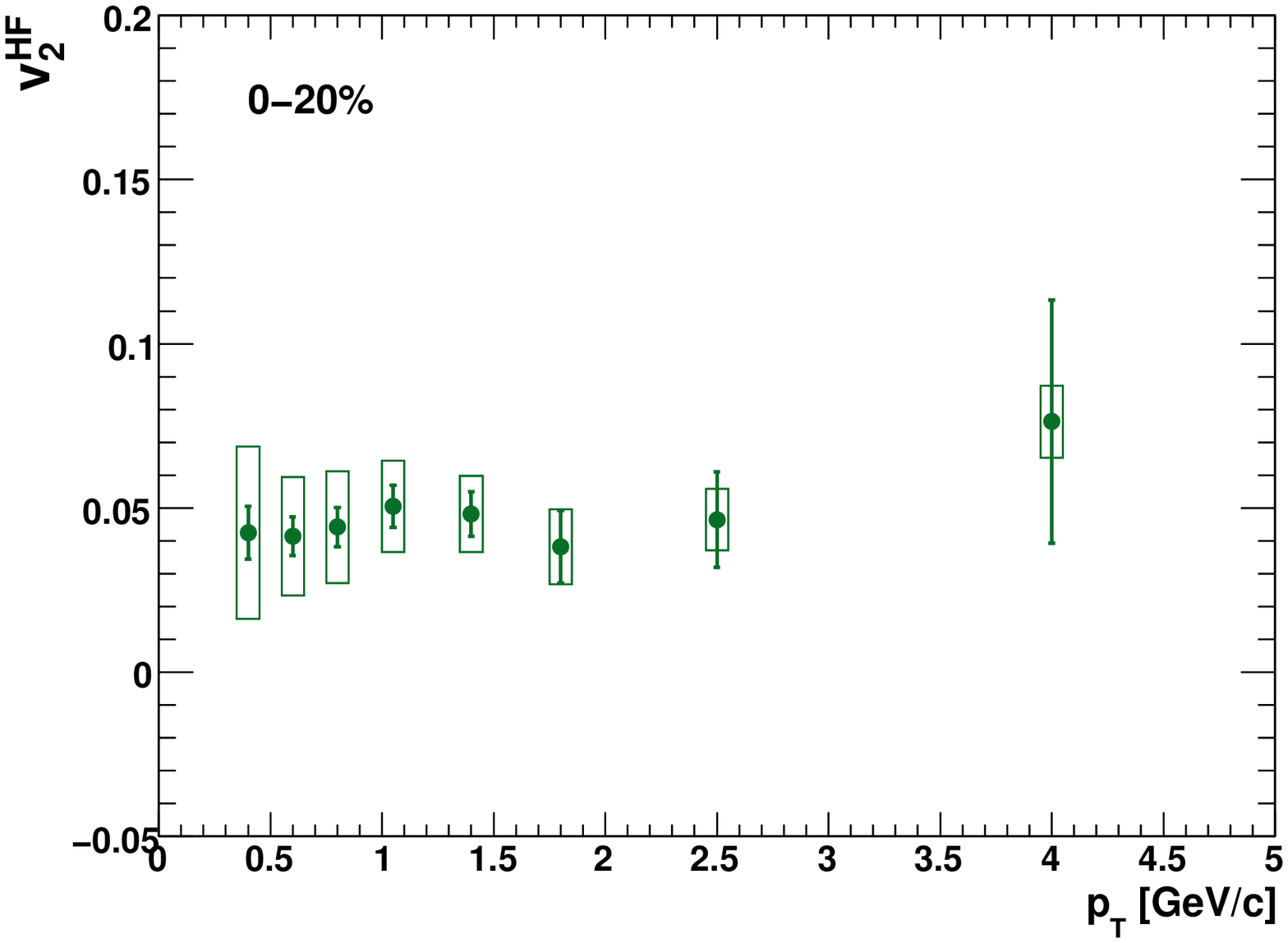}
\includegraphics[width=0.48\linewidth]{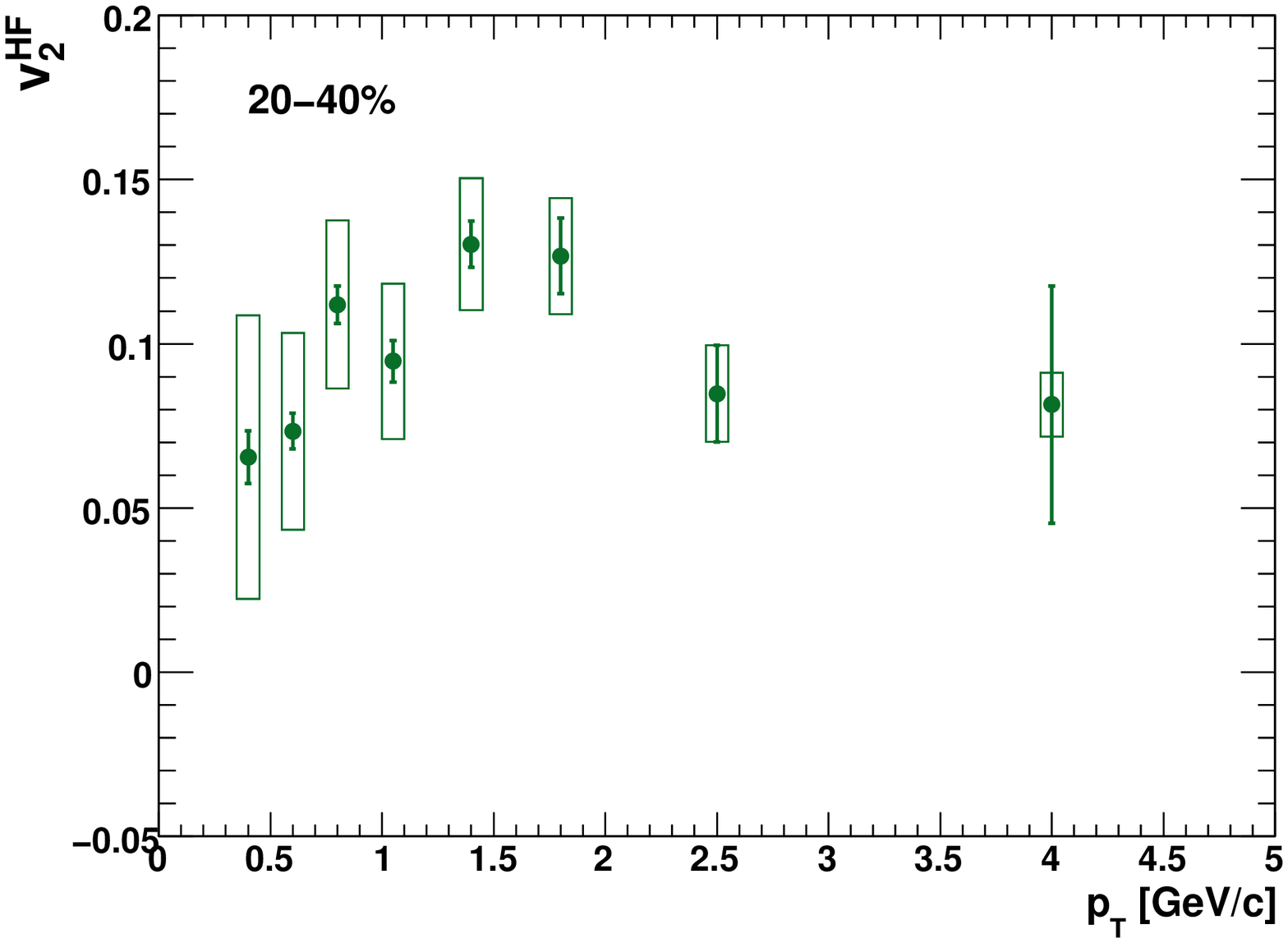}
\includegraphics[width=0.48\linewidth]{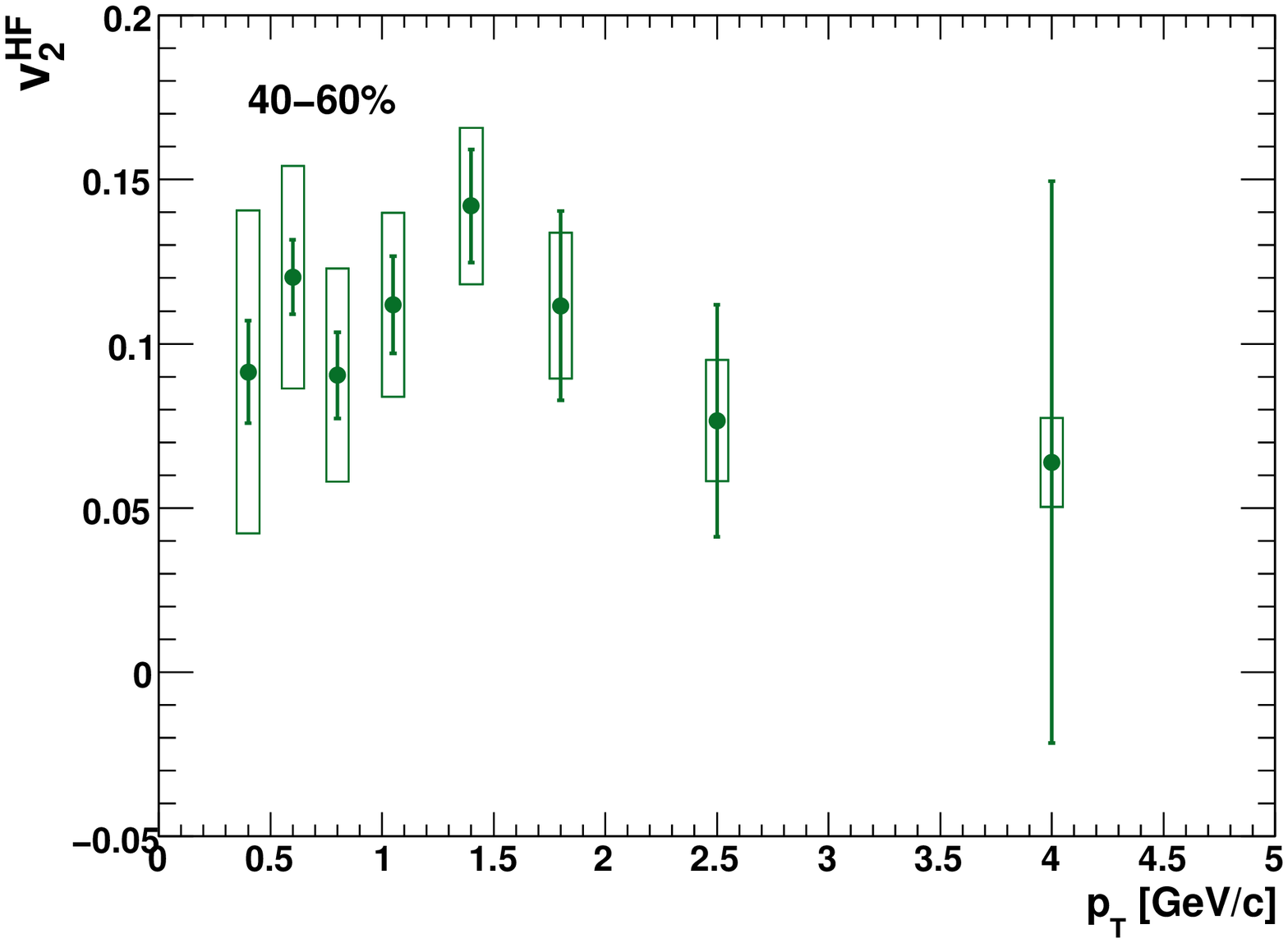}
\includegraphics[width=0.48\linewidth]{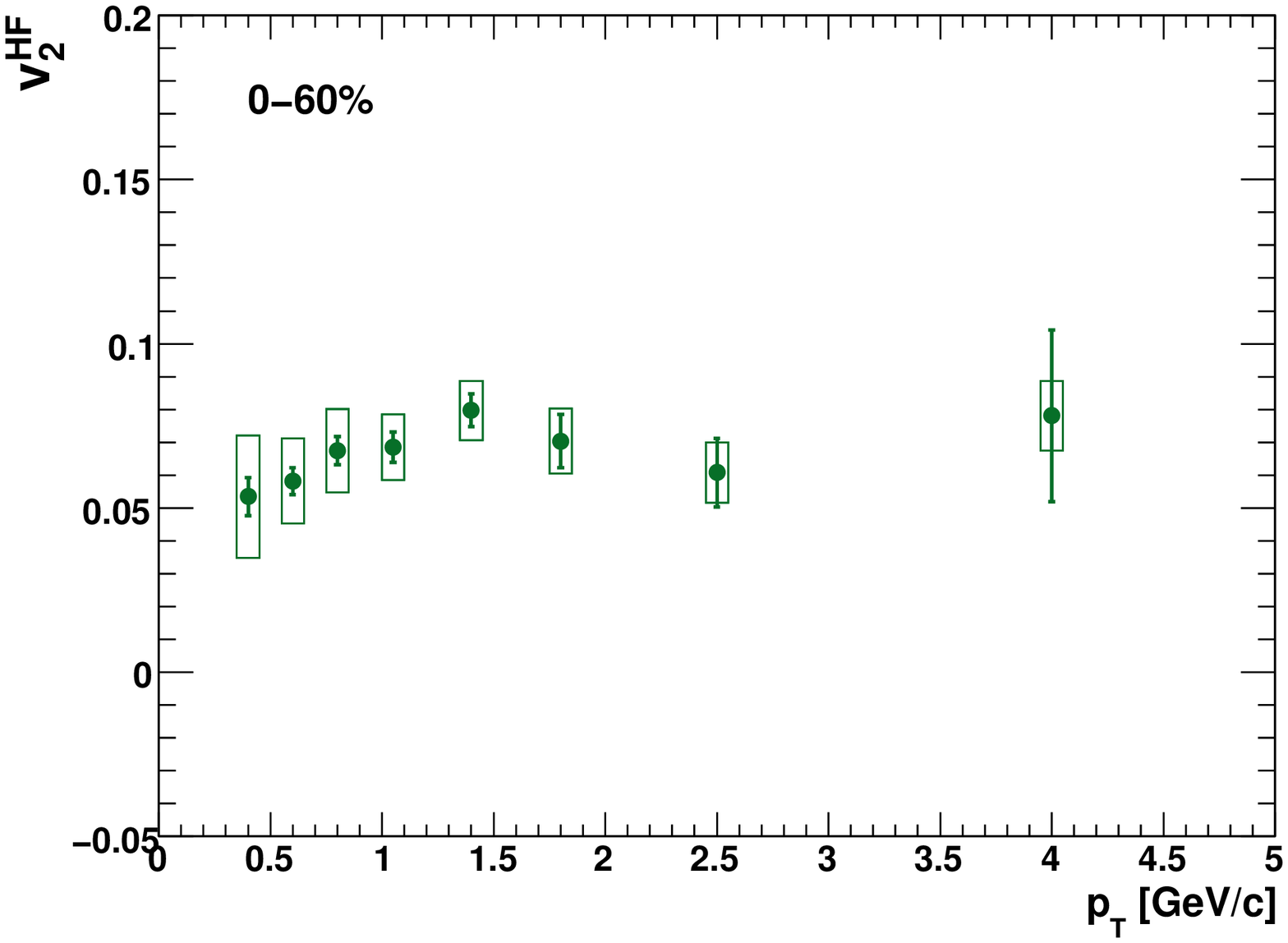}
\caption{(Color online) $v_2^{\rm HF}$ for the indicated centralities.
\label{fig:v2_0_60}}
\end{figure*}

The nonphotonic electron $v_{2}$ for minimum-bias events is shown in 
Fig.~\ref{fig:hqelectronv2}.  The vertical lines are the statistical 
errors and the $1 \sigma$ systematic uncertainties are shown as bands.  
The nonphotonic electron $v_{2}$ increases with $p_{T}$ below 2.0 
GeV/$c$ then it saturates or perhaps decreases.  This trend is similar 
to that of the meson $v_{2}$ ($\pi$ and $K$).  The main source of the 
nonphotonic electrons is semileptonic decays of heavy flavor mesons 
(charm and bottom).  Therefore the nonphotonic electron $v_{2}$ that was 
obtained by subtracting photonic electrons and kaon decays from 
inclusive electrons should be dominated by heavy flavor electrons, which 
reflects the azimuthal anisotropy of heavy quarks.  In this momentum 
region, electrons mainly decay from $D$ mesons.  Therefore the non-zero 
nonphotonic electron $v_{2}$ indicates that the $D$ meson also has a 
non-zero $v_{2}$.

Figure \ref{fig:v2_0_60} show the $v_2^{\rm HF}$ for various centrality 
ranges.  The $p_T$ binning used is coarser than that shown in the 
minimum-bias $v_2^{\rm HF}$ due to lack of statistics.  The $v_2^{\rm 
HF}$ for 60--93\% collisions is not shown as the pion $v_2$ was not 
measured by PHENIX separately for that centrality range.  At high $p_T$, 
the inclusive electron $v_2$ in the 60--93\% centrality range was 
measured to be negative by between 2 and 3 standard deviations in the 
statistical error, and this accounts for the difference between the 
minimum-bias $v_2^{\rm HF}$ and the 0-60\% $v_2^{\rm HF}$.



 
\section{DISCUSSION}\label{sec:discussion}

 
\subsection{Theory comparison $(\raa(\pte))$}

\begin{figure}[htbp]
  \includegraphics[width=1.0\linewidth]{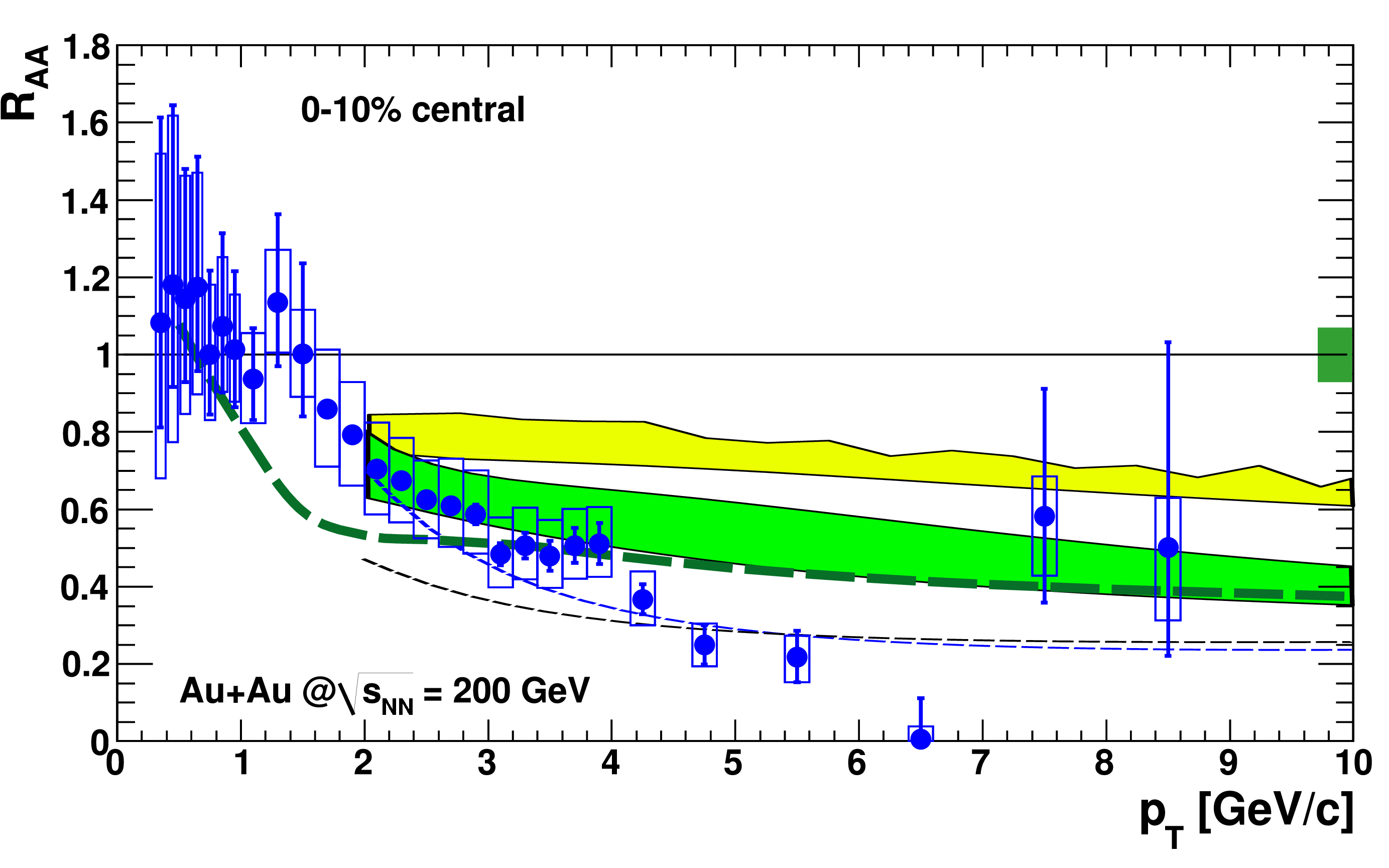}
  \caption{(Color online) $\raa$ in 0--10\% centrality class compared with
energy loss models.  The thick dashed curve is the BDMPS \cite{Armesto:2005mz} 
calculation for electrons from D and B decays.  The  bands are
DGLV \cite{Wicks:2005gt} calculations for electrons from D 
and B decays.  The lower band contains collisional energy loss as well as
radiative energy loss.  The thin dashed curves are DGLV 
calculations for electrons from D decays only.\label{fig:DGLV_BDMPS}}
\end{figure}

Before the collection and analysis of RHIC data, it was generally 
expected that the dominant mechanism for the suppression of heavy quarks 
in the medium would be gluon radiation, and thus the heavy quarks would 
not be suppressed as much as the light quarks due to the dead-cone 
effect.  Predictions from the BDMPS model and DGLV model for radiative 
energy loss can be seen in Fig.~\ref{fig:DGLV_BDMPS} 
\cite{Armesto:2005mz,Wicks:2005gt}.  These models do not account for the 
suppression of the heavy flavor electrons, but they are sensitive to the 
relative admixture of D and B mesons produced in the collisions as D 
mesons are predicted to be suppressed due to gluon radiation more than 
are B mesons.  However, recent measurements by PHENIX \cite{ppg094} 
suggest that the admixture of D and B is comparable to what was used in 
the BDMPS and DGLV calculations.

Mustafa \cite{Mustafa01} found that radiative and elastic scattering 
energy loss for heavy quarks are comparable over a very wide kinematic 
range accessible at RHIC.  Contrary to what was previously thought, 
collisional energy loss should be taken into account in the calculation 
of suppression of heavy flavor mesons in Au+Au collisions.  
Figure~\ref{fig:DGLV_BDMPS} shows the DGLV prediction for suppression 
when collisional energy loss is taken into account in addition to 
radiative energy loss \cite{Wicks:2005gt}.

\begin{figure}[htbp]
  \includegraphics[width=1.0\linewidth]{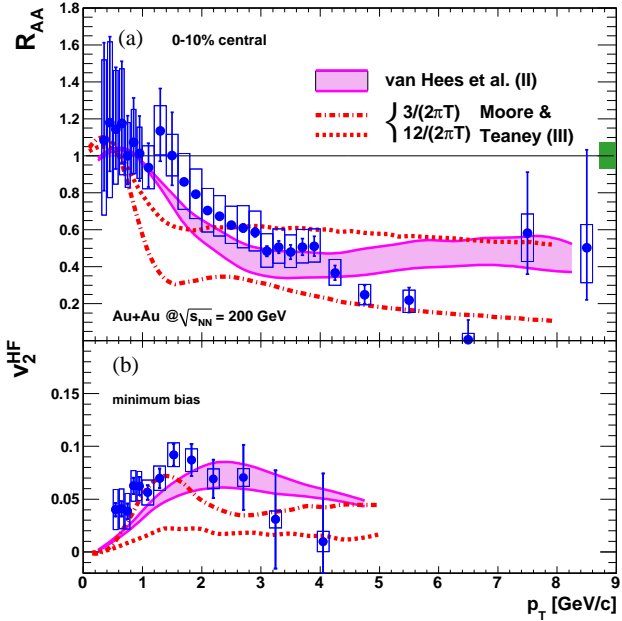}
  \caption{(Color online) Comparison of Langevin-based models from
\cite{Moore:2004tg,vanHees,vanHees:2007a} to the heavy flavor electron $\raa$
for 0--10\% centrality and $v_2$ for minimum-bias collisions.  
\label{fig:Langevin}}
\end{figure}

In \cite{Moore:2004tg}, it was shown that a Langevin-based heavy quark 
transport model can qualitatively explain the large suppression (and 
azimuthal anisotropy) of electrons from heavy flavor decays in Au+Au 
collisions.  The model places a heavy quark into a thermal medium, and 
assumes that the interaction of the heavy quark with the medium can be 
described by uncorrelated momentum kicks.  Contrary to the models 
described above, the interaction in the Langevin model is given 
exclusively by elastic collisions, which is a good approximation for 
quarks which are not ultra-relativistic in the center of mass frame of 
the collision. The parameter which is tuned in this model is the heavy 
quark diffusion coefficient. While the above Langevin model fails to 
simultaneously describe $R_{\rm AA}$ and $v_2$ for a single value of the 
diffusion coefficient, another Langevin-based model 
\cite{vanHees,vanHees:2007a} is in good agreement with both the 
suppression and anisotropy.  In this model, the elastic scattering is 
mediated by resonance excitation of D and B- like states in the medium.  
The theoretical evidence for the existence of such resonance states 
comes from lattice computations.  Figure~\ref{fig:Langevin} shows the 
calculations from these two models.

\begin{figure}[htbp]
  \includegraphics[width=1.0\linewidth]{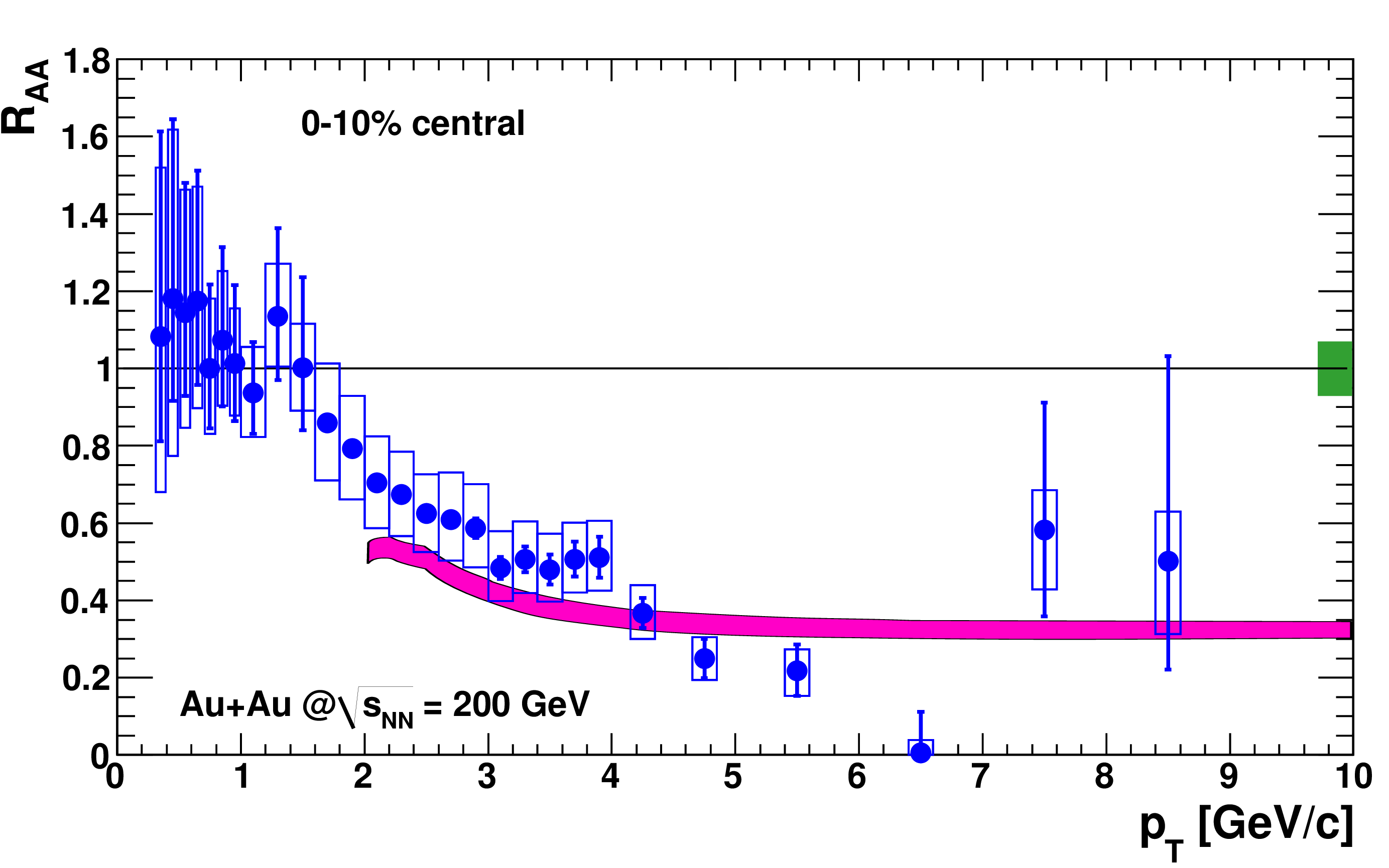}
  \caption{(Color online) $\raa$ in 0--10\% centrality class compared with a
collisional dissociation model \cite{Vitev01} (band) in \auau~collisions.
\label{fig:RAA_theory_03}}
\end{figure}

Recently, it has been suggested that collisional dissociation of heavy 
quarkonia in the quark-gluon plasma \cite{dissociation01} may be a 
possible explanation for suppression of $J/\psi$ production in heavy ion 
collisions.  A.~Adil and I.~Vitev investigated the pQCD dynamics of open 
charm and bottom production and, in the framework of the GLV approach 
extended to composite $q\bar{q}$ systems, derived the medium induced 
dissociation probability for $D$ and $B$ mesons traveling through dense 
nuclear matter~\cite{Vitev01}.  They showed that the effective energy 
loss, which arises from the sequential fragmentation and dissociation of 
heavy quarks and mesons, is sensitive to the interplay between the 
formation times of the hadrons and QGP and the detailed expansion 
dynamics of hot nuclear matter.  Figure \ref{fig:RAA_theory_03} shows 
their result as the band.

Most of the models calculate the nonphotonic electron production 
assuming the same chemical composition of charm and beauty hadrons in 
p+p and Au+Au collisions.  As it has been observed for the light hadrons 
\cite{star_baryon_meson}, one could expect a modification of the charm 
hadron chemical composition in the most central Au+Au collisions.  In 
particular, an enhancement of the $\Lambda_c$ production has been 
predicted \cite{lambda_c_01}.  A $\Lambda_c$ enhancement leads naturally 
to a nonphotonic electron $R_{\rm AA}$ smaller than one due to a smaller 
semileptonic decay branching ratio of charm baryons compared to charm 
mesons and also due to a softer spectrum of the electrons from the charm 
baryon decay.  A $R_{\rm AA}$ of about 0.65 for electrons from charm 
hadron decay is predicted when a charm baryon to charm meson ratio in 
central Au+Au collisions close to one is assumed \cite{lambda_c_02}.

\begin{figure*}[htbp]
\includegraphics[width=0.48\linewidth]{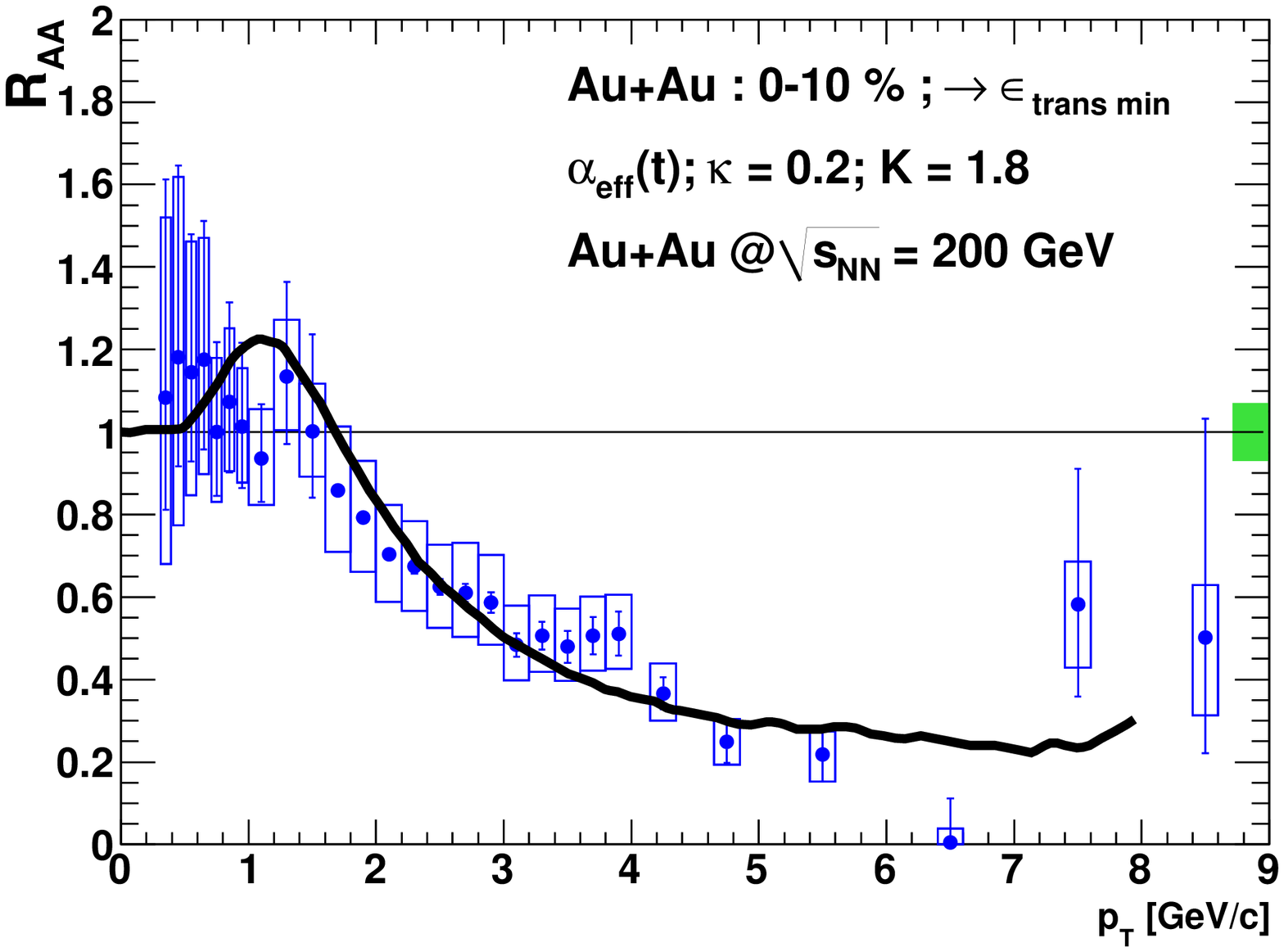}
\includegraphics[width=0.48\linewidth]{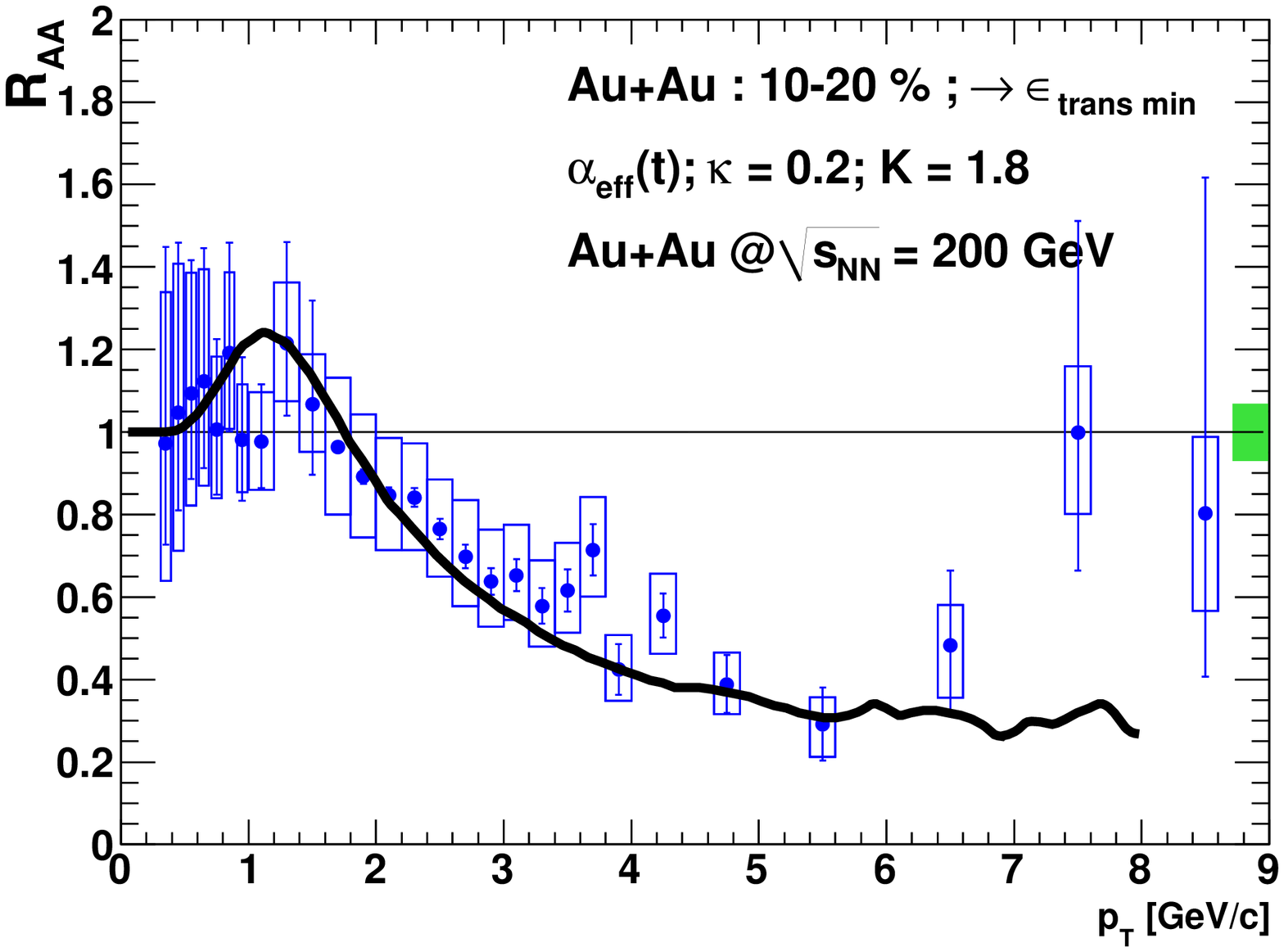}
\includegraphics[width=0.48\linewidth]{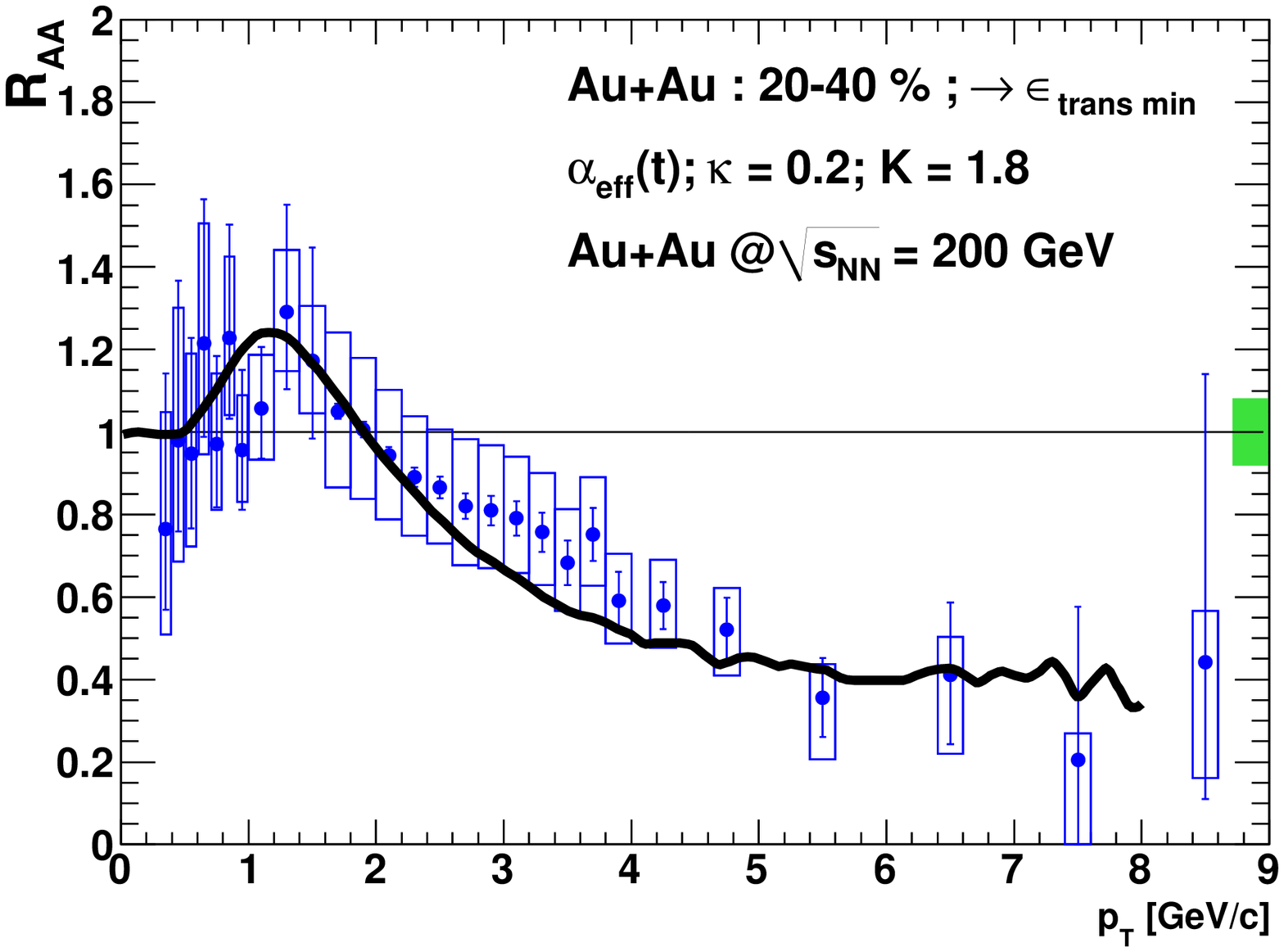}
\includegraphics[width=0.48\linewidth]{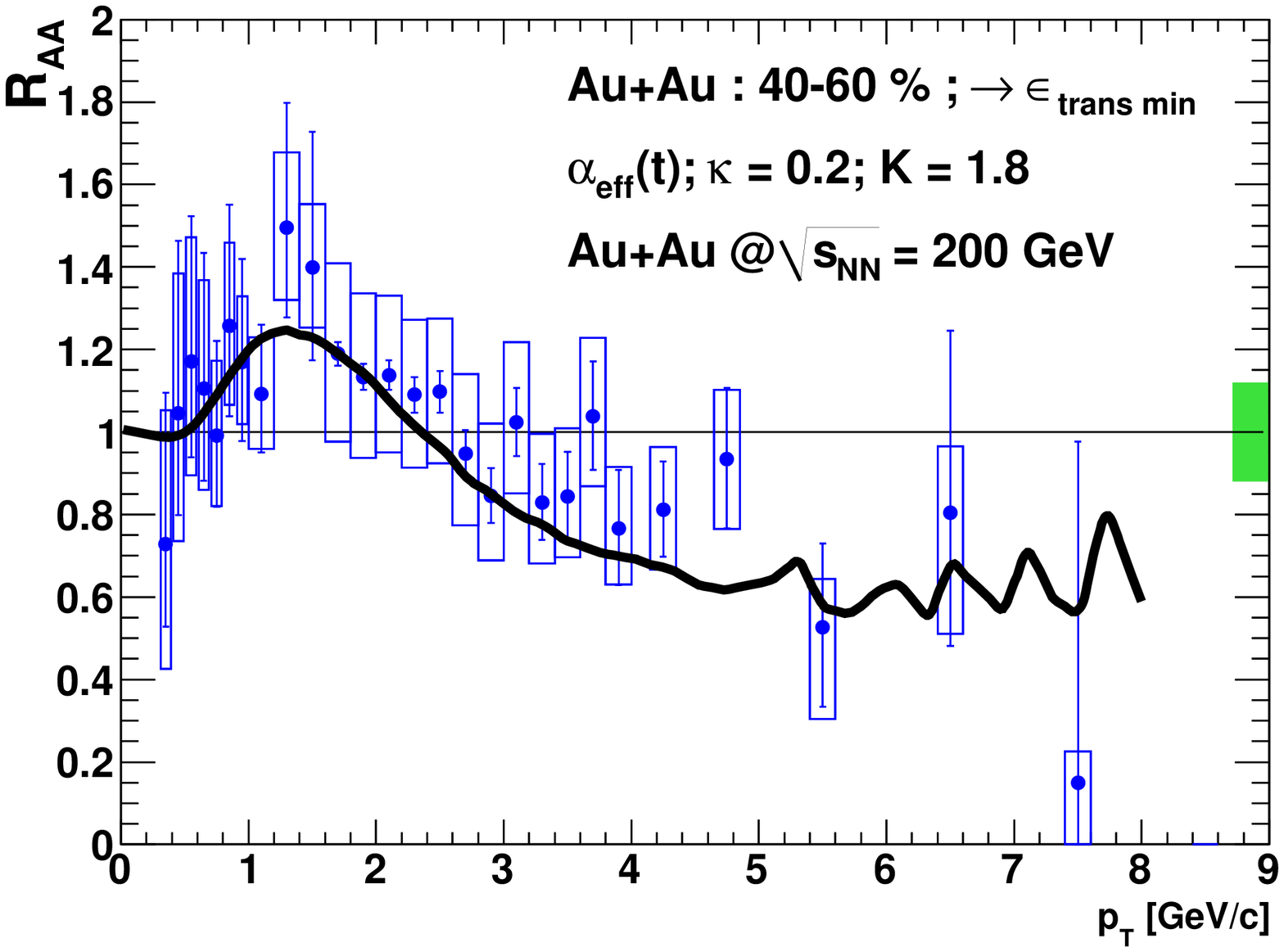}
\includegraphics[width=0.48\linewidth]{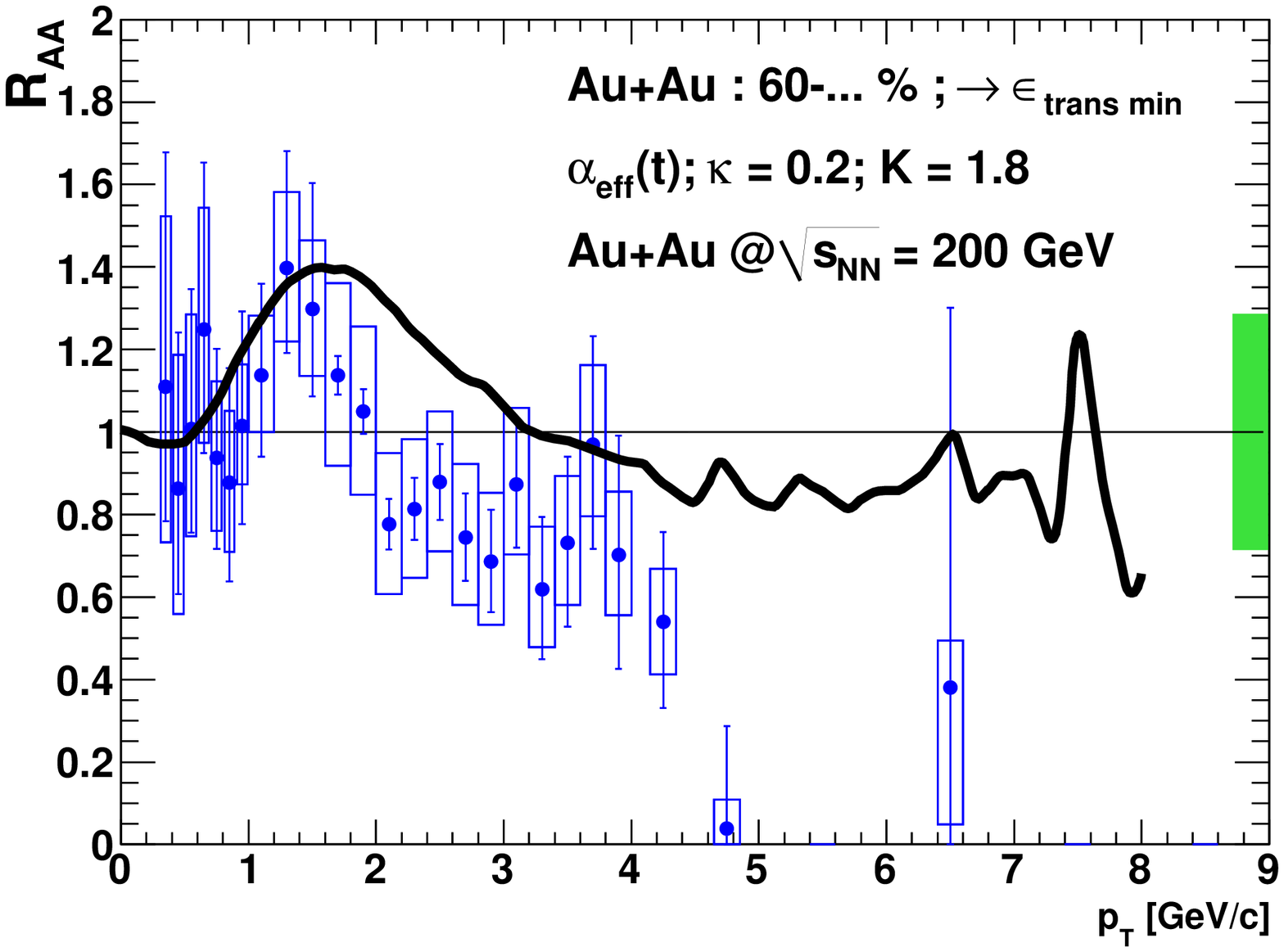}
\includegraphics[width=0.48\linewidth]{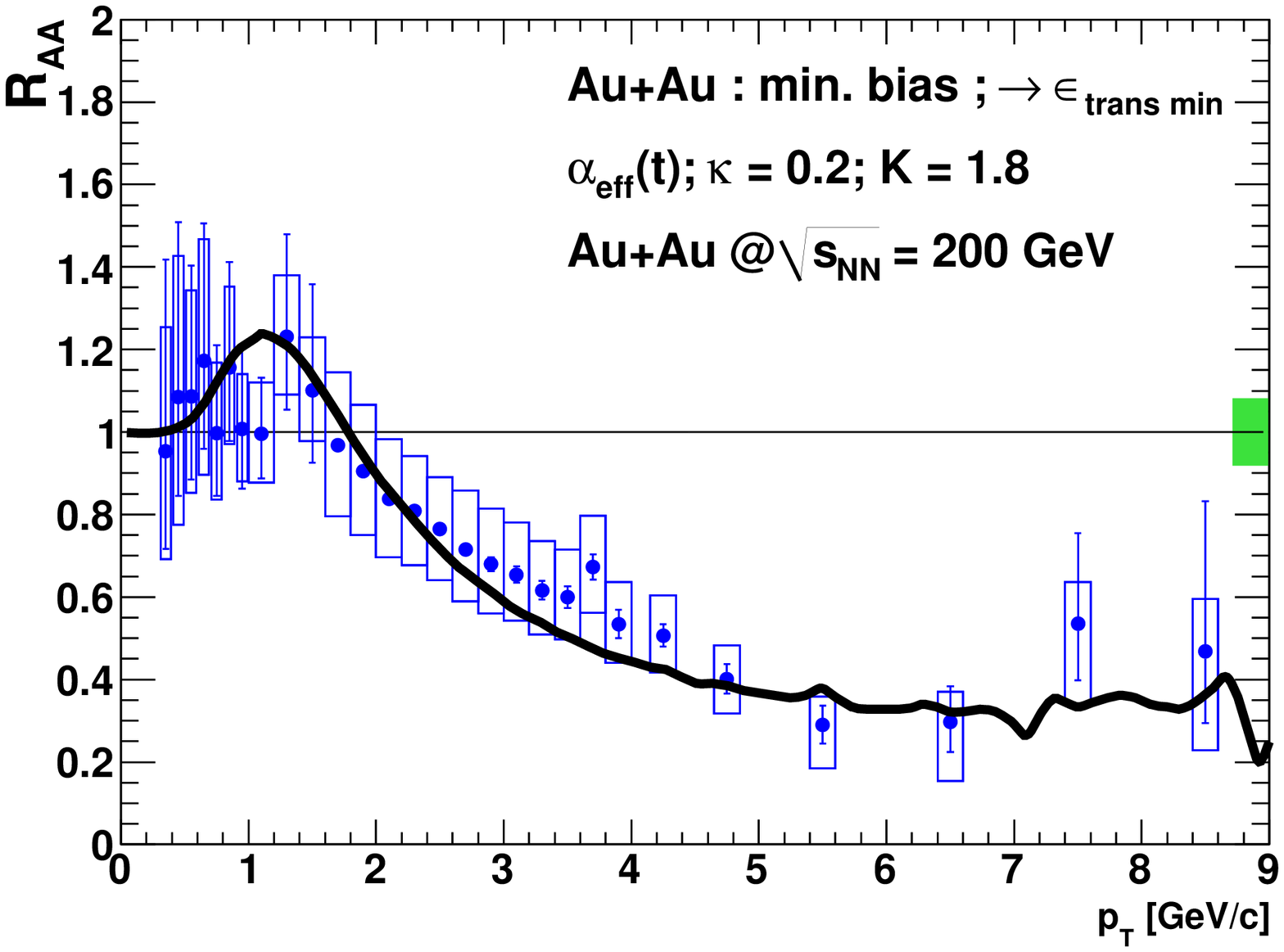}
\caption{(Color online) $\raa$ vs centrality and $p_T$ from Gossiaux and
Aichelin \cite{Pol1,Pol2,Pol3}.  \label{fig:Pol1}}
\end{figure*}

Gossiaux and Aichelin \cite{Pol1,Pol2,Pol3} calculated the $R_{\rm AA}$ from
collisional energy loss in pQCD using a running coupling constant and replacing
the Debye mass with a hard thermal loop calculation.  The model finds a value
close to the experimental $R_{\rm AA}$ for all centralities, while leaving room for
a possible radiative contribution as well.  Figure~\ref{fig:Pol1} shows the
$R_{\rm AA}$ as a function of $p_T$ and centrality from this model.

Kharzeev has predicted a universal bound on the energy of a parton escaping
strongly coupled matter \cite{UniversalBound1} in ${\cal N}=4$ SUSY Yang-Mills
theory, under some assumptions about the evolution of gauge fields in heavy-ion
collisions.  In the model, the $R_{\rm AA}$ at high $p_T$ is given by a constant
times $N_{\rm part}/N_{{\rm coll}}$ with the constant universal for all
particle species.

\subsection{Theory comparison $(v_2)$}

The azimuthal anisotropy of heavy quarks gives information about the 
density of the colliding system.  The collectivity of heavy quarks 
during the early stage of the collision is expected to be developed via 
the strong interaction inside the quark gluon plasma, which will be seen 
in the relatively low $p_T$ region.  On the other hand, the suppression 
and/or modification of the high $p_T$ yield, which is caused by the 
interaction of the heavy quarks with the overlapped almond shape 
geometry and path length of the heavy quark inside the plasma, could 
also create an azimuthal anisotropy at higher $p_T$. Therefore, when 
considering the $p_T$ dependence of the anisotropy parameter $v_2$, it 
is important to understand these different effects as well as the 
relative fraction of bottom quarks compared to charm quarks.  Several 
model predictions are compared to the experimental data in this section.

\begin{figure}[htbp]
  \includegraphics[width=1.0\linewidth]{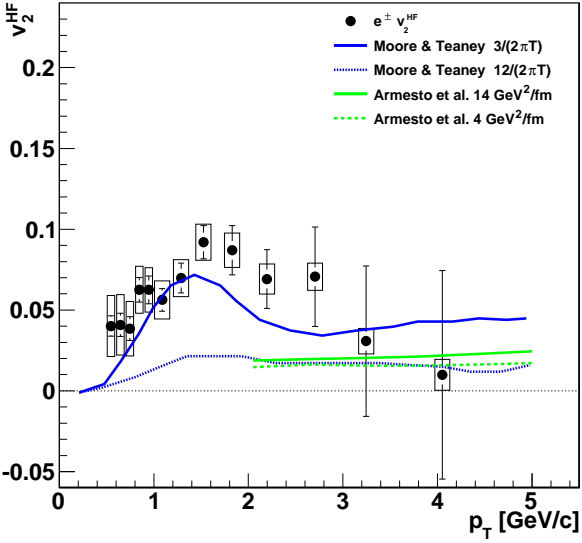}
  \caption{(Color online) Comparison of experimental $v_2$ data with 
  the models of Moore and Teaney \cite{Moore:2004tg} and Armesto et al.
\cite{Armesto:2005mz}.
  \label{fig:comp_v2_fig}}
\end{figure}

Figure~\ref{fig:comp_v2_fig} shows a comparison of two different model 
calculations to experimental data for the nonphotonic electron $v_2$ 
from minimum-bias collisions.  The models shown in this figure were also 
discussed in the previous section as they are also used to calculate 
$R_{\rm AA}$.  In the model of Moore and Teaney \cite{Moore:2004tg}, a 
Langevin-based heavy quark transport calculation is performed in a 
perturbative quark gluon plasma with hydrodynamic simulation.  Two 
different calculations with two different assumptions for the value of 
the diffusion coefficient are shown as a solid curve for D = 
3/(2$\pi$T), which agrees with experimental data, and as a dashed curve 
for D = 12/(2$\pi$T).  In the model of Armesto et al.  
\cite{Armesto:2005mz}, the effect of heavy quark energy loss for $p_T > 
2$ GeV/$c$ is considered without any collective effect in the lower 
$p_T$ region, which gives a lower limit on $v_2$ at high $p_T$ given by 
the energy loss and the geometry of the almond shape, with two values of 
the transport coefficient $\hat{q}$ for the green solid curve $\hat{q}$ 
= 14 GeV$^2$/fm and for the green dashed curve $\hat{q}$ = 4 GeV$^2$/fm.

\begin{figure}[htbp]
  \includegraphics[width=1.0\linewidth]{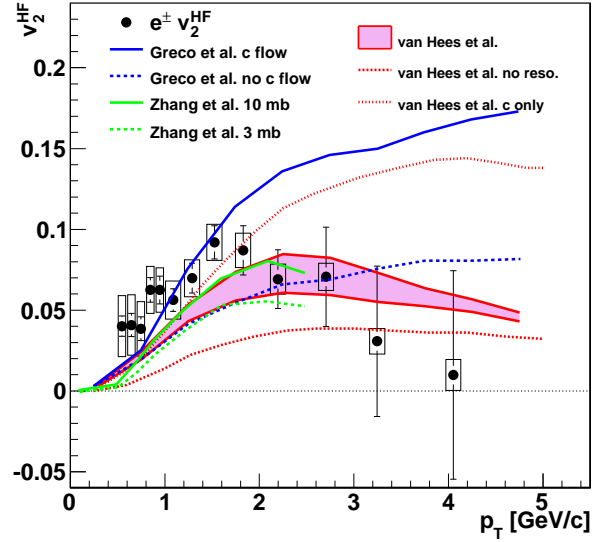}
  \caption {(Color online) Theory comparison of experimental $v_2$ data with 
  the models of Greco et al.~\protect\cite{Greco:2003vf}, 
Zhang et al.~\protect\cite{Zhang:2005pc},
  and van Hees et al.~\protect\cite{vanHees,vanHees:2007a}.
  \label{fig:comp_v2_fig2}}
\end{figure}

Figure~\ref{fig:comp_v2_fig2} shows a set of comparisons of three model 
calculations with the experimental data.  The model of Greco et al.  
\cite{Greco:2003vf} is based on the quark coalescence model with light 
and charm quarks $v_2$, where the light quark $v_2$ is estimated from 
the measured meson $v_2$ and the same amount of radial flow is assumed 
for all quark species with the same saturating $v_2$ for the charm quark 
as for the light quark.  The calculated electron $v_2$ from the D meson 
is labeled as ``Greco et al. c flow".  On the other hand, the line 
labeled as ``Greco et al. no c flow" assumes zero $v_2$ for the charm 
quark, and the predicted $v_2$ is given only by the light quark $v_2$ 
via the quark coalescence mechanism.  The model of Zhang et al. 
\cite{Zhang:2005pc} is a hybrid model starting with HIJING as an initial 
condition, followed by a parton cascade, and then a hadron cascade after 
a hadronization with a quark coalescence model.  This is based on a 
multiphase transport model including rescattering of the charm quark 
with the other partons.  The line labeled ``Zhang et al. 10 mb" is the 
resulting electron $v_2$ from D-meson decay with the charm quark 
parton-scattering cross section of $\sigma_p$ = 10 mb, while the line 
labeled ``Zhang et al. 3 mb" corresponds to $\sigma_p$ = 3 mb.  The 
model of van Hess et al.  \cite{vanHees,vanHees:2007a}, discussed in 
the previous section as it is also used to calculate $R_{\rm AA}$, 
includes resonant interactions in a strongly interacting QGP and parton 
coalescence of both charm and bottom quarks.  The band with shows the 
results of this model, while the result with only charm quarks is shown 
as the line labeled ``van Hees et al. c only", and the result without 
resonances is shown as the line labeled ``van Hees et al. no reso".  
Both the resonance and the coalescence are responsible for the large 
magnitude of the measured electron $v_2$ in the low $p_T$ region, while 
the reduction or flattening of $v_2$ at high $p_T$ requires the bottom 
quark contribution.

\begin{figure}[htbp]
  \includegraphics[width=1.0\linewidth]{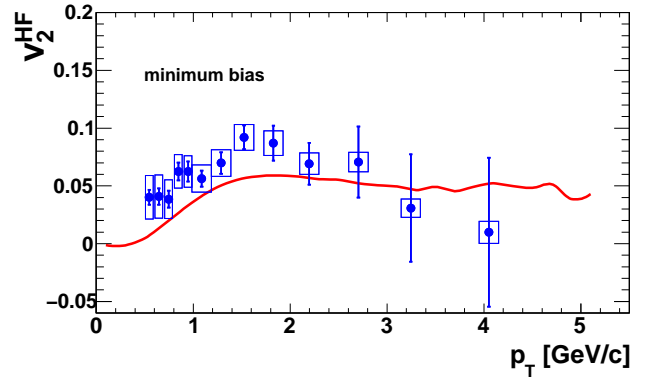}
  \caption {(Color online) 
Minimum-bias $v_2$ compared with a model calculation from Gossiaux
and Aichelin~\protect\cite{Pol1,Pol2,Pol3}.
  \label{fig:pol_v2}
}
\end{figure}

A pQCD-based elastic scattering model \cite{Pol1,Pol2,Pol3} with 
running coupling constant is used to calculate $v_2$ as well as the 
$R_{\rm AA}$.  The minimum-bias $v_2$ from this model is shown in 
Fig.~\ref{fig:pol_v2}.  The $v_2$ is very sensitive to the interaction 
time, and a later freeze-out can produce a larger $v_2$ \cite{Pol1}.

  Various models discussed in this section indicate strong 
interaction of heavy quarks with the high energy density matter created 
in heavy-ion collisions at RHIC and also support quark
coalescence(recombination) as a dominant hadronization mechanism, which would
mean that the heavy quarks are also a part of the quark gluon plasma.  Therefore 
the sizable event anisotropy parameter $v_2$ of electrons from
heavy quarks measured at low $p_T$ should be related with the 
collectivity of the heavy quark.  The reduction or flattening of $v_2$ in the
high $p_T$ region would have to be caused by the change  
of the dominant mechanism from the collectivity of the heavy quark 
to the energy loss of the heavy quark, especially when most of the heavy quarks
are charm quarks.  However the reduction can also be qualitatively understood by
an increase of the relative fraction of bottom quarks as a function of $p_T$ and
a smaller magnitude of event anisotropy for the heavier quark.
 
\begin{figure}[htbp]
  \includegraphics[width=1.0\linewidth]{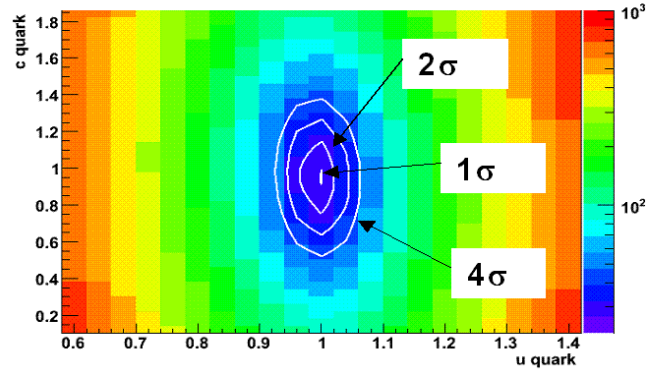}
  \caption {(Color online) 
$\chi^{2}$ map as a function of the light quark
  $v_2$ (horizontal axis) and charm quark $v_2$ (vertical axis), each divided
by the measured light quark $v_2$.
  \label{fig:chi_map}}
\end{figure}

In order to determine the charm quark $v_2$, the measured nonphotonic electron
$v_2$ is compared to that expected from coalescent production of $D$ mesons.  Chi
square values are calculated for different combinations of light and charm quark
$v_2$ values.  The $p_T$ range is restricted to below 2~GeV/c in order to be
sensitive to only the charm quark, and not to the bottom quark.  The quark
coalescence is assumed to occur at similar velocities, therefore it gives the
momentum ratio of about 1/6 for light/charm quarks, which is given by the ratio
of the effective mass of the quarks.  A common shape of $v_2$ vs $p_T$ for both
light and charm quarks is assumed, where this shape is determined by the light
hadron $v_2$.  For given $v_2$ values of light and charm quarks, $\pi$, $K$ and
proton $v_2$'s and electron $v_2$ from D mesons are calculated, then the total
chi-square sum is calculated and plotted in Fig.~\ref{fig:chi_map} as a
function of the light quark $v_2$ (horizontal axis) and charm quark $v_2$
(vertical axis).  Both the horizontal and vertical axes are normalized to the
measured light quark $v_2$, so that unity corresponds to the light quark $v_2$.
The chi-square minimum lies at about equal light quark and heavy quark $v_2$ 
magnitudes, although the relative error in $v_2$ is much larger for the 
charm quark than for the light quark.  This indicates a common quark 
collectivity in the quark gluon plasma phase.

\subsection{Viscosity to entropy ratio $\eta/s$}

The ratio of shear viscosity $\eta$ to entropy density $s$ is a key 
parameter that determines the damping rate in a relativistic system.  At 
temperature $T$, characteristic damping times $\tau$ are of order 
$\eta/sT$~\cite{Policastro:2002se}.

While the apparent success of ideal hydrodynamics in describing particle 
spectra and elliptic flow patterns at RHIC would imply a vanishing value 
of $\eta$, straightforward arguments based on the uncertainty principle 
suggest that the viscosity for any thermal system must be 
non-zero\cite{Danielewicz:1984ww}.  This observation was extended by 
Kovtun, Son and Starinets (KSS)~\cite{Kovtun:2004de}, who demonstrated 
that conformal field theories with gravity duals have a ratio of 
viscosity $\eta$ to entropy density $s$ of $1/4\pi$ (in natural units).  
KSS conjectured that this value is a bound for any relativistic thermal 
field theory, that is, $\eta/s \geq 1/4\pi$.

The results presented in Section~\ref{sec:results} for
$R_{\rm AuAu}(p_T^e)$ and $v_2(p_T^e)$ show that heavy quarks
lose energy in the medium, while acquiring a substantial component of
the medium's collective flow.  Both of these effects may be regarded
as the damping of the initial non-equilibrium dynamics of the
``external'' heavy quark by the medium.
 The simultaneous description of these phenomena by the model of van Hees 
et al.  provides a straightforward, albeit indirect, method to infer the ratio of
viscosity to entropy density.  The resonance model employed in
Ref.~\cite{vanHees} leads to an estimate for the heavy quark spatial diffusion
constant $D_s \sim (4-6)/2\pi T$ for temperatures $T$ in the range $0.2 \
{\rm GeV} < T < 0.4 \ {\rm GeV}$(see also Figure 23 of
Ref.~\cite{Rapp:2008qc}).  Moore and Teaney~\cite{Moore:2004tg} perform a
perturbative calculation of this quantity, and find that for a medium with three
light flavors the ratio of $D_s$ to the hydrodynamic diffusion constant
$\eta/(\epsilon + p)$ for the bulk ($\epsilon$ is the energy density and $p$ the
pressure of the medium) has a value of $\sim 6$ roughly independent of the 
coupling strength $\sim m_D/T$, where $m_D$ is the Debye mass.  They argue 
that the weak variation with coupling strength is to be expected in this 
{\em ratio} of transport coefficients, making it plausible that it 
remains near 6 in the strongly-coupled regime.  In this case, and 
approximating the thermodynamic identity $\epsilon+p = T s + \mu_B n_B 
\approx Ts $ appropriate for the baryon-free central region, one readily 
finds $\eta/s \sim (1.33-2)/4\pi$, that is, a value near the KSS bound 
and consistent with other estimates for the RHIC plasma based on 
flow~\cite{Lacey:2006bc,Drescher:2007cd}, 
fluctuations~\cite{Gavin:2006xd}, entropy 
production~\cite{Dumitru:2007qr} and detailed hydrodynamic 
calculations~\cite{Romatschke:2007mq,Luzum:2008cw}.  It should be noted 
that these various estimates are based on observables of the ``bulk'' 
medium, in flavor channels dominated by $u$, $d$ and $s$ quarks, while 
the result presented here relies explicitly on the coupling of heavy 
flavor to the medium.  The consistency of the derived value of $\eta/s$ 
supports both the strong coupling of heavy flavor to the medium and the 
low value of $\eta/s$ for the RHIC plasma.

 
\section{SUMMARY AND CONCLUSIONS}\label{sec:summary}

This article has detailed the measurement of the yield of single 
electrons from semileptonic decays of heavy flavor mesons at midrapidity 
in $p+p$ and Au+Au collisions at $\sqrt{s_{\rm NN}}$ = 200 GeV, as well as 
the azimuthal anisotropy parameter $v_2$ of such electrons in Au+Au 
collisions.  The unexpectedly large suppression of heavy flavor electrons 
in Au+Au collisions relative to those from $p+p$ collisions and the large 
$v_2$ of heavy flavor electrons has generated much theoretical work.  
In a system as complicated as this medium, confidence from any given
model depends on its ability to describe multiple observables simultaneously.  
The intermediate-mass range dilepton measurement \cite{ppg075} at PHENIX is 
dominated by correlated decays of charm mesons which are not included in 
the calculations of any published model.  In the coming years, PHENIX will
measure the yields and flow of bottom 
and charm mesons separately, which will further constrain the 
interpretation of the data.

\section{Acknowledgements} 

We thank the staff of the Collider-Accelerator and Physics Departments 
at Brookhaven National Laboratory and the staff of the other PHENIX 
participating institutions for their vital contributions.  We 
acknowledge support from the Office of Nuclear Physics in the Office of 
Science of the Department of Energy, the National Science Foundation, a 
sponsored research grant from Renaissance Technologies LLC, Abilene 
Christian University Research Council, Research Foundation of SUNY, and 
Dean of the College of Arts and Sciences, Vanderbilt University (USAA), 
Ministry of Education, Culture, Sports, Science, and Technology and the 
Japan Society for the Promotion of Science (Japan), Conselho Nacional de 
Desenvolvimento Cient\'{\i}fico e Tecnol{\'o}gico and Funda\c c{\~a}o de 
Amparo {\`a} Pesquisa do Estado de S{\~a}o Paulo (Brazil), Natural 
Science Foundation of China (People's Republic of China), Ministry of 
Education, Youth and Sports (Czech Republic), Centre National de la 
Recherche Scientifique, Commissariat {\`a} l'{\'E}nergie Atomique, and 
Institut National de Physique Nucl{\'e}aire et de Physique des 
Particules (France), Ministry of Industry, Science and Tekhnologies, 
Bundesministerium f\"ur Bildung und Forschung, Deutscher Akademischer 
Austausch Dienst, and Alexander von Humboldt Stiftung (Germany), 
Hungarian National Science Fund, OTKA (Hungary), Department of Atomic 
Energy (India), Israel Science Foundation (Israel), Korea Research 
Foundation and Korea Science and Engineering Foundation (Korea), 
Ministry of Education and Science, Rassia Academy of Sciences, Federal 
Agency of Atomic Energy (Russia), VR and the Wallenberg Foundation 
(Sweden), the USA  Civilian Research and Development Foundation for the 
Independent States of the Former Soviet Union, the US-Hungarian Fulbight 
Foundation for Educational Exchange, and the US-Israel Binational 
Science Foundation.

 
\section*{APPENDIX:~DATA TABLES}
 
 

Data are presented in this Appendix for the minimum bias events 
(0--92\%) and for each centrality class (0--10\%, 10--20\%, 20--40\%, 
40--60\%, and 60--92\%) for Au+Au collisions at $\snn = 200$ GeV 
at midrapidity.
Tables ~\ref{tab:inv_0092_0_10}-\ref{tab:inv_40_92} tabulate
the differential invariant yield of heavy flavor electrons.
Tables~\ref{tab:raa_0092_0_10}-\ref{tab:raa_40_92} give 
the nuclear modification factor $\raa$ of heavy flavor electrons.
Table~\ref{tab:v2} shows $v_2$ for heavy flavor electrons.
Table~\ref{tab:inv_pp} gives the differential invariant cross section 
of heavy flavor electrons.


\begingroup \squeezetable

\begin{table*}[htbp]
\caption{Differential invariant yield of electrons ($\aee$) from
heavy-flavor decays 
	 for (upper, minimum bias) 0--92\% and (lower) 0--10\% centrality classes.  
         The $\pte$ is in units of GeV/$c$.  The yield and 
	 corresponding errors are in units of $(\mbox{GeV}/c)^{-2}$ 
\label{tab:inv_0092_0_10}}
\begin{ruledtabular} \begin{tabular}{cccccccc}
Centrality & \quad $\pte$ \quad & \quad Invariant yield \quad & \quad Stat.
error ($+$) \quad & \quad Stat.  error ($-$) \quad & \quad Sys.  error ($+$) \quad
& \quad Sys.  error ($-$) \quad \\
\hline
& 0.30 & $ 7.30 \times 10^{-2}$ & $ 3.01 \times 10^{-3}$ & $ 2.97 \times 
10^{-3}$ & $ 1.94 \times 10^{-2}$ & $ 1.91 \times 10^{-2}$ \\            
& 0.50 & $ 3.74 \times 10^{-2}$ & $ 1.29 \times 10^{-3}$ & $ 1.27 \times 
10^{-3}$ & $ 7.73 \times 10^{-3}$ & $ 7.63 \times 10^{-3}$ \\            
& 0.60 & $ 2.09 \times 10^{-2}$ & $ 6.55 \times 10^{-4}$ & $ 6.47 \times 
10^{-4}$ & $ 3.66 \times 10^{-3}$ & $ 3.61 \times 10^{-3}$ \\            
& 0.70 & $ 1.22 \times 10^{-2}$ & $ 3.86 \times 10^{-4}$ & $ 3.82 \times 
10^{-4}$ & $ 1.94 \times 10^{-3}$ & $ 1.92 \times 10^{-3}$ \\            
& 0.80 & $ 7.46 \times 10^{-3}$ & $ 2.32 \times 10^{-4}$ & $ 2.30 \times 
10^{-4}$ & $ 1.04 \times 10^{-3}$ & $ 1.03 \times 10^{-3}$ \\            
& 0.80 & $ 5.04 \times 10^{-3}$ & $ 1.54 \times 10^{-4}$ & $ 1.52 \times 
10^{-4}$ & $ 6.09 \times 10^{-4}$ & $ 6.03 \times 10^{-4}$ \\
& 0.90 & $ 3.27 \times 10^{-3}$ & $ 1.06 \times 10^{-4}$ & $ 1.05 \times
10^{-4}$ & $ 3.70 \times 10^{-4}$ & $ 3.67 \times 10^{-4}$ \\
& 1.10 & $ 1.84 \times 10^{-3}$ & $ 4.74 \times 10^{-5}$ & $ 4.60 \times
10^{-5}$ & $ 1.95 \times 10^{-4}$ & $ 1.90 \times 10^{-4}$ \\
& 1.30 & $ 8.21 \times 10^{-4}$ & $ 2.04 \times 10^{-5}$ & $ 1.99 \times
10^{-5}$ & $ 9.04 \times 10^{-5}$ & $ 8.81 \times 10^{-5}$ \\
& 1.50 & $ 3.90 \times 10^{-4}$ & $ 8.81 \times 10^{-6}$ & $ 8.62 \times
10^{-6}$ & $ 4.35 \times 10^{-5}$ & $ 4.26 \times 10^{-5}$ \\
& 1.70 & $ 1.87 \times 10^{-4}$ & $ 2.56 \times 10^{-6}$ & $ 2.56 \times
10^{-6}$ & $ 2.87 \times 10^{-5}$ & $ 2.87 \times 10^{-5}$ \\
& 1.90 & $ 9.49 \times 10^{-5}$ & $ 1.35 \times 10^{-6}$ & $ 1.35 \times
10^{-6}$ & $ 1.40 \times 10^{-5}$ & $ 1.40 \times 10^{-5}$ \\
& 2.10 & $ 5.05 \times 10^{-5}$ & $ 7.71 \times 10^{-7}$ & $ 7.71 \times
10^{-7}$ & $ 7.18 \times 10^{-6}$ & $ 7.18 \times 10^{-6}$ \\
0--92\% & 2.30 & $ 2.91 \times 10^{-5}$ & $ 4.83 \times 10^{-7}$ & $ 4.83 \times
10^{-7}$ & $ 3.90 \times 10^{-6}$ & $ 3.90 \times 10^{-6}$ \\
& 2.50 & $ 1.71 \times 10^{-5}$ & $ 3.12 \times 10^{-7}$ & $ 3.12 \times
10^{-7}$ & $ 2.18 \times 10^{-6}$ & $ 2.18 \times 10^{-6}$ \\
& 2.70 & $ 9.98 \times 10^{-6}$ & $ 2.12 \times 10^{-7}$ & $ 2.12 \times
10^{-7}$ & $ 1.26 \times 10^{-6}$ & $ 1.26 \times 10^{-6}$ \\
& 2.90 & $ 6.07 \times 10^{-6}$ & $ 1.50 \times 10^{-7}$ & $ 1.50 \times
10^{-7}$ & $ 7.54 \times 10^{-7}$ & $ 7.54 \times 10^{-7}$ \\
& 3.10 & $ 3.87 \times 10^{-6}$ & $ 1.11 \times 10^{-7}$ & $ 1.11 \times
10^{-7}$ & $ 4.67 \times 10^{-7}$ & $ 4.67 \times 10^{-7}$ \\
& 3.30 & $ 2.45 \times 10^{-6}$ & $ 8.27 \times 10^{-8}$ & $ 8.27 \times
10^{-8}$ & $ 2.93 \times 10^{-7}$ & $ 2.93 \times 10^{-7}$ \\
& 3.50 & $ 1.61 \times 10^{-6}$ & $ 6.39 \times 10^{-8}$ & $ 6.39 \times
10^{-8}$ & $ 1.91 \times 10^{-7}$ & $ 1.91 \times 10^{-7}$ \\
& 3.70 & $ 1.25 \times 10^{-6}$ & $ 5.23 \times 10^{-8}$ & $ 5.23 \times
10^{-8}$ & $ 1.38 \times 10^{-7}$ & $ 1.38 \times 10^{-7}$ \\
& 3.90 & $ 7.05 \times 10^{-7}$ & $ 3.94 \times 10^{-8}$ & $ 3.90 \times
10^{-8}$ & $ 8.40 \times 10^{-8}$ & $ 8.40 \times 10^{-8}$ \\
& 4.20 & $ 3.81 \times 10^{-7}$ & $ 1.71 \times 10^{-8}$ & $ 1.71 \times
10^{-8}$ & $ 4.38 \times 10^{-8}$ & $ 4.38 \times 10^{-8}$ \\
& 4.80 & $ 1.48 \times 10^{-7}$ & $ 1.03 \times 10^{-8}$ & $ 1.02 \times
10^{-8}$ & $ 1.76 \times 10^{-8}$ & $ 1.76 \times 10^{-8}$ \\
& 5.50 & $ 4.04 \times 10^{-8}$ & $ 4.45 \times 10^{-9}$ & $ 4.37 \times
10^{-9}$ & $ 5.61 \times 10^{-9}$ & $ 5.61 \times 10^{-9}$ \\
& 6.50 & $ 1.18 \times 10^{-8}$ & $ 2.22 \times 10^{-9}$ & $ 2.15 \times
10^{-9}$ & $ 1.61 \times 10^{-9}$ & $ 1.61 \times 10^{-9}$ \\
& 7.50 & $ 5.84 \times 10^{-9}$ & $ 1.59 \times 10^{-9}$ & $ 1.53 \times
10^{-9}$ & $ 6.93 \times 10^{-10}$ & $ 6.93 \times 10^{-10}$ \\
& 8.50 & $ 1.93 \times 10^{-9}$ & $ 1.02 \times 10^{-9}$ & $ 6.87 \times
10^{-10}$ & $ 3.37 \times 10^{-10}$ & $ 3.37 \times 10^{-10}$ \\
\hline
& 0.30 & $ 3.35 \times 10^{-1}$ & $ 1.63 \times 10^{-2}$ & $ 1.61 \times
10^{-2}$ & $ 6.70 \times 10^{-2}$ & $ 6.60 \times 10^{-2}$ \\
& 0.50 & $ 1.62 \times 10^{-1}$ & $ 7.05 \times 10^{-3}$ & $ 6.95 \times
10^{-3}$ & $ 2.73 \times 10^{-2}$ & $ 2.69 \times 10^{-2}$ \\
& 0.60 & $ 8.75 \times 10^{-2}$ & $ 3.59 \times 10^{-3}$ & $ 3.55 \times
10^{-3}$ & $ 1.30 \times 10^{-2}$ & $ 1.28 \times 10^{-2}$ \\
& 0.70 & $ 4.83 \times 10^{-2}$ & $ 2.15 \times 10^{-3}$ & $ 2.12 \times
10^{-3}$ & $ 6.87 \times 10^{-3}$ & $ 6.79 \times 10^{-3}$ \\
& 0.80 & $ 2.96 \times 10^{-2}$ & $ 1.31 \times 10^{-3}$ & $ 1.30 \times
10^{-3}$ & $ 3.74 \times 10^{-3}$ & $ 3.70 \times 10^{-3}$ \\
& 0.80 & $ 1.83 \times 10^{-2}$ & $ 8.65 \times 10^{-4}$ & $ 8.57 \times
10^{-4}$ & $ 2.17 \times 10^{-3}$ & $ 2.14 \times 10^{-3}$ \\
& 0.90 & $ 1.29 \times 10^{-2}$ & $ 6.13 \times 10^{-4}$ & $ 6.07 \times
10^{-4}$ & $ 1.37 \times 10^{-3}$ & $ 1.36 \times 10^{-3}$ \\
& 1.10 & $ 6.78 \times 10^{-3}$ & $ 2.74 \times 10^{-4}$ & $ 2.66 \times
10^{-4}$ & $ 7.05 \times 10^{-4}$ & $ 6.83 \times 10^{-4}$ \\
& 1.30 & $ 3.25 \times 10^{-3}$ & $ 7.08 \times 10^{-5}$ & $ 6.89 \times
10^{-5}$ & $ 4.01 \times 10^{-4}$ & $ 3.90 \times 10^{-4}$ \\
& 1.50 & $ 1.50 \times 10^{-3}$ & $ 3.09 \times 10^{-5}$ & $ 3.02 \times
10^{-5}$ & $ 1.91 \times 10^{-4}$ & $ 1.86 \times 10^{-4}$ \\
& 1.70 & $ 6.62 \times 10^{-4}$ & $ 1.05 \times 10^{-5}$ & $ 1.05 \times
10^{-5}$ & $ 9.80 \times 10^{-5}$ & $ 9.80 \times 10^{-5}$ \\
& 1.90 & $ 3.32 \times 10^{-4}$ & $ 5.84 \times 10^{-6}$ & $ 5.84 \times
10^{-6}$ & $ 4.68 \times 10^{-5}$ & $ 4.68 \times 10^{-5}$ \\
0--10\% & 2.10 & $ 1.70 \times 10^{-4}$ & $ 3.51 \times 10^{-6}$ & $ 3.51 \times
10^{-6}$ & $ 2.34 \times 10^{-5}$ & $ 2.34 \times 10^{-5}$ \\
& 2.30 & $ 9.52 \times 10^{-5}$ & $ 2.30 \times 10^{-6}$ & $ 2.30 \times
10^{-6}$ & $ 1.25 \times 10^{-5}$ & $ 1.25 \times 10^{-5}$ \\
&2.5 & $ 5.45 \times 10^{-5}$ & $ 1.55 \times 10^{-6}$ & $ 1.55 \times 10^{-6}$
& $ 6.89 \times 10^{-6}$ & $ 6.89 \times 10^{-6}$ \\
&2.7 & $ 3.32 \times 10^{-5}$ & $ 1.11 \times 10^{-6}$ & $ 1.11 \times 10^{-6}$
& $ 4.06 \times 10^{-6}$ & $ 4.06 \times 10^{-6}$ \\
&2.9 & $ 2.03 \times 10^{-5}$ & $ 8.02 \times 10^{-7}$ & $ 8.02 \times 10^{-7}$
& $ 2.43 \times 10^{-6}$ & $ 2.43 \times 10^{-6}$ \\
&3.1 & $ 1.11 \times 10^{-5}$ & $ 5.84 \times 10^{-7}$ & $ 5.79 \times 10^{-7}$
& $ 1.38 \times 10^{-6}$ & $ 1.38 \times 10^{-6}$ \\
&3.3 & $ 7.73 \times 10^{-6}$ & $ 4.53 \times 10^{-7}$ & $ 4.48 \times 10^{-7}$
& $ 9.14 \times 10^{-7}$ & $ 9.14 \times 10^{-7}$ \\
&3.5 & $ 4.96 \times 10^{-6}$ & $ 3.52 \times 10^{-7}$ & $ 3.48 \times 10^{-7}$
& $ 5.87 \times 10^{-7}$ & $ 5.87 \times 10^{-7}$ \\
&3.7 & $ 3.65 \times 10^{-6}$ & $ 2.82 \times 10^{-7}$ & $ 2.78 \times 10^{-7}$
& $ 4.08 \times 10^{-7}$ & $ 4.08 \times 10^{-7}$ \\
&3.9 & $ 2.56 \times 10^{-6}$ & $ 2.32 \times 10^{-7}$ & $ 2.28 \times 10^{-7}$
& $ 2.84 \times 10^{-7}$ & $ 2.84 \times 10^{-7}$ \\
&4.2 & $ 1.08 \times 10^{-6}$ & $ 9.32 \times 10^{-8}$ & $ 9.19 \times 10^{-8}$
& $ 1.28 \times 10^{-7}$ & $ 1.28 \times 10^{-7}$ \\
&4.8 & $ 3.76 \times 10^{-7}$ & $ 5.51 \times 10^{-8}$ & $ 5.39 \times 10^{-8}$
& $ 4.88 \times 10^{-8}$ & $ 4.88 \times 10^{-8}$ \\
&5.5 & $ 1.22 \times 10^{-7}$ & $ 2.50 \times 10^{-8}$ & $ 2.41 \times 10^{-8}$
& $ 1.72 \times 10^{-8}$ & $ 1.72 \times 10^{-8}$ \\
&6.5 & $ 1.21 \times 10^{-8}$ & $ 1.11 \times 10^{-8}$ & $ 8.26 \times 10^{-9}$
& $ 3.20 \times 10^{-9}$ & $ 3.20 \times 10^{-9}$ \\
&7.5 & $ 2.38 \times 10^{-8}$ & $ 1.15 \times 10^{-8}$ & $ 7.78 \times 10^{-9}$
& $ 2.66 \times 10^{-9}$ & $ 2.66 \times 10^{-9}$ \\
&8.5 & $ 7.78 \times 10^{-9}$ & $ 6.97 \times 10^{-9}$ & $ 3.70 \times 10^{-9}$
& $ 1.25 \times 10^{-9}$ & $ 1.25 \times 10^{-9}$ \\
\end{tabular} \end{ruledtabular}
\end{table*}


\begin{table*}[htbp]
\caption{Differential invariant yield of electrons ($\aee$) from
heavy-flavor decays 
	 for (upper) 10--20\% and (lower) 20--40\% centrality classes.  
         The $\pte$ is in units of GeV/$c$.  The yield and 
	 corresponding errors are in units of $(\mbox{GeV}/c)^{-2}$ 
\label{tab:inv_0092_10_40}}
\begin{ruledtabular} \begin{tabular}{ccccccccc}
Centrality & \quad $\pte$ \quad & \quad Invariant yield \quad & \quad Stat.
error ($+$) \quad & \quad Stat.  error ($-$) \quad & \quad Sys.  error ($+$) \quad
& \quad Sys.  error ($-$) \quad \\
\hline
&0.3 & $ 1.90 \times 10^{-1}$ & $ 1.11 \times 10^{-2}$ & $ 1.09 \times 10^{-2}$ 
& $ 4.48 \times 10^{-2}$ & $ 4.42 \times 10^{-2}$ \\                            
&0.5 & $ 9.11 \times 10^{-2}$ & $ 4.81 \times 10^{-3}$ & $ 4.75 \times 10^{-3}$ 
& $ 1.79 \times 10^{-2}$ & $ 1.77 \times 10^{-2}$ \\                            
&0.6 & $ 5.29 \times 10^{-2}$ & $ 2.52 \times 10^{-3}$ & $ 2.49 \times 10^{-3}$ 
& $ 8.57 \times 10^{-3}$ & $ 8.48 \times 10^{-3}$ \\                            
&0.7 & $ 2.92 \times 10^{-2}$ & $ 1.53 \times 10^{-3}$ & $ 1.51 \times 10^{-3}$ 
& $ 4.55 \times 10^{-3}$ & $ 4.50 \times 10^{-3}$ \\                            
&0.8 & $ 1.88 \times 10^{-2}$ & $ 9.51 \times 10^{-4}$ & $ 9.42 \times 10^{-4}$ 
& $ 2.47 \times 10^{-3}$ & $ 2.45 \times 10^{-3}$ \\                            
&0.8 & $ 1.29 \times 10^{-2}$ & $ 6.42 \times 10^{-4}$ & $ 6.36 \times 10^{-4}$ 
& $ 1.47 \times 10^{-3}$ & $ 1.46 \times 10^{-3}$ \\
&0.9 & $ 7.93 \times 10^{-3}$ & $ 4.50 \times 10^{-4}$ & $ 4.46 \times 10^{-4}$
& $ 8.77 \times 10^{-4}$ & $ 8.70 \times 10^{-4}$ \\
&1.1 & $ 4.47 \times 10^{-3}$ & $ 2.07 \times 10^{-4}$ & $ 2.01 \times 10^{-4}$
& $ 4.69 \times 10^{-4}$ & $ 4.55 \times 10^{-4}$ \\
&1.3 & $ 2.20 \times 10^{-3}$ & $ 5.03 \times 10^{-5}$ & $ 4.90 \times 10^{-5}$
& $ 2.50 \times 10^{-4}$ & $ 2.44 \times 10^{-4}$ \\
&1.5 & $ 1.01 \times 10^{-3}$ & $ 2.24 \times 10^{-5}$ & $ 2.19 \times 10^{-5}$
& $ 1.16 \times 10^{-4}$ & $ 1.13 \times 10^{-4}$ \\
&1.7 & $ 4.69 \times 10^{-4}$ & $ 7.98 \times 10^{-6}$ & $ 7.98 \times 10^{-6}$
& $ 6.73 \times 10^{-5}$ & $ 6.73 \times 10^{-5}$ \\
&1.9 & $ 2.41 \times 10^{-4}$ & $ 4.54 \times 10^{-6}$ & $ 4.54 \times 10^{-6}$
& $ 3.27 \times 10^{-5}$ & $ 3.27 \times 10^{-5}$ \\
10--20\% & 2.10 & $ 1.28 \times 10^{-4}$ & $ 2.81 \times 10^{-6}$ & $ 2.81
\times
10^{-6}$ & $ 1.67 \times 10^{-5}$ & $ 1.67 \times 10^{-5}$ \\
&2.3 & $ 7.54 \times 10^{-5}$ & $ 1.89 \times 10^{-6}$ & $ 1.89 \times 10^{-6}$
& $ 9.24 \times 10^{-6}$ & $ 9.24 \times 10^{-6}$ \\
&2.5 & $ 4.18 \times 10^{-5}$ & $ 1.27 \times 10^{-6}$ & $ 1.27 \times 10^{-6}$
& $ 5.04 \times 10^{-6}$ & $ 5.04 \times 10^{-6}$ \\
&2.7 & $ 2.38 \times 10^{-5}$ & $ 8.92 \times 10^{-7}$ & $ 8.92 \times 10^{-7}$
& $ 2.86 \times 10^{-6}$ & $ 2.86 \times 10^{-6}$ \\
&2.9 & $ 1.40 \times 10^{-5}$ & $ 6.44 \times 10^{-7}$ & $ 6.44 \times 10^{-7}$
& $ 1.66 \times 10^{-6}$ & $ 1.66 \times 10^{-6}$ \\
&3.1 & $ 9.37 \times 10^{-6}$ & $ 4.96 \times 10^{-7}$ & $ 4.91 \times 10^{-7}$
& $ 1.06 \times 10^{-6}$ & $ 1.06 \times 10^{-6}$ \\
&3.3 & $ 5.59 \times 10^{-6}$ & $ 3.69 \times 10^{-7}$ & $ 3.65 \times 10^{-7}$
& $ 6.41 \times 10^{-7}$ & $ 6.41 \times 10^{-7}$ \\
&3.5 & $ 3.98 \times 10^{-6}$ & $ 2.94 \times 10^{-7}$ & $ 2.90 \times 10^{-7}$
& $ 4.39 \times 10^{-7}$ & $ 4.39 \times 10^{-7}$ \\
&3.7 & $ 3.14 \times 10^{-6}$ & $ 2.48 \times 10^{-7}$ & $ 2.44 \times 10^{-7}$
& $ 3.25 \times 10^{-7}$ & $ 3.25 \times 10^{-7}$ \\
&3.9 & $ 1.40 \times 10^{-6}$ & $ 1.71 \times 10^{-7}$ & $ 1.68 \times 10^{-7}$
& $ 1.68 \times 10^{-7}$ & $ 1.68 \times 10^{-7}$ \\
&4.2 & $ 9.77 \times 10^{-7}$ & $ 8.13 \times 10^{-8}$ & $ 8.01 \times 10^{-8}$
& $ 1.05 \times 10^{-7}$ & $ 1.05 \times 10^{-7}$ \\
&4.8 & $ 3.42 \times 10^{-7}$ & $ 4.82 \times 10^{-8}$ & $ 4.70 \times 10^{-8}$
& $ 3.92 \times 10^{-8}$ & $ 3.92 \times 10^{-8}$ \\
&5.5 & $ 9.71 \times 10^{-8}$ & $ 2.10 \times 10^{-8}$ & $ 2.02 \times 10^{-8}$
& $ 1.26 \times 10^{-8}$ & $ 1.26 \times 10^{-8}$ \\
&6.5 & $ 4.00 \times 10^{-8}$ & $ 1.20 \times 10^{-8}$ & $ 1.13 \times 10^{-8}$
& $ 4.53 \times 10^{-9}$ & $ 4.53 \times 10^{-9}$ \\
&7.5 & $ 2.45 \times 10^{-8}$ & $ 1.13 \times 10^{-8}$ & $ 7.40 \times 10^{-9}$
& $ 2.38 \times 10^{-9}$ & $ 2.38 \times 10^{-9}$ \\
&8.5 & $ 7.54 \times 10^{-9}$ & $ 6.78 \times 10^{-9}$ & $ 3.30 \times 10^{-9}$
& $ 1.07 \times 10^{-9}$ & $ 1.07 \times 10^{-9}$ \\
\hline
&0.3 & $ 7.35 \times 10^{-2}$ & $ 5.15 \times 10^{-3}$ & $ 5.09 \times 10^{-3}$
& $ 2.48 \times 10^{-2}$ & $ 2.45 \times 10^{-2}$ \\
&0.5 & $ 4.19 \times 10^{-2}$ & $ 2.30 \times 10^{-3}$ & $ 2.28 \times 10^{-3}$
& $ 9.85 \times 10^{-3}$ & $ 9.74 \times 10^{-3}$ \\
&0.6 & $ 2.25 \times 10^{-2}$ & $ 1.20 \times 10^{-3}$ & $ 1.19 \times 10^{-3}$
& $ 4.66 \times 10^{-3}$ & $ 4.61 \times 10^{-3}$ \\
&0.7 & $ 1.55 \times 10^{-2}$ & $ 7.55 \times 10^{-4}$ & $ 7.47 \times 10^{-4}$
& $ 2.50 \times 10^{-3}$ & $ 2.48 \times 10^{-3}$ \\
&0.8 & $ 8.93 \times 10^{-3}$ & $ 4.63 \times 10^{-4}$ & $ 4.59 \times 10^{-4}$
& $ 1.33 \times 10^{-3}$ & $ 1.31 \times 10^{-3}$ \\
&0.8 & $ 6.50 \times 10^{-3}$ & $ 3.16 \times 10^{-4}$ & $ 3.13 \times 10^{-4}$
& $ 7.83 \times 10^{-4}$ & $ 7.76 \times 10^{-4}$ \\
&0.9 & $ 3.79 \times 10^{-3}$ & $ 2.21 \times 10^{-4}$ & $ 2.19 \times 10^{-4}$
& $ 4.66 \times 10^{-4}$ & $ 4.63 \times 10^{-4}$ \\
&1.1 & $ 2.38 \times 10^{-3}$ & $ 1.03 \times 10^{-4}$ & $ 9.98 \times 10^{-5}$
& $ 2.51 \times 10^{-4}$ & $ 2.44 \times 10^{-4}$ \\
&1.3 & $ 1.15 \times 10^{-3}$ & $ 2.77 \times 10^{-5}$ & $ 2.70 \times 10^{-5}$
& $ 1.26 \times 10^{-4}$ & $ 1.23 \times 10^{-4}$ \\
&1.5 & $ 5.45 \times 10^{-4}$ & $ 1.24 \times 10^{-5}$ & $ 1.21 \times 10^{-5}$
& $ 6.07 \times 10^{-5}$ & $ 5.95 \times 10^{-5}$ \\
&1.7 & $ 2.56 \times 10^{-4}$ & $ 4.27 \times 10^{-6}$ & $ 4.27 \times 10^{-6}$
& $ 3.76 \times 10^{-5}$ & $ 3.76 \times 10^{-5}$ \\
&1.9 & $ 1.31 \times 10^{-4}$ & $ 2.42 \times 10^{-6}$ & $ 2.42 \times 10^{-6}$
& $ 1.85 \times 10^{-5}$ & $ 1.85 \times 10^{-5}$ \\
20--40\% & 2.10 & $ 7.12 \times 10^{-5}$ & $ 1.49 \times 10^{-6}$ &
$ 1.49 \times 10^{-6}$ & $ 9.60 \times 10^{-6}$ & $ 9.60 \times 10^{-6}$ \\
&2.3 & $ 4.00 \times 10^{-5}$ & $ 9.78 \times 10^{-7}$ & $ 9.78 \times 10^{-7}$
& $ 5.15 \times 10^{-6}$ & $ 5.15 \times 10^{-6}$ \\
&2.5 & $ 2.35 \times 10^{-5}$ & $ 6.70 \times 10^{-7}$ & $ 6.70 \times 10^{-7}$
& $ 2.93 \times 10^{-6}$ & $ 2.93 \times 10^{-6}$ \\
&2.7 & $ 1.41 \times 10^{-5}$ & $ 4.73 \times 10^{-7}$ & $ 4.73 \times 10^{-7}$
& $ 1.70 \times 10^{-6}$ & $ 1.70 \times 10^{-6}$ \\
&2.9 & $ 8.86 \times 10^{-6}$ & $ 3.50 \times 10^{-7}$ & $ 3.50 \times 10^{-7}$
& $ 1.03 \times 10^{-6}$ & $ 1.03 \times 10^{-6}$ \\
&3.1 & $ 5.62 \times 10^{-6}$ & $ 2.64 \times 10^{-7}$ & $ 2.64 \times 10^{-7}$
& $ 6.43 \times 10^{-7}$ & $ 6.43 \times 10^{-7}$ \\
&3.3 & $ 3.63 \times 10^{-6}$ & $ 2.02 \times 10^{-7}$ & $ 2.00 \times 10^{-7}$
& $ 4.09 \times 10^{-7}$ & $ 4.09 \times 10^{-7}$ \\
&3.5 & $ 2.25 \times 10^{-6}$ & $ 1.54 \times 10^{-7}$ & $ 1.52 \times 10^{-7}$
& $ 2.54 \times 10^{-7}$ & $ 2.54 \times 10^{-7}$ \\
&3.7 & $ 1.69 \times 10^{-6}$ & $ 1.26 \times 10^{-7}$ & $ 1.25 \times 10^{-7}$
& $ 1.80 \times 10^{-7}$ & $ 1.80 \times 10^{-7}$ \\
&3.9 & $ 9.65 \times 10^{-7}$ & $ 9.42 \times 10^{-8}$ & $ 9.26 \times 10^{-8}$
& $ 1.10 \times 10^{-7}$ & $ 1.10 \times 10^{-7}$ \\
&4.2 & $ 5.37 \times 10^{-7}$ & $ 4.21 \times 10^{-8}$ & $ 4.15 \times 10^{-8}$
& $ 5.84 \times 10^{-8}$ & $ 5.84 \times 10^{-8}$ \\
&4.8 & $ 2.27 \times 10^{-7}$ & $ 2.57 \times 10^{-8}$ & $ 2.52 \times 10^{-8}$
& $ 2.47 \times 10^{-8}$ & $ 2.47 \times 10^{-8}$ \\
&5.5 & $ 6.08 \times 10^{-8}$ & $ 1.11 \times 10^{-8}$ & $ 1.07 \times 10^{-8}$
& $ 7.65 \times 10^{-9}$ & $ 7.65 \times 10^{-9}$ \\
&6.5 & $ 1.91 \times 10^{-8}$ & $ 5.75 \times 10^{-9}$ & $ 5.45 \times 10^{-9}$
& $ 2.28 \times 10^{-9}$ & $ 2.28 \times 10^{-9}$ \\
&7.5 & $ 3.95 \times 10^{-9}$ & $ 4.02 \times 10^{-9}$ & $ 2.32 \times 10^{-9}$
& $ 5.98 \times 10^{-10}$ & $ 5.98 \times 10^{-10}$ \\
&8.5 & $ 2.39 \times 10^{-9}$ & $ 2.85 \times 10^{-9}$ & $ 1.36 \times 10^{-9}$
& $ 3.98 \times 10^{-10}$ & $ 3.98 \times 10^{-10}$ \\
\end{tabular} \end{ruledtabular}
\end{table*}


\begin{table*}[htbp]
\caption{Differential invariant yield of electrons ($\aee$) from
heavy-flavor decays 
	 for (upper) 40--60\% and (lower) 60--92\% centrality classes.  
         The $\pte$ is in units of GeV/$c$.  The yield and 
	 corresponding errors are in units of $(\mbox{GeV}/c)^{-2}$ 
\label{tab:inv_40_92}}
\begin{ruledtabular} \begin{tabular}{ccccccccc}
Centrality & \quad $\pte$ \quad & \quad Invariant yield \quad & \quad Stat.
error ($+$) \quad & \quad Stat.  error ($-$) \quad & \quad Sys.  error ($+$) \quad
& \quad Sys.  error ($-$) \quad \\
\hline
&0.3 & $ 2.13 \times 10^{-2}$ & $ 2.66 \times 10^{-3}$ & $ 2.63 \times 10^{-3}$
& 
$ 9.42 \times 10^{-3}$ & $ 9.31 \times 10^{-3}$ \\                              
 
&0.5 & $ 1.36 \times 10^{-2}$ & $ 1.20 \times 10^{-3}$ & $ 1.18 \times 10^{-3}$
& 
$ 3.69 \times 10^{-3}$ & $ 3.65 \times 10^{-3}$ \\                              
 
&0.6 & $ 8.48 \times 10^{-3}$ & $ 6.44 \times 10^{-4}$ & $ 6.38 \times 10^{-4}$
& 
$ 1.72 \times 10^{-3}$ & $ 1.70 \times 10^{-3}$ \\
&0.7 & $ 4.30 \times 10^{-3}$ & $ 4.00 \times 10^{-4}$ & $ 3.96 \times 10^{-4}$
&
$ 9.02 \times 10^{-4}$ & $ 8.94 \times 10^{-4}$ \\
&0.8 & $ 2.78 \times 10^{-3}$ & $ 2.51 \times 10^{-4}$ & $ 2.49 \times 10^{-4}$
&
$ 4.73 \times 10^{-4}$ & $ 4.69 \times 10^{-4}$ \\
&0.8 & $ 2.03 \times 10^{-3}$ & $ 1.71 \times 10^{-4}$ & $ 1.70 \times 10^{-4}$
&
$ 2.74 \times 10^{-4}$ & $ 2.72 \times 10^{-4}$ \\
&0.9 & $ 1.41 \times 10^{-3}$ & $ 1.23 \times 10^{-4}$ & $ 1.22 \times 10^{-4}$
&
$ 1.68 \times 10^{-4}$ & $ 1.67 \times 10^{-4}$ \\
&1.1 & $ 7.48 \times 10^{-4}$ & $ 5.70 \times 10^{-5}$ & $ 5.55 \times 10^{-5}$
&
$ 8.65 \times 10^{-5}$ & $ 8.42 \times 10^{-5}$ \\
&1.3 & $ 4.05 \times 10^{-4}$ & $ 1.05 \times 10^{-5}$ & $ 1.03 \times 10^{-5}$
&
$ 4.23 \times 10^{-5}$ & $ 4.13 \times 10^{-5}$ \\
&1.5 & $ 1.98 \times 10^{-4}$ & $ 5.00 \times 10^{-6}$ & $ 4.90 \times 10^{-6}$
&
$ 2.10 \times 10^{-5}$ & $ 2.06 \times 10^{-5}$ \\
&1.7 & $ 8.78 \times 10^{-5}$ & $ 2.01 \times 10^{-6}$ & $ 2.01 \times 10^{-6}$
&
$ 1.33 \times 10^{-5}$ & $ 1.33 \times 10^{-5}$ \\
&1.9 & $ 4.64 \times 10^{-5}$ & $ 1.22 \times 10^{-6}$ & $ 1.22 \times 10^{-6}$
&
$ 6.67 \times 10^{-6}$ & $ 6.67 \times 10^{-6}$ \\
40--60\% & 2.10 & $ 2.59 \times 10^{-5}$ & $ 7.91 \times 10^{-7}$ & $ 7.91
\times
10^{-7}$ & $ 3.54 \times 10^{-6}$ & $ 3.54 \times 10^{-6}$ \\
&2.3 & $ 1.50 \times 10^{-5}$ & $ 5.45 \times 10^{-7}$ & $ 5.45 \times 10^{-7}$
&
$ 1.96 \times 10^{-6}$ & $ 1.96 \times 10^{-6}$ \\
&2.5 & $ 9.10 \times 10^{-6}$ & $ 3.85 \times 10^{-7}$ & $ 3.85 \times 10^{-7}$
&
$ 1.13 \times 10^{-6}$ & $ 1.13 \times 10^{-6}$ \\
&2.7 & $ 5.05 \times 10^{-6}$ & $ 2.71 \times 10^{-7}$ & $ 2.71 \times 10^{-7}$
&
$ 6.37 \times 10^{-7}$ & $ 6.37 \times 10^{-7}$ \\
&2.9 & $ 2.90 \times 10^{-6}$ & $ 1.98 \times 10^{-7}$ & $ 1.96 \times 10^{-7}$
&
$ 3.70 \times 10^{-7}$ & $ 3.70 \times 10^{-7}$ \\
&3.1 & $ 2.27 \times 10^{-6}$ & $ 1.60 \times 10^{-7}$ & $ 1.58 \times 10^{-7}$
&
$ 2.54 \times 10^{-7}$ & $ 2.54 \times 10^{-7}$ \\
&3.3 & $ 1.25 \times 10^{-6}$ & $ 1.18 \times 10^{-7}$ & $ 1.16 \times 10^{-7}$
&
$ 1.50 \times 10^{-7}$ & $ 1.50 \times 10^{-7}$ \\
&3.5 & $ 8.68 \times 10^{-7}$ & $ 9.34 \times 10^{-8}$ & $ 9.18 \times 10^{-8}$
&
$ 9.93 \times 10^{-8}$ & $ 9.93 \times 10^{-8}$ \\
&3.7 & $ 7.14 \times 10^{-7}$ & $ 7.86 \times 10^{-8}$ & $ 7.70 \times 10^{-8}$
&
$ 7.43 \times 10^{-8}$ & $ 7.43 \times 10^{-8}$ \\
&3.9 & $ 3.85 \times 10^{-7}$ & $ 5.82 \times 10^{-8}$ & $ 5.67 \times 10^{-8}$
&
$ 4.40 \times 10^{-8}$ & $ 4.40 \times 10^{-8}$ \\
&4.2 & $ 2.30 \times 10^{-7}$ & $ 2.62 \times 10^{-8}$ & $ 2.57 \times 10^{-8}$
&
$ 2.41 \times 10^{-8}$ & $ 2.41 \times 10^{-8}$ \\
&4.8 & $ 1.19 \times 10^{-7}$ & $ 1.75 \times 10^{-8}$ & $ 1.71 \times 10^{-8}$
&
$ 1.17 \times 10^{-8}$ & $ 1.17 \times 10^{-8}$ \\
&5.5 & $ 2.70 \times 10^{-8}$ & $ 7.04 \times 10^{-9}$ & $ 6.71 \times 10^{-9}$
&
$ 3.21 \times 10^{-9}$ & $ 3.21 \times 10^{-9}$ \\
&6.5 & $ 1.03 \times 10^{-8}$ & $ 4.37 \times 10^{-9}$ & $ 3.21 \times 10^{-9}$
&
$ 1.08 \times 10^{-9}$ & $ 1.08 \times 10^{-9}$ \\
&7.5 & $ 1.23 \times 10^{-9}$ & $ 2.74 \times 10^{-9}$ & $ 1.13 \times 10^{-9}$
&
$ 2.04 \times 10^{-10}$ & $ 2.04 \times 10^{-10}$ \\
\hline
&0.3 & $ 5.28 \times 10^{-3}$ & $ 8.56 \times 10^{-4}$ & $ 8.47 \times 10^{-4}$
&
$ 2.16 \times 10^{-3}$ & $ 2.14 \times 10^{-3}$ \\
&0.5 & $ 1.83 \times 10^{-3}$ & $ 3.68 \times 10^{-4}$ & $ 3.65 \times 10^{-4}$
&
$ 8.06 \times 10^{-4}$ & $ 7.97 \times 10^{-4}$ \\
&0.6 & $ 1.18 \times 10^{-3}$ & $ 2.01 \times 10^{-4}$ & $ 1.99 \times 10^{-4}$
&
$ 3.72 \times 10^{-4}$ & $ 3.69 \times 10^{-4}$ \\
&0.7 & $ 7.89 \times 10^{-4}$ & $ 1.26 \times 10^{-4}$ & $ 1.25 \times 10^{-4}$
&
$ 1.93 \times 10^{-4}$ & $ 1.91 \times 10^{-4}$ \\
&0.8 & $ 4.27 \times 10^{-4}$ & $ 7.87 \times 10^{-5}$ & $ 7.81 \times 10^{-5}$
&
$ 9.85 \times 10^{-5}$ & $ 9.77 \times 10^{-5}$ \\
&0.8 & $ 2.30 \times 10^{-4}$ & $ 5.28 \times 10^{-5}$ & $ 5.24 \times 10^{-5}$
&
$ 5.39 \times 10^{-5}$ & $ 5.34 \times 10^{-5}$ \\
&0.9 & $ 1.99 \times 10^{-4}$ & $ 3.81 \times 10^{-5}$ & $ 3.78 \times 10^{-5}$
&
$ 3.29 \times 10^{-5}$ & $ 3.27 \times 10^{-5}$ \\
&1.1 & $ 1.27 \times 10^{-4}$ & $ 1.78 \times 10^{-5}$ & $ 1.74 \times 10^{-5}$
&
$ 1.74 \times 10^{-5}$ & $ 1.69 \times 10^{-5}$ \\
&1.3 & $ 6.15 \times 10^{-5}$ & $ 2.08 \times 10^{-6}$ & $ 2.03 \times 10^{-6}$
&
$ 1.04 \times 10^{-5}$ & $ 1.01 \times 10^{-5}$ \\
&1.5 & $ 2.99 \times 10^{-5}$ & $ 1.08 \times 10^{-6}$ & $ 1.06 \times 10^{-6}$
&
$ 5.21 \times 10^{-6}$ & $ 5.10 \times 10^{-6}$ \\
&1.7 & $ 1.39 \times 10^{-5}$ & $ 5.43 \times 10^{-7}$ & $ 5.43 \times 10^{-7}$
&
$ 2.27 \times 10^{-6}$ & $ 2.27 \times 10^{-6}$ \\
&1.9 & $ 7.09 \times 10^{-6}$ & $ 3.43 \times 10^{-7}$ & $ 3.43 \times 10^{-7}$
&
$ 1.14 \times 10^{-6}$ & $ 1.14 \times 10^{-6}$ \\
60--92\% & 2.10 & $ 3.08 \times 10^{-6}$ & $ 2.18 \times 10^{-7}$ & $ 2.18
\times
10^{-7}$ & $ 5.54 \times 10^{-7}$ & $ 5.54 \times 10^{-7}$ \\
&2.3 & $ 1.91 \times 10^{-6}$ & $ 1.57 \times 10^{-7}$ & $ 1.56 \times 10^{-7}$
&
$ 3.16 \times 10^{-7}$ & $ 3.16 \times 10^{-7}$ \\
&2.5 & $ 1.25 \times 10^{-6}$ & $ 1.15 \times 10^{-7}$ & $ 1.14 \times 10^{-7}$
&
$ 1.89 \times 10^{-7}$ & $ 1.89 \times 10^{-7}$ \\
&2.7 & $ 6.77 \times 10^{-7}$ & $ 8.19 \times 10^{-8}$ & $ 8.06 \times 10^{-8}$
&
$ 1.08 \times 10^{-7}$ & $ 1.08 \times 10^{-7}$ \\
&2.9 & $ 4.09 \times 10^{-7}$ & $ 6.08 \times 10^{-8}$ & $ 5.97 \times 10^{-8}$
&
$ 6.46 \times 10^{-8}$ & $ 6.46 \times 10^{-8}$ \\
&3.1 & $ 3.27 \times 10^{-7}$ & $ 4.95 \times 10^{-8}$ & $ 4.84 \times 10^{-8}$
&
$ 4.40 \times 10^{-8}$ & $ 4.40 \times 10^{-8}$ \\
&3.3 & $ 1.68 \times 10^{-7}$ & $ 3.62 \times 10^{-8}$ & $ 3.52 \times 10^{-8}$
&
$ 2.58 \times 10^{-8}$ & $ 2.58 \times 10^{-8}$ \\
&3.5 & $ 1.30 \times 10^{-7}$ & $ 2.94 \times 10^{-8}$ & $ 2.84 \times 10^{-8}$
&
$ 1.78 \times 10^{-8}$ & $ 1.78 \times 10^{-8}$ \\
&3.7 & $ 1.12 \times 10^{-7}$ & $ 2.52 \times 10^{-8}$ & $ 2.43 \times 10^{-8}$
&
$ 1.33 \times 10^{-8}$ & $ 1.33 \times 10^{-8}$ \\
&3.9 & $ 6.20 \times 10^{-8}$ & $ 1.92 \times 10^{-8}$ & $ 1.84 \times 10^{-8}$
&
$ 8.08 \times 10^{-9}$ & $ 8.08 \times 10^{-9}$ \\
&4.2 & $ 2.90 \times 10^{-8}$ & $ 7.93 \times 10^{-9}$ & $ 7.63 \times 10^{-9}$
&
$ 3.82 \times 10^{-9}$ & $ 3.82 \times 10^{-9}$ \\
&4.8 & $ 5.34 \times 10^{-9}$ & $ 4.10 \times 10^{-9}$ & $ 3.84 \times 10^{-9}$
&
$ 1.14 \times 10^{-9}$ & $ 1.14 \times 10^{-9}$ \\
&5.5 & $ 1.29 \times 10^{-9}$ & $ 2.17 \times 10^{-9}$ & $ 1.37 \times 10^{-9}$
&
$ 3.59 \times 10^{-10}$ & $ 3.59 \times 10^{-10}$ \\
&6.5 & $ 1.12 \times 10^{-9}$ & $ 1.48 \times 10^{-9}$ & $ 7.82 \times 10^{-10}$
& $ 1.51 \times 10^{-10}$ & $ 1.51 \times 10^{-10}$ \\
\end{tabular} \end{ruledtabular}
\end{table*}

\endgroup




\begingroup \squeezetable 

\begin{table*}[htbp]
\caption{$\raa(\pte)$ for (upper, minimum bias) 0--92\% and (lower) 0--10\%
centrality classes.  \label{tab:raa_0092_0_10}}
\begin{ruledtabular} \begin{tabular}{ccccccccccccccccccccccc}
Centrality & \quad $\pte$ \quad & \quad $\raa$ \quad & \quad Stat.  error ($+$)
\quad & \quad Stat.  error ($-$) \quad & \quad Sys.  error ($+$) \quad & \quad
Sys.  error ($-$) \quad \\
\hline
&0.3 & $ 8.74 \times 10^{-1}$ & $ 4.27 \times 10^{-1}$ & $ 2.18 \times 10^{-1}$ 
& $ 3.19 \times 10^{-1}$ & $ 2.89 \times 10^{-1}$ \\                            
&0.5 & $ 1.01 \times 10^{+0}$ & $ 3.96 \times 10^{-1}$ & $ 2.24 \times 10^{-1}$ 
& $ 3.43 \times 10^{-1}$ & $ 3.15 \times 10^{-1}$ \\                            
&0.6 & $ 1.01 \times 10^{+0}$ & $ 2.96 \times 10^{-1}$ & $ 1.89 \times 10^{-1}$ 
& $ 2.65 \times 10^{-1}$ & $ 2.46 \times 10^{-1}$ \\                            
&0.7 & $ 1.10 \times 10^{+0}$ & $ 3.12 \times 10^{-1}$ & $ 2.01 \times 10^{-1}$ 
& $ 2.64 \times 10^{-1}$ & $ 2.46 \times 10^{-1}$ \\                            
&0.8 & $ 9.34 \times 10^{-1}$ & $ 2.01 \times 10^{-1}$ & $ 1.42 \times 10^{-1}$ 
& $ 1.64 \times 10^{-1}$ & $ 1.54 \times 10^{-1}$ \\                            
&0.8 & $ 1.09 \times 10^{+0}$ & $ 2.42 \times 10^{-1}$ & $ 1.70 \times 10^{-1}$ 
& $ 1.77 \times 10^{-1}$ & $ 1.68 \times 10^{-1}$ \\
&0.9 & $ 9.49 \times 10^{-1}$ & $ 1.87 \times 10^{-1}$ & $ 1.37 \times 10^{-1}$
& $ 1.30 \times 10^{-1}$ & $ 1.24 \times 10^{-1}$ \\
&1.1 & $ 9.44 \times 10^{-1}$ & $ 1.29 \times 10^{-1}$ & $ 1.03 \times 10^{-1}$
& $ 1.17 \times 10^{-1}$ & $ 1.12 \times 10^{-1}$ \\
&1.3 & $ 1.06 \times 10^{+0}$ & $ 2.15 \times 10^{-1}$ & $ 1.54 \times 10^{-1}$
& $ 1.35 \times 10^{-1}$ & $ 1.29 \times 10^{-1}$ \\
&1.5 & $ 9.66 \times 10^{-1}$ & $ 2.27 \times 10^{-1}$ & $ 1.55 \times 10^{-1}$
& $ 1.16 \times 10^{-1}$ & $ 1.11 \times 10^{-1}$ \\
&1.7 & $ 9.11 \times 10^{-1}$ & $ 1.25 \times 10^{-2}$ & $ 1.25 \times 10^{-2}$
& $ 1.93 \times 10^{-1}$ & $ 1.76 \times 10^{-1}$ \\
&1.9 & $ 8.55 \times 10^{-1}$ & $ 1.21 \times 10^{-2}$ & $ 1.21 \times 10^{-2}$
& $ 1.78 \times 10^{-1}$ & $ 1.62 \times 10^{-1}$ \\
&2.1 & $ 8.00 \times 10^{-1}$ & $ 1.22 \times 10^{-2}$ & $ 1.22 \times 10^{-2}$
& $ 1.63 \times 10^{-1}$ & $ 1.48 \times 10^{-1}$ \\
&2.3 & $ 7.79 \times 10^{-1}$ & $ 1.29 \times 10^{-2}$ & $ 1.29 \times 10^{-2}$
& $ 1.53 \times 10^{-1}$ & $ 1.38 \times 10^{-1}$ \\
0--92\% & 2.50 & $ 7.47 \times 10^{-1}$ & $ 1.36 \times 10^{-2}$ & $ 1.36 \times
10^{-2}$ & $ 1.45 \times 10^{-1}$ & $ 1.29 \times 10^{-1}$ \\
&2.7 & $ 6.90 \times 10^{-1}$ & $ 1.46 \times 10^{-2}$ & $ 1.46 \times 10^{-2}$
& $ 1.30 \times 10^{-1}$ & $ 1.15 \times 10^{-1}$ \\
&2.9 & $ 6.47 \times 10^{-1}$ & $ 1.60 \times 10^{-2}$ & $ 1.60 \times 10^{-2}$
& $ 1.21 \times 10^{-1}$ & $ 1.07 \times 10^{-1}$ \\
&3.1 & $ 6.23 \times 10^{-1}$ & $ 1.79 \times 10^{-2}$ & $ 1.79 \times 10^{-2}$
& $ 1.15 \times 10^{-1}$ & $ 1.01 \times 10^{-1}$ \\
&3.3 & $ 5.84 \times 10^{-1}$ & $ 1.97 \times 10^{-2}$ & $ 1.97 \times 10^{-2}$
& $ 1.07 \times 10^{-1}$ & $ 9.44 \times 10^{-2}$ \\
&3.5 & $ 5.59 \times 10^{-1}$ & $ 2.21 \times 10^{-2}$ & $ 2.21 \times 10^{-2}$
& $ 1.02 \times 10^{-1}$ & $ 8.99 \times 10^{-2}$ \\
&3.7 & $ 6.18 \times 10^{-1}$ & $ 2.59 \times 10^{-2}$ & $ 2.59 \times 10^{-2}$
& $ 1.10 \times 10^{-1}$ & $ 9.58 \times 10^{-2}$ \\
&3.9 & $ 4.93 \times 10^{-1}$ & $ 2.76 \times 10^{-2}$ & $ 2.73 \times 10^{-2}$
& $ 9.03 \times 10^{-2}$ & $ 7.96 \times 10^{-2}$ \\
&4.2 & $ 4.74 \times 10^{-1}$ & $ 2.12 \times 10^{-2}$ & $ 2.12 \times 10^{-2}$
& $ 8.55 \times 10^{-2}$ & $ 7.50 \times 10^{-2}$ \\
&4.8 & $ 3.95 \times 10^{-1}$ & $ 2.75 \times 10^{-2}$ & $ 2.71 \times 10^{-2}$
& $ 7.24 \times 10^{-2}$ & $ 6.38 \times 10^{-2}$ \\
&5.5 & $ 3.07 \times 10^{-1}$ & $ 3.38 \times 10^{-2}$ & $ 3.32 \times 10^{-2}$
& $ 6.03 \times 10^{-2}$ & $ 5.41 \times 10^{-2}$ \\
&6.5 & $ 3.10 \times 10^{-1}$ & $ 5.85 \times 10^{-2}$ & $ 5.67 \times 10^{-2}$
& $ 6.06 \times 10^{-2}$ & $ 5.43 \times 10^{-2}$ \\
&7.5 & $ 4.68 \times 10^{-1}$ & $ 1.27 \times 10^{-1}$ & $ 1.22 \times 10^{-1}$
& $ 8.56 \times 10^{-2}$ & $ 7.53 \times 10^{-2}$ \\
&8.5 & $ 4.22 \times 10^{-1}$ & $ 2.23 \times 10^{-1}$ & $ 1.50 \times 10^{-1}$
& $ 9.42 \times 10^{-2}$ & $ 8.68 \times 10^{-2}$ \\
\hline
&0.3 & $ 1.08 \times 10^{+0}$ & $ 5.30 \times 10^{-1}$ & $ 2.72 \times 10^{-1}$
& $ 4.37 \times 10^{-1}$ & $ 4.03 \times 10^{-1}$ \\
&0.5 & $ 1.18 \times 10^{+0}$ & $ 4.65 \times 10^{-1}$ & $ 2.64 \times 10^{-1}$
& $ 4.37 \times 10^{-1}$ & $ 4.07 \times 10^{-1}$ \\
&0.6 & $ 1.14 \times 10^{+0}$ & $ 3.36 \times 10^{-1}$ & $ 2.15 \times 10^{-1}$
& $ 3.18 \times 10^{-1}$ & $ 2.98 \times 10^{-1}$ \\
&0.7 & $ 1.17 \times 10^{+0}$ & $ 3.36 \times 10^{-1}$ & $ 2.18 \times 10^{-1}$
& $ 2.96 \times 10^{-1}$ & $ 2.77 \times 10^{-1}$ \\
&0.8 & $ 1.00 \times 10^{+0}$ & $ 2.17 \times 10^{-1}$ & $ 1.55 \times 10^{-1}$
& $ 1.80 \times 10^{-1}$ & $ 1.70 \times 10^{-1}$ \\
&0.8 & $ 1.07 \times 10^{+0}$ & $ 2.40 \times 10^{-1}$ & $ 1.71 \times 10^{-1}$
& $ 1.78 \times 10^{-1}$ & $ 1.69 \times 10^{-1}$ \\
&0.9 & $ 1.01 \times 10^{+0}$ & $ 2.03 \times 10^{-1}$ & $ 1.50 \times 10^{-1}$
& $ 1.42 \times 10^{-1}$ & $ 1.35 \times 10^{-1}$ \\
&1.1 & $ 9.36 \times 10^{-1}$ & $ 1.32 \times 10^{-1}$ & $ 1.06 \times 10^{-1}$
& $ 1.19 \times 10^{-1}$ & $ 1.14 \times 10^{-1}$ \\
&1.3 & $ 1.13 \times 10^{+0}$ & $ 2.29 \times 10^{-1}$ & $ 1.64 \times 10^{-1}$
& $ 1.36 \times 10^{-1}$ & $ 1.30 \times 10^{-1}$ \\
&1.5 & $ 1.00 \times 10^{+0}$ & $ 2.35 \times 10^{-1}$ & $ 1.61 \times 10^{-1}$
& $ 1.15 \times 10^{-1}$ & $ 1.10 \times 10^{-1}$ \\
&1.7 & $ 8.71 \times 10^{-1}$ & $ 1.38 \times 10^{-2}$ & $ 1.38 \times 10^{-2}$
& $ 1.81 \times 10^{-1}$ & $ 1.65 \times 10^{-1}$ \\
&1.9 & $ 8.08 \times 10^{-1}$ & $ 1.42 \times 10^{-2}$ & $ 1.42 \times 10^{-2}$
& $ 1.65 \times 10^{-1}$ & $ 1.49 \times 10^{-1}$ \\
&2.1 & $ 7.29 \times 10^{-1}$ & $ 1.50 \times 10^{-2}$ & $ 1.50 \times 10^{-2}$
& $ 1.46 \times 10^{-1}$ & $ 1.32 \times 10^{-1}$ \\
0--10\% & 2.30 & $ 6.88 \times 10^{-1}$ & $ 1.66 \times 10^{-2}$ & $ 1.66 \times
10^{-2}$ & $ 1.34 \times 10^{-1}$ & $ 1.20 \times 10^{-1}$ \\
&2.5 & $ 6.42 \times 10^{-1}$ & $ 1.83 \times 10^{-2}$ & $ 1.83 \times 10^{-2}$
& $ 1.24 \times 10^{-1}$ & $ 1.11 \times 10^{-1}$ \\
&2.7 & $ 6.18 \times 10^{-1}$ & $ 2.07 \times 10^{-2}$ & $ 2.07 \times 10^{-2}$
& $ 1.15 \times 10^{-1}$ & $ 1.01 \times 10^{-1}$ \\
&2.9 & $ 5.83 \times 10^{-1}$ & $ 2.31 \times 10^{-2}$ & $ 2.31 \times 10^{-2}$
& $ 1.07 \times 10^{-1}$ & $ 9.44 \times 10^{-2}$ \\
&3.1 & $ 4.84 \times 10^{-1}$ & $ 2.54 \times 10^{-2}$ & $ 2.51 \times 10^{-2}$
& $ 9.01 \times 10^{-2}$ & $ 7.97 \times 10^{-2}$ \\
&3.3 & $ 4.97 \times 10^{-1}$ & $ 2.91 \times 10^{-2}$ & $ 2.88 \times 10^{-2}$
& $ 9.07 \times 10^{-2}$ & $ 7.99 \times 10^{-2}$ \\
&3.5 & $ 4.64 \times 10^{-1}$ & $ 3.29 \times 10^{-2}$ & $ 3.25 \times 10^{-2}$
& $ 8.47 \times 10^{-2}$ & $ 7.45 \times 10^{-2}$ \\
&3.7 & $ 4.89 \times 10^{-1}$ & $ 3.77 \times 10^{-2}$ & $ 3.72 \times 10^{-2}$
& $ 8.72 \times 10^{-2}$ & $ 7.62 \times 10^{-2}$ \\
&3.9 & $ 4.83 \times 10^{-1}$ & $ 4.39 \times 10^{-2}$ & $ 4.31 \times 10^{-2}$
& $ 8.60 \times 10^{-2}$ & $ 7.51 \times 10^{-2}$ \\
&4.2 & $ 3.62 \times 10^{-1}$ & $ 3.13 \times 10^{-2}$ & $ 3.08 \times 10^{-2}$
& $ 6.62 \times 10^{-2}$ & $ 5.83 \times 10^{-2}$ \\
&4.8 & $ 2.71 \times 10^{-1}$ & $ 3.97 \times 10^{-2}$ & $ 3.88 \times 10^{-2}$
& $ 5.15 \times 10^{-2}$ & $ 4.59 \times 10^{-2}$ \\
&5.5 & $ 2.49 \times 10^{-1}$ & $ 5.13 \times 10^{-2}$ & $ 4.95 \times 10^{-2}$
& $ 4.94 \times 10^{-2}$ & $ 4.44 \times 10^{-2}$ \\
&6.5 & $ 8.61 \times 10^{-2}$ & $ 7.88 \times 10^{-2}$ & $ 5.88 \times 10^{-2}$
& $ 2.58 \times 10^{-2}$ & $ 2.47 \times 10^{-2}$ \\
&7.5 & $ 5.14 \times 10^{-1}$ & $ 2.49 \times 10^{-1}$ & $ 1.68 \times 10^{-1}$
& $ 9.17 \times 10^{-2}$ & $ 8.02 \times 10^{-2}$ \\
&8.5 & $ 4.60 \times 10^{-1}$ & $ 4.12 \times 10^{-1}$ & $ 2.18 \times 10^{-1}$
& $ 9.77 \times 10^{-2}$ & $ 8.92 \times 10^{-2}$ \\
\end{tabular} \end{ruledtabular}
\end{table*}


\begin{table*}[htbp]
\caption{$\raa(\pte)$ for (upper) 10--20\% and (lower) 20--40\% centrality
classes.  \label{tab:raa_10_40}}
\begin{ruledtabular} \begin{tabular}{ccccccccccccccccccccccc}
Centrality & \quad $\pte$ \quad & \quad $\raa$ \quad & \quad Stat.  error ($+$)
\quad & \quad Stat.  error ($-$) \quad & \quad Sys.  error ($+$) \quad & \quad
Sys.  error ($-$) \quad \\
\hline
&0.3 & $ 9.72 \times 10^{-1}$ & $ 4.76 \times 10^{-1}$ & $ 2.46 \times 10^{-1}$
& 
$ 3.66 \times 10^{-1}$ & $ 3.34 \times 10^{-1}$ \\                              
 
&0.5 & $ 1.05 \times 10^{+0}$ & $ 4.13 \times 10^{-1}$ & $ 2.36 \times 10^{-1}$
& 
$ 3.62 \times 10^{-1}$ & $ 3.33 \times 10^{-1}$ \\                              
 
&0.6 & $ 1.09 \times 10^{+0}$ & $ 3.22 \times 10^{-1}$ & $ 2.08 \times 10^{-1}$
& 
$ 2.92 \times 10^{-1}$ & $ 2.72 \times 10^{-1}$ \\                              
 
&0.7 & $ 1.12 \times 10^{+0}$ & $ 3.22 \times 10^{-1}$ & $ 2.11 \times 10^{-1}$
& 
$ 2.71 \times 10^{-1}$ & $ 2.53 \times 10^{-1}$ \\                              
 
&0.8 & $ 1.01 \times 10^{+0}$ & $ 2.20 \times 10^{-1}$ & $ 1.58 \times 10^{-1}$
& 
$ 1.77 \times 10^{-1}$ & $ 1.67 \times 10^{-1}$ \\                              
 
&0.8 & $ 1.19 \times 10^{+0}$ & $ 2.67 \times 10^{-1}$ & $ 1.91 \times 10^{-1}$
& 
$ 1.95 \times 10^{-1}$ & $ 1.85 \times 10^{-1}$ \\
&0.9 & $ 9.82 \times 10^{-1}$ & $ 1.99 \times 10^{-1}$ & $ 1.48 \times 10^{-1}$
&
$ 1.34 \times 10^{-1}$ & $ 1.28 \times 10^{-1}$ \\
&1.1 & $ 9.76 \times 10^{-1}$ & $ 1.39 \times 10^{-1}$ & $ 1.12 \times 10^{-1}$
&
$ 1.21 \times 10^{-1}$ & $ 1.16 \times 10^{-1}$ \\
&1.3 & $ 1.22 \times 10^{+0}$ & $ 2.45 \times 10^{-1}$ & $ 1.76 \times 10^{-1}$
&
$ 1.48 \times 10^{-1}$ & $ 1.41 \times 10^{-1}$ \\
&1.5 & $ 1.07 \times 10^{+0}$ & $ 2.51 \times 10^{-1}$ & $ 1.72 \times 10^{-1}$
&
$ 1.21 \times 10^{-1}$ & $ 1.16 \times 10^{-1}$ \\
&1.7 & $ 9.74 \times 10^{-1}$ & $ 1.66 \times 10^{-2}$ & $ 1.66 \times 10^{-2}$
&
$ 2.00 \times 10^{-1}$ & $ 1.81 \times 10^{-1}$ \\
&1.9 & $ 9.25 \times 10^{-1}$ & $ 1.75 \times 10^{-2}$ & $ 1.75 \times 10^{-2}$
&
$ 1.85 \times 10^{-1}$ & $ 1.67 \times 10^{-1}$ \\
&2.1 & $ 8.68 \times 10^{-1}$ & $ 1.90 \times 10^{-2}$ & $ 1.90 \times 10^{-2}$
&
$ 1.70 \times 10^{-1}$ & $ 1.52 \times 10^{-1}$ \\
&2.3 & $ 8.61 \times 10^{-1}$ & $ 2.15 \times 10^{-2}$ & $ 2.15 \times 10^{-2}$
&
$ 1.63 \times 10^{-1}$ & $ 1.45 \times 10^{-1}$ \\
10--20\% & 2.50 & $ 7.78 \times 10^{-1}$ & $ 2.37 \times 10^{-2}$ & $ 2.37
\times
10^{-2}$ & $ 1.47 \times 10^{-1}$ & $ 1.30 \times 10^{-1}$ \\
&2.7 & $ 7.00 \times 10^{-1}$ & $ 2.63 \times 10^{-2}$ & $ 2.63 \times 10^{-2}$
&
$ 1.29 \times 10^{-1}$ & $ 1.14 \times 10^{-1}$ \\
&2.9 & $ 6.39 \times 10^{-1}$ & $ 2.93 \times 10^{-2}$ & $ 2.93 \times 10^{-2}$
&
$ 1.17 \times 10^{-1}$ & $ 1.03 \times 10^{-1}$ \\
&3.1 & $ 6.43 \times 10^{-1}$ & $ 3.40 \times 10^{-2}$ & $ 3.37 \times 10^{-2}$
&
$ 1.15 \times 10^{-1}$ & $ 1.01 \times 10^{-1}$ \\
&3.3 & $ 5.67 \times 10^{-1}$ & $ 3.74 \times 10^{-2}$ & $ 3.70 \times 10^{-2}$
&
$ 1.02 \times 10^{-1}$ & $ 8.97 \times 10^{-2}$ \\
&3.5 & $ 5.88 \times 10^{-1}$ & $ 4.34 \times 10^{-2}$ & $ 4.29 \times 10^{-2}$
&
$ 1.04 \times 10^{-1}$ & $ 9.11 \times 10^{-2}$ \\
&3.7 & $ 6.65 \times 10^{-1}$ & $ 5.23 \times 10^{-2}$ & $ 5.16 \times 10^{-2}$
&
$ 1.15 \times 10^{-1}$ & $ 9.98 \times 10^{-2}$ \\
&3.9 & $ 4.19 \times 10^{-1}$ & $ 5.11 \times 10^{-2}$ & $ 5.01 \times 10^{-2}$
&
$ 7.68 \times 10^{-2}$ & $ 6.77 \times 10^{-2}$ \\
&4.2 & $ 5.18 \times 10^{-1}$ & $ 4.31 \times 10^{-2}$ & $ 4.24 \times 10^{-2}$
&
$ 9.10 \times 10^{-2}$ & $ 7.91 \times 10^{-2}$ \\
&4.8 & $ 3.89 \times 10^{-1}$ & $ 5.48 \times 10^{-2}$ & $ 5.35 \times 10^{-2}$
&
$ 7.02 \times 10^{-2}$ & $ 6.15 \times 10^{-2}$ \\
&5.5 & $ 3.14 \times 10^{-1}$ & $ 6.79 \times 10^{-2}$ & $ 6.54 \times 10^{-2}$
&
$ 5.99 \times 10^{-2}$ & $ 5.33 \times 10^{-2}$ \\
&6.5 & $ 4.50 \times 10^{-1}$ & $ 1.35 \times 10^{-1}$ & $ 1.27 \times 10^{-1}$
&
$ 8.07 \times 10^{-2}$ & $ 7.07 \times 10^{-2}$ \\
&7.5 & $ 8.39 \times 10^{-1}$ & $ 3.88 \times 10^{-1}$ & $ 2.53 \times 10^{-1}$
&
$ 1.42 \times 10^{-1}$ & $ 1.22 \times 10^{-1}$ \\
&8.5 & $ 7.03 \times 10^{-1}$ & $ 6.32 \times 10^{-1}$ & $ 3.08 \times 10^{-1}$
&
$ 1.40 \times 10^{-1}$ & $ 1.26 \times 10^{-1}$ \\
\hline
&0.3 & $ 7.65 \times 10^{-1}$ & $ 3.76 \times 10^{-1}$ & $ 1.96 \times 10^{-1}$
& $ 2.82 \times 10^{-1}$ & $ 2.56 \times 10^{-1}$ \\
&0.5 & $ 9.79 \times 10^{-1}$ & $ 3.87 \times 10^{-1}$ & $ 2.21 \times 10^{-1}$
& $ 3.21 \times 10^{-1}$ & $ 2.93 \times 10^{-1}$ \\
&0.6 & $ 9.47 \times 10^{-1}$ & $ 2.80 \times 10^{-1}$ & $ 1.81 \times 10^{-1}$
& $ 2.43 \times 10^{-1}$ & $ 2.24 \times 10^{-1}$ \\
&0.7 & $ 1.21 \times 10^{+0}$ & $ 3.48 \times 10^{-1}$ & $ 2.27 \times 10^{-1}$
& $ 2.89 \times 10^{-1}$ & $ 2.69 \times 10^{-1}$ \\
&0.8 & $ 9.71 \times 10^{-1}$ & $ 2.12 \times 10^{-1}$ & $ 1.53 \times 10^{-1}$
& $ 1.70 \times 10^{-1}$ & $ 1.60 \times 10^{-1}$ \\
&0.8 & $ 1.23 \times 10^{+0}$ & $ 2.75 \times 10^{-1}$ & $ 1.96 \times 10^{-1}$
& $ 1.97 \times 10^{-1}$ & $ 1.86 \times 10^{-1}$ \\
&0.9 & $ 9.56 \times 10^{-1}$ & $ 1.94 \times 10^{-1}$ & $ 1.45 \times 10^{-1}$
& $ 1.32 \times 10^{-1}$ & $ 1.26 \times 10^{-1}$ \\
&1.1 & $ 1.06 \times 10^{+0}$ & $ 1.49 \times 10^{-1}$ & $ 1.21 \times 10^{-1}$
& $ 1.29 \times 10^{-1}$ & $ 1.24 \times 10^{-1}$ \\
&1.3 & $ 1.29 \times 10^{+0}$ & $ 2.61 \times 10^{-1}$ & $ 1.87 \times 10^{-1}$
& $ 1.51 \times 10^{-1}$ & $ 1.43 \times 10^{-1}$ \\
&1.5 & $ 1.17 \times 10^{+0}$ & $ 2.75 \times 10^{-1}$ & $ 1.88 \times 10^{-1}$
& $ 1.33 \times 10^{-1}$ & $ 1.27 \times 10^{-1}$ \\
&1.7 & $ 1.08 \times 10^{+0}$ & $ 1.81 \times 10^{-2}$ & $ 1.81 \times 10^{-2}$
& $ 2.25 \times 10^{-1}$ & $ 2.04 \times 10^{-1}$ \\
&1.9 & $ 1.03 \times 10^{+0}$ & $ 1.89 \times 10^{-2}$ & $ 1.89 \times 10^{-2}$
& $ 2.09 \times 10^{-1}$ & $ 1.90 \times 10^{-1}$ \\
&2.1 & $ 9.81 \times 10^{-1}$ & $ 2.05 \times 10^{-2}$ & $ 2.05 \times 10^{-2}$
& $ 1.95 \times 10^{-1}$ & $ 1.75 \times 10^{-1}$ \\
&2.3 & $ 9.30 \times 10^{-1}$ & $ 2.27 \times 10^{-2}$ & $ 2.27 \times 10^{-2}$
& $ 1.80 \times 10^{-1}$ & $ 1.61 \times 10^{-1}$ \\
20--40\% &2.5 & $ 8.91 \times 10^{-1}$ & $ 2.54 \times 10^{-2}$ & $ 2.54 \times
10^{-2}$ & $ 1.71 \times 10^{-1}$ & $ 1.52 \times 10^{-1}$ \\
&2.7 & $ 8.46 \times 10^{-1}$ & $ 2.84 \times 10^{-2}$ & $ 2.84 \times 10^{-2}$
& $ 1.55 \times 10^{-1}$ & $ 1.37 \times 10^{-1}$ \\
&2.9 & $ 8.20 \times 10^{-1}$ & $ 3.24 \times 10^{-2}$ & $ 3.24 \times 10^{-2}$
& $ 1.49 \times 10^{-1}$ & $ 1.31 \times 10^{-1}$ \\
&3.1 & $ 7.86 \times 10^{-1}$ & $ 3.68 \times 10^{-2}$ & $ 3.68 \times 10^{-2}$
& $ 1.41 \times 10^{-1}$ & $ 1.24 \times 10^{-1}$ \\
&3.3 & $ 7.50 \times 10^{-1}$ & $ 4.17 \times 10^{-2}$ & $ 4.13 \times 10^{-2}$
& $ 1.34 \times 10^{-1}$ & $ 1.17 \times 10^{-1}$ \\
&3.5 & $ 6.77 \times 10^{-1}$ & $ 4.63 \times 10^{-2}$ & $ 4.57 \times 10^{-2}$
& $ 1.21 \times 10^{-1}$ & $ 1.06 \times 10^{-1}$ \\
&3.7 & $ 7.26 \times 10^{-1}$ & $ 5.44 \times 10^{-2}$ & $ 5.37 \times 10^{-2}$
& $ 1.27 \times 10^{-1}$ & $ 1.11 \times 10^{-1}$ \\
&3.9 & $ 5.87 \times 10^{-1}$ & $ 5.73 \times 10^{-2}$ & $ 5.63 \times 10^{-2}$
& $ 1.05 \times 10^{-1}$ & $ 9.24 \times 10^{-2}$ \\
&4.2 & $ 5.80 \times 10^{-1}$ & $ 4.54 \times 10^{-2}$ & $ 4.48 \times 10^{-2}$
& $ 1.02 \times 10^{-1}$ & $ 8.92 \times 10^{-2}$ \\
&4.8 & $ 5.26 \times 10^{-1}$ & $ 5.96 \times 10^{-2}$ & $ 5.84 \times 10^{-2}$
& $ 9.28 \times 10^{-2}$ & $ 8.09 \times 10^{-2}$ \\
&5.5 & $ 4.01 \times 10^{-1}$ & $ 7.31 \times 10^{-2}$ & $ 7.08 \times 10^{-2}$
& $ 7.52 \times 10^{-2}$ & $ 6.67 \times 10^{-2}$ \\
&6.5 & $ 4.36 \times 10^{-1}$ & $ 1.32 \times 10^{-1}$ & $ 1.25 \times 10^{-1}$
& $ 8.01 \times 10^{-2}$ & $ 7.06 \times 10^{-2}$ \\
&7.5 & $ 2.75 \times 10^{-1}$ & $ 2.80 \times 10^{-1}$ & $ 1.62 \times 10^{-1}$
& $ 5.65 \times 10^{-2}$ & $ 5.13 \times 10^{-2}$ \\
&8.5 & $ 4.54 \times 10^{-1}$ & $ 5.42 \times 10^{-1}$ & $ 2.58 \times 10^{-1}$
& $ 9.85 \times 10^{-2}$ & $ 9.03 \times 10^{-2}$ \\
\end{tabular} \end{ruledtabular}
\end{table*}


\begin{table*}[htbp]
\caption{$\raa(\pte)$ for (upper) 40--60\% and (lower) 60--92\% centrality
classes.  \label{tab:raa_40_92}}
\begin{ruledtabular} \begin{tabular}{ccccccccccccccccccccccc}
Centrality & \quad $\pte$ \quad & \quad $\raa$ \quad & \quad Stat.  error ($+$)
\quad & \quad Stat.  error ($-$) \quad & \quad Sys.  error ($+$) \quad & \quad
Sys.  error ($-$) \quad \\
\hline
&0.3 & $ 7.27 \times 10^{-1}$ & $ 3.65 \times 10^{-1}$ & $ 2.01 \times 10^{-1}$
& 
$ 3.23 \times 10^{-1}$ & $ 3.03 \times 10^{-1}$ \\                              
 
&0.5 & $ 1.04 \times 10^{+0}$ & $ 4.18 \times 10^{-1}$ & $ 2.46 \times 10^{-1}$
& 
$ 3.39 \times 10^{-1}$ & $ 3.09 \times 10^{-1}$ \\                              
 
&0.6 & $ 1.17 \times 10^{+0}$ & $ 3.51 \times 10^{-1}$ & $ 2.32 \times 10^{-1}$
& 
$ 2.99 \times 10^{-1}$ & $ 2.76 \times 10^{-1}$ \\
&0.7 & $ 1.10 \times 10^{+0}$ & $ 3.28 \times 10^{-1}$ & $ 2.23 \times 10^{-1}$
&
$ 2.62 \times 10^{-1}$ & $ 2.44 \times 10^{-1}$ \\
&0.8 & $ 9.89 \times 10^{-1}$ & $ 2.28 \times 10^{-1}$ & $ 1.72 \times 10^{-1}$
&
$ 1.80 \times 10^{-1}$ & $ 1.71 \times 10^{-1}$ \\
&0.8 & $ 1.25 \times 10^{+0}$ & $ 2.94 \times 10^{-1}$ & $ 2.18 \times 10^{-1}$
&
$ 2.01 \times 10^{-1}$ & $ 1.90 \times 10^{-1}$ \\
&0.9 & $ 1.17 \times 10^{+0}$ & $ 2.49 \times 10^{-1}$ & $ 1.92 \times 10^{-1}$
&
$ 1.59 \times 10^{-1}$ & $ 1.51 \times 10^{-1}$ \\
&1.1 & $ 1.09 \times 10^{+0}$ & $ 1.68 \times 10^{-1}$ & $ 1.41 \times 10^{-1}$
&
$ 1.38 \times 10^{-1}$ & $ 1.32 \times 10^{-1}$ \\
&1.3 & $ 1.49 \times 10^{+0}$ & $ 3.02 \times 10^{-1}$ & $ 2.17 \times 10^{-1}$
&
$ 1.83 \times 10^{-1}$ & $ 1.75 \times 10^{-1}$ \\
&1.5 & $ 1.40 \times 10^{+0}$ & $ 3.28 \times 10^{-1}$ & $ 2.25 \times 10^{-1}$
&
$ 1.53 \times 10^{-1}$ & $ 1.46 \times 10^{-1}$ \\
&1.7 & $ 1.22 \times 10^{+0}$ & $ 2.79 \times 10^{-2}$ & $ 2.79 \times 10^{-2}$
&
$ 2.57 \times 10^{-1}$ & $ 2.34 \times 10^{-1}$ \\
&1.9 & $ 1.19 \times 10^{+0}$ & $ 3.11 \times 10^{-2}$ & $ 3.11 \times 10^{-2}$
&
$ 2.45 \times 10^{-1}$ & $ 2.22 \times 10^{-1}$ \\
&2.1 & $ 1.17 \times 10^{+0}$ & $ 3.57 \times 10^{-2}$ & $ 3.57 \times 10^{-2}$
&
$ 2.33 \times 10^{-1}$ & $ 2.10 \times 10^{-1}$ \\
&2.3 & $ 1.14 \times 10^{+0}$ & $ 4.15 \times 10^{-2}$ & $ 4.15 \times 10^{-2}$
&
$ 2.22 \times 10^{-1}$ & $ 1.99 \times 10^{-1}$ \\
40--60\% & 2.50 & $ 1.13 \times 10^{+0}$ & $ 4.78 \times 10^{-2}$ & $ 4.78
\times
10^{-2}$ & $ 2.16 \times 10^{-1}$ & $ 1.92 \times 10^{-1}$ \\
&2.7 & $ 9.93 \times 10^{-1}$ & $ 5.34 \times 10^{-2}$ & $ 5.34 \times 10^{-2}$
&
$ 1.86 \times 10^{-1}$ & $ 1.65 \times 10^{-1}$ \\
&2.9 & $ 8.81 \times 10^{-1}$ & $ 6.01 \times 10^{-2}$ & $ 5.95 \times 10^{-2}$
&
$ 1.66 \times 10^{-1}$ & $ 1.48 \times 10^{-1}$ \\
&3.1 & $ 1.04 \times 10^{+0}$ & $ 7.32 \times 10^{-2}$ & $ 7.24 \times 10^{-2}$
&
$ 1.85 \times 10^{-1}$ & $ 1.62 \times 10^{-1}$ \\
&3.3 & $ 8.49 \times 10^{-1}$ & $ 8.00 \times 10^{-2}$ & $ 7.88 \times 10^{-2}$
&
$ 1.56 \times 10^{-1}$ & $ 1.37 \times 10^{-1}$ \\
&3.5 & $ 8.54 \times 10^{-1}$ & $ 9.19 \times 10^{-2}$ & $ 9.03 \times 10^{-2}$
&
$ 1.54 \times 10^{-1}$ & $ 1.35 \times 10^{-1}$ \\
&3.7 & $ 1.01 \times 10^{+0}$ & $ 1.11 \times 10^{-1}$ & $ 1.09 \times 10^{-1}$
&
$ 1.75 \times 10^{-1}$ & $ 1.51 \times 10^{-1}$ \\
&3.9 & $ 7.66 \times 10^{-1}$ & $ 1.16 \times 10^{-1}$ & $ 1.13 \times 10^{-1}$
&
$ 1.38 \times 10^{-1}$ & $ 1.21 \times 10^{-1}$ \\
&4.2 & $ 8.12 \times 10^{-1}$ & $ 9.27 \times 10^{-2}$ & $ 9.09 \times 10^{-2}$
&
$ 1.41 \times 10^{-1}$ & $ 1.23 \times 10^{-1}$ \\
&4.8 & $ 9.03 \times 10^{-1}$ & $ 1.33 \times 10^{-1}$ & $ 1.30 \times 10^{-1}$
&
$ 1.54 \times 10^{-1}$ & $ 1.32 \times 10^{-1}$ \\
&5.5 & $ 5.82 \times 10^{-1}$ & $ 1.52 \times 10^{-1}$ & $ 1.45 \times 10^{-1}$
&
$ 1.07 \times 10^{-1}$ & $ 9.39 \times 10^{-2}$ \\
&6.5 & $ 7.75 \times 10^{-1}$ & $ 3.28 \times 10^{-1}$ & $ 2.40 \times 10^{-1}$
&
$ 1.35 \times 10^{-1}$ & $ 1.17 \times 10^{-1}$ \\
&7.5 & $ 2.79 \times 10^{-1}$ & $ 6.24 \times 10^{-1}$ & $ 2.57 \times 10^{-1}$
&
$ 6.06 \times 10^{-2}$ & $ 5.56 \times 10^{-2}$ \\
\hline
&0.3 & $ 1.11 \times 10^{+0}$ & $ 5.69 \times 10^{-1}$ & $ 3.26 \times 10^{-1}$
& $ 4.14 \times 10^{-1}$ & $ 3.76 \times 10^{-1}$ \\
&0.5 & $ 8.62 \times 10^{-1}$ & $ 3.80 \times 10^{-1}$ & $ 2.56 \times 10^{-1}$
& $ 3.25 \times 10^{-1}$ & $ 3.04 \times 10^{-1}$ \\
&0.6 & $ 1.01 \times 10^{+0}$ & $ 3.39 \times 10^{-1}$ & $ 2.51 \times 10^{-1}$
& $ 2.78 \times 10^{-1}$ & $ 2.59 \times 10^{-1}$ \\
&0.7 & $ 1.25 \times 10^{+0}$ & $ 4.05 \times 10^{-1}$ & $ 3.00 \times 10^{-1}$
& $ 2.95 \times 10^{-1}$ & $ 2.75 \times 10^{-1}$ \\
&0.8 & $ 9.38 \times 10^{-1}$ & $ 2.64 \times 10^{-1}$ & $ 2.21 \times 10^{-1}$
& $ 1.86 \times 10^{-1}$ & $ 1.77 \times 10^{-1}$ \\
&0.8 & $ 8.77 \times 10^{-1}$ & $ 2.78 \times 10^{-1}$ & $ 2.40 \times 10^{-1}$
& $ 1.74 \times 10^{-1}$ & $ 1.68 \times 10^{-1}$ \\
&0.9 & $ 1.01 \times 10^{+0}$ & $ 2.77 \times 10^{-1}$ & $ 2.39 \times 10^{-1}$
& $ 1.49 \times 10^{-1}$ & $ 1.43 \times 10^{-1}$ \\
&1.1 & $ 1.14 \times 10^{+0}$ & $ 2.22 \times 10^{-1}$ & $ 1.97 \times 10^{-1}$
& $ 1.44 \times 10^{-1}$ & $ 1.38 \times 10^{-1}$ \\
&1.3 & $ 1.40 \times 10^{+0}$ & $ 2.84 \times 10^{-1}$ & $ 2.05 \times 10^{-1}$
& $ 1.85 \times 10^{-1}$ & $ 1.78 \times 10^{-1}$ \\
&1.5 & $ 1.30 \times 10^{+0}$ & $ 3.07 \times 10^{-1}$ & $ 2.12 \times 10^{-1}$
& $ 1.67 \times 10^{-1}$ & $ 1.61 \times 10^{-1}$ \\
&1.7 & $ 1.19 \times 10^{+0}$ & $ 4.64 \times 10^{-2}$ & $ 4.64 \times 10^{-2}$
& $ 2.60 \times 10^{-1}$ & $ 2.39 \times 10^{-1}$ \\
&1.9 & $ 1.12 \times 10^{+0}$ & $ 5.43 \times 10^{-2}$ & $ 5.43 \times 10^{-2}$
& $ 2.44 \times 10^{-1}$ & $ 2.24 \times 10^{-1}$ \\
&2.1 & $ 8.56 \times 10^{-1}$ & $ 6.06 \times 10^{-2}$ & $ 6.06 \times 10^{-2}$
& $ 1.98 \times 10^{-1}$ & $ 1.84 \times 10^{-1}$ \\
&2.3 & $ 8.95 \times 10^{-1}$ & $ 7.38 \times 10^{-2}$ & $ 7.31 \times 10^{-2}$
& $ 1.97 \times 10^{-1}$ & $ 1.81 \times 10^{-1}$ \\
60--92\% & 2.50 & $ 9.55 \times 10^{-1}$ & $ 8.84 \times 10^{-2}$ & $ 8.73 \times
10^{-2}$ & $ 2.00 \times 10^{-1}$ & $ 1.82 \times 10^{-1}$ \\
&2.7 & $ 8.21 \times 10^{-1}$ & $ 9.93 \times 10^{-2}$ & $ 9.78 \times 10^{-2}$
& $ 1.73 \times 10^{-1}$ & $ 1.58 \times 10^{-1}$ \\
&2.9 & $ 7.66 \times 10^{-1}$ & $ 1.14 \times 10^{-1}$ & $ 1.12 \times 10^{-1}$
& $ 1.61 \times 10^{-1}$ & $ 1.47 \times 10^{-1}$ \\
&3.1 & $ 9.24 \times 10^{-1}$ & $ 1.40 \times 10^{-1}$ & $ 1.37 \times 10^{-1}$
& $ 1.79 \times 10^{-1}$ & $ 1.60 \times 10^{-1}$ \\
&3.3 & $ 7.01 \times 10^{-1}$ & $ 1.51 \times 10^{-1}$ & $ 1.47 \times 10^{-1}$
& $ 1.45 \times 10^{-1}$ & $ 1.32 \times 10^{-1}$ \\
&3.5 & $ 7.92 \times 10^{-1}$ & $ 1.78 \times 10^{-1}$ & $ 1.73 \times 10^{-1}$
& $ 1.54 \times 10^{-1}$ & $ 1.38 \times 10^{-1}$ \\
&3.7 & $ 9.76 \times 10^{-1}$ & $ 2.19 \times 10^{-1}$ & $ 2.11 \times 10^{-1}$
& $ 1.78 \times 10^{-1}$ & $ 1.57 \times 10^{-1}$ \\
&3.9 & $ 7.62 \times 10^{-1}$ & $ 2.36 \times 10^{-1}$ & $ 2.26 \times 10^{-1}$
& $ 1.45 \times 10^{-1}$ & $ 1.29 \times 10^{-1}$ \\
&4.2 & $ 6.33 \times 10^{-1}$ & $ 1.73 \times 10^{-1}$ & $ 1.66 \times 10^{-1}$
& $ 1.21 \times 10^{-1}$ & $ 1.08 \times 10^{-1}$ \\
&4.8 & $ 2.50 \times 10^{-1}$ & $ 1.92 \times 10^{-1}$ & $ 1.80 \times 10^{-1}$
& $ 6.38 \times 10^{-2}$ & $ 6.00 \times 10^{-2}$ \\
&5.5 & $ 1.72 \times 10^{-1}$ & $ 2.89 \times 10^{-1}$ & $ 1.83 \times 10^{-1}$
& $ 5.35 \times 10^{-2}$ & $ 5.14 \times 10^{-2}$ \\
&6.5 & $ 5.20 \times 10^{-1}$ & $ 6.86 \times 10^{-1}$ & $ 3.62 \times 10^{-1}$
& $ 1.01 \times 10^{-1}$ & $ 8.99 \times 10^{-2}$ \\
\end{tabular} \end{ruledtabular}
\end{table*}
 

\endgroup

\begingroup \squeezetable 


\begin{table*}[htbp]
\caption{$\pte$-integrated nuclear modification factors $\raa$.  
\label{tab:Int_raa_0_092}}
\begin{ruledtabular} \begin{tabular}{ccccccc}
$\pte$ & \quad $\npart$ \quad & \quad $\raa$ \quad & \quad Stat.  error ($+$)
\quad & \quad Stat error ($-$) \quad & \quad Sys.  error ($\pm$) \quad & Common
fractional error \\
(GeV/$c$) & & & & & & from \pp~cross section.  \\
\hline
&14.5 & $ 1.04 \times 10^{+0}$ & $ 7.64 \times 10^{-2}$ & $ 7.55 \times 10^{-2}$
&
$ 3.88 \times 10^{-1}$ &\\
&59.95 & $ 9.86 \times 10^{-1}$ & $ 3.90 \times 10^{-2}$ & $ 3.86 \times
10^{-2}$
& $ 2.53 \times 10^{-1}$ &\\
$ >$ 0.30 &140.4 & $ 9.43 \times 10^{-1}$ & $ 2.29 \times 10^{-2}$ & $ 2.26
\times
10^{-2}$ & $ 2.19 \times 10^{-1}$ &
$ +2.325 \times 10^{-1}-1.587 \times 10^{-1}$ \\
&234.6 & $ 1.03 \times 10^{+0}$ & $ 2.38 \times 10^{-2}$ & $ 2.35 \times
10^{-2}$
& $ 2.43 \times 10^{-1}$ &\\
&325.2 & $ 1.09 \times 10^{+0}$ & $ 2.20 \times 10^{-2}$ & $ 2.17 \times
10^{-2}$
& $ 2.71 \times 10^{-1}$ &\\
\hline
&14.5 & $ 1.08 \times 10^{+0}$ & $ 7.16 \times 10^{-2}$ & $ 7.09 \times 10^{-2}$
&
$ 3.49 \times 10^{-1}$ &\\
&59.95 & $ 1.14 \times 10^{+0}$ & $ 3.71 \times 10^{-2}$ & $ 3.67 \times
10^{-2}$
& $ 2.08 \times 10^{-1}$ &\\
$ >$ 0.60 &140.4 & $ 1.10 \times 10^{+0}$ & $ 2.11 \times 10^{-2}$ & $ 2.09
\times
10^{-2}$ & $ 1.71 \times 10^{-1}$ &
$ +1.309 \times 10^{-1}-1.037 \times 10^{-1}$ \\
&234.6 & $ 1.05 \times 10^{+0}$ & $ 2.11 \times 10^{-2}$ & $ 2.08 \times
10^{-2}$
& $ 1.59 \times 10^{-1}$ &\\
&325.2 & $ 1.03 \times 10^{+0}$ & $ 1.84 \times 10^{-2}$ & $ 1.81 \times
10^{-2}$
& $ 1.59 \times 10^{-1}$ &\\
\hline
&14.5 & $ 1.07 \times 10^{+0}$ & $ 7.27 \times 10^{-2}$ & $ 7.17 \times 10^{-2}$
&
$ 3.33 \times 10^{-1}$ &\\
&59.95 & $ 1.20 \times 10^{+0}$ & $ 3.82 \times 10^{-2}$ & $ 3.77 \times
10^{-2}$
& $ 1.97 \times 10^{-1}$ &\\
$ >0.8$ &140.4 & $ 1.09 \times 10^{+0}$ & $ 2.14 \times 10^{-2}$ & $ 2.11 \times
10^{-2}$ & $ 1.49 \times 10^{-1}$ &
$ +1.096 \times 10^{-1}-8.989 \times 10^{-2}$ \\
&234.6 & $ 1.04 \times 10^{+0}$ & $ 2.13 \times 10^{-2}$ & $ 2.09 \times
10^{-2}$
& $ 1.35 \times 10^{-1}$ &\\
&325.2 & $ 9.77 \times 10^{-1}$ & $ 1.81 \times 10^{-2}$ & $ 1.78 \times
10^{-2}$
& $ 1.29 \times 10^{-1}$ &\\
\hline
&14.5 & $ 7.78 \times 10^{-1}$ & $ 2.96 \times 10^{-2}$ & $ 2.92 \times 10^{-2}$
&
$ 2.57 \times 10^{-1}$ &\\
&59.95 & $ 9.91 \times 10^{-1}$ & $ 1.66 \times 10^{-2}$ & $ 1.66 \times
10^{-2}$
& $ 1.77 \times 10^{-1}$ &\\
$ >2.0$ &140.4 & $ 8.13 \times 10^{-1}$ & $ 9.13 \times 10^{-3}$ & $ 9.12 \times
10^{-3}$ & $ 1.25 \times 10^{-1}$ &
$ +1.096 \times 10^{-1}-8.989 \times 10^{-2}$ \\
&234.6 & $ 7.10 \times 10^{-1}$ & $ 8.52 \times 10^{-3}$ & $ 8.50 \times
10^{-3}$
& $ 1.03 \times 10^{-1}$ &\\
&325.2 & $ 5.89 \times 10^{-1}$ & $ 6.64 \times 10^{-3}$ & $ 6.63 \times
10^{-3}$
& $ 8.88 \times 10^{-2}$ &\\
\hline
&14.5 & $ 7.04 \times 10^{-1}$ & $ 6.39 \times 10^{-2}$ & $ 5.98 \times 10^{-2}$
&
$ 2.24 \times 10^{-1}$ &\\
&59.95 & $ 8.55 \times 10^{-1}$ & $ 3.35 \times 10^{-2}$ & $ 3.25 \times
10^{-2}$
& $ 1.40 \times 10^{-1}$ &\\
$ >$ 3.0&140.4 & $ 6.52 \times 10^{-1}$ & $ 1.67 \times 10^{-2}$ &
$ 1.64 \times 10^{-2}$ & $ 9.09 \times 10^{-2}$ &
$ +9.049 \times 10^{-2}-7.662 \times 10^{-2}$ \\
&234.6 & $ 5.40 \times 10^{-1}$ & $ 1.56 \times 10^{-2}$ & $ 1.52 \times
10^{-2}$
& $ 7.12 \times 10^{-2}$ &\\
&325.2 & $ 4.24 \times 10^{-1}$ & $ 1.17 \times 10^{-2}$ & $ 1.14 \times
10^{-2}$
& $ 5.90 \times 10^{-2}$ &\\
\hline
&14.5 & $ 3.99 \times 10^{-1}$ & $ 1.22 \times 10^{-1}$ & $ 9.72 \times 10^{-2}$
&
$ 1.31 \times 10^{-1}$ &\\
&59.95 & $ 7.59 \times 10^{-1}$ & $ 6.73 \times 10^{-2}$ & $ 6.17 \times
10^{-2}$
& $ 1.22 \times 10^{-1}$ &\\
$ >4.0$ &140.4 & $ 5.11 \times 10^{-1}$ & $ 3.16 \times 10^{-2}$ & $ 3.00 \times
10^{-2}$ & $ 7.14 \times 10^{-2}$ &
$ +0.066 \times 10^{-1}-8.787 \times 10^{-2}$ \\
&234.6 & $ 4.51 \times 10^{-1}$ & $ 3.09 \times 10^{-2}$ & $ 2.88 \times
10^{-2}$
& $ 5.96 \times 10^{-2}$ &\\
&325.2 & $ 3.03 \times 10^{-1}$ & $ 2.19 \times 10^{-2}$ & $ 2.05 \times
10^{-2}$
& $ 4.42 \times 10^{-2}$ &\\
\end{tabular} \end{ruledtabular}
\end{table*}
 
 
\endgroup
\begingroup \squeezetable




\begin{table}[tbh]
\caption{Heavy-flavor $e^\pm$ $v_2$ from Au+Au collisions, for the centralities
indicated.  
The $p_T$ is in units of GeV/c.  \label{tab:v2}}
\begin{ruledtabular} \begin{tabular}{cccccc}
Centrality & $p_T$&$v_2$&stat error&syst error \\ 
\hline
& 0.546 & 0.0401 & 0.00636 & 0.0189 \\
& 0.646 & 0.0408 & 0.00712 & 0.0188 \\
& 0.746 & 0.0385 & 0.00742 & 0.0168 \\
& 0.847 & 0.0626 & 0.00767 & 0.0145 \\
& 0.947 & 0.0625 & 0.00861 & 0.0137 \\
& 1.09 & 0.0563 & 0.00699 & 0.0118 \\
0--92\% & 1.29 & 0.0698 & 0.00923 & 0.0115 \\
& 1.52 & 0.092 & 0.0103 & 0.0112 \\
& 1.83 & 0.087 & 0.0152 & 0.0107 \\
& 2.20 & 0.0692 & 0.0181 & 0.00931 \\
& 2.70 & 0.0706 & 0.0308 & 0.00842 \\
& 3.24 & 0.0308 & 0.0465 & 0.00789 \\
& 4.05 & 0.00986 & 0.0645 & 0.00957 \\
\hline
& 0.40 & 5.13 $ \times 10^{-2} $ & 1.16 $ \times 10^{-2} $ 
  & 9.44 $ \times 10^{-3} $ \\
& 0.60 & 3.52 $ \times 10^{-2} $ & 8.82 $ \times 10^{-3} $ 
  & 6.77 $ \times 10^{-3} $ \\
& 0.80 & 4.13 $ \times 10^{-2} $ & 9.13 $ \times 10^{-3} $ 
  & 8.52 $ \times 10^{-3} $ \\
0--10\% & 1.05 & 4.20 $ \times 10^{-2} $ & 9.65 $ \times 10^{-3} $ 
  & 5.45 $ \times 10^{-3} $ \\
& 1.40 & 4.08 $ \times 10^{-2} $ & 1.02 $ \times 10^{-2} $ 
  & 4.96 $ \times 10^{-3} $ \\
& 1.80 & 7.05 $ \times 10^{-3} $ & 1.67 $ \times 10^{-2} $ 
  & 7.54 $ \times 10^{-3} $ \\
& 2.50 & 9.81 $ \times 10^{-3} $ & 2.22 $ \times 10^{-2} $ 
  & 6.51 $ \times 10^{-3} $ \\
& 4.00 & 1.09 $ \times 10^{-1} $ & 5.83 $ \times 10^{-2} $ 
  & 1.31 $ \times 10^{-2} $ \\
\hline
& 0.40 & 2.90 $ \times 10^{-2} $ & 9.59 $ \times 10^{-3} $ 
  & 2.03 $ \times 10^{-2} $ \\
& 0.60 & 5.01 $ \times 10^{-2} $ & 6.52 $ \times 10^{-3} $ 
  & 1.13 $ \times 10^{-2} $ \\
& 0.80 & 4.78 $ \times 10^{-2} $ & 6.59 $ \times 10^{-3} $ 
  & 1.07 $ \times 10^{-2} $ \\
10--20\% & 1.05 & 6.28 $ \times 10^{-2} $ & 7.04 $ \times 10^{-3} $ 
  & 8.54 $ \times 10^{-3} $ \\
& 1.40 & 5.89 $ \times 10^{-2} $ & 7.65 $ \times 10^{-3} $ 
  & 8.19 $ \times 10^{-3} $ \\
& 1.80 & 8.25 $ \times 10^{-2} $ & 1.25 $ \times 10^{-2} $ 
  & 7.07 $ \times 10^{-3} $ \\
& 2.50 & 9.60 $ \times 10^{-2} $ & 1.63 $ \times 10^{-2} $ & 6.34 $ \times
10^{-3} $ \\
& 4.00 & 4.04 $ \times 10^{-2} $ & 4.06 $ \times 10^{-2} $ & 5.44 $ \times
10^{-3} $ \\
\hline
& 0.40 & 6.55 $ \times 10^{-2} $ & 8.01 $ \times 10^{-3} $ & 1.82 $ \times
10^{-2} $ \\
& 0.60 & 7.34 $ \times 10^{-2} $ & 5.44 $ \times 10^{-3} $ & 1.36 $ \times
10^{-2} $ \\
& 0.80 & 1.12 $ \times 10^{-1} $ & 5.73 $ \times 10^{-3} $ & 1.05 $ \times
10^{-2} $ \\
& 1.05 & 9.47 $ \times 10^{-2} $ & 6.30 $ \times 10^{-3} $ & 1.01 $ \times
10^{-2} $ \\
20--40\% & 1.40 & 1.30 $ \times 10^{-1} $ & 7.08 $ \times 10^{-3} $ & 8.69 $
\times 10^{-3} $ \\
& 1.80 & 1.27 $ \times 10^{-1} $ & 1.14 $ \times 10^{-2} $ & 8.36 $ \times
10^{-3} $ \\
& 2.50 & 8.48 $ \times 10^{-2} $ & 1.47 $ \times 10^{-2} $ & 8.17 $ \times
10^{-3} $ \\
& 4.00 & 8.15 $ \times 10^{-2} $ & 3.62 $ \times 10^{-2} $ & 6.29 $ \times
10^{-3} $ \\
\hline
& 0.40 & 9.14 $ \times 10^{-2} $ & 1.57 $ \times 10^{-2} $ & 2.10 $ \times
10^{-2} $ \\
& 0.60 & 1.20 $ \times 10^{-1} $ & 1.13 $ \times 10^{-2} $ & 1.49 $ \times
10^{-2} $ \\
& 0.80 & 9.05 $ \times 10^{-2} $ & 1.31 $ \times 10^{-2} $ & 1.63 $ \times
10^{-2} $ \\
& 1.05 & 1.12 $ \times 10^{-1} $ & 1.49 $ \times 10^{-2} $ & 1.25 $ \times
10^{-2} $ \\
40--60\% & 1.40 & 1.42 $ \times 10^{-1} $ & 1.73 $ \times 10^{-2} $ & 1.07 $
\times 10^{-2} $ \\
& 1.80 & 1.12 $ \times 10^{-1} $ & 2.88 $ \times 10^{-2} $ & 1.19 $ \times
10^{-2} $ \\
& 2.50 & 7.66 $ \times 10^{-2} $ & 3.53 $ \times 10^{-2} $ & 1.20 $ \times
10^{-2} $ \\
& 4.00 & 6.39 $ \times 10^{-2} $ & 8.55 $ \times 10^{-2} $ & 1.03 $ \times
10^{-2} $ \\
\end{tabular} \end{ruledtabular}
\end{table}
 

\endgroup
\begingroup \squeezetable
 


\begin{table}[htbp]
\caption{Differential invariant cross section of electrons ($\aee$)
from heavy-flavor decays for 200 GeV $p+p$ collisions at midrapidity.  The
$\pte$ is in units of GeV/$c$.  The cross section and 
 corresponding errors are in units of mb \label{tab:inv_pp}}
\begin{ruledtabular} \begin{tabular}{cccc}
 \quad $\pte$ \quad & \quad Invariant yield \quad & \quad Stat.  error \quad &
\quad Sys.  error \quad \\
\hline
0.350&1.36$\times 10^{-2}$&4.45$\times 10^{-3}$&5.95$\times 10^{-3}$ \\
0.400&6.05$\times 10^{-3}$&1.70$\times 10^{-3}$&2.39$\times 10^{-3}$ \\
0.550&3.36$\times 10^{-3}$&7.56$\times 10^{-4}$&1.04$\times 10^{-3}$ \\
0.650&1.81$\times 10^{-3}$&3.98$\times 10^{-4}$&5.12$\times 10^{-4}$ \\
0.750&1.30$\times 10^{-3}$&2.28$\times 10^{-4}$&2.59$\times 10^{-4}$ \\
0.850&7.50$\times 10^{-4}$&1.35$\times 10^{-4}$&1.35$\times 10^{-4}$ \\
0.950&5.61$\times 10^{-4}$&9.14$\times 10^{-5}$&7.85$\times 10^{-5}$ \\
1.10&3.18$\times 10^{-4}$&3.77$\times 10^{-5}$&3.64$\times 10^{-5}$ \\
1.30&1.26$\times 10^{-4}$&2.10$\times 10^{-5}$&1.50$\times 10^{-5}$ \\
1.50&6.58$\times 10^{-5}$&1.25$\times 10^{-5}$&6.56$\times 10^{-6}$ \\
1.70&3.97$\times 10^{-5}$&2.07$\times 10^{-6}$&3.32$\times 10^{-6}$ \\
1.90&1.99$\times 10^{-5}$&1.18$\times 10^{-6}$&1.65$\times 10^{-6}$ \\
2.10&1.14$\times 10^{-5}$&7.44$\times 10^{-7}$&8.96$\times 10^{-7}$ \\
2.30&6.83$\times 10^{-6}$&4.96$\times 10^{-7}$&5.09$\times 10^{-7}$ \\
2.50&3.98$\times 10^{-6}$&3.63$\times 10^{-7}$&3.05$\times 10^{-7}$ \\
2.70&2.44$\times 10^{-6}$&6.49$\times 10^{-8}$&2.80$\times 10^{-7}$ \\
2.90&1.63$\times 10^{-6}$&4.77$\times 10^{-8}$&1.77$\times 10^{-7}$ \\
3.10&1.05$\times 10^{-6}$&3.53$\times 10^{-8}$&1.10$\times 10^{-7}$ \\
3.30&7.21$\times 10^{-7}$&2.73$\times 10^{-8}$&7.31$\times 10^{-8}$ \\
3.50&5.04$\times 10^{-7}$&2.15$\times 10^{-8}$&4.89$\times 10^{-8}$ \\
3.70&3.45$\times 10^{-7}$&1.70$\times 10^{-8}$&3.30$\times 10^{-8}$ \\
3.90&2.37$\times 10^{-7}$&1.36$\times 10^{-8}$&2.25$\times 10^{-8}$ \\
4.25&1.33$\times 10^{-7}$&5.95$\times 10^{-9}$&1.22$\times 10^{-8}$ \\
4.75&6.13$\times 10^{-8}$&3.73$\times 10^{-9}$&5.45$\times 10^{-9}$ \\
5.50&2.06$\times 10^{-8}$&1.79$\times 10^{-9}$&1.85$\times 10^{-9}$ \\
6.50&6.62$\times 10^{-9}$&9.26$\times 10^{-10}$&5.72$\times 10^{-10}$ \\
7.50&1.70$\times 10^{-9}$&5.73$\times 10^{-10}$&1.60$\times 10^{-10}$ \\
8.50&9.72$\times 10^{-10}$&5.16$\times 10^{-10}$&8.07$\times 10^{-11}$ \\
\end{tabular} \end{ruledtabular}
\end{table}

\endgroup


\clearpage


\end{document}